\newif \iffull     \fullfalse
\newif \ifveryfull \veryfullfalse
\newif \ifdraft    \draftfalse
\newif \ifpopl     \popltrue

\makeatletter \@input{texdirectives} \makeatother

\iffull \poplfalse \else \popltrue \fi

\iffull
  \documentclass[9pt]{extarticle}
  \usepackage{fullpage}
\else
  \ifdraft
    \documentclass[preprint]{sigplanconf}
  \else
    \documentclass{sigplanconf}
  \fi
\fi

\usepackage{floatrow}
\usepackage{amsmath}
\usepackage{mathtools}

\usepackage{amssymb,amsmath}
\usepackage{supertabular}
\usepackage{bcptheorem}

\usepackage{natbib}
\usepackage{enumitem}
\setitemize{wide,noitemsep}
\setlist[itemize,1]{label=--}

\usepackage{hyperref}
\hypersetup{
    pdftitle={Space-Efficient Manifest Contracts},
    pdfauthor={Michael Greenberg},
    bookmarksnumbered=true,     
    bookmarksopen=true,         
    bookmarksopenlevel=1,       
    hidelinks,
    naturalnames=true,
    pdfstartview=Fit,           
    pdfpagemode=UseOutlines,
    pdfpagelayout=SinglePage
}

\usepackage{local}

\usepackage{tikz}
\usetikzlibrary{matrix,arrows}

\tikzset{align at top/.style={baseline=(current bounding box.north)}}

\tikzset{
    F/.style={
     shorten >=.25em,#1-to,
     to path={-- node[inner sep=0pt,at end,sloped] {${}_{ \mathsf{F} }^{\phantom{*}}$}
              (\tikztotarget) \tikztonodes}
    },
    F/.default=
}
\tikzset{
    F*/.style={
     shorten >=.25em,#1-to,
     to path={-- node[inner sep=0pt,at end,sloped] {${}_{ \mathsf{F} }^*$}
              (\tikztotarget) \tikztonodes}
    },
    F*/.default=
}
\tikzset{
    H/.style={
     shorten >=.25em,#1-to,
     to path={-- node[inner sep=0pt,at end,sloped] {${}_{ \mathsf{H} }^{\phantom{*}}$}
              (\tikztotarget) \tikztonodes}
    },
    H/.default=
}
\tikzset{
    H*/.style={
     shorten >=.25em,#1-to,
     to path={-- node[inner sep=0pt,at end,sloped] {${}_{ \mathsf{H} }^*$}
              (\tikztotarget) \tikztonodes}
    },
    H*/.default=
}
\tikzset{
    E/.style={
     shorten >=.25em,#1-to,
     to path={-- node[inner sep=0pt,at end,sloped] {${}_{ \mathsf{E} }^{\phantom{*}}$}
              (\tikztotarget) \tikztonodes}
    },
    E/.default=
}
\tikzset{
    E*/.style={
     shorten >=.25em,#1-to,
     to path={-- node[inner sep=0pt,at end,sloped] {${}_{ \mathsf{E} }^*$}
              (\tikztotarget) \tikztonodes}
    },
    E*/.default=
}

\newcommand{\reflr}{\ifpopl\ref{fig:eideticlr}\else\ref{fig:lr}\fi}

\ifdraft
\fi

\newcommand{\ottdrule}[4][]{{\displaystyle\frac{\begin{array}{l}#2\end{array}}{#3}\quad\ottdrulename{#4}}}
\newcommand{\ottusedrule}[1]{\[#1\]}
\newcommand{\ottpremise}[1]{ #1 \\}
\newenvironment{ottdefnblock}[3][]{ \framebox{\mbox{#2}} \quad #3 \\[0pt]}{}

\newcommand{\ottnt}[1]{\mathit{#1}}
\newcommand{\ottmv}[1]{\mathit{#1}}

\newcommand{\ottsym}[1]{#1}

\newcommand{\ottdrulename}[1]{\textsc{#1}}

\newcommand{\ottdruleWFXXEmpty}[1]{\ottdrule[#1]{%
}{
 \mathord{  \vdash _{ \ottnt{m} } }~ \emptyset }{%
{\ottdrulename{WF\_Empty}}{}%
}}

\newcommand{\ottdruleWFXXExtend}[1]{\ottdrule[#1]{%
\ottpremise{  \mathord{  \vdash _{ \ottnt{m} } }~ \Gamma   \quad   \mathord{  \vdash _{ \ottnt{m} } }~ \ottnt{T}  }%
}{
 \mathord{  \vdash _{ \ottnt{m} } }~  \Gamma , \mathit{x} \mathord{:} \ottnt{T}  }{%
{\ottdrulename{WF\_Extend}}{}%
}}

\newcommand{\ottdruleWFXXFun}[1]{\ottdrule[#1]{%
\ottpremise{  \mathord{  \vdash _{ \ottnt{m} } }~ \ottnt{T_{{\mathrm{1}}}}   \quad   \mathord{  \vdash _{ \ottnt{m} } }~ \ottnt{T_{{\mathrm{2}}}}  }%
}{
 \mathord{  \vdash _{ \ottnt{m} } }~  \ottnt{T_{{\mathrm{1}}}} \mathord{ \rightarrow } \ottnt{T_{{\mathrm{2}}}}  }{%
{\ottdrulename{WF\_Fun}}{}%
}}

\newcommand{\ottdruleWFXXBase}[1]{\ottdrule[#1]{%
}{
 \mathord{  \vdash _{ \ottnt{m} } }~  \{ \mathit{x} \mathord{:} \ottnt{B} \mathrel{\mid}  \mathsf{true}  \}  }{%
{\ottdrulename{WF\_Base}}{}%
}}

\newcommand{\ottdruleWFXXRefine}[1]{\ottdrule[#1]{%
\ottpremise{  \mathit{x} \mathord{:}  \{ \mathit{x} \mathord{:} \ottnt{B} \mathrel{\mid}  \mathsf{true}  \}     \vdash _{ \ottnt{m} }  \ottnt{e}  :   \{ \mathit{x} \mathord{:}  \mathsf{Bool}  \mathrel{\mid}  \mathsf{true}  \}  }%
}{
 \mathord{  \vdash _{ \ottnt{m} } }~  \{ \mathit{x} \mathord{:} \ottnt{B} \mathrel{\mid} \ottnt{e} \}  }{%
{\ottdrulename{WF\_Refine}}{}%
}}

\newcommand{\ottdruleAXXNone}[1]{\ottdrule[#1]{%
\ottpremise{  \vdash  \ottnt{T_{{\mathrm{1}}}}  \mathrel{\parallel}  \ottnt{T_{{\mathrm{2}}}}  \quad     \mathord{  \vdash _{ \ottnt{m} } }~ \ottnt{T_{{\mathrm{1}}}}   \quad   \mathord{  \vdash _{ \ottnt{m} } }~ \ottnt{T_{{\mathrm{2}}}}     }%
}{
 \mathord{  \vdash _{ \ottnt{m} } }~ \bullet   \mathrel{\parallel}   \ottnt{T_{{\mathrm{1}}}}  \Rightarrow  \ottnt{T_{{\mathrm{2}}}} }{%
{\ottdrulename{A\_None}}{}%
}}

\newcommand{\ottdruleAXXTypeSet}[1]{\ottdrule[#1]{%
\ottpremise{ \vdash  \ottnt{T_{{\mathrm{1}}}}  \mathrel{\parallel}  \ottnt{T_{{\mathrm{2}}}}  \quad     \mathord{  \vdash _{  \mathsf{H}  } }~ \ottnt{T_{{\mathrm{1}}}}   \quad   \mathord{  \vdash _{  \mathsf{H}  } }~ \ottnt{T_{{\mathrm{2}}}}    }%
\ottpremise{ \forall   \ottnt{T}  \in  \mathcal{S}   . ~     \mathord{  \vdash _{  \mathsf{H}  } }~ \ottnt{T}   \quad  \vdash  \ottnt{T}  \mathrel{\parallel}  \ottnt{T_{{\mathrm{1}}}}   }%
}{
 \mathord{  \vdash _{  \mathsf{H}  } }~ \mathcal{S}   \mathrel{\parallel}   \ottnt{T_{{\mathrm{1}}}}  \Rightarrow  \ottnt{T_{{\mathrm{2}}}} }{%
{\ottdrulename{A\_TypeSet}}{}%
}}

\newcommand{\ottdruleAXXRefine}[1]{\ottdrule[#1]{%
\ottpremise{   \mathord{  \vdash _{  \mathsf{E}  } }~  \{ \mathit{x} \mathord{:} \ottnt{B} \mathrel{\mid} \ottnt{e_{{\mathrm{1}}}} \}    \quad   \mathord{  \vdash _{  \mathsf{E}  } }~  \{ \mathit{x} \mathord{:} \ottnt{B} \mathrel{\mid} \ottnt{e_{{\mathrm{2}}}} \}    }%
\ottpremise{   \forall   \{ \mathit{x} \mathord{:} \ottnt{B} \mathrel{\mid} \ottnt{e} \} \in  \ottnt{r}   . ~   \mathord{  \vdash _{  \mathsf{E}  } }~  \{ \mathit{x} \mathord{:} \ottnt{B} \mathrel{\mid} \ottnt{e} \}      \quad   \text{no duplicates in $ \ottnt{r} $}  }%
\ottpremise{ \exists   \{ \mathit{x} \mathord{:} \ottnt{B} \mathrel{\mid} \ottnt{e} \}   \in  \ottnt{r}  . ~   \{ \mathit{x} \mathord{:} \ottnt{B} \mathrel{\mid} \ottnt{e} \}  \, \supset \,  \{ \mathit{x} \mathord{:} \ottnt{B} \mathrel{\mid} \ottnt{e_{{\mathrm{2}}}} \}  }%
}{
 \mathord{  \vdash _{  \mathsf{E}  } }~ \ottnt{r}   \mathrel{\parallel}    \{ \mathit{x} \mathord{:} \ottnt{B} \mathrel{\mid} \ottnt{e_{{\mathrm{1}}}} \}   \Rightarrow   \{ \mathit{x} \mathord{:} \ottnt{B} \mathrel{\mid} \ottnt{e_{{\mathrm{2}}}} \}  }{%
{\ottdrulename{A\_Refine}}{}%
}}

\newcommand{\ottdruleAXXFun}[1]{\ottdrule[#1]{%
\ottpremise{  \mathord{  \vdash _{  \mathsf{E}  } }~ \ottnt{c_{{\mathrm{1}}}}   \mathrel{\parallel}   \ottnt{T_{{\mathrm{21}}}}  \Rightarrow  \ottnt{T_{{\mathrm{11}}}}   \quad   \mathord{  \vdash _{  \mathsf{E}  } }~ \ottnt{c_{{\mathrm{2}}}}   \mathrel{\parallel}   \ottnt{T_{{\mathrm{12}}}}  \Rightarrow  \ottnt{T_{{\mathrm{22}}}}  }%
}{
 \mathord{  \vdash _{  \mathsf{E}  } }~ \ottnt{c_{{\mathrm{1}}}}  \mapsto  \ottnt{c_{{\mathrm{2}}}}   \mathrel{\parallel}    (  \ottnt{T_{{\mathrm{11}}}} \mathord{ \rightarrow } \ottnt{T_{{\mathrm{12}}}}  )   \Rightarrow   (  \ottnt{T_{{\mathrm{21}}}} \mathord{ \rightarrow } \ottnt{T_{{\mathrm{22}}}}  )  }{%
{\ottdrulename{A\_Fun}}{}%
}}

\newcommand{\ottdruleSXXRefine}[1]{\ottdrule[#1]{%
}{
\vdash   \{ \mathit{x} \mathord{:} \ottnt{B} \mathrel{\mid} \ottnt{e_{{\mathrm{1}}}} \}   \mathrel{\parallel}   \{ \mathit{x} \mathord{:} \ottnt{B} \mathrel{\mid} \ottnt{e_{{\mathrm{2}}}} \} }{%
{\ottdrulename{S\_Refine}}{}%
}}

\newcommand{\ottdruleSXXFun}[1]{\ottdrule[#1]{%
\ottpremise{ \vdash  \ottnt{T_{{\mathrm{11}}}}  \mathrel{\parallel}  \ottnt{T_{{\mathrm{21}}}}  \quad  \vdash  \ottnt{T_{{\mathrm{12}}}}  \mathrel{\parallel}  \ottnt{T_{{\mathrm{22}}}} }%
}{
\vdash   \ottnt{T_{{\mathrm{11}}}} \mathord{ \rightarrow } \ottnt{T_{{\mathrm{12}}}}   \mathrel{\parallel}   \ottnt{T_{{\mathrm{21}}}} \mathord{ \rightarrow } \ottnt{T_{{\mathrm{22}}}} }{%
{\ottdrulename{S\_Fun}}{}%
}}

\newcommand{\ottdruleTXXVar}[1]{\ottdrule[#1]{%
\ottpremise{  \mathord{  \vdash _{ \ottnt{m} } }~ \Gamma   \quad   \mathit{x}  \mathord{:}  \ottnt{T}  \in  \Gamma  }%
}{
 \Gamma   \vdash _{ \ottnt{m} }  \mathit{x}  :  \ottnt{T} }{%
{\ottdrulename{T\_Var}}{}%
}}

\newcommand{\ottdruleTXXConst}[1]{\ottdrule[#1]{%
\ottpremise{    \mathord{  \vdash _{ \ottnt{m} } }~ \Gamma   \quad   \mathord{  \vdash _{ \ottnt{m} } }~  \{ \mathit{x} \mathord{:} \ottnt{B} \mathrel{\mid} \ottnt{e} \}      \quad     \mathsf{ty} ( \ottnt{k} )   \ottsym{=}  \ottnt{B}  \quad   \ottnt{e}  [  \ottnt{k} / \mathit{x}  ]  \,  \longrightarrow ^{*}_{ \ottnt{m} }  \,  \mathsf{true}    }%
}{
 \Gamma   \vdash _{ \ottnt{m} }  \ottnt{k}  :   \{ \mathit{x} \mathord{:} \ottnt{B} \mathrel{\mid} \ottnt{e} \}  }{%
{\ottdrulename{T\_Const}}{}%
}}

\newcommand{\ottdruleTXXAbs}[1]{\ottdrule[#1]{%
\ottpremise{  \mathord{  \vdash _{ \ottnt{m} } }~ \ottnt{T_{{\mathrm{1}}}}   \quad    \Gamma , \mathit{x} \mathord{:} \ottnt{T_{{\mathrm{1}}}}    \vdash _{ \ottnt{m} }  \ottnt{e_{{\mathrm{12}}}}  :  \ottnt{T_{{\mathrm{2}}}}  }%
}{
 \Gamma   \vdash _{ \ottnt{m} }   \lambda \mathit{x} \mathord{:} \ottnt{T_{{\mathrm{1}}}} .~  \ottnt{e_{{\mathrm{12}}}}   :   \ottnt{T_{{\mathrm{1}}}} \mathord{ \rightarrow } \ottnt{T_{{\mathrm{2}}}}  }{%
{\ottdrulename{T\_Abs}}{}%
}}

\newcommand{\ottdruleTXXOp}[1]{\ottdrule[#1]{%
\ottpremise{  \mathsf{ty} (\mathord{ \ottnt{op} })   \ottsym{=}   {}  \ottnt{T_{{\mathrm{1}}}}  \rightarrow \, ... \, \rightarrow  \ottnt{T_{\ottmv{n}}}  {} \mathord{ \rightarrow } \ottnt{T}   \quad   \Gamma   \vdash _{ \ottnt{m} }  \ottnt{e_{\ottmv{i}}}  :  \ottnt{T_{\ottmv{i}}}  }%
}{
 \Gamma   \vdash _{ \ottnt{m} }  \ottnt{op}  \ottsym{(}  \ottnt{e_{{\mathrm{1}}}}  \ottsym{,}  \dots  \ottsym{,}  \ottnt{e_{\ottmv{n}}}  \ottsym{)}  :  \ottnt{T} }{%
{\ottdrulename{T\_Op}}{}%
}}

\newcommand{\ottdruleTXXApp}[1]{\ottdrule[#1]{%
\ottpremise{  \Gamma   \vdash _{ \ottnt{m} }  \ottnt{e_{{\mathrm{1}}}}  :   (  \ottnt{T_{{\mathrm{1}}}} \mathord{ \rightarrow } \ottnt{T_{{\mathrm{2}}}}  )    \quad   \Gamma   \vdash _{ \ottnt{m} }  \ottnt{e_{{\mathrm{2}}}}  :  \ottnt{T_{{\mathrm{1}}}}  }%
}{
 \Gamma   \vdash _{ \ottnt{m} }   \ottnt{e_{{\mathrm{1}}}} ~ \ottnt{e_{{\mathrm{2}}}}   :  \ottnt{T_{{\mathrm{2}}}} }{%
{\ottdrulename{T\_App}}{}%
}}

\newcommand{\ottdruleTXXCast}[1]{\ottdrule[#1]{%
\ottpremise{  \mathord{  \vdash _{ \ottnt{m} } }~ \ottnt{a}   \mathrel{\parallel}   \ottnt{T_{{\mathrm{1}}}}  \Rightarrow  \ottnt{T_{{\mathrm{2}}}}   \quad   \Gamma   \vdash _{ \ottnt{m} }  \ottnt{e}  :  \ottnt{T_{{\mathrm{1}}}}  }%
}{
 \Gamma   \vdash _{ \ottnt{m} }   \langle  \ottnt{T_{{\mathrm{1}}}}  \mathord{ \overset{ \ottnt{a} }{\Rightarrow} }  \ottnt{T_{{\mathrm{2}}}}  \rangle^{ \ottnt{l} } ~  \ottnt{e}   :  \ottnt{T_{{\mathrm{2}}}} }{%
{\ottdrulename{T\_Cast}}{}%
}}

\newcommand{\ottdruleTXXBlame}[1]{\ottdrule[#1]{%
\ottpremise{  \mathord{  \vdash _{ \ottnt{m} } }~ \Gamma   \quad   \mathord{  \vdash _{ \ottnt{m} } }~ \ottnt{T}  }%
}{
 \Gamma   \vdash _{ \ottnt{m} }   \mathord{\Uparrow}  \ottnt{l}   :  \ottnt{T} }{%
{\ottdrulename{T\_Blame}}{}%
}}

\newcommand{\ottdruleTXXCheck}[1]{\ottdrule[#1]{%
\ottpremise{      \mathord{  \vdash _{ \ottnt{m} } }~ \Gamma   \quad   \mathord{  \vdash _{ \ottnt{m} } }~  \{ \mathit{x} \mathord{:} \ottnt{B} \mathrel{\mid} \ottnt{e_{{\mathrm{1}}}} \}      \quad     \mathsf{ty} ( \ottnt{k} )   \ottsym{=}  \ottnt{B}  \quad   \emptyset   \vdash _{ \ottnt{m} }  \ottnt{e_{{\mathrm{2}}}}  :   \{ \mathit{x} \mathord{:}  \mathsf{Bool}  \mathrel{\mid}  \mathsf{true}  \}        \quad   \ottnt{e_{{\mathrm{1}}}}  [  \ottnt{k} / \mathit{x}  ]  \,  \longrightarrow ^{*}_{ \ottnt{m} }  \, \ottnt{e_{{\mathrm{2}}}} }%
}{
 \Gamma   \vdash _{ \ottnt{m} }   \langle   \{ \mathit{x} \mathord{:} \ottnt{B} \mathrel{\mid} \ottnt{e_{{\mathrm{1}}}} \}  ,  \ottnt{e_{{\mathrm{2}}}} ,  \ottnt{k}  \rangle^{ \ottnt{l} }   :   \{ \mathit{x} \mathord{:} \ottnt{B} \mathrel{\mid} \ottnt{e_{{\mathrm{1}}}} \}  }{%
{\ottdrulename{T\_Check}}{}%
}}

\newcommand{\ottdruleTXXStack}[1]{\ottdrule[#1]{%
\ottpremise{      \mathord{  \vdash _{  \mathsf{E}  } }~ \Gamma   \quad   \mathord{  \vdash _{  \mathsf{E}  } }~  \{ \mathit{x} \mathord{:} \ottnt{B} \mathrel{\mid} \ottnt{e_{{\mathrm{1}}}} \}      \quad     \mathsf{ty} ( \ottnt{k} )   \ottsym{=}  \ottnt{B}  \quad   \emptyset   \vdash _{  \mathsf{E}  }  \ottnt{e_{{\mathrm{2}}}}  :   \{ \mathit{x} \mathord{:} \ottnt{B} \mathrel{\mid} \ottnt{e_{{\mathrm{3}}}} \}        \quad   \forall   \{ \mathit{x} \mathord{:} \ottnt{B} \mathrel{\mid} \ottnt{e} \} \in  \ottnt{r}   . ~   \mathord{  \vdash _{  \mathsf{E}  } }~  \{ \mathit{x} \mathord{:} \ottnt{B} \mathrel{\mid} \ottnt{e} \}    }%
\ottpremise{\ottnt{s}  \ottsym{=}   \mathord{\checkmark}  \, \text{implies} \,   \ottnt{e_{{\mathrm{1}}}}  [  \ottnt{k} / \mathit{x}  ]  \,  \longrightarrow ^{*}_{  \mathsf{E}  }  \,  \mathsf{true}  \, \vee \,  (  \exists   \{ \mathit{x} \mathord{:} \ottnt{B} \mathrel{\mid} \ottnt{e} \}   . ~   \exists  \ottnt{l}  . ~    \{ \mathit{x} \mathord{:} \ottnt{B} \mathrel{\mid} \ottnt{e} \}  \, \supset \,  \{ \mathit{x} \mathord{:} \ottnt{B} \mathrel{\mid} \ottnt{e_{{\mathrm{1}}}} \}  \, \wedge \,  \langle   \{ \mathit{x} \mathord{:} \ottnt{B} \mathrel{\mid} \ottnt{e} \}  ,   \ottnt{e}  [  \ottnt{k} / \mathit{x}  ]  ,  \ottnt{k}  \rangle^{ \ottnt{l} }  \,  \longrightarrow ^{*}_{  \mathsf{E}  }  \, \ottnt{e_{{\mathrm{2}}}}    )  }%
\ottpremise{\ottnt{s}  \ottsym{=}   \mathord{?}  \, \text{implies} \,  (  \exists   \{ \mathit{x} \mathord{:} \ottnt{B} \mathrel{\mid} \ottnt{e} \}   \in  \ottnt{r}  . ~   \{ \mathit{x} \mathord{:} \ottnt{B} \mathrel{\mid} \ottnt{e} \}  \, \supset \,  \{ \mathit{x} \mathord{:} \ottnt{B} \mathrel{\mid} \ottnt{e_{{\mathrm{1}}}} \}   ) }%
}{
 \Gamma   \vdash _{  \mathsf{E}  }   \langle   \{ \mathit{x} \mathord{:} \ottnt{B} \mathrel{\mid} \ottnt{e_{{\mathrm{1}}}} \}  ,  \ottnt{s} ,  \ottnt{r} ,  \ottnt{k} ,  \ottnt{e_{{\mathrm{2}}}}  \rangle^{\bullet}   :   \{ \mathit{x} \mathord{:} \ottnt{B} \mathrel{\mid} \ottnt{e_{{\mathrm{1}}}} \}  }{%
{\ottdrulename{T\_Stack}}{}%
}}

\newcommand{\ottdruleVXXConst}[1]{\ottdrule[#1]{%
}{
 \mathsf{val} _{ \ottnt{m} }~ \ottnt{k} }{%
{\ottdrulename{V\_Const}}{}%
}}

\newcommand{\ottdruleVXXAbs}[1]{\ottdrule[#1]{%
}{
 \mathsf{val} _{ \ottnt{m} }~  \lambda \mathit{x} \mathord{:} \ottnt{T} .~  \ottnt{e}  }{%
{\ottdrulename{V\_Abs}}{}%
}}

\newcommand{\ottdruleVXXProxyC}[1]{\ottdrule[#1]{%
\ottpremise{ \mathsf{val} _{  \mathsf{C}  }~ \ottnt{e} }%
}{
 \mathsf{val} _{  \mathsf{C}  }~  \langle   \ottnt{T_{{\mathrm{11}}}} \mathord{ \rightarrow } \ottnt{T_{{\mathrm{12}}}}   \mathord{ \overset{ \bullet }{\Rightarrow} }   \ottnt{T_{{\mathrm{21}}}} \mathord{ \rightarrow } \ottnt{T_{{\mathrm{22}}}}   \rangle^{ \ottnt{l} } ~  \ottnt{e}  }{%
{\ottdrulename{V\_ProxyC}}{}%
}}

\newcommand{\ottdruleVXXProxyF}[1]{\ottdrule[#1]{%
}{
 \mathsf{val} _{  \mathsf{F}  }~  \langle   \ottnt{T_{{\mathrm{11}}}} \mathord{ \rightarrow } \ottnt{T_{{\mathrm{12}}}}   \mathord{ \overset{ \emptyset }{\Rightarrow} }   \ottnt{T_{{\mathrm{21}}}} \mathord{ \rightarrow } \ottnt{T_{{\mathrm{22}}}}   \rangle^{ \ottnt{l} } ~   \lambda \mathit{x} \mathord{:} \ottnt{T} .~  \ottnt{e}   }{%
{\ottdrulename{V\_ProxyF}}{}%
}}

\newcommand{\ottdruleVXXProxyH}[1]{\ottdrule[#1]{%
}{
 \mathsf{val} _{  \mathsf{H}  }~  \langle   \ottnt{T_{{\mathrm{11}}}} \mathord{ \rightarrow } \ottnt{T_{{\mathrm{12}}}}   \mathord{ \overset{ \mathcal{S} }{\Rightarrow} }   \ottnt{T_{{\mathrm{21}}}} \mathord{ \rightarrow } \ottnt{T_{{\mathrm{22}}}}   \rangle^{ \ottnt{l} } ~   \lambda \mathit{x} \mathord{:} \ottnt{T} .~  \ottnt{e}   }{%
{\ottdrulename{V\_ProxyH}}{}%
}}

\newcommand{\ottdruleVXXProxyE}[1]{\ottdrule[#1]{%
}{
 \mathsf{val} _{  \mathsf{E}  }~  \langle   \ottnt{T_{{\mathrm{11}}}} \mathord{ \rightarrow } \ottnt{T_{{\mathrm{12}}}}   \mathord{ \overset{ \ottnt{c_{{\mathrm{1}}}}  \mapsto  \ottnt{c_{{\mathrm{2}}}} }{\Rightarrow} }   \ottnt{T_{{\mathrm{21}}}} \mathord{ \rightarrow } \ottnt{T_{{\mathrm{22}}}}   \rangle^{\bullet} ~   \lambda \mathit{x} \mathord{:} \ottnt{T} .~  \ottnt{e}   }{%
{\ottdrulename{V\_ProxyE}}{}%
}}

\newcommand{\ottdruleRXXVal}[1]{\ottdrule[#1]{%
\ottpremise{ \mathsf{val} _{ \ottnt{m} }~ \ottnt{e} }%
}{
 \mathsf{result} _{ \ottnt{m} }~ \ottnt{e} }{%
{\ottdrulename{R\_Val}}{}%
}}

\newcommand{\ottdruleRXXBlame}[1]{\ottdrule[#1]{%
}{
 \mathsf{result} _{ \ottnt{m} }~  \mathord{\Uparrow}  \ottnt{l}  }{%
{\ottdrulename{R\_Blame}}{}%
}}

\newcommand{\ottdruleEXXBeta}[1]{\ottdrule[#1]{%
\ottpremise{ \mathsf{val} _{ \ottnt{m} }~ \ottnt{e_{{\mathrm{2}}}} }%
}{
  (  \lambda \mathit{x} \mathord{:} \ottnt{T} .~  \ottnt{e_{{\mathrm{12}}}}  )  ~ \ottnt{e_{{\mathrm{2}}}}  \,  \longrightarrow _{ \ottnt{m} }  \,  \ottnt{e_{{\mathrm{12}}}}  [  \ottnt{e_{{\mathrm{2}}}} / \mathit{x}  ] }{%
{\ottdrulename{E\_Beta}}{}%
}}

\newcommand{\ottdruleEXXOp}[1]{\ottdrule[#1]{%
\ottpremise{   \mathsf{val} _{ \ottnt{m} }~ \ottnt{e_{{\mathrm{1}}}}    ~...~    \mathsf{val} _{ \ottnt{m} }~ \ottnt{e_{\ottmv{n}}}   }%
}{
\ottnt{op}  \ottsym{(}  \ottnt{e_{{\mathrm{1}}}}  \ottsym{,} \, ... \, \ottsym{,}  \ottnt{e_{\ottmv{n}}}  \ottsym{)} \,  \longrightarrow _{ \ottnt{m} }  \, \denot{ op } \, \ottsym{(}  \ottnt{e_{{\mathrm{1}}}}  \ottsym{,} \, ... \, \ottsym{,}  \ottnt{e_{\ottmv{n}}}  \ottsym{)}}{%
{\ottdrulename{E\_Op}}{}%
}}

\newcommand{\ottdruleEXXUnwrap}[1]{\ottdrule[#1]{%
\ottpremise{  \mathsf{val} _{ \ottnt{m} }~  \langle   \ottnt{T_{{\mathrm{11}}}} \mathord{ \rightarrow } \ottnt{T_{{\mathrm{12}}}}   \mathord{ \overset{ \ottnt{a} }{\Rightarrow} }   \ottnt{T_{{\mathrm{21}}}} \mathord{ \rightarrow } \ottnt{T_{{\mathrm{22}}}}   \rangle^{ \ottnt{l} } ~  \ottnt{e_{{\mathrm{1}}}}    \quad   \mathsf{val} _{ \ottnt{m} }~ \ottnt{e_{{\mathrm{2}}}}  }%
}{
  (  \langle   \ottnt{T_{{\mathrm{11}}}} \mathord{ \rightarrow } \ottnt{T_{{\mathrm{12}}}}   \mathord{ \overset{ \ottnt{a} }{\Rightarrow} }   \ottnt{T_{{\mathrm{21}}}} \mathord{ \rightarrow } \ottnt{T_{{\mathrm{22}}}}   \rangle^{ \ottnt{l} } ~  \ottnt{e_{{\mathrm{1}}}}  )  ~ \ottnt{e_{{\mathrm{2}}}}  \,  \longrightarrow _{ \ottnt{m} }  \,  \langle  \ottnt{T_{{\mathrm{12}}}}  \mathord{ \overset{  \mathsf{cod} ( \ottnt{a} )  }{\Rightarrow} }  \ottnt{T_{{\mathrm{22}}}}  \rangle^{ \ottnt{l} } ~   (  \ottnt{e_{{\mathrm{1}}}} ~  (  \langle  \ottnt{T_{{\mathrm{21}}}}  \mathord{ \overset{  \mathsf{dom} ( \ottnt{a} )  }{\Rightarrow} }  \ottnt{T_{{\mathrm{11}}}}  \rangle^{ \ottnt{l} } ~  \ottnt{e_{{\mathrm{2}}}}  )   )  }{%
{\ottdrulename{E\_Unwrap}}{}%
}}

\newcommand{\ottdruleEXXCheckNoneC}[1]{\ottdrule[#1]{%
}{
 \langle   \{ \mathit{x} \mathord{:} \ottnt{B} \mathrel{\mid} \ottnt{e_{{\mathrm{1}}}} \}   \mathord{ \overset{ \bullet }{\Rightarrow} }   \{ \mathit{x} \mathord{:} \ottnt{B} \mathrel{\mid} \ottnt{e_{{\mathrm{2}}}} \}   \rangle^{ \ottnt{l} } ~  \ottnt{k}  \,  \longrightarrow _{  \mathsf{C}  }  \,  \langle   \{ \mathit{x} \mathord{:} \ottnt{B} \mathrel{\mid} \ottnt{e_{{\mathrm{2}}}} \}  ,   \ottnt{e_{{\mathrm{2}}}}  [  \ottnt{k} / \mathit{x}  ]  ,  \ottnt{k}  \rangle^{ \ottnt{l} } }{%
{\ottdrulename{E\_CheckNoneC}}{}%
}}

\newcommand{\ottdruleEXXCheckOK}[1]{\ottdrule[#1]{%
}{
 \langle   \{ \mathit{x} \mathord{:} \ottnt{B} \mathrel{\mid} \ottnt{e} \}  ,   \mathsf{true}  ,  \ottnt{k}  \rangle^{ \ottnt{l} }  \,  \longrightarrow _{ \ottnt{m} }  \, \ottnt{k}}{%
{\ottdrulename{E\_CheckOK}}{}%
}}

\newcommand{\ottdruleEXXCheckFail}[1]{\ottdrule[#1]{%
}{
 \langle   \{ \mathit{x} \mathord{:} \ottnt{B} \mathrel{\mid} \ottnt{e} \}  ,   \mathsf{false}  ,  \ottnt{k}  \rangle^{ \ottnt{l} }  \,  \longrightarrow _{ \ottnt{m} }  \,  \mathord{\Uparrow}  \ottnt{l} }{%
{\ottdrulename{E\_CheckFail}}{}%
}}

\newcommand{\ottdruleEXXAppL}[1]{\ottdrule[#1]{%
\ottpremise{\ottnt{e_{{\mathrm{1}}}} \,  \longrightarrow _{ \ottnt{m} }  \, \ottnt{e'_{{\mathrm{1}}}}}%
}{
 \ottnt{e_{{\mathrm{1}}}} ~ \ottnt{e_{{\mathrm{2}}}}  \,  \longrightarrow _{ \ottnt{m} }  \,  \ottnt{e'_{{\mathrm{1}}}} ~ \ottnt{e_{{\mathrm{2}}}} }{%
{\ottdrulename{E\_AppL}}{}%
}}

\newcommand{\ottdruleEXXAppR}[1]{\ottdrule[#1]{%
\ottpremise{  \mathsf{val} _{ \ottnt{m} }~ \ottnt{e_{{\mathrm{1}}}}   \quad  \ottnt{e_{{\mathrm{2}}}} \,  \longrightarrow _{ \ottnt{m} }  \, \ottnt{e'_{{\mathrm{2}}}} }%
}{
 \ottnt{e_{{\mathrm{1}}}} ~ \ottnt{e_{{\mathrm{2}}}}  \,  \longrightarrow _{ \ottnt{m} }  \,  \ottnt{e_{{\mathrm{1}}}} ~ \ottnt{e'_{{\mathrm{2}}}} }{%
{\ottdrulename{E\_AppR}}{}%
}}

\newcommand{\ottdruleEXXOpInner}[1]{\ottdrule[#1]{%
\ottpremise{    \mathsf{val} _{ \ottnt{m} }~ \ottnt{e_{{\mathrm{1}}}}   ~...~   \mathsf{val} _{ \ottnt{m} }~ \ottnt{e_{{\ottmv{i}-1}}}     \quad  \ottnt{e_{\ottmv{i}}} \,  \longrightarrow _{ \ottnt{m} }  \, \ottnt{e'_{\ottmv{i}}} }%
}{
\ottnt{op}  \ottsym{(}  \ottnt{e_{{\mathrm{1}}}}  \ottsym{,}  \dots  \ottsym{,}  \ottnt{e_{{\ottmv{i}-1}}}  \ottsym{,}  \ottnt{e_{\ottmv{i}}}  \ottsym{,}  \dots  \ottsym{,}  \ottnt{e_{\ottmv{n}}}  \ottsym{)} \,  \longrightarrow _{ \ottnt{m} }  \, \ottnt{op}  \ottsym{(}  \ottnt{e_{{\mathrm{1}}}}  \ottsym{,}  \dots  \ottsym{,}  \ottnt{e_{{\ottmv{i}-1}}}  \ottsym{,}  \ottnt{e'_{\ottmv{i}}}  \ottsym{,}  \dots  \ottsym{,}  \ottnt{e_{\ottmv{n}}}  \ottsym{)}}{%
{\ottdrulename{E\_OpInner}}{}%
}}

\newcommand{\ottdruleEXXCheckInner}[1]{\ottdrule[#1]{%
\ottpremise{\ottnt{e_{{\mathrm{2}}}} \,  \longrightarrow _{ \ottnt{m} }  \, \ottnt{e'_{{\mathrm{2}}}}}%
}{
 \langle   \{ \mathit{x} \mathord{:} \ottnt{B} \mathrel{\mid} \ottnt{e_{{\mathrm{1}}}} \}  ,  \ottnt{e_{{\mathrm{2}}}} ,  \ottnt{k}  \rangle^{ \ottnt{l} }  \,  \longrightarrow _{ \ottnt{m} }  \,  \langle   \{ \mathit{x} \mathord{:} \ottnt{B} \mathrel{\mid} \ottnt{e_{{\mathrm{1}}}} \}  ,  \ottnt{e'_{{\mathrm{2}}}} ,  \ottnt{k}  \rangle^{ \ottnt{l} } }{%
{\ottdrulename{E\_CheckInner}}{}%
}}

\newcommand{\ottdruleEXXAppRaiseL}[1]{\ottdrule[#1]{%
}{
  \mathord{\Uparrow}  \ottnt{l}  ~ \ottnt{e_{{\mathrm{2}}}}  \,  \longrightarrow _{ \ottnt{m} }  \,  \mathord{\Uparrow}  \ottnt{l} }{%
{\ottdrulename{E\_AppRaiseL}}{}%
}}

\newcommand{\ottdruleEXXAppRaiseR}[1]{\ottdrule[#1]{%
\ottpremise{ \mathsf{val} _{ \ottnt{m} }~ \ottnt{e_{{\mathrm{1}}}} }%
}{
 \ottnt{e_{{\mathrm{1}}}} ~  \mathord{\Uparrow}  \ottnt{l}   \,  \longrightarrow _{ \ottnt{m} }  \,  \mathord{\Uparrow}  \ottnt{l} }{%
{\ottdrulename{E\_AppRaiseR}}{}%
}}

\newcommand{\ottdruleEXXCastRaise}[1]{\ottdrule[#1]{%
}{
 \langle  \ottnt{T_{{\mathrm{1}}}}  \mathord{ \overset{ \mathcal{S} }{\Rightarrow} }  \ottnt{T_{{\mathrm{2}}}}  \rangle^{ \ottnt{l} } ~   \mathord{\Uparrow}  \ottnt{l'}   \,  \longrightarrow _{ \ottnt{m} }  \,  \mathord{\Uparrow}  \ottnt{l'} }{%
{\ottdrulename{E\_CastRaise}}{}%
}}

\newcommand{\ottdruleEXXOpRaise}[1]{\ottdrule[#1]{%
\ottpremise{  \mathsf{val} _{ \ottnt{m} }~ \ottnt{e_{{\mathrm{1}}}}   ~...~   \mathsf{val} _{ \ottnt{m} }~ \ottnt{e_{{\ottmv{i}-1}}}  }%
}{
\ottnt{op}  \ottsym{(}  \ottnt{e_{{\mathrm{1}}}}  \ottsym{,}  \dots  \ottsym{,}  \ottnt{e_{{\ottmv{i}-1}}}  \ottsym{,}   \mathord{\Uparrow}  \ottnt{l}   \ottsym{,}  \dots  \ottsym{,}  \ottnt{e_{\ottmv{n}}}  \ottsym{)} \,  \longrightarrow _{ \ottnt{m} }  \,  \mathord{\Uparrow}  \ottnt{l} }{%
{\ottdrulename{E\_OpRaise}}{}%
}}

\newcommand{\ottdruleEXXCheckRaise}[1]{\ottdrule[#1]{%
}{
 \langle   \{ \mathit{x} \mathord{:} \ottnt{B} \mathrel{\mid} \ottnt{e} \}  ,   \mathord{\Uparrow}  \ottnt{l}  ,  \ottnt{k}  \rangle^{ \ottnt{l'} }  \,  \longrightarrow _{ \ottnt{m} }  \,  \mathord{\Uparrow}  \ottnt{l} }{%
{\ottdrulename{E\_CheckRaise}}{}%
}}

\newcommand{\ottdruleEXXCastInnerC}[1]{\ottdrule[#1]{%
\ottpremise{\ottnt{e} \,  \longrightarrow _{  \mathsf{C}  }  \, \ottnt{e'}}%
}{
 \langle  \ottnt{T_{{\mathrm{1}}}}  \mathord{ \overset{ \bullet }{\Rightarrow} }  \ottnt{T_{{\mathrm{2}}}}  \rangle^{ \ottnt{l} } ~  \ottnt{e}  \,  \longrightarrow _{  \mathsf{C}  }  \,  \langle  \ottnt{T_{{\mathrm{1}}}}  \mathord{ \overset{ \bullet }{\Rightarrow} }  \ottnt{T_{{\mathrm{2}}}}  \rangle^{ \ottnt{l} } ~  \ottnt{e'} }{%
{\ottdrulename{E\_CastInnerC}}{}%
}}

\newcommand{\ottdruleEXXCastInner}[1]{\ottdrule[#1]{%
\ottpremise{ \ottnt{m}  \neq   \mathsf{C}   \quad    \ottnt{e_{{\mathrm{2}}}} \,  \longrightarrow _{ \ottnt{m} }  \, \ottnt{e'_{{\mathrm{2}}}}  \quad  \ottnt{e_{{\mathrm{2}}}}  \neq   \langle  \ottnt{T_{{\mathrm{1}}}}  \mathord{ \overset{ \ottnt{a'} }{\Rightarrow} }  \ottnt{T_{{\mathrm{2}}}}  \rangle^{ \ottnt{l'} } ~  \ottnt{e''_{{\mathrm{2}}}}    }%
}{
 \langle  \ottnt{T_{{\mathrm{2}}}}  \mathord{ \overset{ \ottnt{a} }{\Rightarrow} }  \ottnt{T_{{\mathrm{3}}}}  \rangle^{ \ottnt{l} } ~  \ottnt{e_{{\mathrm{2}}}}  \,  \longrightarrow _{ \ottnt{m} }  \,  \langle  \ottnt{T_{{\mathrm{2}}}}  \mathord{ \overset{ \ottnt{a} }{\Rightarrow} }  \ottnt{T_{{\mathrm{3}}}}  \rangle^{ \ottnt{l} } ~  \ottnt{e'_{{\mathrm{2}}}} }{%
{\ottdrulename{E\_CastInner}}{}%
}}

\newcommand{\ottdruleEXXCastMerge}[1]{\ottdrule[#1]{%
\ottpremise{\ottnt{a_{{\mathrm{3}}}}  \ottsym{=}   \mathsf{merge} _{ \ottnt{m} }( \ottnt{T_{{\mathrm{1}}}} , \ottnt{a_{{\mathrm{1}}}} , \ottnt{T_{{\mathrm{2}}}} , \ottnt{a_{{\mathrm{2}}}} , \ottnt{T_{{\mathrm{3}}}} ) }%
}{
 \langle  \ottnt{T_{{\mathrm{2}}}}  \mathord{ \overset{ \ottnt{a_{{\mathrm{2}}}} }{\Rightarrow} }  \ottnt{T_{{\mathrm{3}}}}  \rangle^{ \ottnt{l} } ~   (  \langle  \ottnt{T_{{\mathrm{1}}}}  \mathord{ \overset{ \ottnt{a_{{\mathrm{1}}}} }{\Rightarrow} }  \ottnt{T_{{\mathrm{2}}}}  \rangle^{ \ottnt{l'} } ~  \ottnt{e_{{\mathrm{2}}}}  )   \,  \longrightarrow _{ \ottnt{m} }  \,  \langle  \ottnt{T_{{\mathrm{1}}}}  \mathord{ \overset{ \ottnt{a_{{\mathrm{3}}}} }{\Rightarrow} }  \ottnt{T_{{\mathrm{3}}}}  \rangle^{ \ottnt{l} } ~  \ottnt{e_{{\mathrm{2}}}} }{%
{\ottdrulename{E\_CastMerge}}{}%
}}

\newcommand{\ottdruleEXXCastInnerE}[1]{\ottdrule[#1]{%
\ottpremise{ \ottnt{e_{{\mathrm{2}}}} \,  \longrightarrow _{  \mathsf{E}  }  \, \ottnt{e'_{{\mathrm{2}}}}  \quad  \ottnt{e_{{\mathrm{2}}}}  \neq   \langle  \ottnt{T_{{\mathrm{1}}}}  \mathord{ \overset{ \ottnt{a'} }{\Rightarrow} }  \ottnt{T_{{\mathrm{2}}}}  \rangle^{ \ottnt{l'} } ~  \ottnt{e''_{{\mathrm{2}}}}  }%
}{
 \langle  \ottnt{T_{{\mathrm{2}}}}  \mathord{ \overset{ \ottnt{a} }{\Rightarrow} }  \ottnt{T_{{\mathrm{3}}}}  \rangle^{ \ottnt{l} } ~  \ottnt{e_{{\mathrm{2}}}}  \,  \longrightarrow _{  \mathsf{E}  }  \,  \langle  \ottnt{T_{{\mathrm{2}}}}  \mathord{ \overset{ \ottnt{a} }{\Rightarrow} }  \ottnt{T_{{\mathrm{3}}}}  \rangle^{ \ottnt{l} } ~  \ottnt{e'_{{\mathrm{2}}}} }{%
{\ottdrulename{E\_CastInnerE}}{}%
}}

\newcommand{\ottdruleEXXCastMergeE}[1]{\ottdrule[#1]{%
\ottpremise{\ottnt{c_{{\mathrm{3}}}}  \ottsym{=}   \mathsf{join} ( \ottnt{c_{{\mathrm{1}}}} , \ottnt{c_{{\mathrm{2}}}} ) }%
}{
 \langle  \ottnt{T_{{\mathrm{2}}}}  \mathord{ \overset{ \ottnt{c_{{\mathrm{2}}}} }{\Rightarrow} }  \ottnt{T_{{\mathrm{3}}}}  \rangle^{ \ottnt{l} } ~   (  \langle  \ottnt{T_{{\mathrm{1}}}}  \mathord{ \overset{ \ottnt{c_{{\mathrm{1}}}} }{\Rightarrow} }  \ottnt{T_{{\mathrm{2}}}}  \rangle^{ \ottnt{l'} } ~  \ottnt{e_{{\mathrm{2}}}}  )   \,  \longrightarrow _{  \mathsf{E}  }  \,  \langle  \ottnt{T_{{\mathrm{1}}}}  \mathord{ \overset{ \ottnt{c_{{\mathrm{3}}}} }{\Rightarrow} }  \ottnt{T_{{\mathrm{3}}}}  \rangle^{ \ottnt{l} } ~  \ottnt{e_{{\mathrm{2}}}} }{%
{\ottdrulename{E\_CastMergeE}}{}%
}}

\newcommand{\ottdruleEXXTypeSet}[1]{\ottdrule[#1]{%
}{
 \langle  \ottnt{T_{{\mathrm{1}}}}  \mathord{ \overset{\bullet}{\Rightarrow} }  \ottnt{T_{{\mathrm{2}}}}  \rangle^{ \ottnt{l} } ~  \ottnt{e}  \,  \longrightarrow _{  \mathsf{H}  }  \,  \langle  \ottnt{T_{{\mathrm{1}}}}  \mathord{ \overset{ \emptyset }{\Rightarrow} }  \ottnt{T_{{\mathrm{2}}}}  \rangle^{ \ottnt{l} } ~  \ottnt{e} }{%
{\ottdrulename{E\_TypeSet}}{}%
}}

\newcommand{\ottdruleEXXCheckEmpty}[1]{\ottdrule[#1]{%
}{
\begin{array}{@{}l@{} }   \langle   \{ \mathit{x} \mathord{:} \ottnt{B} \mathrel{\mid} \ottnt{e_{{\mathrm{1}}}} \}   \mathord{ \overset{ \emptyset }{\Rightarrow} }   \{ \mathit{x} \mathord{:} \ottnt{B} \mathrel{\mid} \ottnt{e_{{\mathrm{2}}}} \}   \rangle^{ \ottnt{l} } ~  \ottnt{k}  \,  \longrightarrow _{  \mathsf{H}  }  \, {} \\   \langle   \{ \mathit{x} \mathord{:} \ottnt{B} \mathrel{\mid} \ottnt{e_{{\mathrm{2}}}} \}  ,   \ottnt{e_{{\mathrm{2}}}}  [  \ottnt{k} / \mathit{x}  ]  ,  \ottnt{k}  \rangle^{ \ottnt{l} }   \end{array}}{%
{\ottdrulename{E\_CheckEmpty}}{}%
}}

\newcommand{\ottdruleEXXCheckSet}[1]{\ottdrule[#1]{%
\ottpremise{ \mathsf{choose} ( \mathcal{S} )   \ottsym{=}   \{ \mathit{x} \mathord{:} \ottnt{B} \mathrel{\mid} \ottnt{e_{{\mathrm{2}}}} \} }%
}{
 \begin{array}{@{}l@{} }   \langle   \{ \mathit{x} \mathord{:} \ottnt{B} \mathrel{\mid} \ottnt{e_{{\mathrm{1}}}} \}   \mathord{ \overset{ \mathcal{S} }{\Rightarrow} }   \{ \mathit{x} \mathord{:} \ottnt{B} \mathrel{\mid} \ottnt{e_{{\mathrm{3}}}} \}   \rangle^{ \ottnt{l} } ~  \ottnt{k}   \,  \longrightarrow _{  \mathsf{H}  }  \, {} \\    \langle   \{ \mathit{x} \mathord{:} \ottnt{B} \mathrel{\mid} \ottnt{e_{{\mathrm{2}}}} \}   \mathord{ \overset{  \mathcal{S}  \setminus   \{ \mathit{x} \mathord{:} \ottnt{B} \mathrel{\mid} \ottnt{e_{{\mathrm{2}}}} \}   }{\Rightarrow} }   \{ \mathit{x} \mathord{:} \ottnt{B} \mathrel{\mid} \ottnt{e_{{\mathrm{3}}}} \}   \rangle^{ \ottnt{l} } ~  {} \\  \quad   \langle   \{ \mathit{x} \mathord{:} \ottnt{B} \mathrel{\mid} \ottnt{e_{{\mathrm{2}}}} \}  ,   \ottnt{e_{{\mathrm{2}}}}  [  \ottnt{k} / \mathit{x}  ]  ,  \ottnt{k}  \rangle^{ \ottnt{l} }     \end{array}}{%
{\ottdrulename{E\_CheckSet}}{}%
}}

\newcommand{\ottdruleEXXCoerce}[1]{\ottdrule[#1]{%
}{
 \langle  \ottnt{T_{{\mathrm{1}}}}  \mathord{ \overset{\bullet}{\Rightarrow} }  \ottnt{T_{{\mathrm{2}}}}  \rangle^{ \ottnt{l} } ~  \ottnt{e}  \,  \longrightarrow _{  \mathsf{E}  }  \,  \langle  \ottnt{T_{{\mathrm{1}}}}  \mathord{ \overset{  \mathsf{coerce} ( \ottnt{T_{{\mathrm{1}}}} , \ottnt{T_{{\mathrm{2}}}} , \ottnt{l} )  }{\Rightarrow} }  \ottnt{T_{{\mathrm{2}}}}  \rangle^{\bullet} ~  \ottnt{e} }{%
{\ottdrulename{E\_Coerce}}{}%
}}

\newcommand{\ottdruleEXXCoerceStack}[1]{\ottdrule[#1]{%
}{
 \langle   \{ \mathit{x} \mathord{:} \ottnt{B} \mathrel{\mid} \ottnt{e_{{\mathrm{1}}}} \}   \mathord{ \overset{ \ottnt{r} }{\Rightarrow} }   \{ \mathit{x} \mathord{:} \ottnt{B} \mathrel{\mid} \ottnt{e_{{\mathrm{2}}}} \}   \rangle^{\bullet} ~  \ottnt{k}  \,  \longrightarrow _{  \mathsf{E}  }  \,  \langle   \{ \mathit{x} \mathord{:} \ottnt{B} \mathrel{\mid} \ottnt{e_{{\mathrm{2}}}} \}  ,   \mathord{?}  ,  \ottnt{r} ,  \ottnt{k} ,  \ottnt{k}  \rangle^{\bullet} }{%
{\ottdrulename{E\_CoerceStack}}{}%
}}

\newcommand{\ottdruleEXXStackDone}[1]{\ottdrule[#1]{%
}{
 \langle   \{ \mathit{x} \mathord{:} \ottnt{B} \mathrel{\mid} \ottnt{e} \}  ,   \mathord{\checkmark}  ,  \mathsf{nil} ,  \ottnt{k} ,  \ottnt{k}  \rangle^{\bullet}  \,  \longrightarrow _{  \mathsf{E}  }  \, \ottnt{k}}{%
{\ottdrulename{E\_StackDone}}{}%
}}

\newcommand{\ottdruleEXXStackPop}[1]{\ottdrule[#1]{%
}{
 \langle   \{ \mathit{x} \mathord{:} \ottnt{B} \mathrel{\mid} \ottnt{e} \}  ,  \ottnt{s} ,  \ottsym{(}    \{ \mathit{x} \mathord{:} \ottnt{B} \mathrel{\mid} \ottnt{e'} \}^{ \ottnt{l} }  , \ottnt{r}   \ottsym{)} ,  \ottnt{k} ,  \ottnt{k}  \rangle^{\bullet}  \,  \longrightarrow _{  \mathsf{E}  }  \,  \langle   \{ \mathit{x} \mathord{:} \ottnt{B} \mathrel{\mid} \ottnt{e} \}  ,   \ottnt{s}  \vee ( \ottnt{e}  =  \ottnt{e'} )  ,  \ottnt{r} ,  \ottnt{k} ,   \langle   \{ \mathit{x} \mathord{:} \ottnt{B} \mathrel{\mid} \ottnt{e'} \}  ,   \ottnt{e'}  [  \ottnt{k} / \mathit{x}  ]  ,  \ottnt{k}  \rangle^{ \ottnt{l} }   \rangle^{\bullet} }{%
{\ottdrulename{E\_StackPop}}{}%
}}

\newcommand{\ottdruleEXXStackInner}[1]{\ottdrule[#1]{%
\ottpremise{\ottnt{e'} \,  \longrightarrow _{  \mathsf{E}  }  \, \ottnt{e''}}%
}{
 \langle   \{ \mathit{x} \mathord{:} \ottnt{B} \mathrel{\mid} \ottnt{e} \}  ,  \ottnt{s} ,  \ottnt{r} ,  \ottnt{k} ,  \ottnt{e'}  \rangle^{\bullet}  \,  \longrightarrow _{  \mathsf{E}  }  \,  \langle   \{ \mathit{x} \mathord{:} \ottnt{B} \mathrel{\mid} \ottnt{e} \}  ,  \ottnt{s} ,  \ottnt{r} ,  \ottnt{k} ,  \ottnt{e''}  \rangle^{\bullet} }{%
{\ottdrulename{E\_StackInner}}{}%
}}

\newcommand{\ottdruleEXXStackRaise}[1]{\ottdrule[#1]{%
}{
 \langle   \{ \mathit{x} \mathord{:} \ottnt{B} \mathrel{\mid} \ottnt{e} \}  ,  \ottnt{s} ,  \ottnt{r} ,  \ottnt{k} ,   \mathord{\Uparrow}  \ottnt{l'}   \rangle^{\bullet}  \,  \longrightarrow _{  \mathsf{E}  }  \,  \mathord{\Uparrow}  \ottnt{l'} }{%
{\ottdrulename{E\_StackRaise}}{}%
}}

\newcommand{\ottdruleSWFXXRefine}[1]{\ottdrule[#1]{%
\ottpremise{ \forall  \ottnt{k}  . ~    \mathsf{ty} ( \ottnt{k} )   \ottsym{=}  \ottnt{B} \, \text{implies} \,  \ottnt{e}  [  \ottnt{k} / \mathit{x}  ]  \, \in \,  \denot{  \{ \mathit{x} \mathord{:}  \mathsf{Bool}  \mathrel{\mid}  \mathsf{true}  \}  }   }%
}{
\models   \{ \mathit{x} \mathord{:} \ottnt{B} \mathrel{\mid} \ottnt{e} \} }{%
{\ottdrulename{SWF\_Refine}}{}%
}}

\newcommand{\ottdruleSWFXXFun}[1]{\ottdrule[#1]{%
\ottpremise{ \models  \ottnt{T_{{\mathrm{1}}}}  \quad  \models  \ottnt{T_{{\mathrm{2}}}} }%
}{
\models   \ottnt{T_{{\mathrm{1}}}} \mathord{ \rightarrow } \ottnt{T_{{\mathrm{2}}}} }{%
{\ottdrulename{SWF\_Fun}}{}%
}}

\newcommand{\ottdruleSWFXXTypeSet}[1]{\ottdrule[#1]{%
\ottpremise{  \vdash  \ottnt{T_{{\mathrm{1}}}}  \mathrel{\parallel}  \ottnt{T_{{\mathrm{2}}}}  \quad    \models  \ottnt{T_{{\mathrm{1}}}}  \quad  \models  \ottnt{T_{{\mathrm{2}}}}    }%
\ottpremise{ \forall   \ottnt{T}  \in  \mathcal{S}   . ~    \models  \ottnt{T}  \quad  \vdash  \ottnt{T}  \mathrel{\parallel}  \ottnt{T_{{\mathrm{1}}}}   }%
}{
\models  \mathcal{S}  \mathrel{\parallel}  \ottnt{T_{{\mathrm{1}}}}  \Rightarrow  \ottnt{T_{{\mathrm{2}}}}}{%
{\ottdrulename{SWF\_TypeSet}}{}%
}}

\newcommand{\ottdruleSubXXNaiveRefine}[1]{\ottdrule[#1]{%
\ottpremise{ \{ \mathit{x} \mathord{:} \ottnt{B} \mathrel{\mid} \ottnt{e_{{\mathrm{1}}}} \}  \, \supset \,  \{ \mathit{x} \mathord{:} \ottnt{B} \mathrel{\mid} \ottnt{e_{{\mathrm{2}}}} \} }%
}{
 \{ \mathit{x} \mathord{:} \ottnt{B} \mathrel{\mid} \ottnt{e_{{\mathrm{1}}}} \}  \,  <: _{ \ottmv{n} }  \,  \{ \mathit{x} \mathord{:} \ottnt{B} \mathrel{\mid} \ottnt{e_{{\mathrm{2}}}} \} }{%
{\ottdrulename{Sub\_NaiveRefine}}{}%
}}

\newcommand{\ottdruleSubXXNaiveFun}[1]{\ottdrule[#1]{%
\ottpremise{ \ottnt{T_{{\mathrm{21}}}} \,  <: _{ \ottmv{n} }  \, \ottnt{T_{{\mathrm{11}}}}  \quad  \ottnt{T_{{\mathrm{12}}}} \,  <: _{ \ottmv{n} }  \, \ottnt{T_{{\mathrm{22}}}} }%
}{
 \ottnt{T_{{\mathrm{11}}}} \mathord{ \rightarrow } \ottnt{T_{{\mathrm{12}}}}  \,  <: _{ \ottmv{n} }  \,  \ottnt{T_{{\mathrm{21}}}} \mathord{ \rightarrow } \ottnt{T_{{\mathrm{22}}}} }{%
{\ottdrulename{Sub\_NaiveFun}}{}%
}}

\begin{document}

\numbertheoremsfalse

\ifpopl
\conferenceinfo{POPL '15}{January 15--17, 2015, Mumbai, India} 
\copyrightyear{2015} 
\copyrightdata{978-1-4503-3300-9/15/01} 
\doi{2676726.2676967}
\fi

% Uncomment one of the following two, if you are not going for the 
% traditional copyright transfer agreement.

\iffull\else
\exclusivelicense                 % ACM gets exclusive license to publish, 
                                   % you retain copyright
\fi
%\permissiontopublish             % ACM gets nonexclusive license to publish
                                 % (paid open-access papers, 
                                 % short abstracts)

{\iffull\else
\ifdraft
\titlebanner{DRAFT---do not distribute}        % These are ignored unless
\preprintfooter{DRAFT}   % 'preprint' option specified.
\fi
\fi}

\title{Space-Efficient Manifest Contracts}
%\subtitle{Subtitle Text, if any}

{\iffull
\author{Michael Greenberg \\ Princeton University}
\else
\authorinfo{Michael Greenberg}
           {Princeton University}
           {mg19@cs.princeton.edu}
\fi}

\maketitle

\begin{abstract}
  The standard algorithm for higher-order contract checking can lead
  to unbounded space consumption and can destroy tail recursion,
  altering a program's asymptotic space complexity.
  While space efficiency for gradual types---contracts mediating
  untyped and typed code---is well studied, sound space efficiency for
  manifest contracts---contracts that check stronger properties than
  simple types, e.g., ``is a natural'' instead of ``is an
  integer''---remains an open problem.

  We show how to achieve sound space efficiency for manifest contracts
  with strong predicate contracts. The essential trick is
  breaking the contract checking down into \textit{coercions}:
  structured, blame-annotated lists of checks. By carefully preventing
  duplicate coercions from appearing, we can restore space efficiency
  while keeping the same observable behavior.

  {\iffull Along the way, we define a framework for space efficiency,
    traversing the design space with three different space-efficient
    manifest calculi. We examine the diverse correctness criteria for
    contract semantics; we conclude with a coercion-based language
    whose contracts enjoy (galactically) bounded, \textit{sound} space
    consumption---they are observationally equivalent to the standard,
    space-inefficient semantics.  

    \textit{This is an extended version of \citet{Greenberg15space}, with a
    great deal of material that does not appear in the conference
    paper: an exploration of the design space with two other
    space-efficient calculi and complete proofs.}  \fi}
\end{abstract}

\iffull\else

% terms are not compulsory anymore, 
% you may leave them out
%\terms
%term1, term2

\fi

\section{Introduction}
\label{sec:intro}

Types are an extremely successful form of lightweight specification:
programmers can state their intent---e.g., plus is a function that
takes two numbers and returns another number---and then type checkers
can ensure that a program conforms to the programmer's intent.
Types can only go so far though: division is, like addition, a
function that takes two numbers and returns another number... so long
as the second number isn't zero. Conventional type systems do a good
job of stopping many kinds of errors, but most type systems cannot
protect partial operations like division and array indexing.
Advanced techniques---singleton and dependent types, for example---can
cover many of these cases, allowing programmers to use types like
``non-zero number'' or ``index within bounds'' to specify the domains
on which partial operations are safe. Such techniques are demanding:
they can be difficult to understand, they force certain programming
idioms, and they place heavy constraints on the programming language,
requiring purity or even strong normalization.

\textit{Contracts} are a popular compromise: programmers write
type-like contracts of the form $ \mathsf{Int}   \rightarrow   \{ \mathit{x} \mathord{:}  \mathsf{Int}  \mathrel{\mid}  \mathit{x}  \mathrel{\ne}  \ottsym{0}  \} 
 \rightarrow   \mathsf{Int} $, where the predicates $ \mathit{x}  \mathrel{\ne}  \ottsym{0} $ are written in
code. These type-like specifications can then be checked at
runtime~\cite{Findler02contracts}.
Models of contract calculi have taken two forms: latent and
manifest~\cite{Greenberg12contracts}. We take the manifest approach
here, which means checking contracts with \textit{casts}, written
$ \langle  \ottnt{T_{{\mathrm{1}}}}  \mathord{ \overset{    }{\Rightarrow} }  \ottnt{T_{{\mathrm{2}}}}  \rangle^{ \ottnt{l} } ~  \ottnt{e} $.
Checking a \textit{predicate contract} (also called a
\textit{refinement type}, though that term is overloaded) like
$ \{ \mathit{x} \mathord{:}  \mathsf{Int}  \mathrel{\mid}  \mathit{x}  \mathrel{\ne}  \ottsym{0}  \} $ on a number $\ottmv{n}$ involves running the
predicate $ \mathit{x}  \mathrel{\ne}  \ottsym{0} $ with $\ottmv{n}$ for $\mathit{x}$. Casts from one
predicate contract to another, $ \langle   \{ \mathit{x} \mathord{:} \ottnt{B} \mathrel{\mid} \ottnt{e_{{\mathrm{1}}}} \}   \mathord{ \overset{    }{\Rightarrow} }   \{ \mathit{x} \mathord{:} \ottnt{B} \mathrel{\mid} \ottnt{e_{{\mathrm{2}}}} \}   \rangle^{ \ottnt{l} } ~     $\!\!, take a constant $\ottnt{k}$ and check to see that $ \ottnt{e_{{\mathrm{2}}}}  [  \ottnt{k} / \mathit{x}  ]  \,  \longrightarrow ^{*}_{    }  \,  \mathsf{true} $.
It's hard to know what to do with function casts at runtime: in
$ \langle   \ottnt{T_{{\mathrm{11}}}} \mathord{ \rightarrow } \ottnt{T_{{\mathrm{12}}}}   \mathord{ \overset{    }{\Rightarrow} }   \ottnt{T_{{\mathrm{21}}}} \mathord{ \rightarrow } \ottnt{T_{{\mathrm{22}}}}   \rangle^{ \ottnt{l} } ~  \ottnt{e} $, we know that $\ottnt{e}$ is a
$ \ottnt{T_{{\mathrm{11}}}} \mathord{ \rightarrow } \ottnt{T_{{\mathrm{12}}}} $, but what does that tell us about treating $\ottnt{e}$ as a
$ \ottnt{T_{{\mathrm{21}}}} \mathord{ \rightarrow } \ottnt{T_{{\mathrm{22}}}} $? Findler and Felleisen's insight is that we must defer
checking, waiting until the cast value $\ottnt{e}$ gets an
argument~\cite{Findler02contracts}. These deferred checks are recorded
on the value by means of a \textit{function proxy}, i.e., $ \langle   \ottnt{T_{{\mathrm{11}}}} \mathord{ \rightarrow } \ottnt{T_{{\mathrm{12}}}}   \mathord{ \overset{    }{\Rightarrow} }   \ottnt{T_{{\mathrm{21}}}} \mathord{ \rightarrow } \ottnt{T_{{\mathrm{22}}}}   \rangle^{ \ottnt{l} } ~  \ottnt{e} $ is a value when $\ottnt{e}$ is a value; applying
a function proxy unwraps it contravariantly. We check the domain
contract $\ottnt{T_{{\mathrm{1}}}}$ on $\ottnt{e}$, run the original function $\mathit{f}$ on the
result, and then check that result against the codomain contract
$\ottnt{T_{{\mathrm{2}}}}$:
\[ \begin{array}{@{}l@{}}
    (  \langle   \ottnt{T_{{\mathrm{11}}}} \mathord{ \rightarrow } \ottnt{T_{{\mathrm{12}}}}   \mathord{ \overset{    }{\Rightarrow} }   \ottnt{T_{{\mathrm{21}}}} \mathord{ \rightarrow } \ottnt{T_{{\mathrm{22}}}}   \rangle^{ \ottnt{l} } ~  \ottnt{e_{{\mathrm{1}}}}  )  ~ \ottnt{e_{{\mathrm{2}}}}  \,  \longrightarrow _{    }  \, {} \\   \langle  \ottnt{T_{{\mathrm{12}}}}  \mathord{ \overset{    }{\Rightarrow} }  \ottnt{T_{{\mathrm{22}}}}  \rangle^{ \ottnt{l} } ~   (  \ottnt{e_{{\mathrm{1}}}} ~  (  \langle  \ottnt{T_{{\mathrm{21}}}}  \mathord{ \overset{    }{\Rightarrow} }  \ottnt{T_{{\mathrm{11}}}}  \rangle^{ \ottnt{l} } ~  \ottnt{e_{{\mathrm{2}}}}  )   )  
\end{array}
\]
Findler and Felleisen neatly designed a system for contract checking
in a higher-order world, but there is a problem: contract checking is space
inefficient~\cite{Herman07space}.

Contract checking's space inefficiency can be summed up as follows:
\textbf{function proxies break tail calls}. Calls to an unproxied
function from a tail position can be optimized to not allocate stack
frames. Proxied functions, however, will unwrap to have codomain
contracts---breaking tail calls.
We discuss other sources of space inefficiency below, but breaking
tail calls is the most severe.
Consider factorial written in accumulator passing style. The developer
may believe that the following can be compiled to use tail calls:
\[ \begin{array}{r@{~}l}
   \mathsf{fact}  : &    \{ \mathit{x} \mathord{:}  \mathsf{Int}  \mathrel{\mid}  \mathit{x}  \mathrel{\ge}  \ottsym{0}  \}  \mathord{ \rightarrow }  \{ \mathit{x} \mathord{:}  \mathsf{Int}  \mathrel{\mid}  \mathit{x}  \mathrel{\ge}  \ottsym{0}  \}   \mathord{ \rightarrow }  \{ \mathit{x} \mathord{:}  \mathsf{Int}  \mathrel{\mid}  \mathit{x}  \mathrel{\ge}  \ottsym{0}  \}   \\
           = &    \lambda \mathit{x} \mathord{:}  \{ \mathit{x} \mathord{:}  \mathsf{Int}  \mathrel{\mid}  \mathsf{true}  \}  .~   \lambda \mathit{y} \mathord{:}  \{ \mathit{y} \mathord{:}  \mathsf{Int}  \mathrel{\mid}  \mathsf{true}  \}  .~  {} \\  &  \quad   \mathsf{if}~  \mathit{x}  \mathrel{=}  \ottsym{0}  ~\mathsf{then}~ \mathit{y} ~\mathsf{else}~  \mathsf{fact}     ~  ( \mathit{x} \,  -  \, \ottsym{1} )   ~  (  \mathit{x}  \mathrel{*}  \mathit{y}  )  
\end{array} \]
A cast insertion algorithm~\cite{Swamy09coercions} might produce the
following non-tail recursive function:
\[ \begin{array}{r@{}l}
  \multicolumn{2}{l}{ \mathsf{fact}  = } \\
  ~~~ &  \langle    \{ \mathit{x} \mathord{:}  \mathsf{Int}  \mathrel{\mid}  \mathsf{true}  \}  \mathord{ \rightarrow }   \{ \mathit{y} \mathord{:}  \mathsf{Int}  \mathrel{\mid}  \mathsf{true}  \}  \mathord{ \rightarrow }  \{ \mathit{z} \mathord{:}  \mathsf{Int}  \mathrel{\mid}  \mathsf{true}  \}     \mathord{ \overset{    }{\Rightarrow} }    {} \\  &  \phantom{\langle}   \{ \mathit{x} \mathord{:}  \mathsf{Int}  \mathrel{\mid}  \mathit{x}  \mathrel{\ge}  \ottsym{0}  \}  \mathord{ \rightarrow }  \{ \mathit{y} \mathord{:}  \mathsf{Int}  \mathrel{\mid}  \mathit{y}  \mathrel{\ge}  \ottsym{0}  \}   \mathord{ \rightarrow }  \{ \mathit{z} \mathord{:}  \mathsf{Int}  \mathrel{\mid}  \mathit{x}  \mathrel{\ge}  \ottsym{0}  \}    \rangle^{  l_{ \mathsf{fact} }  } ~  {} \\  &   \lambda \mathit{x} \mathord{:}  \{ \mathit{x} \mathord{:}  \mathsf{Int}  \mathrel{\mid}  \mathsf{true}  \}  .~   \lambda \mathit{y} \mathord{:}  \{ \mathit{y} \mathord{:}  \mathsf{Int}  \mathrel{\mid}  \mathsf{true}  \}  .~  {} \\  &   ~~   \mathsf{if}~  \mathit{x}  \mathrel{=}  \ottsym{0}  ~\mathsf{then}~ \mathit{y} ~\mathsf{else}~ {} \\  &  \quad   (  \langle   \{ \mathit{x} \mathord{:}  \mathsf{Int}  \mathrel{\mid}  \mathit{x}  \mathrel{\ge}  \ottsym{0}  \}   \mathord{ \overset{    }{\Rightarrow} }   \{ \mathit{x} \mathord{:}  \mathsf{Int}  \mathrel{\mid}  \mathsf{true}  \}   \rangle^{  l_{ \mathsf{fact} }  } ~   (   \mathsf{fact}  ~ \dots  )   )      
\end{array} \]
\finishlater{this function evaluating in classic, eidetic}
Tail-call optimization is essential for usable functional
languages. Space inefficiency has been one of two
significant obstacles for pervasive use of higher-order contract
checking. (The other is state, which we do not treat here.)

\medskip

In this work, we show how to achieve semantics-preserving space
efficiency for non-dependent contract checking.
Our approach is inspired by work on \textit{gradual
  typing}~\cite{Siek06gradual}, a form of (manifest) contracts
designed to mediate dynamic and simple typing---that is, gradual typing
(a) allows the dynamic type, and (b) restricts the predicates in
contracts to checks on type tags.
\Citet{Herman07space} developed the first space-efficient gradually
typed system, using Henglein's coercions~\cite{Henglein94dynamic};
\citet{Siek10threesomes} devised a related system supporting blame.
The essence of the solution is to allow casts to merge: given two
adjacent casts $ \langle  \ottnt{T_{{\mathrm{2}}}}  \mathord{ \overset{    }{\Rightarrow} }  \ottnt{T_{{\mathrm{3}}}}  \rangle^{ \ottnt{l_{{\mathrm{2}}}} } ~   (  \langle  \ottnt{T_{{\mathrm{1}}}}  \mathord{ \overset{    }{\Rightarrow} }  \ottnt{T_{{\mathrm{2}}}}  \rangle^{ \ottnt{l_{{\mathrm{1}}}} } ~  \ottnt{e}  )  $, we
must somehow combine them into a single cast.
\Citeauthor{Siek10threesomes} annotate their casts with an
intermediate type representing the greatest lower bound of the types
encountered.
Such a trick doesn't work in our more general setting: simple types
plus dynamic form a straightforward lattice using type precision as
the ordering, but it's less clear what to do when we have arbitrary
predicate contracts.

{\ifpopl
We define two modes of a single calculus, \lambdah. The \textit{\underline{c}lassic} mode is just the conventional, inefficient semantics; 
\else
We offer three \textit{modes} of space-efficiency; all of the modes
are defined in a single calculus which we call \lambdah. Each mode
enjoys varying levels of soundness with respect to the standard,
space-inefficient semantics of classic \lambdah.
We sketch here the mode-indexed rules for combining annotations on
casts---the key rules for space efficiency.

The \textit{\underline{f}orgetful} mode uses empty annotations, $ \bullet $; we combine two
casts by dropping intermediate types:
\[  \langle  \ottnt{T_{{\mathrm{2}}}}  \mathord{ \overset{ \bullet }{\Rightarrow} }  \ottnt{T_{{\mathrm{3}}}}  \rangle^{ \ottnt{l_{{\mathrm{2}}}} } ~   (  \langle  \ottnt{T_{{\mathrm{1}}}}  \mathord{ \overset{ \bullet }{\Rightarrow} }  \ottnt{T_{{\mathrm{2}}}}  \rangle^{ \ottnt{l_{{\mathrm{1}}}} } ~  \ottnt{e}  )   \,  \longrightarrow _{  \mathsf{F}  }  \,  \langle  \ottnt{T_{{\mathrm{1}}}}  \mathord{ \overset{ \bullet }{\Rightarrow} }  \ottnt{T_{{\mathrm{3}}}}  \rangle^{ \ottnt{l_{{\mathrm{2}}}} } ~  \ottnt{e}  \]
Surprisingly, this evaluation rule is type safe and somewhat sound
with respect to the classic mode, as discovered by
\citet{Greenberg13thesis}: if classic \lambdah produces a value, so
does forgetful \lambdah.

The \textit{\underline{h}eedful} mode uses sets of types $\mathcal{S}_{\ottmv{i}}$ as its
annotations, making sure to save the intermediate type:
\[  \langle  \ottnt{T_{{\mathrm{2}}}}  \mathord{ \overset{ \mathcal{S}_{{\mathrm{2}}} }{\Rightarrow} }  \ottnt{T_{{\mathrm{3}}}}  \rangle^{ \ottnt{l_{{\mathrm{2}}}} } ~   (  \langle  \ottnt{T_{{\mathrm{1}}}}  \mathord{ \overset{ \mathcal{S}_{{\mathrm{1}}} }{\Rightarrow} }  \ottnt{T_{{\mathrm{2}}}}  \rangle^{ \ottnt{l_{{\mathrm{1}}}} } ~  \ottnt{e}  )   \,  \longrightarrow _{  \mathsf{H}  }  \,  \langle  \ottnt{T_{{\mathrm{1}}}}  \mathord{ \overset{   \mathcal{S}_{{\mathrm{1}}}  \cup  \mathcal{S}_{{\mathrm{2}}}   \cup   \set{  \ottnt{T_{{\mathrm{2}}}}  }   }{\Rightarrow} }  \ottnt{T_{{\mathrm{3}}}}  \rangle^{ \ottnt{l_{{\mathrm{2}}}} } ~  \ottnt{e}  \]
In Siek and Wadler's terms, we use the powerset lattice for
annotations, while they use pointed types.
Heedful and classic \lambdah are almost identical, except sometimes
they blame different labels. 

Finally,\fi} the \textit{\underline{e}idetic} mode annotates casts with
\textit{refinement lists} and \textit{function coercions}---a new form
of coercion inspired by \citet{Greenberg13thesis}. The coercions keep
track of checking so well that the type indices and blame labels on casts are
unnecessary:
\[  \langle  \ottnt{T_{{\mathrm{2}}}}  \mathord{ \overset{ \ottnt{c_{{\mathrm{2}}}} }{\Rightarrow} }  \ottnt{T_{{\mathrm{3}}}}  \rangle^{ \bullet } ~   (  \langle  \ottnt{T_{{\mathrm{1}}}}  \mathord{ \overset{ \ottnt{c_{{\mathrm{1}}}} }{\Rightarrow} }  \ottnt{T_{{\mathrm{2}}}}  \rangle^{ \bullet } ~  \ottnt{e}  )   \,  \longrightarrow _{  \mathsf{E}  }  \,  \langle  \ottnt{T_{{\mathrm{1}}}}  \mathord{ \overset{  \mathsf{join} ( \ottnt{c_{{\mathrm{1}}}} , \ottnt{c_{{\mathrm{2}}}} )  }{\Rightarrow} }  \ottnt{T_{{\mathrm{3}}}}  \rangle^{ \bullet } ~  \ottnt{e}  \]
These coercions form a skew lattice: refinement lists have ordering
constraints that break commutativity.
Eidetic \lambdah is space efficient and observationally equivalent to
the classic mode.

{\iffull
Since eidetic and classic \lambdah behave the same, why bother with
forgetful and heedful?
First and foremost, the calculi offer insights into the semantics of
contracts\iffull: the soundness of forgetful \lambdah depends on a certain
philosophy of contracts; heedful \lambdah relates to threesomes
without blame~\cite{Siek10threesomes}\fi.
Second, we offer them as alternative points in the design space.
Finally and perhaps cynically, they are strawmen---warm up exercises
for eidetic \lambdah.
\fi}

\medskip

\iffull
We claim two contributions:
\begin{enumerate}
\item \fi Eidetic \lambdah is the first manifest contract calculus that is
  both \textit{sound} and \textit{space efficient} with respect to the
  classic semantics---a result contrary to \citet{Greenberg13thesis},
  who conjectured that such a result is impossible. We believe that
  space efficiency is a critical step towards the implementation of
  practical languages with manifest contracts.

\iffull
\item A framework for defining space efficiency in manifest contract
  systems, with an exploration of the design space. We identify common
  structures and methods in the operational semantics as well as in
  the proofs of type soundness, soundness with regard to the classic
  framework, and space bounds.
\end{enumerate} \fi
We do not prove a blame theorem~\cite{Tobin-Hochstadt06interlanguage}, since we lack
the clear separation of dynamic and static typing found in gradual
typing. We conjecture that such a theorem could be proved\iffull for classic
and eidetic \lambdah---but perhaps not for forgetful and heedful
\lambdah, which skip checks and change blame labels\fi.
Our model has two limits worth mentioning: we do not handle
dependency, a common and powerful feature in manifest systems; and,
our bounds for space efficiency are \textit{galactic}---they establish
that contracts consume constant space, but do nothing to reduce that
constant~\cite{Lipton10galactic}. Our contribution is showing that
sound space efficiency is \textit{possible} where it was believed to
be impossible~\cite{Greenberg13thesis}; we leave evidence
that it is \textit{practicable} for future work.

Our proofs are available in the extended
version~\cite{Greenberg14spacetr},
Appendices~\ref{app:typesoundness}--\ref{app:bounds}.

Readers who are very familiar with this topic can read
Figures~\ref{fig:syntax},~\ref{fig:opsem}, and~\ref{fig:typing} and
then skip directly to Section~\ifpopl\ref{sec:eidetic}\else\ref{sec:overview}\fi.
Readers who understand the space inefficiency of contracts but aren't
particularly familiar with manifest contracts can skip
Section~\ref{sec:inefficiency} and proceed to
Section~\ref{sec:classic}.

\section{Function proxies}
\label{sec:inefficiency}

Space inefficient contract checking breaks tail recursion---a
showstopping problem for realistic implementations of pervasive
contract use. 
Racket's contract system~\cite{RacketContracts}, the most widely
used higher-order contract system, takes a ``macro'' approach to
contracts: contracts typically appear only on module interfaces, and
aren't checked within a module. Their approach comes partly out of a
philosophy of breaking invariants inside modules but not out of them,
but also partly out of a need to retain tail recursion within
modules. Space inefficiency has shaped the way their contract system
has developed. They do not use our ``micro'' approach, wherein
annotations and casts permeate the code.

Tail recursion aside, there is another important source of space
inefficiency: the unbounded number of function proxies. Hierarchies of
libraries are a typical example: consider a list library and a set
library built using increasingly sorted lists. We might have:
\[ \begin{array}{r@{~:~}l@{~=~}l}
   \mathsf{null}     &    \alpha  ~\mathsf{List}  \mathord{ \rightarrow }  \{ \mathit{x} \mathord{:}  \mathsf{Bool}  \mathrel{\mid}  \mathsf{true}  \}   & ... \\
   \mathsf{head}     &   \{ \mathit{x} \mathord{:}   \alpha  ~\mathsf{List}  \mathrel{\mid}   \mathsf{not}  ~  (   \mathsf{null}  ~ \mathit{x}  )   \}  \mathord{ \rightarrow }  \alpha   & ... \\[6pt]
   \mathsf{empty}  &    \alpha  ~\mathsf{Set}  \mathord{ \rightarrow }  \{ \mathit{x} \mathord{:}  \mathsf{Bool}  \mathrel{\mid}  \mathsf{true}  \}   &  \mathsf{null}  \\
   \mathsf{min}  &   \{ \mathit{x} \mathord{:}   \alpha  ~\mathsf{Set}  \mathrel{\mid}   \mathsf{not}  ~  (   \mathsf{empty}  ~ \mathit{x}  )   \}  \mathord{ \rightarrow }  \alpha   &  \mathsf{head} 
\end{array} \]
Our code reuse comes with a price: even though the precondition on
$ \mathsf{min} $ is effectively the same as that on $ \mathsf{head} $, we must
have two function proxies, and the non-emptiness of the list
representing the set is checked twice: first by checking
$ \mathsf{empty} $, and again by checking $ \mathsf{null} $ (which is the same
function).
Blame systems like those in Racket encourage modules to declare
contracts to avoid being blamed, which can result in redundant
checking like the above when libraries requirements imply
sub-libraries' requirements.

Or consider a library of drawing primitives based
around painters, functions of type $  \mathsf{Canvas}  \mathord{ \rightarrow }  \mathsf{Canvas}  $.
An underlying graphics library offers basic functions for manipulating
canvases and functions over canvases, e.g., $ \mathsf{primFlipH} $ is a
painter transformer---of type
$  (   \mathsf{Canvas}  \mathord{ \rightarrow }  \mathsf{Canvas}   )  \mathord{ \rightarrow }  (   \mathsf{Canvas}  \mathord{ \rightarrow }  \mathsf{Canvas}   )  $---that flips the generated
images horizontally.
A wrapper library may add derived functions while re-exporting the
underlying functions with refinement types specifying a canvas's square
dimensions, where $ \mathsf{SquareCanvas}   \ottsym{=}   \{ \mathit{x} \mathord{:}  \mathsf{Canvas}  \mathrel{\mid}   \mathsf{width}   \ottsym{(}  \mathit{x}  \ottsym{)}  \mathrel{=}   \mathsf{height}   \ottsym{(}  \mathit{x}  \ottsym{)}  \} $:
\[ \begin{array}{@{}r@{~}l}
    \mathsf{flipH}  ~ \mathit{p}  =&  \langle    \mathsf{Canvas}  \mathord{ \rightarrow }  \mathsf{Canvas}    \mathord{ \overset{    }{\Rightarrow} }   {} \\  &  \phantom{\langle}   \mathsf{SquareCanvas}  \mathord{ \rightarrow }  \mathsf{SquareCanvas}    \rangle^{ \ottnt{l} } ~  {} \\  &  \quad   (   \mathsf{primFlipH}  ~ {} \\  &  \qquad   (  \langle    \mathsf{SquareCanvas}  \mathord{ \rightarrow }  \mathsf{SquareCanvas}    \mathord{ \overset{    }{\Rightarrow} }   {} \\  &  \qquad  \phantom{(}  \phantom{\langle}   \mathsf{Canvas}  \mathord{ \rightarrow }  \mathsf{Canvas}    \rangle^{ \ottnt{l} } ~  \mathit{p}  )   )  
\end{array}
\]
The wrapper library only accepts painters with appropriately refined
types, but must strip away these refinements before calling the
underlying implementation---which demands $  \mathsf{Canvas}  \mathord{ \rightarrow }  \mathsf{Canvas}  $
painters. The wrapper library then has to cast these modified
functions \textit{back} to the refined types. Calling $  \mathsf{flipH}  ~  (   \mathsf{flipH}  ~ \mathit{p}  )  $ yields:
\[\begin{array}{@{}l@{}}
   \langle    \mathsf{Canvas}  \mathord{ \rightarrow }  \mathsf{Canvas}    \mathord{ \overset{    }{\Rightarrow} }    \mathsf{SquareCanvas}  \mathord{ \rightarrow }  \mathsf{SquareCanvas}    \rangle^{ \ottnt{l} } ~  {} \\  \quad   (   \mathsf{primFlipH}  ~ {} \\  \qquad   (  \langle    \mathsf{SquareCanvas}  \mathord{ \rightarrow }  \mathsf{SquareCanvas}    \mathord{ \overset{    }{\Rightarrow} }    \mathsf{Canvas}  \mathord{ \rightarrow }  \mathsf{Canvas}    \rangle^{ \ottnt{l} } ~  {} \\  \qquad  \quad   (  \langle    \mathsf{Canvas}  \mathord{ \rightarrow }  \mathsf{Canvas}    \mathord{ \overset{    }{\Rightarrow} }    \mathsf{SquareCanvas}  \mathord{ \rightarrow }  \mathsf{SquareCanvas}    \rangle^{ \ottnt{l} } ~  {} \\  \qquad  \qquad   (   \mathsf{primFlipH}  ~ {} \\  \qquad  \qquad  \quad   (  \langle    \mathsf{SquareCanvas}  \mathord{ \rightarrow }  \mathsf{SquareCanvas}    \mathord{ \overset{    }{\Rightarrow} }   {} \\  \qquad  \qquad  \quad  \phantom{(}  \phantom{\langle}   \mathsf{Canvas}  \mathord{ \rightarrow }  \mathsf{Canvas}    \rangle^{ \ottnt{l} } ~  \mathit{p}  )   )   )   )   )  
\end{array}\]
That is, we first cast $\mathit{p}$ to a plain painter and return a new
painter $\mathit{p'}$. We then cast $\mathit{p'}$ into and then immediately out
of the refined type, before continuing on to flip $\mathit{p'}$.
All the while, we are accumulating many function proxies beyond the
wrapping done by the underlying implementation of $ \mathsf{primFlipH} $.
Redundant wrapping can become quite extreme, especially for
continuation-passing programs.
Function proxies are the essential problem: nothing bounds their
accumulation. Unfolding unboundedly many function proxies creates
stacks of unboundedly many checks---which breaks tail calls.
{\iffull If we can have a constant number of function proxies that
  produce stacks of checks of a constant size, then we can have tail
  call optimization. \fi}
A space-efficient scheme for manifest contracts bounds the number of
function proxies that can accumulate.

\finishlater{game objects, re-exported with partial applications}

{\iffull

  We can also adapt Herman et al.'s mutually recursive even and odd
  functions, writing them with type ascriptions in the following
  apparently tail recursive program consuming constant space:
\[ \begin{array}{@{}r@{~}l} 
   \mathsf{odd}  = &   \lambda \mathit{x} \mathord{:}  \{ \mathit{x} \mathord{:}  \mathsf{Int}  \mathrel{\mid}  \mathsf{true}  \}  .~   \mathsf{if}~  (  \mathit{x}  \mathrel{=}  \ottsym{0}  )   \\ & \mathsf{then}~   \mathsf{false}   :   \{ \mathit{b} \mathord{:}  \mathsf{Bool}  \mathrel{\mid}  \mathit{b}  \mathrel{\vee}   (  \mathit{x} \,  \mathsf{mod}  \, \ottsym{2}  \mathrel{=}  \ottsym{0}  )   \}    \\ & \mathsf{else}~  \mathsf{even}    ~  ( \mathit{x} \,  -  \, \ottsym{1} )   \\
   \mathsf{even}  : &   \{ \mathit{x} \mathord{:}  \mathsf{Int}  \mathrel{\mid}  \mathsf{true}  \}  \mathord{ \rightarrow }  \{ \mathit{b} \mathord{:}  \mathsf{Bool}  \mathrel{\mid}  \mathit{b}  \mathrel{\vee}   (  \mathit{x} \,  -  \, \ottsym{1} \,  \mathsf{mod}  \, \ottsym{2}  \mathrel{=}  \ottsym{0}  )   \}   = {} \\
  \multicolumn{2}{l}{\quad   \lambda \mathit{x} \mathord{:}  \{ \mathit{x} \mathord{:}  \mathsf{Int}  \mathrel{\mid}  \mathsf{true}  \}  .~   \mathsf{if}~  (  \mathit{x}  \mathrel{=}  \ottsym{0}  )  ~\mathsf{then}~  \mathsf{true}  ~\mathsf{else}~  \mathsf{odd}    ~  ( \mathit{x} \,  -  \, \ottsym{1} )  }
\end{array} \]
A cast insertion algorithm~\cite{Swamy09coercions} might produce:
\[ \begin{array}{@{}r@{~}l} 
   \mathsf{odd}  = &   \lambda \mathit{x} \mathord{:}  \{ \mathit{x} \mathord{:}  \mathsf{Int}  \mathrel{\mid}  \mathsf{true}  \}  .~   \mathsf{if}~  (  \mathit{x}  \mathrel{=}  \ottsym{0}  )   \\ & \mathsf{then}~  \langle   \{ \mathit{b} \mathord{:}  \mathsf{Bool}  \mathrel{\mid}  \mathsf{true}  \}   \mathord{ \overset{    }{\Rightarrow} }  {} \\  &  \phantom{\mathsf{then}~}  \phantom{\langle}   \{ \mathit{b} \mathord{:}  \mathsf{Bool}  \mathrel{\mid}  \mathit{b}  \mathrel{\vee}   (  \mathit{x} \,  \mathsf{mod}  \, \ottsym{2}  \mathrel{=}  \ottsym{0}  )   \}   \rangle^{ \ottnt{l_{{\mathrm{1}}}} } ~   \mathsf{false}    \\ & \mathsf{else}~  \mathsf{even}    ~  ( \mathit{x} \,  -  \, \ottsym{1} )   \\
   \mathsf{even}  = &  \langle    \{ \mathit{x} \mathord{:}  \mathsf{Int}  \mathrel{\mid}  \mathsf{true}  \}  \mathord{ \rightarrow }  \{ \mathit{b} \mathord{:}  \mathsf{Bool}  \mathrel{\mid}  \mathsf{true}  \}    \mathord{ \overset{    }{\Rightarrow} }   {} \\  &  \phantom{\langle}   \{ \mathit{x} \mathord{:}  \mathsf{Int}  \mathrel{\mid}  \mathsf{true}  \}  \mathord{ \rightarrow } {} \\  &  \phantom{\langle}   \{ \mathit{b} \mathord{:}  \mathsf{Bool}  \mathrel{\mid}  \mathit{b}  \mathrel{\vee}   (   (  \mathit{x}  \mathrel{+}  \ottsym{1}  )  \,  \mathsf{mod}  \, \ottsym{2}  \mathrel{=}  \ottsym{0}  )   \}    \rangle^{ \ottnt{l_{{\mathrm{2}}}} } ~  {} \\  &   (  \lambda \mathit{x} \mathord{:}  \{ \mathit{x} \mathord{:}  \mathsf{Int}  \mathrel{\mid}  \mathsf{true}  \}  .~   \mathsf{if}~  (  \mathit{x}  \mathrel{=}  \ottsym{0}  )   \\ & \mathsf{then}~  \mathsf{true}   \\ & \mathsf{else}~  \langle   \{ \mathit{b} \mathord{:}  \mathsf{Bool}  \mathrel{\mid}  \mathit{b}  \mathrel{\vee}   (  \mathit{x} \,  -  \, \ottsym{1} \,  \mathsf{mod}  \, \ottsym{2}  \mathrel{=}  \ottsym{0}  )   \}   \mathord{ \overset{    }{\Rightarrow} }  {} \\  &  \phantom{\mathsf{else}~}  \phantom{\langle}   \{ \mathit{b} \mathord{:}  \mathsf{Bool}  \mathrel{\mid}  \mathsf{true}  \}   \rangle^{ \ottnt{l_{{\mathrm{3}}}} } ~   (   \mathsf{odd}  ~  ( \mathit{x} \,  -  \, \ottsym{1} )   )     )  
\end{array} \]
Note that the casts labeled $\ottnt{l_{{\mathrm{2}}}}$ and $\ottnt{l_{{\mathrm{3}}}}$ both break tail
recursion: the former leads to unwrapping whenever $ \mathsf{even} $ is
called, while the latter moves the call to $ \mathsf{odd} $ from tail
position.
Breaking tail-call optimization is very bad: we went from constant
space to \textit{linear} space. That is, inserting contract checks can
change a program's asymptotic space efficiency!

\fi}

\section{Classic manifest contracts}
\label{sec:classic}

\begin{figure}[t]
  \[\begin{array}{r@{~~~}c@{~~~}l}
    \multicolumn{3}{l}{\textbf{Modes}} \\
    \ottnt{m} &::=&  \mathsf{C}  \qquad\text{\underline{c}lassic \lambdah; Section~\ref{sec:classic}} \\ 
\iffull      &\BNFALT&  \mathsf{F}  \qquad\text{\underline{f}orgetful \lambdah; Section~\ref{sec:forgetful}} \\
      &\BNFALT&  \mathsf{H}  \qquad\text{\underline{h}eedful \lambdah; Section~\ref{sec:heedful}} \\ \fi
      &\BNFALT&  \mathsf{E}  \qquad\text{\underline{e}idetic \lambdah; Section~\ref{sec:eidetic}} \\
    && \\
    \multicolumn{3}{l}{\textbf{Types}} \\
    \ottnt{B} &::=&  \mathsf{Bool}  \BNFALT  \dots  \\
    \ottnt{T} &::=&  \{ \mathit{x} \mathord{:} \ottnt{B} \mathrel{\mid} \ottnt{e} \}  \BNFALT  \ottnt{T_{{\mathrm{1}}}} \mathord{ \rightarrow } \ottnt{T_{{\mathrm{2}}}}  \\

    \multicolumn{3}{l}{\textbf{Terms}} \\
    \ottnt{e} &::=& \mathit{x} \BNFALT \ottnt{k} \BNFALT  \lambda \mathit{x} \mathord{:} \ottnt{T} .~  \ottnt{e}  \BNFALT 
        \ottnt{e_{{\mathrm{1}}}} ~ \ottnt{e_{{\mathrm{2}}}}  \BNFALT \ottnt{op}  \ottsym{(}  \ottnt{e_{{\mathrm{1}}}}  \ottsym{,}  \dots  \ottsym{,}  \ottnt{e_{\ottmv{n}}}  \ottsym{)} \BNFALT \\
    &&  \highlight{  \langle  \ottnt{T_{{\mathrm{1}}}}  \mathord{ \overset{ \ottnt{a} }{\Rightarrow} }  \ottnt{T_{{\mathrm{2}}}}  \rangle^{ \ottnt{l} } ~  \ottnt{e}  }  \BNFALT  \highlight{  \langle   \{ \mathit{x} \mathord{:} \ottnt{B} \mathrel{\mid} \ottnt{e_{{\mathrm{1}}}} \}  ,  \ottnt{e_{{\mathrm{2}}}} ,  \ottnt{k}  \rangle^{ \ottnt{l} }  }  \BNFALT  \highlight{  \mathord{\Uparrow}  \ottnt{l}  }  \BNFALT \\
    &&  \highlight{  \langle   \{ \mathit{x} \mathord{:} \ottnt{B} \mathrel{\mid} \ottnt{e_{{\mathrm{1}}}} \}  ,  \ottnt{s} ,  \ottnt{r} ,  \ottnt{k} ,  \ottnt{e}  \rangle^{\bullet}  }  \\
    \multicolumn{3}{l}{\textbf{Annotations: type set, coercions, and refinement lists}} \\
    \ottnt{a} &::=&  \bullet  \iffull \BNFALT \mathcal{S} \fi \BNFALT \ottnt{c} \\
\iffull    \mathcal{S} &::=&  \emptyset  \BNFALT \set{\ottnt{T_{{\mathrm{1}}}},...,\ottnt{T_{\ottmv{n}}}} \\ \fi
    \ottnt{c} &::=& \ottnt{r} \BNFALT \ottnt{c_{{\mathrm{1}}}}  \mapsto  \ottnt{c_{{\mathrm{2}}}} \\
    \ottnt{r} &::=&  \mathsf{nil}  \BNFALT   \{ \mathit{x} \mathord{:} \ottnt{B} \mathrel{\mid} \ottnt{e} \}^{ \ottnt{l} }  , \ottnt{r}  \\
    \multicolumn{3}{l}{\textbf{Statuses}} \\
    \ottnt{s} &::=&  \mathord{\checkmark}  \BNFALT  \mathord{?}  \\
    \multicolumn{3}{l}{\textbf{Locations}} \\
    \ottnt{l} &::=&  \bullet  \BNFALT \ottnt{l_{{\mathrm{1}}}} \BNFALT ...
  \end{array}\]
%  \vspace*{-10pt}
  \caption{Syntax of \lambdah}
  \label{fig:syntax}
\end{figure}

The standard manifest contract calculus, \lambdah, is originally due
to Flanagan~\cite{Flanagan06hybrid}. We give the syntax for the
non-dependent fragment in Figure~\ref{fig:syntax}. We have highlighted
in $\highlight{\text{yellow}}$ the four syntactic forms relevant to
contract checking.
This paper paper discusses \ifpopl two \else four \fi modes of
\lambdah: classic \lambdah, mode $ \mathsf{C} $\ifpopl,\else; forgetful
\lambdah, mode $ \mathsf{F} $; heedful \lambdah, mode $ \mathsf{H} $;\fi and eidetic
\lambdah, mode $ \mathsf{E} $. Each of these languages uses the syntax of
Figure~\ref{fig:syntax}, while the typing rules and operational
semantics are indexed by the mode $\ottnt{m}$. The proofs and metatheory
are also mode-indexed.
In an extended version of this work, we develop two additional modes
with slightly different properties from eidetic \lambdah, filling out
a ``framework'' for space-efficient manifest
contracts~\cite{Greenberg14spacetr}. We omit the other two modes here
to save space for eidetic \lambdah, which is the only mode that is
sound with respect to classic \lambdah.
{\iffull We summarize how each of these modes differ in
  Section~\ref{sec:example}---but first we (laconically) explain the
  syntax and the interesting bits of the operational semantics.\fi}

The metavariable $\ottnt{B}$ is used for base types, of which at least
$ \mathsf{Bool} $ must be present. There are two kinds of types. First,
\textit{predicate contracts} $ \{ \mathit{x} \mathord{:} \ottnt{B} \mathrel{\mid} \ottnt{e} \} $, also called
\textit{refinements of base types} or just \textit{refinement types},
denotes constants $\ottnt{k}$ of base type $\ottnt{B}$ such that $ \ottnt{e}  [  \ottnt{k} / \mathit{x}  ] $
holds---that is, such that $ \ottnt{e}  [  \ottnt{k} / \mathit{x}  ]  \,  \longrightarrow ^{*}_{ \ottnt{m} }  \,  \mathsf{true} $ for any mode
$\ottnt{m}$. Function types $ \ottnt{T_{{\mathrm{1}}}} \mathord{ \rightarrow } \ottnt{T_{{\mathrm{2}}}} $ are standard.

The terms of \lambdah are largely those of the simply-typed lambda
calculus: variables, constants $\ottnt{k}$, abstractions, applications,
and operations should all be familiar. 
The first distinguishing feature of \lambdah's terms is the
\textit{cast}, written $ \langle  \ottnt{T_{{\mathrm{1}}}}  \mathord{ \overset{ \ottnt{a} }{\Rightarrow} }  \ottnt{T_{{\mathrm{2}}}}  \rangle^{ \ottnt{l} } ~  \ottnt{e} $. Here $\ottnt{e}$ is a term of
type $\ottnt{T_{{\mathrm{1}}}}$; the cast checks whether $\ottnt{e}$ can be treated as a
$\ottnt{T_{{\mathrm{2}}}}$---if $\ottnt{e}$ doesn't cut it, the cast will use its label
$\ottnt{l}$ to raise the uncatchable exception $ \mathord{\Uparrow}  \ottnt{l} $, read
``blame $\ottnt{l}$''.
Our casts also have annotations $\ottnt{a}$. Classic \ifpopl doesn't \else
and forgetful \lambdah don't \fi need annotations---we write
$ \bullet $ and say ``none''. \iffull Heedful \lambdah uses type sets $\mathcal{S}$ to
track space-efficiently pending checks.\fi Eidetic \lambdah uses
coercions $\ottnt{c}$, based on coercions in \citet{Henglein94dynamic}. We
explain coercions in greater detail in Section~\ref{sec:eidetic}, but
they amount to lists of blame-annotated refinement types $\ottnt{r}$ and
function coercions.

The three remaining forms---active checks, blame, and coercion
stacks---only occur as the program evaluates.
Casts between refinement types are checked by \textit{active checks}
$ \langle   \{ \mathit{x} \mathord{:} \ottnt{B} \mathrel{\mid} \ottnt{e_{{\mathrm{1}}}} \}  ,  \ottnt{e_{{\mathrm{2}}}} ,  \ottnt{k}  \rangle^{ \ottnt{l} } $. The first term is the type being
checked---necessary for the typing rule. The second term is the
current status of the check; it is an invariant that $ \ottnt{e_{{\mathrm{1}}}}  [  \ottnt{k} / \mathit{x}  ]  \,  \longrightarrow ^{*}_{ \ottnt{m} }  \, \ottnt{e_{{\mathrm{2}}}}$. The final term is the constant being checked, which is
returned wholesale if the check succeeds.
When checks fail, the program raises \textit{blame}, an uncatchable
exception written $ \mathord{\Uparrow}  \ottnt{l} $.
A \textit{coercion stack} $ \langle   \{ \mathit{x} \mathord{:} \ottnt{B} \mathrel{\mid} \ottnt{e_{{\mathrm{1}}}} \}  ,  \ottnt{s} ,  \ottnt{r} ,  \ottnt{k} ,  \ottnt{e}  \rangle^{\bullet} $ represents the
state of checking a coercion; we only use it in eidetic \lambdah, so
we postpone discussing it until Section~\ref{sec:eidetic}.

\subsection{Core operational semantics}
\label{sec:opsem}

\colorlet{salmon}{orange!15}
\colorlet{periwinkle}{blue!35}

\begin{bigfigure}
  \hdr{Values and results}{\qquad \fbox{$ \mathsf{val} _{ \ottnt{m} }~ \ottnt{e} $} \qquad \fbox{$ \mathsf{result} _{ \ottnt{m} }~ \ottnt{e} $}}

  \iffull
  \threesidebyside[.25][.25][.4]
    {\ottusedrule{\ottdruleVXXConst{}}}
    {\ottusedrule{\ottdruleVXXAbs{}}}
    {\ottusedrule{\highlight[salmon]{\ottdruleVXXProxyC{}}}}
  \sidebyside
      {\ottusedrule{\ottdruleRXXVal{}}}
      {\ottusedrule{\ottdruleRXXBlame{}}}
  \else
  \threesidebysidesqueeze[.15][.17][.33][.01]
    {\ottusedrule{\ottdruleVXXConst{}}}
    {\ottusedrule{\ottdruleVXXAbs{}}}
    {\ottusedrule{\highlight[salmon]{\ottdruleVXXProxyC{}}}}
  {\sidebysidesqueeze[.15][.16][.01]
      {\ottusedrule{\ottdruleRXXVal{}}}
      {\ottusedrule{\ottdruleRXXBlame{}}}}
  \fi

  \hdr{Shared operational semantics}{\qquad \fbox{$\ottnt{e_{{\mathrm{1}}}} \,  \longrightarrow _{ \ottnt{m} }  \, \ottnt{e_{{\mathrm{2}}}}$}}

  \sidebyside
    {\ottusedrule{\ottdruleEXXBeta{}}}
    {\ottusedrule{\ottdruleEXXOp{}}}
  {\ottusedrule{\ottdruleEXXUnwrap{}}}
  \sidebyside
    {\[\begin{array}{rcl}
         \mathsf{dom} ( \bullet )  &=&  \bullet  \\
\iffull         \mathsf{dom} ( \mathcal{S} )  &=& \bigcup_{ \ottnt{T}  \in  \mathcal{S} }  \mathsf{dom} ( \ottnt{T} )  \\ \fi
         \mathsf{dom} ( \ottnt{c_{{\mathrm{1}}}}  \mapsto  \ottnt{c_{{\mathrm{2}}}} )  &=& \ottnt{c_{{\mathrm{1}}}}
      \end{array} \]}
    {\[\begin{array}{rcl}
         \mathsf{cod} ( \bullet )  &=&  \bullet  \\
\iffull         \mathsf{cod} ( \mathcal{S} )  &=& \bigcup_{ \ottnt{T}  \in  \mathcal{S} }  \mathsf{cod} ( \ottnt{T} )  \\ \fi
         \mathsf{cod} ( \ottnt{c_{{\mathrm{1}}}}  \mapsto  \ottnt{c_{{\mathrm{2}}}} )  &=& \ottnt{c_{{\mathrm{2}}}}
      \end{array} \]}
  {\ottusedrule{\highlight[salmon]{\ottdruleEXXCheckNoneC{}}}}
  \sidebyside
    {\ottusedrule{\ottdruleEXXCheckOK{}}}
    {\ottusedrule{\ottdruleEXXCheckFail{}}}
  \sidebyside
    {\ottusedrule{\ottdruleEXXAppL{}}}
    {\ottusedrule{\ottdruleEXXAppR{}}}
  {\ottusedrule{\ottdruleEXXOpInner{}}}
  \sidebyside[.45][.5]
    {\ottusedrule{\highlight[salmon]{\ottdruleEXXCastInnerC{}}}}
    {\ottusedrule{\ottdruleEXXCheckInner{}}}
  \iffull
    {\ottusedrule{\highlight[periwinkle]{\ottdruleEXXCastInner{}}}}
    {\ottusedrule{\highlight[periwinkle]{\ottdruleEXXCastMerge{}}}}  
  \else\sidebyside[.45][.48]
    {\ottusedrule{\highlight[periwinkle]{\ottdruleEXXCastInnerE{}}}}
    {\ottusedrule{\highlight[periwinkle]{\ottdruleEXXCastMergeE{}}}}
  \fi
  \threesidebyside[.28][.28][.36]
    {\ottusedrule{\ottdruleEXXAppRaiseL{}}}
    {\ottusedrule{\ottdruleEXXAppRaiseR{}}}
    {\ottusedrule{\ottdruleEXXCastRaise{}}}
  \sidebyside
    {\ottusedrule{\ottdruleEXXOpRaise{}}}
    {\ottusedrule{\ottdruleEXXCheckRaise{}}}
% 
%  \[ \highlight{ \mathsf{merge} _{ \ottnt{m} }( \ottnt{T_{{\mathrm{1}}}} , \ottnt{a_{{\mathrm{1}}}} , \ottnt{T_{{\mathrm{2}}}} , \ottnt{a_{{\mathrm{2}}}} , \ottnt{T_{{\mathrm{3}}}} )  : \ottnt{a_{{\mathrm{3}}}}} \]
%
    \vspace*{-1em}
    \caption{Core operational semantics of \lambdah; classic \lambdah
      rules are $\highlight[salmon]{\text{salmon}}$; space-efficient rules are
      $\highlight[periwinkle]{\text{periwinkle}}$}
  \label{fig:opsem}
\end{bigfigure}

\newcommand{\ECheckNone}{\ifpopl\E{CheckNoneC}\else\E{CheckNone}\fi}
\newcommand{\ECastInner}{\ifpopl\E{CastInnerE}\else\E{CastInner}\fi}
\newcommand{\ECastMerge}{\ifpopl\E{CastMergeE}\else\E{CastMerge}\fi}

Our mode-indexed operational semantics for our manifest calculi
comprise three relations: $ \mathsf{val} _{ \ottnt{m} }~ \ottnt{e} $ identifies terms that are
values in mode $\ottnt{m}$ (or $\ottnt{m}$-values), $ \mathsf{result} _{ \ottnt{m} }~ \ottnt{e} $
identifies $\ottnt{m}$-results, and $\ottnt{e_{{\mathrm{1}}}} \,  \longrightarrow _{ \ottnt{m} }  \, \ottnt{e_{{\mathrm{2}}}}$ is the small-step
reduction relation for mode $\ottnt{m}$.
{\iffull
It is more conventional to fix values as a syntactic subset, but that
approach would be confusing here: we would need three different
metavariables to unambiguously refer to values from each language. The
mode-indexed value and result relations neatly avoid any potential
confusion between metavariables.
Each mode defines its own value rule for function
proxies. 
\fi}
Figure~\ref{fig:opsem} defines the core rules. The
rules for classic \lambdah ($\ottnt{m}  \ottsym{=}   \mathsf{C} $) are in
$\highlight[salmon]{\text{salmon}}$; the shared space-efficient
rules are in $\highlight[periwinkle]{\text{periwinkle}}$.
To save space, we pass over standard rules.

The mode-agnostic value rules are straightforward: constants are
always values (\V{Const}), as are lambdas (\V{Abs}). Each mode defines
its own value rule for function proxies, \V{Proxy$\ottnt{m}$}. The classic
rule, \V{ProxyC}, says that a function proxy  \[  \langle   \ottnt{T_{{\mathrm{11}}}} \mathord{ \rightarrow } \ottnt{T_{{\mathrm{12}}}}   \mathord{ \overset{ \bullet }{\Rightarrow} }   \ottnt{T_{{\mathrm{21}}}} \mathord{ \rightarrow } \ottnt{T_{{\mathrm{22}}}}   \rangle^{ \ottnt{l} } ~  \ottnt{e}  \] is a $ \mathsf{C} $-value when $\ottnt{e}$ is a
$ \mathsf{C} $-value. That is, function proxies can wrap lambda abstractions
and other function proxies alike.
\ifpopl Eidetic \lambdah only allows \else Other modes only allow \fi
lambda abstractions to be proxied\iffull while requiring that the
annotations are appropriate\fi.
All of the space-efficient calculi in the literature take our
approach, where a function cast applied to a value \textit{is} a
value; some space inefficient ones do,
too~\cite{Findler02contracts,Gronski07unifying,Greenberg12contracts}.
In other formulations of \lambdah in the literature, function proxies
are implemented by introducing a new lambda as a wrapper \`a la
Findler and Felleisen's $\overline{wrap}$
operator~\cite{Findler02contracts,Flanagan06hybrid,Siek06gradual,Belo11fh}. Such
an $\eta$-expansion semantics is convenient, since then applications
only ever reduce by $\beta$-reduction. But it wouldn't suit our
purposes at all: space efficiency demands that we combine function
proxies.
We can also imagine a third, ungainly semantics that looks into closures rather
than having explicit function proxies.
Results don't depend on the mode: $\ottnt{m}$-values are
always $\ottnt{m}$-results (\R{Val}); blame is always a result, too
(\R{Blame}).

\E{Beta} applies lambda abstractions via substitution, using a
call-by-value rule. Note that $\beta$ reduction in mode $\ottnt{m}$
requires that the argument is an $\ottnt{m}$-value.
The reduction rule for operations (\E{Op}) defers to operations'
denotations, $ \denot{ op } $; since these may be partial (e.g.,
division), we assign types to operations that guarantee totality (see
Section~\ref{sec:typing}).
That is, partial operations are a potential source of stuckness, and
the types assigned to operations must guarantee the absence of
stuckness.
Robin Milner famously stated that ``well typed expressions don't go
wrong''~\cite{Milner78atheory}; his programs could go wrong by (a)
applying a boolean like a function or (b) conditioning on a function
like a boolean. Systems with more base types can go wrong in more
ways, some of which are hard to capture in standard type
systems. Contracts allow us to bridge that gap.
Letting operations get stuck is a philosophical stance---contracts
expand the notion of ``wrong''\iffull---that supports our forgetful semantics
(Section~\ref{sec:forgetful})\fi. \finishlater{contributions?}

\E{Unwrap} applies function proxies to values, contravariantly in the
domain and covariantly in the codomain. We also split up each cast's
annotation, using $ \mathsf{dom} ( \ottnt{a} ) $ and $ \mathsf{cod} ( \ottnt{a} ) $\ifpopl---each mode is
discussed in its respective section\iffull (type sets in
Section~\ref{sec:heedful} and coercions in
Section~\ref{sec:eidetic})\fi\fi.
\ECheckNone\ turns a cast between refinement types into
an active check with the same blame label. We discard the source
type---we already know that $\ottnt{k}$ is a $ \{ \mathit{x} \mathord{:} \ottnt{B} \mathrel{\mid} \ottnt{e_{{\mathrm{1}}}} \} $---and
substitute the scrutinee into the target type, $ \ottnt{e_{{\mathrm{2}}}}  [  \ottnt{k} / \mathit{x}  ] $, as the
current state of checking. We must also hold onto the scrutinee, in
case the check succeeds.
We are careful to not apply this rule in \ifpopl eidietc \lambdah\else
heedful and eidetic modes\fi, which must generate annotations before
running checks\iffull; we discuss these more in those modes'
sections\fi.
Active checks evaluate by the congruence rule \E{CheckInner} until one
of three results adheres: the predicate returns $ \mathsf{true} $, so the
whole active check returns the scrutinee (\E{CheckOK}); the predicate
returns $ \mathsf{false} $, so the whole active check raises blame using the
label on the chceck (\E{CheckFail}); or blame was raised during
checking, and we propagate it via \E{CheckRaise}.  
\ifpopl Checks in eidetic \lambdah \else heedful and eidetic \fi use
slightly different forms, described in \ifpopl
Section~\ref{sec:eidetic}\else their respective sections\fi.

The core semantics includes several other congruence rules: \E{AppL},
\E{AppR}, and \E{OpInner}.
Since space bounds rely not only on limiting the number of function
proxies but also on accumulation of casts on the stack, the core
semantics doesn't include a cast congruence rule. The congruence rule
for casts in classic \lambdah, \E{CastInnerC}, allows for free use of
congruence. In the space-efficient \ifpopl{}calculus\else{}calculi\fi, the use of congruence is
instead limited by the rules \ECastInner\ and \ECastMerge. Cast
arguments only take congruent steps when they aren't casts
themselves. A cast applied to another cast \textit{merges}, using the
\iffull $ \mathsf{merge} $ \else $ \mathsf{join} $ \fi function.
Each space-efficient calculus uses a different annotation scheme, so
each one has a different merge function. \iffull We deliberately leave
$ \mathsf{merge} $ undefined sometimes---heedful and eidetic \lambdah must
control when \E{CastMerge} can apply.
Note that we don't need to specify $\ottnt{m}  \neq   \mathsf{C} $ in \ECastMerge---we
just don't define a merge operator for classic \lambdah.  \else We are
careful to define $ \mathsf{join} $ only over coercions, so \ECastMerge\ won't
apply on the empty annotation, $ \bullet $ (read ``none''). \fi
We have \ECastMerge\ arbitrarily retain the label of the outer
cast. \ifpopl This choice is ultimately irrelevant, since eidetic
\lambdah won't need to keep track of blame labels on casts themselves
(Section~\ref{sec:eidetic}). \else No choice is ``right'' here---we
discuss this issue further in Section~\ref{sec:heedful}. \fi
In addition to congruence rules, there are blame propagation rules,
which are universal: \E{AppRaiseL}, \E{AppRaiseR}, \E{CastRaise},
\E{OpRaise}. These rules propagate the uncatchable exception
$ \mathord{\Uparrow}  \ottnt{l} $ while obeying call-by-value rules.

{\iffull
\subsection{Cast merges by example}
\label{sec:example}

\ifpopl We'll explain eidetic \lambdah's semantics in detail later\else
Each mode's section explains its semantics in detail\fi, but we can
summarize the cast merging rules here by example. Consider the
following term:
\[ \begin{array}{@{}l@{} }  \ottnt{e}  \ottsym{=}   \langle   \{ \mathit{x} \mathord{:}  \mathsf{Int}  \mathrel{\mid}  \mathit{x} \,  \mathsf{mod}  \, \ottsym{2}  \mathrel{=}  \ottsym{0}  \}   \mathord{ \overset{\bullet}{\Rightarrow} }   \{ \mathit{x} \mathord{:}  \mathsf{Int}  \mathrel{\mid}  \mathit{x}  \mathrel{\ne}  \ottsym{0}  \}   \rangle^{ \ottnt{l_{{\mathrm{3}}}} } ~  {} \\  \qquad  \quad   (  \langle   \{ \mathit{x} \mathord{:}  \mathsf{Int}  \mathrel{\mid}  \mathit{x}  \mathrel{\ge}  \ottsym{0}  \}   \mathord{ \overset{\bullet}{\Rightarrow} }   \{ \mathit{x} \mathord{:}  \mathsf{Int}  \mathrel{\mid}  \mathit{x} \,  \mathsf{mod}  \, \ottsym{2}  \mathrel{=}  \ottsym{0}  \}   \rangle^{ \ottnt{l_{{\mathrm{2}}}} } ~  {} \\  \qquad  \qquad   (  \langle   \{ \mathit{x} \mathord{:}  \mathsf{Int}  \mathrel{\mid}  \mathsf{true}  \}   \mathord{ \overset{\bullet}{\Rightarrow} }   \{ \mathit{x} \mathord{:}  \mathsf{Int}  \mathrel{\mid}  \mathit{x}  \mathrel{\ge}  \ottsym{0}  \}   \rangle^{ \ottnt{l_{{\mathrm{1}}}} } ~   {-1}   )   )    \end{array} \]
Here $\ottnt{e}$ runs three checks on integer $ {-1} $: first for
non-negativity (blaming $\ottnt{l_{{\mathrm{1}}}}$ on failure), then for evenness
(blaming $\ottnt{l_{{\mathrm{2}}}}$ on failure), and then for non-zeroness (blaming
$\ottnt{l_{{\mathrm{3}}}}$ on failure).
Classic and eidetic \lambdah both blame $\ottnt{l_{{\mathrm{1}}}}$; heedful \lambdah
also raises blame, though it blames a different label, $\ottnt{l_{{\mathrm{3}}}}$;
forgetful \lambdah actually \textit{accepts} the value, returning
$ {-1} $. We discuss the operational rules for modes other than
$ \mathsf{C} $ in detail in each mode's section; for now, we repeat the
derived rules for merging casts from Section~\ref{sec:intro}.

Classic \lambdah evaluates the casts step-by-step: first it checks
whether $ {-1} $ is positive, which fails, so $\ottnt{e} \,  \longrightarrow ^{*}_{  \mathsf{C}  }  \,  \mathord{\Uparrow}  \ottnt{l_{{\mathrm{1}}}} $ (see Figure~\ref{fig:classicexample}).
\begin{figure}
\[ \begin{array}{r@{~}l}
  \ottnt{e} = &  \langle   \{ \mathit{x} \mathord{:}  \mathsf{Int}  \mathrel{\mid}  \mathit{x} \,  \mathsf{mod}  \, \ottsym{2}  \mathrel{=}  \ottsym{0}  \}   \mathord{ \overset{\bullet}{\Rightarrow} }   \{ \mathit{x} \mathord{:}  \mathsf{Int}  \mathrel{\mid}  \mathit{x}  \mathrel{\ne}  \ottsym{0}  \}   \rangle^{ \ottnt{l_{{\mathrm{3}}}} } ~  {} \\  &  \quad   (  \langle   \{ \mathit{x} \mathord{:}  \mathsf{Int}  \mathrel{\mid}  \mathit{x}  \mathrel{\ge}  \ottsym{0}  \}   \mathord{ \overset{\bullet}{\Rightarrow} }   \{ \mathit{x} \mathord{:}  \mathsf{Int}  \mathrel{\mid}  \mathit{x} \,  \mathsf{mod}  \, \ottsym{2}  \mathrel{=}  \ottsym{0}  \}   \rangle^{ \ottnt{l_{{\mathrm{2}}}} } ~  {} \\  &  \qquad   (  \langle   \{ \mathit{x} \mathord{:}  \mathsf{Int}  \mathrel{\mid}  \mathsf{true}  \}   \mathord{ \overset{\bullet}{\Rightarrow} }   \{ \mathit{x} \mathord{:}  \mathsf{Int}  \mathrel{\mid}  \mathit{x}  \mathrel{\ge}  \ottsym{0}  \}   \rangle^{ \ottnt{l_{{\mathrm{1}}}} } ~   {-1}   )   )   \\
 \multicolumn{2}{r}{(\E{CastInnerC}/\ECheckNone)} \\
  \longrightarrow _{  \mathsf{C}  }  &  \langle   \{ \mathit{x} \mathord{:}  \mathsf{Int}  \mathrel{\mid}  \mathit{x}  \mathrel{\ge}  \ottsym{0}  \}   \mathord{ \overset{\bullet}{\Rightarrow} }   \{ \mathit{x} \mathord{:}  \mathsf{Int}  \mathrel{\mid}  \mathit{x}  \mathrel{\ne}  \ottsym{0}  \}   \rangle^{ \ottnt{l_{{\mathrm{3}}}} } ~  {} \\  &  \quad   (  \langle   \{ \mathit{x} \mathord{:}  \mathsf{Int}  \mathrel{\mid}  \mathit{x}  \mathrel{\ge}  \ottsym{0}  \}   \mathord{ \overset{\bullet}{\Rightarrow} }   \{ \mathit{x} \mathord{:}  \mathsf{Int}  \mathrel{\mid}  \mathit{x} \,  \mathsf{mod}  \, \ottsym{2}  \mathrel{=}  \ottsym{0}  \}   \rangle^{ \ottnt{l_{{\mathrm{2}}}} } ~  {} \\  &  \qquad   \langle   \{ \mathit{x} \mathord{:}  \mathsf{Int}  \mathrel{\mid}  \mathit{x}  \mathrel{\ge}  \ottsym{0}  \}  ,    {-1}   \mathrel{\ge}  \ottsym{0}  ,   {-1}   \rangle^{ \ottnt{l_{{\mathrm{1}}}} }   )   \\
 \multicolumn{2}{r}{(\E{CastInnerC}/\E{CheckInner}/\E{Op})} \\
  \longrightarrow _{  \mathsf{C}  }  &  \dots ~  \langle   \{ \mathit{x} \mathord{:}  \mathsf{Int}  \mathrel{\mid}  \mathit{x}  \mathrel{\ge}  \ottsym{0}  \}  ,   \mathsf{false}  ,   {-1}   \rangle^{ \ottnt{l_{{\mathrm{1}}}} }   \\
 \multicolumn{2}{r}{(\E{CastInnerC}/\E{CheckFail})} \\
  \longrightarrow _{  \mathsf{C}  }  &  \dots ~  \mathord{\Uparrow}  \ottnt{l_{{\mathrm{1}}}}   \qquad\qquad\qquad\qquad\qquad (\E{CastRaise}) \\ 
  \longrightarrow ^{*}_{  \mathsf{C}  }  &  \mathord{\Uparrow}  \ottnt{l_{{\mathrm{1}}}}  \\
\end{array} \]
\caption{An example of classic \lambdah}
\label{fig:classicexample}
\end{figure}
We first step by \ECheckNone, starting checking at the innermost
cast. Using congruence rules, we run \E{Op} to reduce the contract's
predicate, finding $ \mathsf{false} $. \E{CheckFail} then raises blame and
\E{CastRaise} propagates it.
Forgetful \lambdah doesn't use annotations at all---it just forgets
the intermediate casts, effectively using the following rule:
\[  \langle  \ottnt{T_{{\mathrm{2}}}}  \mathord{ \overset{ \bullet }{\Rightarrow} }  \ottnt{T_{{\mathrm{3}}}}  \rangle^{ \ottnt{l_{{\mathrm{2}}}} } ~   (  \langle  \ottnt{T_{{\mathrm{1}}}}  \mathord{ \overset{ \bullet }{\Rightarrow} }  \ottnt{T_{{\mathrm{2}}}}  \rangle^{ \ottnt{l_{{\mathrm{1}}}} } ~  \ottnt{e}  )   \,  \longrightarrow _{  \mathsf{F}  }  \,  \langle  \ottnt{T_{{\mathrm{1}}}}  \mathord{ \overset{ \bullet }{\Rightarrow} }  \ottnt{T_{{\mathrm{3}}}}  \rangle^{ \ottnt{l_{{\mathrm{2}}}} } ~  \ottnt{e}  \]
It never checks for non-negativity or evenness, skipping straight to
the check that $ {-1} $ is non-zero.
\[ \begin{array}{r@{~}l}
  \ottnt{e}  \longrightarrow _{  \mathsf{F}  }  &  \langle   \{ \mathit{x} \mathord{:}  \mathsf{Int}  \mathrel{\mid}  \mathit{x}  \mathrel{\ge}  \ottsym{0}  \}   \mathord{ \overset{\bullet}{\Rightarrow} }   \{ \mathit{x} \mathord{:}  \mathsf{Int}  \mathrel{\mid}  \mathit{x}  \mathrel{\ne}  \ottsym{0}  \}   \rangle^{ \ottnt{l_{{\mathrm{3}}}} } ~  {} \\  &  \quad   (  \langle   \{ \mathit{x} \mathord{:}  \mathsf{Int}  \mathrel{\mid}  \mathsf{true}  \}   \mathord{ \overset{\bullet}{\Rightarrow} }   \{ \mathit{x} \mathord{:}  \mathsf{Int}  \mathrel{\mid}  \mathit{x}  \mathrel{\ge}  \ottsym{0}  \}   \rangle^{ \ottnt{l_{{\mathrm{1}}}} } ~   {-1}   )   \\
         \longrightarrow _{  \mathsf{F}  }  &  \langle   \{ \mathit{x} \mathord{:}  \mathsf{Int}  \mathrel{\mid}  \mathsf{true}  \}   \mathord{ \overset{\bullet}{\Rightarrow} }   \{ \mathit{x} \mathord{:}  \mathsf{Int}  \mathrel{\mid}  \mathit{x}  \mathrel{\ne}  \ottsym{0}  \}   \rangle^{ \ottnt{l_{{\mathrm{3}}}} } ~   {-1}   \\
        \longrightarrow ^{*}_{  \mathsf{F}  }  &  {-1}      
\end{array} \]
Heedful \lambdah works by annotating casts with a set of intermediate
types, effectively using the rule:
\[  \langle  \ottnt{T_{{\mathrm{2}}}}  \mathord{ \overset{ \mathcal{S}_{{\mathrm{2}}} }{\Rightarrow} }  \ottnt{T_{{\mathrm{3}}}}  \rangle^{ \ottnt{l_{{\mathrm{2}}}} } ~   (  \langle  \ottnt{T_{{\mathrm{1}}}}  \mathord{ \overset{ \mathcal{S}_{{\mathrm{1}}} }{\Rightarrow} }  \ottnt{T_{{\mathrm{2}}}}  \rangle^{ \ottnt{l_{{\mathrm{1}}}} } ~  \ottnt{e}  )   \,  \longrightarrow _{  \mathsf{H}  }  \,  \langle  \ottnt{T_{{\mathrm{1}}}}  \mathord{ \overset{   \mathcal{S}_{{\mathrm{1}}}  \cup  \mathcal{S}_{{\mathrm{2}}}   \cup   \set{  \ottnt{T_{{\mathrm{2}}}}  }   }{\Rightarrow} }  \ottnt{T_{{\mathrm{3}}}}  \rangle^{ \ottnt{l_{{\mathrm{2}}}} } ~  \ottnt{e}  \]
Every type in a type set needs to be checked, but the order is
essentially nondeterministic: heedful \lambdah checks that $ {-1} $ is
non-negative and even in \textit{some} order. Whichever one is checked
first fails; both cases raise $ \mathord{\Uparrow}  \ottnt{l_{{\mathrm{3}}}} $.
\[ \begin{array}{r@{~}l}
  \ottnt{e}  \longrightarrow ^{*}_{  \mathsf{H}  }  &  \langle   \{ \mathit{x} \mathord{:}  \mathsf{Int}  \mathrel{\mid}  \mathit{x}  \mathrel{\ge}  \ottsym{0}  \}   \mathord{ \overset{  \set{   \{ \mathit{x} \mathord{:}  \mathsf{Int}  \mathrel{\mid}  \mathit{x} \,  \mathsf{mod}  \, \ottsym{2}  \mathrel{=}  \ottsym{0}  \}   }  }{\Rightarrow} }   \{ \mathit{x} \mathord{:}  \mathsf{Int}  \mathrel{\mid}  \mathit{x}  \mathrel{\ne}  \ottsym{0}  \}   \rangle^{ \ottnt{l_{{\mathrm{3}}}} } ~  {} \\  &  \quad   (  \langle   \{ \mathit{x} \mathord{:}  \mathsf{Int}  \mathrel{\mid}  \mathsf{true}  \}   \mathord{ \overset{\bullet}{\Rightarrow} }   \{ \mathit{x} \mathord{:}  \mathsf{Int}  \mathrel{\mid}  \mathit{x}  \mathrel{\ge}  \ottsym{0}  \}   \rangle^{ \ottnt{l_{{\mathrm{1}}}} } ~   {-1}   )   \\
         \longrightarrow ^{*}_{  \mathsf{H}  }  &  \langle   \{ \mathit{x} \mathord{:}  \mathsf{Int}  \mathrel{\mid}  \mathsf{true}  \}   \mathord{ \overset{ \mathcal{S} }{\Rightarrow} }   \{ \mathit{x} \mathord{:}  \mathsf{Int}  \mathrel{\mid}  \mathit{x}  \mathrel{\ne}  \ottsym{0}  \}   \rangle^{ \ottnt{l_{{\mathrm{3}}}} } ~   {-1}   \\
        \text{where} & \mathcal{S}  \ottsym{=}   \set{   \{ \mathit{x} \mathord{:}  \mathsf{Int}  \mathrel{\mid}  \mathit{x} \,  \mathsf{mod}  \, \ottsym{2}  \mathrel{=}  \ottsym{0}  \}  ,   \{ \mathit{x} \mathord{:}  \mathsf{Int}  \mathrel{\mid}  \mathit{x}  \mathrel{\ge}  \ottsym{0}  \}   } 
\end{array} \]

Finally, eidetic \lambdah uses coercions as its annotations; coercions
$\ottnt{c}$ are detailed checking plans for running checks in the same
order as classic \lambdah while skipping redundant checks.
As we will see in Section~\ref{sec:eidetic}, eidetic \lambdah
generates coercions and then drops blame labels, giving us the rule:
\[  \langle  \ottnt{T_{{\mathrm{2}}}}  \mathord{ \overset{ \ottnt{c_{{\mathrm{2}}}} }{\Rightarrow} }  \ottnt{T_{{\mathrm{3}}}}  \rangle^{ \bullet } ~   (  \langle  \ottnt{T_{{\mathrm{1}}}}  \mathord{ \overset{ \ottnt{c_{{\mathrm{1}}}} }{\Rightarrow} }  \ottnt{T_{{\mathrm{2}}}}  \rangle^{ \bullet } ~  \ottnt{e}  )   \,  \longrightarrow _{  \mathsf{E}  }  \,  \langle  \ottnt{T_{{\mathrm{1}}}}  \mathord{ \overset{  \mathsf{join} ( \ottnt{c_{{\mathrm{1}}}} , \ottnt{c_{{\mathrm{2}}}} )  }{\Rightarrow} }  \ottnt{T_{{\mathrm{3}}}}  \rangle^{ \bullet } ~  \ottnt{e}  \]
There are no redundant checks in the example term $\ottnt{e}$, so eidetic
\lambdah does \textit{exactly} the same checking as classic, finding
$\ottnt{e} \,  \longrightarrow ^{*}_{  \mathsf{E}  }  \,  \mathord{\Uparrow}  \ottnt{l_{{\mathrm{1}}}} $.
\fi}

\subsection{Type system}
\label{sec:typing}

\begin{bigfigure}
  \hdr{Context and type well formedness} 
      {\qquad \fbox{$ \mathord{  \vdash _{ \ottnt{m} } }~ \Gamma $} 
       \qquad \fbox{$ \mathord{  \vdash _{ \ottnt{m} } }~ \ottnt{T} $}}

  \sidebyside
    {\ottusedrule{\ottdruleWFXXEmpty{}}}
    {\ottusedrule{\ottdruleWFXXExtend{}}}
  \threesidebysidesqueeze[.23][.43][.28][.02]
    {\ottusedrule{\ottdruleWFXXBase{}}}
    {\ottusedrule{\ottdruleWFXXRefine{}}}
    {\ottusedrule{\ottdruleWFXXFun{}}}

  \hdr{Type compatibility and annotation well formedness}
    {\qquad \fbox{$\vdash  \ottnt{T_{{\mathrm{1}}}}  \mathrel{\parallel}  \ottnt{T_{{\mathrm{2}}}}$} \qquad \fbox{$ \mathord{  \vdash _{ \ottnt{m} } }~ \ottnt{a}   \mathrel{\parallel}   \ottnt{T_{{\mathrm{1}}}}  \Rightarrow  \ottnt{T_{{\mathrm{2}}}} $}}

  \threesidebyside
    {\ottusedrule{\ottdruleSXXRefine{}}}
    {\ottusedrule{\ottdruleSXXFun{}}}
    {\ottusedrule{\ottdruleAXXNone{}}}

  \hdr{Expression typing}{\qquad \fbox{$ \Gamma   \vdash _{ \ottnt{m} }  \ottnt{e}  :  \ottnt{T} $}}

  \threesidebyside
    {\ottusedrule{\ottdruleTXXVar{}}}
    {\ottusedrule{\ottdruleTXXAbs{}}}
    {\ottusedrule{\ottdruleTXXBlame{}}}
  \sidebyside[.5]
    {\ottusedrule{\ottdruleTXXConst{}}}
    {\ottusedrule{\ottdruleTXXOp{}}}
  \sidebyside
    {\ottusedrule{\ottdruleTXXApp{}}}
    {\ottusedrule{\ottdruleTXXCast{}}}
  {\ottusedrule{\ottdruleTXXCheck{}}}

  \caption{Universal typing rules of \lambdah}
  \label{fig:typing}
\end{bigfigure}

All modes share a type system, given in
Figure~\ref{fig:typing}.
All judgments are universal and simply thread the mode
through---except for annotation well formedness $ \mathord{  \vdash _{ \ottnt{m} } }~ \ottnt{a}   \mathrel{\parallel}   \ottnt{T_{{\mathrm{1}}}}  \Rightarrow  \ottnt{T_{{\mathrm{2}}}} $, which is mode specific, and a single eidetic-specific rule
given in Figure~\ref{fig:eideticsemantics}\iffull in
Section~\ref{sec:eidetic}\fi.
The type system comprises several relations:
context well formedness $ \mathord{  \vdash _{ \ottnt{m} } }~ \Gamma $ and type well formedness $ \mathord{  \vdash _{ \ottnt{m} } }~ \ottnt{T} $; type compatibility $\vdash  \ottnt{T_{{\mathrm{1}}}}  \mathrel{\parallel}  \ottnt{T_{{\mathrm{2}}}}$, a mode-less comparison of
the \textit{skeleton} of two types; annotation well formedness $ \mathord{  \vdash _{ \ottnt{m} } }~ \ottnt{a}   \mathrel{\parallel}   \ottnt{T_{{\mathrm{1}}}}  \Rightarrow  \ottnt{T_{{\mathrm{2}}}} $; and term typing $ \Gamma   \vdash _{ \ottnt{m} }  \ottnt{e}  :  \ottnt{T} $.

Context well formedness is entirely straightforward; type well
formedness requires some care to get base types off the ground.
  We establish as an axiom that the \textit{raw} type $ \{ \mathit{x} \mathord{:} \ottnt{B} \mathrel{\mid}  \mathsf{true}  \} $
  is well formed for every base type $\ottnt{B}$ (\WF{Base}); we then use
  raw types to check that refinements are well formed: $ \{ \mathit{x} \mathord{:} \ottnt{B} \mathrel{\mid} \ottnt{e} \} $
  is well formed in mode $\ottnt{m}$ if $\ottnt{e}$ is well typed as a boolean
  in mode $\ottnt{m}$ when $\mathit{x}$ is a value of type $\ottnt{B}$
  (\WF{Refine}). Without \WF{Base}, \WF{Refine} wouldn't have a well
  formed context. Function types are well formed in mode $\ottnt{m}$ when
  their domains and codomains are well formed in mode $\ottnt{m}$. (Unlike
  many recent formulations, our functions are not dependent---we leave
  dependency as future work.) 
Type compatibility $\vdash  \ottnt{T_{{\mathrm{1}}}}  \mathrel{\parallel}  \ottnt{T_{{\mathrm{2}}}}$ identifies types which can be
cast to each other: the types must have the same ``skeleton''. It is
reasonable to try to cast a non-zero integer $ \{ \mathit{x} \mathord{:}  \mathsf{Int}  \mathrel{\mid}  \mathit{x}  \mathrel{\ne}  \ottsym{0}  \} $ to a
positive integer $ \{ \mathit{x} \mathord{:}  \mathsf{Int}  \mathrel{\mid}  \mathit{x}  \mathrel{>} \ottsym{0}  \} $, but it is senseless to cast it
to a boolean $ \{ \mathit{x} \mathord{:}  \mathsf{Bool}  \mathrel{\mid}  \mathsf{true}  \} $ or to a function type $ \ottnt{T_{{\mathrm{1}}}} \mathord{ \rightarrow } \ottnt{T_{{\mathrm{2}}}} $.
Every cast must be between compatible types; at their core, \lambdah
programs are simply typed lambda calculus programs.
Type compatibility is reflexive, symmetric, and transitive; i.e., it
is an equivalence relation.

Our family of calculi use different annotations. All source programs
(defined below) begin without annotations---we write the empty
annotation $ \bullet $, read ``none''. The universal annotation well formedness rule
just defers to type compatibility (\A{None}); it is an invariant that
$ \mathord{  \vdash _{ \ottnt{m} } }~ \ottnt{a}   \mathrel{\parallel}   \ottnt{T_{{\mathrm{1}}}}  \Rightarrow  \ottnt{T_{{\mathrm{2}}}} $ implies $\vdash  \ottnt{T_{{\mathrm{1}}}}  \mathrel{\parallel}  \ottnt{T_{{\mathrm{2}}}}$.

As for term typing, the \T{Var}, \T{Abs}, \T{Op}, and \T{App} rules
are entirely conventional. \T{Blame} types blame at any (well formed)
type. 
A constant $\ottnt{k}$ can be typed by \T{Const} at any type $ \{ \mathit{x} \mathord{:} \ottnt{B} \mathrel{\mid} \ottnt{e} \} $
in mode $\ottnt{m}$ if: (a) $\ottnt{k}$ is a $\ottnt{B}$, i.e., $ \mathsf{ty} ( \ottnt{k} )   \ottsym{=}  \ottnt{B}$;
(b) the type in question is well formed in $\ottnt{m}$; and (c), if
$ \ottnt{e}  [  \ottnt{k} / \mathit{x}  ]  \,  \longrightarrow ^{*}_{ \ottnt{m} }  \,  \mathsf{true} $.
As an immediate consequence, we can derive the following rule typing
constants at their raw type, since $ \mathsf{true}  \,  \longrightarrow ^{*}_{ \ottnt{m} }  \,  \mathsf{true} $ in all modes
and raw types are well formed in all modes (\WF{Base}):
\[ \ottdrule[]{ \mathord{  \vdash _{ \ottnt{m} } }~ \Gamma  \qquad  \mathsf{ty} ( \ottnt{k} )   \ottsym{=}  \ottnt{B}}{ \Gamma   \vdash _{ \ottnt{m} }  \ottnt{k}  :   \{ \mathit{x} \mathord{:} \ottnt{B} \mathrel{\mid}  \mathsf{true}  \}  }{}{} \]
This approach to typing constants in a manifest calculus is novel: it
offers a great deal of latitude with typing, while avoiding the
subtyping of some formulations~\cite{Greenberg12contracts,
  Flanagan06hybrid, Knowles10hybrid, Knowles06sage} and the extra rule
of others~\cite{Belo11fh}.
We assume that $ \mathsf{ty} ( \ottnt{k} )   \ottsym{=}   \mathsf{Bool} $ iff $\ottnt{k} \in \set{ \mathsf{true} ,
   \mathsf{false} }$.

We require in \T{Op} that $ \mathsf{ty} (\mathord{ \ottnt{op} }) $ only produces well formed first-order
types, i.e., types of the form $ \mathord{  \vdash _{ \ottnt{m} } }~ {}   \{ \mathit{x} \mathord{:} \ottnt{B_{{\mathrm{1}}}} \mathrel{\mid} \ottnt{e_{{\mathrm{1}}}} \}   \rightarrow \, ... \, \rightarrow   \{ \mathit{x} \mathord{:} \ottnt{B_{\ottmv{n}}} \mathrel{\mid} \ottnt{e_{\ottmv{n}}} \}   {} $. We require that the type is consistent with the operation's
denotation: $\denot{ op } \, \ottsym{(}  \ottnt{k_{{\mathrm{1}}}}  \ottsym{,} \, ... \, \ottsym{,}  \ottnt{k_{\ottmv{n}}}  \ottsym{)}$ is defined iff $ \ottnt{e_{\ottmv{i}}}  [  \ottnt{k_{\ottmv{i}}} / \mathit{x}  ]  \,  \longrightarrow ^{*}_{ \ottnt{m} }  \,  \mathsf{true} $ for all $\ottnt{m}$.
For this evaluation to hold for every system we consider, the types
assigned to operations can't involve casts that both (a) stack and (b)
can fail\iffull---because forgetful \lambdah may skip them, leading to
different typings\fi. We believe this is not so stringent a requirement:
the types for operations ought to be simple, e.g. $ \mathsf{ty} (\mathord{  \mathsf{div}  })  =
   \{ \mathit{x} \mathord{:}  \mathsf{Real}  \mathrel{\mid}  \mathsf{true}  \}  \mathord{ \rightarrow }  \{ \mathit{y} \mathord{:}  \mathsf{Real}  \mathrel{\mid}  \mathit{y}  \mathrel{\ne}  \ottsym{0}  \}   \mathord{ \rightarrow }  \{ \mathit{z} \mathord{:}  \mathsf{Real}  \mathrel{\mid}  \mathsf{true}  \}  $, and stacked casts
only arise in stack-free terms due to function proxies.
In general, it is interesting to ask what refinement types to assign
to constants, as careless assignments can lead to circular checking
(e.g., if division has a codomain cast checking its work with
multiplication and vice versa).

The typing rule for casts, \T{Cast}, relies on the annotation well
formedness rule: $ \langle  \ottnt{T_{{\mathrm{1}}}}  \mathord{ \overset{ \ottnt{a} }{\Rightarrow} }  \ottnt{T_{{\mathrm{2}}}}  \rangle^{ \ottnt{l} } ~  \ottnt{e} $ is well formed in mode $\ottnt{m}$
when $ \mathord{  \vdash _{ \ottnt{m} } }~ \ottnt{a}   \mathrel{\parallel}   \ottnt{T_{{\mathrm{1}}}}  \Rightarrow  \ottnt{T_{{\mathrm{2}}}} $ and $\ottnt{e}$ is a $\ottnt{T_{{\mathrm{1}}}}$.
Allowing any cast between compatible base types is conservative: a
cast from $ \{ \mathit{x} \mathord{:}  \mathsf{Int}  \mathrel{\mid}  \mathit{x}  \mathrel{>} \ottsym{0}  \} $ to $ \{ \mathit{x} \mathord{:}  \mathsf{Int}  \mathrel{\mid}  \mathit{x}  \mathrel{\le}  \ottsym{0}  \} $ always
fails. Earlier work has used SMT solvers to try to statically reject
certain casts and eliminate those that are guaranteed to
succeed~\cite{Flanagan06hybrid,Knowles06sage,Bierman10smt}; we omit
these checks, as we view them as secondary---a static analysis
offering bug-finding and optimization, and not the essence of the
system.

The final rule, \T{Check}, is used for checking active checks, which
should only occur at runtime. In fact, they should only ever be
applied to closed terms; the rule allows for any well formed context
as a technical device for weakening\iffull
(Lemma~\ref{lem:weakening})\fi.

Active checks $ \langle   \{ \mathit{x} \mathord{:} \ottnt{B} \mathrel{\mid} \ottnt{e_{{\mathrm{1}}}} \}  ,  \ottnt{e_{{\mathrm{2}}}} ,  \ottnt{k}  \rangle^{ \ottnt{l} } $ arise as the result of casts
between refined base types, as in the following classic \lambdah
evaluation of a successful cast:
\[ \begin{array}{r@{~}c@{~}l}
   \langle   \{ \mathit{x} \mathord{:} \ottnt{B} \mathrel{\mid} \ottnt{e} \}   \mathord{ \overset{ \bullet }{\Rightarrow} }   \{ \mathit{x} \mathord{:} \ottnt{B} \mathrel{\mid} \ottnt{e'} \}   \rangle^{ \ottnt{l} } ~  \ottnt{k}  & \longrightarrow _{  \mathsf{C}  } &  \langle   \{ \mathit{x} \mathord{:} \ottnt{B} \mathrel{\mid} \ottnt{e'} \}  ,   \ottnt{e'}  [  \ottnt{k} / \mathit{x}  ]  ,  \ottnt{k}  \rangle^{ \ottnt{l} }  \\
  & \longrightarrow ^{*}_{  \mathsf{C}  } &  \langle   \{ \mathit{x} \mathord{:} \ottnt{B} \mathrel{\mid} \ottnt{e'} \}  ,   \mathsf{true}  ,  \ottnt{k}  \rangle^{ \ottnt{l} }  \\
  & \longrightarrow _{  \mathsf{C}  } & \ottnt{k}
\end{array}\]
If we are going to prove type soundness via syntactic
methods~\cite{Wright94syntactic}, we must have enough information to
type $\ottnt{k}$ at $ \{ \mathit{x} \mathord{:} \ottnt{B} \mathrel{\mid} \ottnt{e'} \} $. For this reason, \T{Check} requires
that $ \ottnt{e_{{\mathrm{1}}}}  [  \ottnt{k} / \mathit{x}  ]  \,  \longrightarrow ^{*}_{ \ottnt{m} }  \, \ottnt{e_{{\mathrm{2}}}}$; this way, we know that $ \ottnt{e'}  [  \ottnt{k} / \mathit{x}  ]  \,  \longrightarrow ^{*}_{ \ottnt{m} }  \,  \mathsf{true} $ at the end of the previous derivation, which is enough to
apply \T{Const}. The other premises of \T{Check} ensure that the types
all match up: that the target refinement type is well formed; that
$\ottnt{k}$ has the base type in question; and that $\ottnt{e_{{\mathrm{2}}}}$, the current
state of the active check, is also well formed.

To truly say that our languages share a syntax and a type system, we
highlight a subset of type derivations as \textit{source program} type
derivations.  We show that source programs well typed in one mode are
well typed in the all modes\ifpopl~\cite{Greenberg14spacetr}\else~(Appendix~\ref{app:typesoundness})\fi.
\numbertheoremstrue
\begin{definition}[Source program]
  \label{def:sourceprogram}
A source program type derivation obeys
the following rules:
\begin{itemize}
\item \T{Const} only ever assigns the type $ \{ \mathit{x} \mathord{:}  \mathsf{ty} ( \ottnt{k} )  \mathrel{\mid}  \mathsf{true}  \} $.
  Variations in each mode's evaluation aren't reflected in the (source
  program) type system. (We could soundly relax this requirement to
  allow $ \{ \mathit{x} \mathord{:}  \mathsf{ty} ( \ottnt{k} )  \mathrel{\mid} \ottnt{e} \} $ such that $ \ottnt{e}  [  \ottnt{k} / \mathit{x}  ]  \,  \longrightarrow ^{*}_{ \ottnt{m} }  \,  \mathsf{true} $ for any
  mode $\ottnt{m}$.)
\item Casts have empty annotations $\ottnt{a}  \ottsym{=}  \bullet$. Casts also have
  blame labels, and not empty blame (also written $ \bullet $).
\item \T{Check}, \T{Stack} (Section~\ref{sec:eidetic}), and
  \T{Blame} are not used---these are for runtime only.
\end{itemize}
\end{definition}
\numbertheoremsfalse
Note that source programs don't use any of the typing rules that defer
to the evaluation relation (\T{Check} and \T{Stack}), so we can
maintain a clear phase distinction between type checking programs and
running them.

\subsection{Metatheory}
\label{sec:metatheory}

One distinct advantage of having a single syntax with parameterized
semantics is that some of the metatheory can be done once for all
modes.
Each mode proves its own canonical forms lemma---since each mode has a
unique notion of value---and its own progress and preservation lemmas
for syntactic type soundness~\cite{Wright94syntactic}. But other
standard metatheoretical machinery---weakening, substitution, and
regularity---can be proved for all modes at once (see
Section~\ref{app:genericmetatheory}).
To wit, we prove syntactic type soundness in
Appendix~\ref{app:classicsoundness} for classic \lambdah in just three
mode-specific lemmas: canonical forms, progress, and preservation. In
every theorem statement, we include a reference to the lemma number
where it is proved in the appendix. In PDF versions, this reference is
hyperlinked.

{\ifpopl
\begin{lemma}[Classic canonical forms (\ref{lem:classiccanonicalforms})]
  If $ \emptyset   \vdash _{  \mathsf{C}  }  \ottnt{e}  :  \ottnt{T} $ and $ \mathsf{val} _{  \mathsf{C}  }~ \ottnt{e} $ then:
  \begin{itemize}
  \item If $\ottnt{T}  \ottsym{=}   \{ \mathit{x} \mathord{:} \ottnt{B} \mathrel{\mid} \ottnt{e'} \} $, then $\ottnt{e}  \ottsym{=}  \ottnt{k}$ and $ \mathsf{ty} ( \ottnt{k} )   \ottsym{=}  \ottnt{B}$
    and $ \ottnt{e'}  [  \ottnt{e} / \mathit{x}  ]  \,  \longrightarrow ^{*}_{  \mathsf{C}  }  \,  \mathsf{true} $.
  \item If $\ottnt{T}  \ottsym{=}   \ottnt{T_{{\mathrm{1}}}} \mathord{ \rightarrow } \ottnt{T_{{\mathrm{2}}}} $, then either $\ottnt{e}  \ottsym{=}   \lambda \mathit{x} \mathord{:} \ottnt{T} .~  \ottnt{e'} $ or $\ottnt{e}  \ottsym{=}   \langle   \ottnt{T_{{\mathrm{11}}}} \mathord{ \rightarrow } \ottnt{T_{{\mathrm{12}}}}   \mathord{ \overset{\bullet}{\Rightarrow} }   \ottnt{T_{{\mathrm{21}}}} \mathord{ \rightarrow } \ottnt{T_{{\mathrm{22}}}}   \rangle^{ \ottnt{l} } ~  \ottnt{e'} $.
  \end{itemize}
\end{lemma}

\begin{lemma}[Classic progress (\ref{lem:classicprogress})]
  If $ \emptyset   \vdash _{  \mathsf{C}  }  \ottnt{e}  :  \ottnt{T} $, then either:
  \begin{enumerate}
  \item $ \mathsf{result} _{  \mathsf{C}  }~ \ottnt{e} $, i.e., $\ottnt{e}  \ottsym{=}   \mathord{\Uparrow}  \ottnt{l} $ or $ \mathsf{val} _{  \mathsf{C}  }~ \ottnt{e} $; or
  \item there exists an $\ottnt{e'}$ such that $\ottnt{e} \,  \longrightarrow _{  \mathsf{C}  }  \, \ottnt{e'}$.
  \end{enumerate}
\end{lemma}

\begin{lemma}[Classic preservation (\ref{lem:classicpreservation})]
  If $ \emptyset   \vdash _{  \mathsf{C}  }  \ottnt{e}  :  \ottnt{T} $ and $\ottnt{e} \,  \longrightarrow _{  \mathsf{C}  }  \, \ottnt{e'}$, then $ \emptyset   \vdash _{  \mathsf{C}  }  \ottnt{e'}  :  \ottnt{T} $.
\end{lemma}
\fi}

{\iffull
\subsection{Overview}
\label{sec:overview}

In the rest of this paper, we give the semantics for three
space-efficient modes for \lambdah, relating the languages' behavior
on source programs (Definition~\ref{def:sourceprogram}).  The
forgetful mode is space efficient without annotations, converging to a
value more often than classic \lambdah ($\ottnt{m}  \ottsym{=}   \mathsf{F} $;
Section~\ref{sec:forgetful}).
The heedful mode is space efficient and uses type sets to converge to
a value exactly when classic \lambdah does; it may blame different
labels, though ($\ottnt{m}  \ottsym{=}   \mathsf{H} $; Section~\ref{sec:heedful}).
The eidetic mode is space efficient and uses coercions to track
pending checks; it behaves exactly like classic \lambdah ($\ottnt{m}  \ottsym{=}   \mathsf{E} $;
Section~\ref{sec:eidetic}).
We show that source programs that are well typed in one mode are well
typed in all of them
(Lemmas~\ref{lem:forgetfulsource},~\ref{lem:heedfulsource},
and~\ref{lem:eideticsource}).
We relate these space-efficient modes back to classic \lambdah in
Section~\ref{sec:soundness}.

One may wonder why we even bother to mention forgetful and heedful
\lambdah, if eidetic \lambdah is soundly space efficient with respect
to classic \lambdah.
These two `intermediate' modes are interesting as an exploration of
the design space---but also in their own right.

Forgetful \lambdah takes a radical approach that involves
skipping checks---its soundness is rather surprising and offers
insights into the semantics of contracts.
Contracts have been used for more than avoiding wrongness, though:
they have been used in Racket for abstraction and information
hiding~\cite{Racket,RacketContracts}.
Forgetful \lambdah can't use contracts for information hiding. Suppose
we implement user records as functions from strings to strings. We
would like to pass a user record to an untrusted component, hiding
some fields but not others.
We can achieve this by specifying a white- or blacklist in a contract,
e.g., $  \{ \mathit{f} \mathord{:}  \mathsf{String}  \mathrel{\mid}  \mathit{f}  \mathrel{\ne}   \mathtt{``password''}   \}  \mathord{ \rightarrow }  \{ \mathit{v} \mathord{:}  \mathsf{String}  \mathrel{\mid}  \mathsf{true}  \}  $. Wrapping a
function in this contract introduces a function proxy... which can be
overwritten by \ECastMerge! To really get information hiding, the
programmer must explicitly $\eta$-expand the function proxy, writing
$ (  \lambda \mathit{f} \mathord{:}  \{ \mathit{f} \mathord{:}  \mathsf{String}  \mathrel{\mid}  \mathit{f}  \mathrel{\ne}   \mathtt{``password''}   \}  .~  \dots  ) $. Forgetful \lambdah's
contracts can't enforce abstractions.\footnote{This
  observation is due to Sam Tobin-Hochstadt.}

While \citet{Siek10threesomes} uses the lattice of type precision in
their threesomes without blame, our heedful \lambdah uses the powerset
lattice of types.
Just as \citeauthor{Siek10threesomes} use labeled types and meet-like
composition for threesomes \textit{with} blame, we may be able to
derive something similar for heedful and eidetic \lambdah: in a
(non-commutative) skew lattice, heedful uses a potentially re-ordering
conjunction while eidetic preserves order. A lattice-theoretic account
of casts, coercions, and blame may be possible.

\section{Forgetful space efficiency}
\label{sec:forgetful}

In forgetful \lambdah, we offer a simple solution to space-inefficient
casts: just forget about them. Function proxies only ever wrap
lambda abstractions; trying to cast a function proxy simply throws
away the inner proxy.
Just the same, when accumulating casts on the stack, we throw away all but the
last cast.
Readers may wonder: how can this ever be sound? Several factors work
together to make forgetful \lambdah a sound calculus. In short, the
key ingredients are call-by-value evaluation and the observation that
type safety only talks about reduction to values in this setting.

In this section, our mode $\ottnt{m}  \ottsym{=}   \mathsf{F} $: our evaluation relation is $ \longrightarrow _{  \mathsf{F}  } $
and we use typing judgments of the form, e.g. $ \Gamma   \vdash _{  \mathsf{F}  }  \ottnt{e}  :  \ottnt{T} $.
Forgetful \lambdah is the simplest of the space-efficient calculi: it
just uses the standard typing rules from Figure~\ref{fig:typing} and
the space-efficient reduction rules from Figure~\ref{fig:opsem}. 
We give the new operational definitions for $\ottnt{m}  \ottsym{=}   \mathsf{F} $ in
Figure~\ref{fig:forgetfulopsem}: a new value rule and the definition
of the $ \mathsf{merge} $ operator.
First, \V{ProxyF} says that function proxies in forgetful \lambdah are
only values when the proxied value is a lambda (and not another
function proxy).  Limiting the number of
function proxies is critical for establishing space bounds, as we do
in Section~\ref{sec:bounds}.
Forgetful casts don't use annotations, so they just use \A{None}.
The forgetful merge operator just \textit{forgets} the intermediate
type $\ottnt{T_{{\mathrm{2}}}}$.

\begin{figure}[t]
  \hdr{Values and merging}{\qquad \fbox{$ \mathsf{val} _{  \mathsf{F}  }~ \ottnt{e} $}}

  {\iffull
    {\ottusedrule{\ottdruleVXXConst{}}}
    {\ottusedrule{\ottdruleVXXAbs{}}}
  \fi}

  {\ottusedrule{\ottdruleVXXProxyF{}}}
  {\[  \mathsf{merge} _{  \mathsf{F}  }( \ottnt{T_{{\mathrm{1}}}} , \bullet , \ottnt{T_{{\mathrm{2}}}} , \bullet , \ottnt{T_{{\mathrm{3}}}} )   \ottsym{=}  \bullet \]}
  \vspace*{-10pt}
  \caption{Operational semantics of forgetful \lambdah}
  \label{fig:forgetfulopsem}
\end{figure}

\begin{figure}
\[ \begin{array}{r@{~}l}
  \ottnt{e} = &  \langle   \{ \mathit{x} \mathord{:}  \mathsf{Int}  \mathrel{\mid}  \mathit{x} \,  \mathsf{mod}  \, \ottsym{2}  \mathrel{=}  \ottsym{0}  \}   \mathord{ \overset{\bullet}{\Rightarrow} }   \{ \mathit{x} \mathord{:}  \mathsf{Int}  \mathrel{\mid}  \mathit{x}  \mathrel{\ne}  \ottsym{0}  \}   \rangle^{ \ottnt{l_{{\mathrm{3}}}} } ~  {} \\  &  \quad   (  \langle   \{ \mathit{x} \mathord{:}  \mathsf{Int}  \mathrel{\mid}  \mathit{x}  \mathrel{\ge}  \ottsym{0}  \}   \mathord{ \overset{\bullet}{\Rightarrow} }   \{ \mathit{x} \mathord{:}  \mathsf{Int}  \mathrel{\mid}  \mathit{x} \,  \mathsf{mod}  \, \ottsym{2}  \mathrel{=}  \ottsym{0}  \}   \rangle^{ \ottnt{l_{{\mathrm{2}}}} } ~  {} \\  &  \qquad   (  \langle   \{ \mathit{x} \mathord{:}  \mathsf{Int}  \mathrel{\mid}  \mathsf{true}  \}   \mathord{ \overset{\bullet}{\Rightarrow} }   \{ \mathit{x} \mathord{:}  \mathsf{Int}  \mathrel{\mid}  \mathit{x}  \mathrel{\ge}  \ottsym{0}  \}   \rangle^{ \ottnt{l_{{\mathrm{1}}}} } ~   {-1}   )   )   \\
 \multicolumn{2}{r}{(\E{CastMerge})} \\
  \longrightarrow _{  \mathsf{F}  }  &  \langle   \{ \mathit{x} \mathord{:}  \mathsf{Int}  \mathrel{\mid}  \mathit{x}  \mathrel{\ge}  \ottsym{0}  \}   \mathord{ \overset{\bullet}{\Rightarrow} }   \{ \mathit{x} \mathord{:}  \mathsf{Int}  \mathrel{\mid}  \mathit{x}  \mathrel{\ne}  \ottsym{0}  \}   \rangle^{ \ottnt{l_{{\mathrm{3}}}} } ~  {} \\  &  \quad   (  \langle   \{ \mathit{x} \mathord{:}  \mathsf{Int}  \mathrel{\mid}  \mathsf{true}  \}   \mathord{ \overset{\bullet}{\Rightarrow} }   \{ \mathit{x} \mathord{:}  \mathsf{Int}  \mathrel{\mid}  \mathit{x}  \mathrel{\ge}  \ottsym{0}  \}   \rangle^{ \ottnt{l_{{\mathrm{1}}}} } ~   {-1}   )   \\
 \multicolumn{2}{r}{(\E{CastMerge})} \\
  \longrightarrow _{  \mathsf{F}  }  &  \langle   \{ \mathit{x} \mathord{:}  \mathsf{Int}  \mathrel{\mid}  \mathsf{true}  \}   \mathord{ \overset{\bullet}{\Rightarrow} }   \{ \mathit{x} \mathord{:}  \mathsf{Int}  \mathrel{\mid}  \mathit{x}  \mathrel{\ne}  \ottsym{0}  \}   \rangle^{ \ottnt{l_{{\mathrm{3}}}} } ~   {-1}   \\
 \multicolumn{2}{r}{(\ECheckNone)} \\
  \longrightarrow _{  \mathsf{F}  }  &  \langle   \{ \mathit{x} \mathord{:}  \mathsf{Int}  \mathrel{\mid}  \mathit{x}  \mathrel{\ne}  \ottsym{0}  \}  ,    {-1}   \mathrel{\ne}  \ottsym{0}  ,   {-1}   \rangle^{ \ottnt{l_{{\mathrm{3}}}} }  \\
 \multicolumn{2}{r}{(\E{CheckInner}/\E{Op})} \\
  \longrightarrow _{  \mathsf{F}  }  &  \langle   \{ \mathit{x} \mathord{:}  \mathsf{Int}  \mathrel{\mid}  \mathit{x}  \mathrel{\ne}  \ottsym{0}  \}  ,   \mathsf{true}  ,   {-1}   \rangle^{ \ottnt{l_{{\mathrm{3}}}} }  \qquad\qquad (\E{CheckOK}) \\
  \longrightarrow _{  \mathsf{F}  }  &  {-1} 
\end{array} \]
  \caption{Example of forgetful \lambdah}
  \label{fig:forgetfulexample}
\end{figure}
We demonstrate this semantics on the example from
Section~\ref{sec:opsem} in Figure~\ref{fig:forgetfulexample}.
We first step by merging casts, forgetting the intermediate type. Then
contract checking proceeds as normal for the target type; since
$ {-1} $ is non-zero, the check succeeds and returns its scrutinee by
\E{CheckOK}.

The type soundness property typically has two parts:
(a) well typed programs don't go `wrong' (for us, getting stuck), and 
(b) well typed programs reduce to programs that are well typed at the
same type.
How could a forgetful \lambdah program go wrong, violating property
(a)? The general ``skeletal'' structure of types means we never have
to worry about errors caught by simple type systems, such as trying to
apply a non-function. Our semantics can get stuck by trying to apply
an operator to an input that isn't in its domain, e.g., trying to
divide by zero.
To guarantee that we avoid stuck operators, \lambdah generally relies
on subject reduction, property (b). Operators are assigned types that
avoid stuckness, i.e., $ \mathsf{ty} (\mathord{ \ottnt{op} }) $ and $ \denot{ op } $ agree. Some
earlier systems have done this~\cite{Flanagan06hybrid,Knowles10hybrid}
while others haven't~\cite{Greenberg12contracts,Belo11fh}. We view it
as a critical component of contract calculi.
So for, say, integer division, $ \mathsf{ty} (\mathord{  \mathsf{div}  })   \ottsym{=}     \{ \mathit{x} \mathord{:}  \mathsf{Int}  \mathrel{\mid}  \mathsf{true}  \}  \mathord{ \rightarrow }  \{ \mathit{y} \mathord{:}  \mathsf{Int}  \mathrel{\mid}  \mathit{y}  \mathrel{\ne}  \ottsym{0}  \}   \mathord{ \rightarrow }  \{ \mathit{z} \mathord{:}  \mathsf{Int}  \mathrel{\mid}  \mathsf{true}  \}  $. To actually use $ \mathsf{div} $
in a program, the second argument must be typed as a non-zero
integer---by a non-source typing with \T{Const} directly (see
Definition~\ref{def:sourceprogram}) or by casting (\T{Cast}).
It may seem dangerous: casts protect operators from improper values,
preventing stuckness; forgetful \lambdah eliminates some casts. But
consider the cast eliminated by \E{CastMerge}:
\[  \langle  \ottnt{T_{{\mathrm{2}}}}  \mathord{ \overset{\bullet}{\Rightarrow} }  \ottnt{T_{{\mathrm{3}}}}  \rangle^{ \ottnt{l} } ~   (  \langle  \ottnt{T_{{\mathrm{1}}}}  \mathord{ \overset{\bullet}{\Rightarrow} }  \ottnt{T_{{\mathrm{2}}}}  \rangle^{ \ottnt{l'} } ~  \ottnt{e}  )   \,  \longrightarrow _{  \mathsf{F}  }  \,  \langle  \ottnt{T_{{\mathrm{1}}}}  \mathord{ \overset{\bullet}{\Rightarrow} }  \ottnt{T_{{\mathrm{3}}}}  \rangle^{ \ottnt{l} } ~  \ottnt{e}  \]
While the program tried to cast $\ottnt{e}$ to a $\ottnt{T_{{\mathrm{2}}}}$, it immediately
cast it back out---no operation relies on $\ottnt{e}$ being a
$\ottnt{T_{{\mathrm{2}}}}$. Skipping the check doesn't risk stuckness.
Since \lambdah is call-by-value, we can use the same reasoning to
allow functions to assume that their inputs inhabit their types---a
critical property for programmer reasoning.

Forgetful \lambdah enjoys soundness via a standard syntactic proof of
progress and preservation, reusing the theorems from
Section~\ref{app:genericmetatheory}.  What's more, source programs are
well typed in classic \lambdah iff they are well typed in forgetful
\lambdah: both languages can run the same terms.
Proofs are in Appendix~\ref{app:forgetfulsoundness}.

\section{Heedful space efficiency}
\label{sec:heedful}

\begin{figure}
  \hdr{Type set well formedness}
    {\qquad \fbox{$ \mathord{  \vdash _{ \ottnt{m} } }~ \mathcal{S}   \mathrel{\parallel}   \ottnt{T_{{\mathrm{1}}}}  \Rightarrow  \ottnt{T_{{\mathrm{2}}}} $}}

    {\ottusedrule{\ottdruleAXXTypeSet{}}}

  \hdr{Values and operational semantics}{\qquad \fbox{$ \mathsf{val} _{  \mathsf{H}  }~ \ottnt{e} $} \qquad \fbox{$\ottnt{e_{{\mathrm{1}}}} \,  \longrightarrow _{  \mathsf{H}  }  \, \ottnt{e_{{\mathrm{2}}}}$}}

  {\ottusedrule{\ottdruleVXXProxyH{}}}
  {\ottusedrule{\ottdruleEXXTypeSet{}}} 
  {\ottusedrule{\ottdruleEXXCheckEmpty{}}}
  {\ottusedrule{\ottdruleEXXCheckSet{}}}
  {\[  \mathsf{choose} ( \mathcal{S} )  \in \mathcal{S} \text{ when } \mathcal{S} \ne  \emptyset  \]}
  {\[  \mathsf{merge} _{  \mathsf{H}  }( \ottnt{T_{{\mathrm{1}}}} , \mathcal{S}_{{\mathrm{1}}} , \ottnt{T_{{\mathrm{2}}}} , \mathcal{S}_{{\mathrm{2}}} , \ottnt{T_{{\mathrm{3}}}} )   \ottsym{=}    \mathcal{S}_{{\mathrm{1}}}  \cup  \mathcal{S}_{{\mathrm{2}}}   \cup   \set{  \ottnt{T_{{\mathrm{2}}}}  }   \]}
  {\[ \begin{array}{rcl}
       \mathsf{dom} ( \mathcal{S} )  &=& \bigcup_{ \ottnt{T}  \in  \mathcal{S} }  \mathsf{dom} ( \ottnt{T} )  \\
       \mathsf{cod} ( \mathcal{S} )  &=& \bigcup_{ \ottnt{T}  \in  \mathcal{S} }  \mathsf{cod} ( \ottnt{T} ) 
    \end{array} \]}

  \caption{Annotation typing and operational semantics of heedful \lambdah}
  \label{fig:heedfulsemantics}
\end{figure}

Heedful \lambdah ($\ottnt{m}  \ottsym{=}   \mathsf{H} $) takes the cast
merging strategy from forgetful \lambdah, but uses \textit{type sets}
on casts and function proxies to avoid dropping casts.
Space efficiency for heedful \lambdah rests on the use of sets:
classic \lambdah allows for arbitrary lists of function proxies and
casts on the stack to accumulate. Restricting this accumulation to a
set gives us a straightforward bound on the amount of accumulation: a
program of fixed size can only have so many types at each size. We
discuss this idea further in Section~\ref{sec:bounds}.

We extend the typing rules and operational semantics in
Figure~\ref{fig:heedfulsemantics}.
Up until this point, we haven't used annotations. Heedful \lambdah
collects type sets as casts merge to record the types that must be
checked.
The \A{TypeSet} annotation well formedness rule extends the premises
of \A{None} with the requirement that if $ \mathord{  \vdash _{  \mathsf{H}  } }~ \mathcal{S}   \mathrel{\parallel}   \ottnt{T_{{\mathrm{1}}}}  \Rightarrow  \ottnt{T_{{\mathrm{2}}}} $,
then all the types in $\mathcal{S}$ are well formed and compatible with
$\ottnt{T_{{\mathrm{1}}}}$ and $\ottnt{T_{{\mathrm{2}}}}$.
Type set compatibility is stable under removing elements from the set
$\mathcal{S}$, and it is symmetric and transitive with respective to its
type indices (since compatibility itself is symmetric and transitive).

One might expect the types in type sets to carry blame labels---might
we then be able to have \textit{sound} space efficiency? It turns out
that just having labels in the sets isn't enough---we actually need to
keep track of the ordering of checks. Eidetic \lambdah
(Section~\ref{sec:eidetic}) does exactly this tracking. Consider this calculus a warmup.

\finishlater{example of why this doesn't work?}

\begin{figure}[t]
\[ \begin{array}{r@{~}l}
  \ottnt{e} = &  \langle   \{ \mathit{x} \mathord{:}  \mathsf{Int}  \mathrel{\mid}  \mathit{x} \,  \mathsf{mod}  \, \ottsym{2}  \mathrel{=}  \ottsym{0}  \}   \mathord{ \overset{\bullet}{\Rightarrow} }   \{ \mathit{x} \mathord{:}  \mathsf{Int}  \mathrel{\mid}  \mathit{x}  \mathrel{\ne}  \ottsym{0}  \}   \rangle^{ \ottnt{l_{{\mathrm{3}}}} } ~  {} \\  &  \quad   (  \langle   \{ \mathit{x} \mathord{:}  \mathsf{Int}  \mathrel{\mid}  \mathit{x}  \mathrel{\ge}  \ottsym{0}  \}   \mathord{ \overset{\bullet}{\Rightarrow} }   \{ \mathit{x} \mathord{:}  \mathsf{Int}  \mathrel{\mid}  \mathit{x} \,  \mathsf{mod}  \, \ottsym{2}  \mathrel{=}  \ottsym{0}  \}   \rangle^{ \ottnt{l_{{\mathrm{2}}}} } ~  {} \\  &  \qquad   (  \langle   \{ \mathit{x} \mathord{:}  \mathsf{Int}  \mathrel{\mid}  \mathsf{true}  \}   \mathord{ \overset{\bullet}{\Rightarrow} }   \{ \mathit{x} \mathord{:}  \mathsf{Int}  \mathrel{\mid}  \mathit{x}  \mathrel{\ge}  \ottsym{0}  \}   \rangle^{ \ottnt{l_{{\mathrm{1}}}} } ~   {-1}   )   )   \\
 \multicolumn{2}{r}{(\E{TypeSet}, \E{CastInner}/\E{TypeSet})} \\
  \longrightarrow ^{*}_{  \mathsf{H}  }  &  \langle   \{ \mathit{x} \mathord{:}  \mathsf{Int}  \mathrel{\mid}  \mathit{x} \,  \mathsf{mod}  \, \ottsym{2}  \mathrel{=}  \ottsym{0}  \}   \mathord{ \overset{ \emptyset }{\Rightarrow} }   \{ \mathit{x} \mathord{:}  \mathsf{Int}  \mathrel{\mid}  \mathit{x}  \mathrel{\ne}  \ottsym{0}  \}   \rangle^{ \ottnt{l_{{\mathrm{3}}}} } ~  {} \\  &  \quad   (  \langle   \{ \mathit{x} \mathord{:}  \mathsf{Int}  \mathrel{\mid}  \mathit{x}  \mathrel{\ge}  \ottsym{0}  \}   \mathord{ \overset{ \emptyset }{\Rightarrow} }   \{ \mathit{x} \mathord{:}  \mathsf{Int}  \mathrel{\mid}  \mathit{x} \,  \mathsf{mod}  \, \ottsym{2}  \mathrel{=}  \ottsym{0}  \}   \rangle^{ \ottnt{l_{{\mathrm{2}}}} } ~  {} \\  &  \qquad   (  \langle   \{ \mathit{x} \mathord{:}  \mathsf{Int}  \mathrel{\mid}  \mathsf{true}  \}   \mathord{ \overset{\bullet}{\Rightarrow} }   \{ \mathit{x} \mathord{:}  \mathsf{Int}  \mathrel{\mid}  \mathit{x}  \mathrel{\ge}  \ottsym{0}  \}   \rangle^{ \ottnt{l_{{\mathrm{1}}}} } ~   {-1}   )   )   \\
 \multicolumn{2}{r}{(\E{CastMerge})} \\ 
  \longrightarrow _{  \mathsf{H}  }  &  \langle   \{ \mathit{x} \mathord{:}  \mathsf{Int}  \mathrel{\mid}  \mathit{x}  \mathrel{\ge}  \ottsym{0}  \}   \mathord{ \overset{  \set{   \{ \mathit{x} \mathord{:}  \mathsf{Int}  \mathrel{\mid}  \mathit{x} \,  \mathsf{mod}  \, \ottsym{2}  \mathrel{=}  \ottsym{0}  \}   }  }{\Rightarrow} }   \{ \mathit{x} \mathord{:}  \mathsf{Int}  \mathrel{\mid}  \mathit{x}  \mathrel{\ne}  \ottsym{0}  \}   \rangle^{ \ottnt{l_{{\mathrm{3}}}} } ~  {} \\  &  \qquad   (  \langle   \{ \mathit{x} \mathord{:}  \mathsf{Int}  \mathrel{\mid}  \mathsf{true}  \}   \mathord{ \overset{\bullet}{\Rightarrow} }   \{ \mathit{x} \mathord{:}  \mathsf{Int}  \mathrel{\mid}  \mathit{x}  \mathrel{\ge}  \ottsym{0}  \}   \rangle^{ \ottnt{l_{{\mathrm{1}}}} } ~   {-1}   )   \\
 \multicolumn{2}{r}{(\E{CastInner}/\E{TypeSet})} \\ 
  \longrightarrow _{  \mathsf{H}  }  &  \langle   \{ \mathit{x} \mathord{:}  \mathsf{Int}  \mathrel{\mid}  \mathit{x}  \mathrel{\ge}  \ottsym{0}  \}   \mathord{ \overset{  \set{   \{ \mathit{x} \mathord{:}  \mathsf{Int}  \mathrel{\mid}  \mathit{x} \,  \mathsf{mod}  \, \ottsym{2}  \mathrel{=}  \ottsym{0}  \}   }  }{\Rightarrow} }   \{ \mathit{x} \mathord{:}  \mathsf{Int}  \mathrel{\mid}  \mathit{x}  \mathrel{\ne}  \ottsym{0}  \}   \rangle^{ \ottnt{l_{{\mathrm{3}}}} } ~  {} \\  &  \qquad   (  \langle   \{ \mathit{x} \mathord{:}  \mathsf{Int}  \mathrel{\mid}  \mathsf{true}  \}   \mathord{ \overset{ \emptyset }{\Rightarrow} }   \{ \mathit{x} \mathord{:}  \mathsf{Int}  \mathrel{\mid}  \mathit{x}  \mathrel{\ge}  \ottsym{0}  \}   \rangle^{ \ottnt{l_{{\mathrm{1}}}} } ~   {-1}   )   \\
 \multicolumn{2}{r}{(\E{CastMerge})} \\
  \longrightarrow _{  \mathsf{H}  }  &  \langle   \{ \mathit{x} \mathord{:}  \mathsf{Int}  \mathrel{\mid}  \mathsf{true}  \}   \mathord{ \overset{ \mathcal{S} }{\Rightarrow} }   \{ \mathit{x} \mathord{:}  \mathsf{Int}  \mathrel{\mid}  \mathit{x}  \mathrel{\ne}  \ottsym{0}  \}   \rangle^{ \ottnt{l_{{\mathrm{3}}}} } ~   {-1}   \\
 \multicolumn{2}{r}{\text{where } \mathcal{S}  \ottsym{=}   \set{   \{ \mathit{x} \mathord{:}  \mathsf{Int}  \mathrel{\mid}  \mathit{x} \,  \mathsf{mod}  \, \ottsym{2}  \mathrel{=}  \ottsym{0}  \}  ,   \{ \mathit{x} \mathord{:}  \mathsf{Int}  \mathrel{\mid}  \mathit{x}  \mathrel{\ge}  \ottsym{0}  \}   } } \\
 \multicolumn{2}{r}{(\E{CheckSet},  \mathsf{choose} ( \mathcal{S} )   \ottsym{=}   \{ \mathit{x} \mathord{:}  \mathsf{Int}  \mathrel{\mid}  \mathit{x} \,  \mathsf{mod}  \, \ottsym{2}  \mathrel{=}  \ottsym{0}  \} )} \\
  \longrightarrow _{  \mathsf{H}  }  &  \langle   \{ \mathit{x} \mathord{:}  \mathsf{Int}  \mathrel{\mid}  \mathit{x} \,  \mathsf{mod}  \, \ottsym{2}  \mathrel{=}  \ottsym{0}  \}   \mathord{ \overset{ \mathcal{S}' }{\Rightarrow} }   \{ \mathit{x} \mathord{:}  \mathsf{Int}  \mathrel{\mid}  \mathit{x}  \mathrel{\ne}  \ottsym{0}  \}   \rangle^{ \ottnt{l_{{\mathrm{3}}}} } ~  {} \\  &  \quad   \langle   \{ \mathit{x} \mathord{:}  \mathsf{Int}  \mathrel{\mid}  \mathit{x} \,  \mathsf{mod}  \, \ottsym{2}  \mathrel{=}  \ottsym{0}  \}  ,    {-1}  \,  \mathsf{mod}  \, \ottsym{2}  \mathrel{=}  \ottsym{0}  ,   {-1}   \rangle^{ \ottnt{l_{{\mathrm{3}}}} }   \\
 \multicolumn{2}{r}{\text{where } \mathcal{S}'  \ottsym{=}   \set{   \{ \mathit{x} \mathord{:}  \mathsf{Int}  \mathrel{\mid}  \mathit{x}  \mathrel{\ge}  \ottsym{0}  \}   } } \\
 \multicolumn{2}{r}{(\E{CastInner}/\E{CheckInner}/\E{Op})} \\
  \longrightarrow _{  \mathsf{H}  }  &  \langle   \{ \mathit{x} \mathord{:}  \mathsf{Int}  \mathrel{\mid}  \mathit{x} \,  \mathsf{mod}  \, \ottsym{2}  \mathrel{=}  \ottsym{0}  \}   \mathord{ \overset{ \mathcal{S}' }{\Rightarrow} }   \{ \mathit{x} \mathord{:}  \mathsf{Int}  \mathrel{\mid}  \mathit{x}  \mathrel{\ne}  \ottsym{0}  \}   \rangle^{ \ottnt{l_{{\mathrm{3}}}} } ~  {} \\  &  \quad   \langle   \{ \mathit{x} \mathord{:}  \mathsf{Int}  \mathrel{\mid}  \mathit{x} \,  \mathsf{mod}  \, \ottsym{2}  \mathrel{=}  \ottsym{0}  \}  ,   \ottsym{1}  \mathrel{=}  \ottsym{0}  ,   {-1}   \rangle^{ \ottnt{l_{{\mathrm{3}}}} }   \\
 \multicolumn{2}{r}{(\E{CastInner}/\E{CheckInner}/\E{Op})} \\
  \longrightarrow _{  \mathsf{H}  }  &  \langle   \{ \mathit{x} \mathord{:}  \mathsf{Int}  \mathrel{\mid}  \mathit{x} \,  \mathsf{mod}  \, \ottsym{2}  \mathrel{=}  \ottsym{0}  \}   \mathord{ \overset{ \mathcal{S}' }{\Rightarrow} }   \{ \mathit{x} \mathord{:}  \mathsf{Int}  \mathrel{\mid}  \mathit{x}  \mathrel{\ne}  \ottsym{0}  \}   \rangle^{ \ottnt{l_{{\mathrm{3}}}} } ~  {} \\  &  \quad   \langle   \{ \mathit{x} \mathord{:}  \mathsf{Int}  \mathrel{\mid}  \mathit{x} \,  \mathsf{mod}  \, \ottsym{2}  \mathrel{=}  \ottsym{0}  \}  ,   \mathsf{false}  ,   {-1}   \rangle^{ \ottnt{l_{{\mathrm{3}}}} }   \\
 \multicolumn{2}{r}{(\E{CastInner}/\E{CheckFail})} \\
  \longrightarrow _{  \mathsf{H}  }  &  \langle   \{ \mathit{x} \mathord{:}  \mathsf{Int}  \mathrel{\mid}  \mathit{x}  \mathrel{\ge}  \ottsym{0}  \}   \mathord{ \overset{ \emptyset }{\Rightarrow} }   \{ \mathit{x} \mathord{:}  \mathsf{Int}  \mathrel{\mid}  \mathit{x}  \mathrel{\ne}  \ottsym{0}  \}   \rangle^{ \ottnt{l_{{\mathrm{3}}}} } ~   \mathord{\Uparrow}  \ottnt{l_{{\mathrm{3}}}}  \\
 \multicolumn{2}{r}{(\E{CastRaise})} \\
  \longrightarrow _{  \mathsf{H}  }  &  \mathord{\Uparrow}  \ottnt{l_{{\mathrm{3}}}}  
\end{array} \]
\caption{Example of heedful \lambdah}
\label{fig:heedfulexample}
\end{figure}

Heedful \lambdah adds some evaluation rules to the universal ones
found in Figure~\ref{fig:opsem}.
First \E{TypeSet} takes a source program cast without an annotation
and annotates it with an empty set.
\E{CheckEmpty} is exactly like \ECheckNone, though we separate the
two to avoid conflating the empty annotation $ \bullet $ and the empty
set $ \emptyset $.
In \E{CheckSet}, we use an essentially unspecified function
$ \mathsf{choose} $ to pick a type from a type set to check. Using the
$ \mathsf{choose} $ function is theoretically expedient, as it hides all of
heedful \lambdah's nondeterminism. Nothing is inherently problematic
with this nondeterminism, but putting it in the reduction relation
itself complicates the proof of strong normalization that is necessary
for the proof relating classic and heedful \lambdah (Section~\ref{sec:soundness}).

For function types, we define $ \mathsf{dom} ( \mathcal{S} ) $ and $ \mathsf{cod} ( \mathcal{S} ) $ by mapping
the underlying function on types over the set. Note that this may
shrink the size of the set $\mathcal{S}$, but never grow it---there can't be
more unique (co)domain types in $\mathcal{S}$ than there are types.

The merge operator, used in \E{CastMerge}, merges two sets by unioning
the two type sets with the intermediate type, i.e.:
\[  \langle  \ottnt{T_{{\mathrm{2}}}}  \mathord{ \overset{ \mathcal{S}_{{\mathrm{2}}} }{\Rightarrow} }  \ottnt{T_{{\mathrm{3}}}}  \rangle^{ \ottnt{l_{{\mathrm{2}}}} } ~   (  \langle  \ottnt{T_{{\mathrm{1}}}}  \mathord{ \overset{ \mathcal{S}_{{\mathrm{1}}} }{\Rightarrow} }  \ottnt{T_{{\mathrm{2}}}}  \rangle^{ \ottnt{l_{{\mathrm{1}}}} } ~  \ottnt{e}  )   \,  \longrightarrow _{  \mathsf{H}  }  \,  \langle  \ottnt{T_{{\mathrm{1}}}}  \mathord{ \overset{   \mathcal{S}_{{\mathrm{1}}}  \cup  \mathcal{S}_{{\mathrm{2}}}   \cup   \set{  \ottnt{T_{{\mathrm{2}}}}  }   }{\Rightarrow} }  \ottnt{T_{{\mathrm{3}}}}  \rangle^{ \ottnt{l_{{\mathrm{2}}}} } ~  \ottnt{e}  \]
There are some subtle interactions here between the different
annotations: we won't merge casts that haven't yet stepped by
\E{TypeSet} because $ \mathsf{merge} _{  \mathsf{H}  }( \ottnt{T_{{\mathrm{1}}}} , \bullet , \ottnt{T_{{\mathrm{2}}}} , \bullet , \ottnt{T_{{\mathrm{3}}}} ) $ isn't
defined.

We demonstrate the heedful semantics by returning to the example from
Section~\ref{sec:opsem} in Figure~\ref{fig:heedfulexample}.
To highlight the difference between classic and heedful \lambdah, we
select a $ \mathsf{choose} $ function that has heedful check the refinements
out of order, failing on the check for evenness rather than the check
for positivity.
The real source of difference, however, is that \E{CastMerge} takes
the second blame label of the two casts it merges.
Taking the first wouldn't be right, either: suppose that the target
type of the $\ottnt{l_{{\mathrm{1}}}}$ cast wasn't $ \{ \mathit{x} \mathord{:}  \mathsf{Int}  \mathrel{\mid}  \mathit{x}  \mathrel{\ge}  \ottsym{0}  \} $, but some other
type that $ {-1} $ inhabits. Then classic \lambdah would blame
$\ottnt{l_{{\mathrm{2}}}}$, but heedful \lambdah would have held onto $\ottnt{l_{{\mathrm{1}}}}$.
The solution to this blame tracking problem is to hold onto blame
labels in annotations---which, again, is exactly what we do in eidetic
\lambdah.

\finishlater{another example, with function proxies?}

The syntactic proof of type soundness for heedful \lambdah appears in
an appendix in Appendix~\ref{app:heedfulsoundness}.
We also have ``source typing'' for heedful \lambdah: source programs
are well typed when $\ottnt{m}  \ottsym{=}   \mathsf{C} $ iff they are well typed when
$\ottnt{m}  \ottsym{=}   \mathsf{H} $. As a corollary, source programs are well typed in $ \mathsf{F} $
if and only if they are well typed in $ \mathsf{H} $.

\fi}

\section{Eidetic space efficiency}
\label{sec:eidetic}

Eidetic \lambdah uses \textit{coercions}\iffull, a more refined system of
annotations than heedful \lambdah's type sets\fi.
Coercions do two critical things\iffull{} that type sets don't\fi: they retain check order,
and they track blame. 
Our coercions are ultimately inspired by those of
\citet{Henglein94dynamic}; we discuss the relationship between
our coercions and his in related work (Section~\ref{sec:related}).
Recall the syntax of coercions from Figure~\ref{fig:syntax}:
\[ \begin{array}{r@{~}c@{~}l}
    \ottnt{c} &::=& \ottnt{r} \BNFALT \ottnt{c_{{\mathrm{1}}}}  \mapsto  \ottnt{c_{{\mathrm{2}}}} \\
    \ottnt{r} &::=&  \mathsf{nil}  \BNFALT   \{ \mathit{x} \mathord{:} \ottnt{B} \mathrel{\mid} \ottnt{e} \}^{ \ottnt{l} }  , \ottnt{r}  \\
\end{array}\]
Coercions come in two flavors: blame-annotated refinement lists
$\ottnt{r}$---zero or more refinement types, each annotated with a blame
label---and function coercions $\ottnt{c_{{\mathrm{1}}}}  \mapsto  \ottnt{c_{{\mathrm{2}}}}$. We write them as comma
separated lists, omitting the empty refinement list $ \mathsf{nil} $ when the
refinement list is non-empty.
We define the coercion well formedness rules, an additional typing
rule, and reduction rules for eidetic \lambdah in
Figure~\ref{fig:eideticsemantics}.
To ease the exposition, our explanation doesn't mirror the rule
groupings in the figure.
\begin{bigfigure}
  \hdr{Coercion implication predicate: axioms}{\qquad \fbox{$ \{ \mathit{x} \mathord{:} \ottnt{B} \mathrel{\mid} \ottnt{e_{{\mathrm{1}}}} \}  \, \supset \,  \{ \mathit{x} \mathord{:} \ottnt{B} \mathrel{\mid} \ottnt{e_{{\mathrm{2}}}} \} $}}

  ~

  \begin{minipage}[t]{.75\linewidth}
  \begin{enumerate}
  \item[(\textbf{Reflexivity})] \labelasm{implrefl} If $ \mathord{  \vdash _{  \mathsf{E}  } }~  \{ \mathit{x} \mathord{:} \ottnt{B} \mathrel{\mid} \ottnt{e} \}  $ then $ \{ \mathit{x} \mathord{:} \ottnt{B} \mathrel{\mid} \ottnt{e} \}  \, \supset \,  \{ \mathit{x} \mathord{:} \ottnt{B} \mathrel{\mid} \ottnt{e} \} $.
  \item[(\textbf{Transitivity})] \labelasm{impltrans} If $ \{ \mathit{x} \mathord{:} \ottnt{B} \mathrel{\mid} \ottnt{e_{{\mathrm{1}}}} \}  \, \supset \,  \{ \mathit{x} \mathord{:} \ottnt{B} \mathrel{\mid} \ottnt{e_{{\mathrm{2}}}} \} $ and $ \{ \mathit{x} \mathord{:} \ottnt{B} \mathrel{\mid} \ottnt{e_{{\mathrm{2}}}} \}  \, \supset \,  \{ \mathit{x} \mathord{:} \ottnt{B} \mathrel{\mid} \ottnt{e_{{\mathrm{3}}}} \} $ then $ \{ \mathit{x} \mathord{:} \ottnt{B} \mathrel{\mid} \ottnt{e_{{\mathrm{1}}}} \}  \, \supset \,  \{ \mathit{x} \mathord{:} \ottnt{B} \mathrel{\mid} \ottnt{e_{{\mathrm{3}}}} \} $.
  \item[(\textbf{Adequacy})] \labelasm{impladeq} If $ \{ \mathit{x} \mathord{:} \ottnt{B} \mathrel{\mid} \ottnt{e_{{\mathrm{1}}}} \}  \, \supset \,  \{ \mathit{x} \mathord{:} \ottnt{B} \mathrel{\mid} \ottnt{e_{{\mathrm{2}}}} \} $ then $ \forall   \ottnt{k}  \in \mathcal{K}_{ \ottnt{B} }   . ~   \ottnt{e_{{\mathrm{1}}}}  [  \ottnt{k} / \mathit{x}  ]  \,  \longrightarrow ^{*}_{  \mathsf{E}  }  \,  \mathsf{true}   \, \text{implies} \,  \ottnt{e_{{\mathrm{2}}}}  [  \ottnt{k} / \mathit{x}  ]  \,  \longrightarrow ^{*}_{  \mathsf{E}  }  \,  \mathsf{true} $.
  \item[(\textbf{Decidability})] \labelasm{impldec} For all $ \mathord{  \vdash _{  \mathsf{E}  } }~  \{ \mathit{x} \mathord{:} \ottnt{B} \mathrel{\mid} \ottnt{e_{{\mathrm{1}}}} \}  $ and $ \mathord{  \vdash _{  \mathsf{E}  } }~  \{ \mathit{x} \mathord{:} \ottnt{B} \mathrel{\mid} \ottnt{e_{{\mathrm{2}}}} \}  $, it is decidable whether
    $ \{ \mathit{x} \mathord{:} \ottnt{B} \mathrel{\mid} \ottnt{e_{{\mathrm{1}}}} \}  \, \supset \,  \{ \mathit{x} \mathord{:} \ottnt{B} \mathrel{\mid} \ottnt{e_{{\mathrm{2}}}} \} $.
  \end{enumerate}
  \end{minipage}

  \hdr{Coercion well formedness and term typing}
    {\qquad \fbox{$ \mathord{  \vdash _{ \ottnt{m} } }~ \ottnt{c}   \mathrel{\parallel}   \ottnt{T_{{\mathrm{1}}}}  \Rightarrow  \ottnt{T_{{\mathrm{2}}}} $} \qquad \fbox{$ \Gamma   \vdash _{ \ottnt{m} }  \ottnt{e}  :  \ottnt{T} $}}

  \sidebyside[.57][.39]
    {\ottusedrule{\ottdruleAXXRefine{}}}
    {\ottusedrule{\ottdruleAXXFun{}}}

  {\ottusedrule{\ottdruleTXXStack{}}}

  \hdr{Values and operational semantics}{\qquad \fbox{$ \mathsf{val} _{  \mathsf{E}  }~ \ottnt{e} $} \qquad \fbox{$\ottnt{e_{{\mathrm{1}}}} \,  \longrightarrow _{  \mathsf{E}  }  \, \ottnt{e_{{\mathrm{2}}}}$}}

%  {\iffull \sidebyside
%    {\ottusedrule{\ottdruleVXXConst{}}}
%    {\ottusedrule{\ottdruleVXXAbs{}}} \fi}
  {\ottusedrule{\ottdruleVXXProxyE{}}}
  \sidebysidesqueeze[.42][.56][.01][t]
    {\ottusedrule{\ottdruleEXXCoerce{}}}
    {\ottusedrule{\ottdruleEXXCoerceStack{}}}
  {\ottusedrule{\ottdruleEXXStackPop{}}}
  \sidebyside[.51][.45]
    {\ottusedrule{\ottdruleEXXStackInner{}}}
    {\ottusedrule{\ottdruleEXXStackRaise{}}}
  {\ottusedrule{\ottdruleEXXStackDone{}}}

  \hdr{Cast translation and coercion operations}{}
  \iffull
  \vspace*{-4pt}
  \fi
  \sidebyside[.33][.6][t]
    {\[ \begin{array}{rcl}
\iffull  \mathsf{merge} _{  \mathsf{E}  }( \ottnt{T_{{\mathrm{1}}}} , \ottnt{c_{{\mathrm{1}}}} , \ottnt{T_{{\mathrm{2}}}} , \ottnt{c_{{\mathrm{2}}}} , \ottnt{T_{{\mathrm{3}}}} )  &=&  \mathsf{join} ( \ottnt{c_{{\mathrm{1}}}} , \ottnt{c_{{\mathrm{2}}}} )  \\ \fi
         \mathsf{dom} ( \ottnt{c_{{\mathrm{1}}}}  \mapsto  \ottnt{c_{{\mathrm{2}}}} )  &=& \ottnt{c_{{\mathrm{1}}}} \\
         \mathsf{cod} ( \ottnt{c_{{\mathrm{1}}}}  \mapsto  \ottnt{c_{{\mathrm{2}}}} )  &=& \ottnt{c_{{\mathrm{2}}}}
        \end{array} \]}
    {\[ \begin{array}{rcl}
       \mathsf{coerce} (  \{ \mathit{x} \mathord{:} \ottnt{B} \mathrel{\mid} \ottnt{e_{{\mathrm{1}}}} \}  ,  \{ \mathit{x} \mathord{:} \ottnt{B} \mathrel{\mid} \ottnt{e_{{\mathrm{2}}}} \}  , \ottnt{l} )  &=&  \{ \mathit{x} \mathord{:} \ottnt{B} \mathrel{\mid} \ottnt{e_{{\mathrm{2}}}} \}^{ \ottnt{l} }  \\
       \mathsf{coerce} (  \ottnt{T_{{\mathrm{11}}}} \mathord{ \rightarrow } \ottnt{T_{{\mathrm{12}}}}  ,  \ottnt{T_{{\mathrm{21}}}} \mathord{ \rightarrow } \ottnt{T_{{\mathrm{22}}}}  , \ottnt{l} )  &=& 
         \mathsf{coerce} ( \ottnt{T_{{\mathrm{21}}}} , \ottnt{T_{{\mathrm{11}}}} , \ottnt{l} )   \mapsto   \mathsf{coerce} ( \ottnt{T_{{\mathrm{12}}}} , \ottnt{T_{{\mathrm{22}}}} , \ottnt{l} )  \\
      \end{array} \]}
  \sidebyside[.45][.45][t]
  {\[ \begin{array}{rcl}
     \mathsf{join} (  \{ \mathit{x} \mathord{:} \ottnt{B} \mathrel{\mid} \ottnt{e} \}^{ \ottnt{l} }  , \mathsf{nil} )  &=&  \{ \mathit{x} \mathord{:} \ottnt{B} \mathrel{\mid} \ottnt{e} \}^{ \ottnt{l} }  \\
     \mathsf{join} (  \{ \mathit{x} \mathord{:} \ottnt{B} \mathrel{\mid} \ottnt{e} \}^{ \ottnt{l} }  , \ottnt{r} )  &=&   \{ \mathit{x} \mathord{:} \ottnt{B} \mathrel{\mid} \ottnt{e} \}^{ \ottnt{l} }  , \mathsf{drop} \, \ottsym{(}  \ottnt{r}  \ottsym{,}   \{ \mathit{x} \mathord{:} \ottnt{B} \mathrel{\mid} \ottnt{e} \}   \ottsym{)}  \\
  \end{array} \]}
  {\[ \begin{array}{rcl}
     \mathsf{join} ( \mathsf{nil} , \ottnt{r_{{\mathrm{2}}}} )  &=& \ottnt{r_{{\mathrm{2}}}} \\
     \mathsf{join} ( \ottsym{(}    \{ \mathit{x} \mathord{:} \ottnt{B} \mathrel{\mid} \ottnt{e} \}^{ \ottnt{l} }  , \ottnt{r_{{\mathrm{1}}}}   \ottsym{)} , \ottnt{r_{{\mathrm{2}}}} )  &=&  \mathsf{join} (  \{ \mathit{x} \mathord{:} \ottnt{B} \mathrel{\mid} \ottnt{e} \}^{ \ottnt{l} }  ,  \mathsf{join} ( \ottnt{r_{{\mathrm{1}}}} , \ottnt{r_{{\mathrm{2}}}} )  )  \\
     \mathsf{join} ( \ottnt{c_{{\mathrm{11}}}}  \mapsto  \ottnt{c_{{\mathrm{12}}}} , \ottnt{c_{{\mathrm{21}}}}  \mapsto  \ottnt{c_{{\mathrm{22}}}} )  &=&  \mathsf{join} ( \ottnt{c_{{\mathrm{21}}}} , \ottnt{c_{{\mathrm{11}}}} )   \mapsto   \mathsf{join} ( \ottnt{c_{{\mathrm{12}}}} , \ottnt{c_{{\mathrm{22}}}} ) 
  \end{array} \]}
  \sidebyside[.7][.26][t]
  {\[ \begin{array}{rcl}
    \mathsf{drop} \, \ottsym{(}  \mathsf{nil}  \ottsym{,}   \{ \mathit{x} \mathord{:} \ottnt{B} \mathrel{\mid} \ottnt{e} \}   \ottsym{)} &=&  \mathsf{nil}  \\
    \mathsf{drop} \, \ottsym{(}  \ottsym{(}    \{ \mathit{x} \mathord{:} \ottnt{B} \mathrel{\mid} \ottnt{e_{{\mathrm{1}}}} \}^{ \ottnt{l} }  , \ottnt{r}   \ottsym{)}  \ottsym{,}   \{ \mathit{x} \mathord{:} \ottnt{B} \mathrel{\mid} \ottnt{e} \}   \ottsym{)} &=& \begin{cases}
      \mathsf{drop} \, \ottsym{(}  \ottnt{r}  \ottsym{,}   \{ \mathit{x} \mathord{:} \ottnt{B} \mathrel{\mid} \ottnt{e} \}   \ottsym{)} &  \{ \mathit{x} \mathord{:} \ottnt{B} \mathrel{\mid} \ottnt{e} \}  \, \supset \,  \{ \mathit{x} \mathord{:} \ottnt{B} \mathrel{\mid} \ottnt{e_{{\mathrm{1}}}} \}  \\
        \{ \mathit{x} \mathord{:} \ottnt{B} \mathrel{\mid} \ottnt{e_{{\mathrm{1}}}} \}^{ \ottnt{l} }  , \mathsf{drop} \, \ottsym{(}  \ottnt{r}  \ottsym{,}   \{ \mathit{x} \mathord{:} \ottnt{B} \mathrel{\mid} \ottnt{e} \}   \ottsym{)}  &  \{ \mathit{x} \mathord{:} \ottnt{B} \mathrel{\mid} \ottnt{e} \}  \, \not \supset \,  \{ \mathit{x} \mathord{:} \ottnt{B} \mathrel{\mid} \ottnt{e_{{\mathrm{1}}}} \} 
    \end{cases}
  \end{array} \]}
  {\[ \begin{array}{@{\qquad}rcl}
      \mathord{\checkmark}   \vee ( \ottnt{e_{{\mathrm{1}}}}  =  \ottnt{e_{{\mathrm{2}}}} )  &=&  \mathord{\checkmark}  \\
      \mathord{?}   \vee ( \ottnt{e_{{\mathrm{1}}}}  =  \ottnt{e_{{\mathrm{2}}}} )  &=& \begin{cases}
       \mathord{\checkmark}  & \ottnt{e_{{\mathrm{1}}}}  \ottsym{=}  \ottnt{e_{{\mathrm{2}}}} \\
       \mathord{?}  & \text{otherwise}
    \end{cases}
  \end{array} \]}
  \vspace*{-10pt}
  \caption{Typing rules and operational semantics for eidetic \lambdah}
  \label{fig:eideticsemantics}
\end{bigfigure}

As a general intuition, coercions are plans for checking: they contain
precisely those types to be checked.
Refinement lists are well formed for casts between $ \{ \mathit{x} \mathord{:} \ottnt{B} \mathrel{\mid} \ottnt{e_{{\mathrm{1}}}} \} $ and
$ \{ \mathit{x} \mathord{:} \ottnt{B} \mathrel{\mid} \ottnt{e_{{\mathrm{2}}}} \} $ when: (a) every type in the list is a blame-annotated,
well formed refinement of $\ottnt{B}$, i.e., all the types are of the form
$ \{ \mathit{x} \mathord{:} \ottnt{B} \mathrel{\mid} \ottnt{e} \}^{ \ottnt{l} } $ and are therefore similar to the indices; (b) there
are no duplicated types in the list; and (c) the target type
$ \{ \mathit{x} \mathord{:} \ottnt{B} \mathrel{\mid} \ottnt{e_{{\mathrm{2}}}} \} $ is implied by some other type in the list.
Note that the input type for all refinement lists can be any well
formed refinement---this corresponds to the intuition that base types
have no negative parts, i.e., casts between refinements ignore the
type on the left.
Finally, we simply write ``no duplicates in $\ottnt{r}$''---it is an
invariant during the evaluation of source programs.
Function coercions, on the other hand, have a straightforward
(contravariant) well formedness rule.

The \E{Coerce} rule translates source-program casts to coercions:
$ \mathsf{coerce} ( \ottnt{T_{{\mathrm{1}}}} , \ottnt{T_{{\mathrm{2}}}} , \ottnt{l} ) $ is a coercion representing exactly the
checking done by the cast $ \langle  \ottnt{T_{{\mathrm{1}}}}  \mathord{ \overset{\bullet}{\Rightarrow} }  \ottnt{T_{{\mathrm{2}}}}  \rangle^{ \ottnt{l} } ~     $\!\!. All of the refinement
types in $ \mathsf{coerce} ( \ottnt{T_{{\mathrm{1}}}} , \ottnt{T_{{\mathrm{2}}}} , \ottnt{l} ) $ are annotated with the blame label
$\ottnt{l}$, since that's the label that would be blamed if the cast
failed at that type.
Since a coercion is a complete plan for checking, a coercion
annotation obviates the need for type indices and blame labels. To
this end, \E{Coerce} drops the blame label from the cast, replacing it
with an empty label. We keep the type indices so that we can reuse
\ECastMerge\ from the universal semantics, and also as a technical
device in the preservation proof.

The actual checking of coercions rests on the treatment of refinement
lists: function coercions are expanded as functions are applied by
\E{Unwrap}, so they don't need much special treatment beyond a
definition for $ \mathsf{dom} $ and $ \mathsf{cod} $.
Eidetic \lambdah uses \textit{coercion stacks}
$ \langle   \{ \mathit{x} \mathord{:} \ottnt{B} \mathrel{\mid} \ottnt{e_{{\mathrm{1}}}} \}  ,  \ottnt{s} ,  \ottnt{r} ,  \ottnt{k} ,  \ottnt{e_{{\mathrm{2}}}}  \rangle^{\bullet} $ to evaluate refinement lists. Coercion
stacks are type checked by \T{Stack} (in
Figure~\ref{fig:eideticsemantics}). We explain the operational
semantics before explaining the typing rule.
Coercion stacks are runtime-only entities comprising five parts: a
target type, a status, a pending refinement list, a constant
scrutinee, and a checking term.
We keep the target type of the coercion for preservation's sake.
The status bit $\ottnt{s}$ is either $ \mathord{\checkmark} $ or $ \mathord{?} $: when the status
is $ \mathord{\checkmark} $, we are currently checking or have already checked the
target type $ \{ \mathit{x} \mathord{:} \ottnt{B} \mathrel{\mid} \ottnt{e_{{\mathrm{1}}}} \} $; when it is $ \mathord{?} $, we haven't.
The pending refinement list $\ottnt{r}$ holds those checks not yet
done. When $\ottnt{s}  \ottsym{=}   \mathord{?} $, the target type is still in $\ottnt{r}$.
The scrutinee $\ottnt{k}$ is the constant we're checking; the checking
term $\ottnt{e_{{\mathrm{2}}}}$ is \textit{either} the scrutinee $\ottnt{k}$ itself, or it is
an active check on $\ottnt{k}$.

The evaluation of a coercion stack proceeds as follows.
First, \E{CoerceStack} starts a coercion stack when a cast between
refinements meets a constant, recording the target type, setting the
status to $ \mathord{?} $, and setting the checking term to $\ottnt{k}$.
Then \E{StackPop} starts an active check on the first type in the
refinement list, using its blame label on the active check---possibly
updating the status if the type being popped from the list is the
target type. The active check runs by the congruence rule \E{StackInner},
eventually returning $\ottnt{k}$ itself or blame. In the latter case,
\E{StackRaise} propagates the blame. If not, then the scrutinee is
$\ottnt{k}$ once more and \E{StackPop} can fire again.
Eventually, the refinement list is exhausted, and \E{StackDone}
returns $\ottnt{k}$.

Now we can explain \T{Stack}'s many jobs. It must recapitulate
\A{Refine}, but not exactly---since eventually the target type will be
checked and no longer appear in $\ottnt{r}$. The status $\ottnt{s}$ differentiates
what our requirement is: when $\ottnt{s}  \ottsym{=}   \mathord{?} $, the target type is in
$\ottnt{r}$. When $\ottnt{s}  \ottsym{=}   \mathord{\checkmark} $, we either know that $\ottnt{k}$ inhabits the
target type or that we are currently checking the target type (i.e.,
an active check of the target type at some blame label reduces to our
current checking term).

\iffull
Finally, we must define a merge operator, $ \mathsf{merge} _{ \mathsf{E} }$. We
define it in terms of
\else
Finally, we need to define how to merge casts. We use
\fi the $ \mathsf{join} $ operator, which is very nearly
concatenation on refinement lists and a contravariant homomorphism on
function coercions. It's not concatenation because it uses an
implication predicate, the pre-order $ \supset $, to eliminate
duplicates (because $ \supset $ is reflexive\refasm{implrefl}) and hide
subsumed types (because $ \supset $ is adequate\refasm{impladeq}). We
read $ \{ \mathit{x} \mathord{:} \ottnt{B} \mathrel{\mid} \ottnt{e_{{\mathrm{1}}}} \}  \, \supset \,  \{ \mathit{x} \mathord{:} \ottnt{B} \mathrel{\mid} \ottnt{e_{{\mathrm{2}}}} \} $ as ``$ \{ \mathit{x} \mathord{:} \ottnt{B} \mathrel{\mid} \ottnt{e_{{\mathrm{1}}}} \} $ implies
$ \{ \mathit{x} \mathord{:} \ottnt{B} \mathrel{\mid} \ottnt{e_{{\mathrm{2}}}} \} $''.
When eliminating types, $ \mathsf{join} $ always chooses the leftmost blame
label. Contravariance means that $ \mathsf{join} ( \ottnt{c_{{\mathrm{1}}}} , \ottnt{c_{{\mathrm{2}}}} ) $ takes leftmost labels
in positive positions and rightmost labels in negative ones.
The $ \mathsf{coerce} $ metafunction and $ \mathsf{join} $ operator work together to
make sure that the refinement lists are correctly ordered.
As we show below, `correctly ordered' means the positive parts take
older labels and negative parts take newer ones.
\iffull As for heedful \lambdah, \fi \ECastMerge\ is slightly subtle---we never
merge casts with $ \bullet $ as an annotation because such merges
aren't defined.

In Figure~\ref{fig:eideticsemantics}, we only give the axioms for
$ \supset $: it must be an adequate, decidable pre-order. Syntactic
type equality is the simplest implementation of the $ \supset $ predicate, but
the reflexive transitive closure of any adequate decidable relation
would work.

By way of example, consider a cast from $\ottnt{T_{{\mathrm{1}}}}  \ottsym{=}    \{ \mathit{x} \mathord{:}  \mathsf{Int}  \mathrel{\mid}  \mathit{x}  \mathrel{\ge}  \ottsym{0}  \}  \mathord{ \rightarrow }  \{ \mathit{x} \mathord{:}  \mathsf{Int}  \mathrel{\mid}  \mathit{x}  \mathrel{\ge}  \ottsym{0}  \}  $ to $\ottnt{T_{{\mathrm{2}}}}  \ottsym{=}    \{ \mathit{x} \mathord{:}  \mathsf{Int}  \mathrel{\mid}  \mathsf{true}  \}  \mathord{ \rightarrow }  \{ \mathit{x} \mathord{:}  \mathsf{Int}  \mathrel{\mid}  \mathit{x}  \mathrel{>} \ottsym{0}  \}  $. 
For brevity, we refer to the domains as $\ottnt{T_{\ottmv{i}\,{\mathrm{1}}}}$ and the codomains as $\ottnt{T_{\ottmv{i}\,{\mathrm{2}}}}$.
We find that $  (  \langle  \ottnt{T_{{\mathrm{1}}}}  \mathord{ \overset{\bullet}{\Rightarrow} }  \ottnt{T_{{\mathrm{2}}}}  \rangle^{ \ottnt{l} } ~  \mathit{v_{{\mathrm{1}}}}  )  ~ \mathit{v_{{\mathrm{2}}}} $ steps in classic \lambdah to:
\[ \begin{array}{@{}l@{}} 
   \langle   \{ \mathit{x} \mathord{:}  \mathsf{Int}  \mathrel{\mid}  \mathit{x}  \mathrel{\ge}  \ottsym{0}  \}   \mathord{ \overset{\bullet}{\Rightarrow} }   \{ \mathit{x} \mathord{:}  \mathsf{Int}  \mathrel{\mid}  \mathit{x}  \mathrel{>} \ottsym{0}  \}   \rangle^{ \ottnt{l} } ~  {} \\  \quad   (  \mathit{v_{{\mathrm{1}}}} ~  (  \langle   \{ \mathit{x} \mathord{:}  \mathsf{Int}  \mathrel{\mid}  \mathsf{true}  \}   \mathord{ \overset{\bullet}{\Rightarrow} }   \{ \mathit{x} \mathord{:}  \mathsf{Int}  \mathrel{\mid}  \mathit{x}  \mathrel{\ge}  \ottsym{0}  \}   \rangle^{ \ottnt{l} } ~  \mathit{v_{{\mathrm{2}}}}  )   )  
\end{array} \]
Note that $\ottnt{T_{{\mathrm{1}}}}$'s domain is checked but its codomain isn't; the
reverse is true for $\ottnt{T_{{\mathrm{2}}}}$. When looking at a cast, we can read off
which refinements are checked by looking at the positive parts of the
target type and the negative parts of the source type. The
relationship between casts and polarity is not a new
one~\cite{Findler06projections,Gronski07unifying,Herman07space,Wadler09blame,Greenberg13thesis}.
Unlike casts, coercions directly express the sequence of checks to be
performed. Consider the coercion generated from the cast above,
recalling that $\ottnt{T_{\ottmv{i}\,{\mathrm{1}}}}$ and $\ottnt{T_{\ottmv{i}\,{\mathrm{2}}}}$ are the domains and codomains
of $\ottnt{T_{{\mathrm{1}}}}$ and $\ottnt{T_{{\mathrm{2}}}}$:
\[ \begin{array}{rl} 
           &   (  \langle  \ottnt{T_{{\mathrm{1}}}}  \mathord{ \overset{\bullet}{\Rightarrow} }  \ottnt{T_{{\mathrm{2}}}}  \rangle^{ \ottnt{l} } ~  \mathit{v_{{\mathrm{1}}}}  )  ~ \mathit{v_{{\mathrm{2}}}}  \\
   \longrightarrow _{  \mathsf{E}  }  &   (  \langle  \ottnt{T_{{\mathrm{1}}}}  \mathord{ \overset{ \ottnt{c} }{\Rightarrow} }  \ottnt{T_{{\mathrm{2}}}}  \rangle^{\bullet} ~  \mathit{v_{{\mathrm{1}}}}  )  ~ \mathit{v_{{\mathrm{2}}}}  \\
  \multicolumn{2}{r}{\text{where } \ottnt{c}  \ottsym{=}   \{ \mathit{x} \mathord{:}  \mathsf{Int}  \mathrel{\mid}  \mathit{x}  \mathrel{\ge}  \ottsym{0}  \}^{ \ottnt{l} }   \mapsto   \{ \mathit{x} \mathord{:}  \mathsf{Int}  \mathrel{\mid}  \mathit{x}  \mathrel{>} \ottsym{0}  \}^{ \ottnt{l} } } \\
   \longrightarrow _{  \mathsf{E}  }  &   (  \langle   \ottnt{T_{{\mathrm{11}}}} \mathord{ \rightarrow } \ottnt{T_{{\mathrm{12}}}}   \mathord{ \overset{ \ottnt{c} }{\Rightarrow} }   \ottnt{T_{{\mathrm{21}}}} \mathord{ \rightarrow } \ottnt{T_{{\mathrm{22}}}}   \rangle^{\bullet} ~  \mathit{v_{{\mathrm{1}}}}  )  ~ \mathit{v_{{\mathrm{2}}}}  \\ 
   \longrightarrow _{  \mathsf{E}  }  &  \langle  \ottnt{T_{{\mathrm{12}}}}  \mathord{ \overset{  \{ \mathit{x} \mathord{:}  \mathsf{Int}  \mathrel{\mid}  \mathit{x}  \mathrel{>} \ottsym{0}  \}^{ \ottnt{l} }  }{\Rightarrow} }  \ottnt{T_{{\mathrm{22}}}}  \rangle^{\bullet} ~   (  \mathit{v_{{\mathrm{1}}}} ~  (  \langle  \ottnt{T_{{\mathrm{21}}}}  \mathord{ \overset{  \{ \mathit{x} \mathord{:}  \mathsf{Int}  \mathrel{\mid}  \mathit{x}  \mathrel{\ge}  \ottsym{0}  \}^{ \ottnt{l} }  }{\Rightarrow} }  \ottnt{T_{{\mathrm{11}}}}  \rangle^{\bullet} ~  \mathit{v_{{\mathrm{2}}}}  )   )  
\end{array} \]
In this example, there is only a single blame label, $\ottnt{l}$. Tracking
blame labels is critical for exactly matching classic \lambdah's
behavior. The examples rely on $ \supset $ being reflexive.
\begin{figure}[t]
\[ \begin{array}{@{}r@{~}l}
  \ottnt{e} = &  \langle   \{ \mathit{x} \mathord{:}  \mathsf{Int}  \mathrel{\mid}  \mathit{x} \,  \mathsf{mod}  \, \ottsym{2}  \mathrel{=}  \ottsym{0}  \}   \mathord{ \overset{\bullet}{\Rightarrow} }   \{ \mathit{x} \mathord{:}  \mathsf{Int}  \mathrel{\mid}  \mathit{x}  \mathrel{\ne}  \ottsym{0}  \}   \rangle^{ \ottnt{l_{{\mathrm{3}}}} } ~  {} \\  &   ~~   (  \langle   \{ \mathit{x} \mathord{:}  \mathsf{Int}  \mathrel{\mid}  \mathit{x}  \mathrel{\ge}  \ottsym{0}  \}   \mathord{ \overset{\bullet}{\Rightarrow} }   \{ \mathit{x} \mathord{:}  \mathsf{Int}  \mathrel{\mid}  \mathit{x} \,  \mathsf{mod}  \, \ottsym{2}  \mathrel{=}  \ottsym{0}  \}   \rangle^{ \ottnt{l_{{\mathrm{2}}}} } ~  {} \\  &  \quad   (  \langle   \{ \mathit{x} \mathord{:}  \mathsf{Int}  \mathrel{\mid}  \mathsf{true}  \}   \mathord{ \overset{\bullet}{\Rightarrow} }   \{ \mathit{x} \mathord{:}  \mathsf{Int}  \mathrel{\mid}  \mathit{x}  \mathrel{\ge}  \ottsym{0}  \}   \rangle^{ \ottnt{l_{{\mathrm{1}}}} } ~   {-1}   )   )    \\
 \multicolumn{2}{r}{(\E{Coerce})} \\
  \longrightarrow _{  \mathsf{E}  }  &  \langle   \{ \mathit{x} \mathord{:}  \mathsf{Int}  \mathrel{\mid}  \mathit{x} \,  \mathsf{mod}  \, \ottsym{2}  \mathrel{=}  \ottsym{0}  \}   \mathord{ \overset{  \{ \mathit{x} \mathord{:}  \mathsf{Int}  \mathrel{\mid}  \mathit{x}  \mathrel{\ne}  \ottsym{0}  \}^{ \ottnt{l_{{\mathrm{3}}}} }  }{\Rightarrow} }   \{ \mathit{x} \mathord{:}  \mathsf{Int}  \mathrel{\mid}  \mathit{x}  \mathrel{\ne}  \ottsym{0}  \}   \rangle^{ \ottnt{l_{{\mathrm{3}}}} } ~  {} \\  &   ~~   (  \langle   \{ \mathit{x} \mathord{:}  \mathsf{Int}  \mathrel{\mid}  \mathit{x}  \mathrel{\ge}  \ottsym{0}  \}   \mathord{ \overset{\bullet}{\Rightarrow} }   \{ \mathit{x} \mathord{:}  \mathsf{Int}  \mathrel{\mid}  \mathit{x} \,  \mathsf{mod}  \, \ottsym{2}  \mathrel{=}  \ottsym{0}  \}   \rangle^{ \ottnt{l_{{\mathrm{2}}}} } ~  {} \\  &  \quad   (  \langle   \{ \mathit{x} \mathord{:}  \mathsf{Int}  \mathrel{\mid}  \mathsf{true}  \}   \mathord{ \overset{\bullet}{\Rightarrow} }   \{ \mathit{x} \mathord{:}  \mathsf{Int}  \mathrel{\mid}  \mathit{x}  \mathrel{\ge}  \ottsym{0}  \}   \rangle^{ \ottnt{l_{{\mathrm{1}}}} } ~   {-1}   )   )    \\
 \multicolumn{2}{r}{(\E{CastInner}/\E{Coerce})} \\ 
  \longrightarrow _{  \mathsf{E}  }  &  \langle   \{ \mathit{x} \mathord{:}  \mathsf{Int}  \mathrel{\mid}  \mathit{x} \,  \mathsf{mod}  \, \ottsym{2}  \mathrel{=}  \ottsym{0}  \}   \mathord{ \overset{  \{ \mathit{x} \mathord{:}  \mathsf{Int}  \mathrel{\mid}  \mathit{x}  \mathrel{\ne}  \ottsym{0}  \}^{ \ottnt{l_{{\mathrm{3}}}} }  }{\Rightarrow} }   \{ \mathit{x} \mathord{:}  \mathsf{Int}  \mathrel{\mid}  \mathit{x}  \mathrel{\ne}  \ottsym{0}  \}   \rangle^{ \ottnt{l_{{\mathrm{3}}}} } ~  {} \\  &   ~~   (  \langle   \{ \mathit{x} \mathord{:}  \mathsf{Int}  \mathrel{\mid}  \mathit{x}  \mathrel{\ge}  \ottsym{0}  \}   \mathord{ \overset{  \{ \mathit{x} \mathord{:}  \mathsf{Int}  \mathrel{\mid}  \mathit{x} \,  \mathsf{mod}  \, \ottsym{2}  \mathrel{=}  \ottsym{0}  \}^{ \ottnt{l_{{\mathrm{2}}}} }  }{\Rightarrow} }   \{ \mathit{x} \mathord{:}  \mathsf{Int}  \mathrel{\mid}  \mathit{x} \,  \mathsf{mod}  \, \ottsym{2}  \mathrel{=}  \ottsym{0}  \}   \rangle^{ \ottnt{l_{{\mathrm{2}}}} } ~  {} \\  &  \quad   (  \langle   \{ \mathit{x} \mathord{:}  \mathsf{Int}  \mathrel{\mid}  \mathsf{true}  \}   \mathord{ \overset{\bullet}{\Rightarrow} }   \{ \mathit{x} \mathord{:}  \mathsf{Int}  \mathrel{\mid}  \mathit{x}  \mathrel{\ge}  \ottsym{0}  \}   \rangle^{ \ottnt{l_{{\mathrm{1}}}} } ~   {-1}   )   )    \\
 \multicolumn{2}{r}{(\ECastMerge)} \\ 
  \longrightarrow _{  \mathsf{E}  }  &  \langle   \{ \mathit{x} \mathord{:}  \mathsf{Int}  \mathrel{\mid}  \mathit{x}  \mathrel{\ge}  \ottsym{0}  \}   \mathord{ \overset{ \ottnt{r'} }{\Rightarrow} }   \{ \mathit{x} \mathord{:}  \mathsf{Int}  \mathrel{\mid}  \mathit{x}  \mathrel{\ne}  \ottsym{0}  \}   \rangle^{ \ottnt{l_{{\mathrm{3}}}} } ~  {} \\  &   ~~   (  \langle   \{ \mathit{x} \mathord{:}  \mathsf{Int}  \mathrel{\mid}  \mathsf{true}  \}   \mathord{ \overset{\bullet}{\Rightarrow} }   \{ \mathit{x} \mathord{:}  \mathsf{Int}  \mathrel{\mid}  \mathit{x}  \mathrel{\ge}  \ottsym{0}  \}   \rangle^{ \ottnt{l_{{\mathrm{1}}}} } ~   {-1}   )    \\
 \multicolumn{2}{l}{\text{where } \ottnt{r'}  \ottsym{=}    \{ \mathit{x} \mathord{:}  \mathsf{Int}  \mathrel{\mid}  \mathit{x} \,  \mathsf{mod}  \, \ottsym{2}  \mathrel{=}  \ottsym{0}  \}^{ \ottnt{l_{{\mathrm{2}}}} }  ,  \{ \mathit{x} \mathord{:}  \mathsf{Int}  \mathrel{\mid}  \mathit{x}  \mathrel{\ne}  \ottsym{0}  \}^{ \ottnt{l_{{\mathrm{3}}}} }  } \\
 \multicolumn{2}{r}{(\E{CastInner}/\E{Coerce})} \\ 
  \longrightarrow _{  \mathsf{E}  }  &  \langle   \{ \mathit{x} \mathord{:}  \mathsf{Int}  \mathrel{\mid}  \mathit{x}  \mathrel{\ge}  \ottsym{0}  \}   \mathord{ \overset{ \ottnt{r'} }{\Rightarrow} }   \{ \mathit{x} \mathord{:}  \mathsf{Int}  \mathrel{\mid}  \mathit{x}  \mathrel{\ne}  \ottsym{0}  \}   \rangle^{ \ottnt{l_{{\mathrm{3}}}} } ~  {} \\  &   ~~   (  \langle   \{ \mathit{x} \mathord{:}  \mathsf{Int}  \mathrel{\mid}  \mathsf{true}  \}   \mathord{ \overset{  \{ \mathit{x} \mathord{:}  \mathsf{Int}  \mathrel{\mid}  \mathit{x}  \mathrel{\ge}  \ottsym{0}  \}^{ \ottnt{l_{{\mathrm{1}}}} }  }{\Rightarrow} }   \{ \mathit{x} \mathord{:}  \mathsf{Int}  \mathrel{\mid}  \mathit{x}  \mathrel{\ge}  \ottsym{0}  \}   \rangle^{ \ottnt{l_{{\mathrm{1}}}} } ~   {-1}   )    \\
 \multicolumn{2}{r}{(\ECastMerge)} \\ 
  \longrightarrow _{  \mathsf{E}  }  &  \langle   \{ \mathit{x} \mathord{:}  \mathsf{Int}  \mathrel{\mid}  \mathsf{true}  \}   \mathord{ \overset{ \ottnt{r} }{\Rightarrow} }   \{ \mathit{x} \mathord{:}  \mathsf{Int}  \mathrel{\mid}  \mathit{x}  \mathrel{\ne}  \ottsym{0}  \}   \rangle^{ \ottnt{l_{{\mathrm{3}}}} } ~   {-1}   \\
 \multicolumn{2}{l}{\text{where } \ottnt{r}  \ottsym{=}    \{ \mathit{x} \mathord{:}  \mathsf{Int}  \mathrel{\mid}  \mathit{x}  \mathrel{\ge}  \ottsym{0}  \}^{ \ottnt{l_{{\mathrm{1}}}} }  , \ottnt{r'} } \\
 \multicolumn{2}{r}{(\E{CoerceStack})} \\ 
  \longrightarrow _{  \mathsf{E}  }  &  \langle   \{ \mathit{x} \mathord{:}  \mathsf{Int}  \mathrel{\mid}  \mathit{x}  \mathrel{\ne}  \ottsym{0}  \}  ,   \mathord{?}  ,  \ottnt{r} ,   {-1}  ,   {-1}   \rangle^{\bullet}  \\
 \multicolumn{2}{r}{(\E{StackPop})} \\ 
  \longrightarrow _{  \mathsf{E}  }  &  \langle   \{ \mathit{x} \mathord{:}  \mathsf{Int}  \mathrel{\mid}  \mathit{x}  \mathrel{\ne}  \ottsym{0}  \}  ,   \mathord{?}  ,  \ottnt{r'} ,   {-1}  ,  {} \\  &  \phantom{\langle}   \langle   \{ \mathit{x} \mathord{:}  \mathsf{Int}  \mathrel{\mid}  \mathit{x}  \mathrel{\ge}  \ottsym{0}  \}  ,    {-1}   \mathrel{\ge}  \ottsym{0}  ,   {-1}   \rangle^{ \ottnt{l_{{\mathrm{1}}}} }   \rangle^{\bullet}  \\
  \longrightarrow ^{*}_{  \mathsf{E}  }  &  \mathord{\Uparrow}  \ottnt{l_{{\mathrm{1}}}}  
\end{array} \]
\caption{Example of eidetic \lambdah}
\label{fig:eideticexample}
\end{figure}
First, we return to our example from before in
Figure~\ref{fig:eideticexample}. Throughout the merging, each
refinement type retains its own original blame label, allowing eidetic
\lambdah to behave just like classic \lambdah.

We offer a final pair of examples, showing how
coercions with redundant types are merged. The intuition here is that
positive positions are checked covariantly---oldest (innermost) cast
first---while negative positions are checked contravariantly---newest
(outermost) cast first. Consider the classic \lambdah term:
\[ \begin{array}{r@{~}l}
  \ottnt{T_{{\mathrm{1}}}} = &   \{ \mathit{x} \mathord{:}  \mathsf{Int}  \mathrel{\mid} \ottnt{e_{{\mathrm{11}}}} \}  \mathord{ \rightarrow }  \{ \mathit{x} \mathord{:}  \mathsf{Int}  \mathrel{\mid} \ottnt{e_{{\mathrm{21}}}} \}   \\
  \ottnt{T_{{\mathrm{2}}}} = &   \{ \mathit{x} \mathord{:}  \mathsf{Int}  \mathrel{\mid} \ottnt{e_{{\mathrm{12}}}} \}  \mathord{ \rightarrow }  \{ \mathit{x} \mathord{:}  \mathsf{Int}  \mathrel{\mid} \ottnt{e_{{\mathrm{22}}}} \}   \\
  \ottnt{T_{{\mathrm{3}}}} = &   \{ \mathit{x} \mathord{:}  \mathsf{Int}  \mathrel{\mid} \ottnt{e_{{\mathrm{13}}}} \}  \mathord{ \rightarrow }  \{ \mathit{x} \mathord{:}  \mathsf{Int}  \mathrel{\mid} \ottnt{e_{{\mathrm{22}}}} \}   \\
   \ottnt{e} = &  \langle  \ottnt{T_{{\mathrm{2}}}}  \mathord{ \overset{\bullet}{\Rightarrow} }  \ottnt{T_{{\mathrm{3}}}}  \rangle^{ \ottnt{l_{{\mathrm{2}}}} } ~   (  \langle  \ottnt{T_{{\mathrm{1}}}}  \mathord{ \overset{\bullet}{\Rightarrow} }  \ottnt{T_{{\mathrm{2}}}}  \rangle^{ \ottnt{l_{{\mathrm{1}}}} } ~  \mathit{v}  )   
\end{array}
\]
Note that the casts run inside-out, from old to new in the positive
position, but they run from the outside-in, new to old, in the
negative position.
\[ \begin{array}{@{}r@{~}l@{}}
    \ottnt{e} ~ \mathit{v'}   \longrightarrow _{  \mathsf{C}  }  &  \langle   \{ \mathit{x} \mathord{:}  \mathsf{Int}  \mathrel{\mid} \ottnt{e_{{\mathrm{22}}}} \}   \mathord{ \overset{\bullet}{\Rightarrow} }   \{ \mathit{x} \mathord{:}  \mathsf{Int}  \mathrel{\mid} \ottnt{e_{{\mathrm{22}}}} \}   \rangle^{ \ottnt{l_{{\mathrm{2}}}} } ~  {} \\  &  \quad   (  \langle   \{ \mathit{x} \mathord{:}  \mathsf{Int}  \mathrel{\mid} \ottnt{e_{{\mathrm{21}}}} \}   \mathord{ \overset{\bullet}{\Rightarrow} }   \{ \mathit{x} \mathord{:}  \mathsf{Int}  \mathrel{\mid} \ottnt{e_{{\mathrm{22}}}} \}   \rangle^{ \ottnt{l_{{\mathrm{1}}}} } ~  {} \\  &  \qquad   (  \mathit{v} ~  (  \langle   \{ \mathit{x} \mathord{:}  \mathsf{Int}  \mathrel{\mid} \ottnt{e_{{\mathrm{12}}}} \}   \mathord{ \overset{\bullet}{\Rightarrow} }   \{ \mathit{x} \mathord{:}  \mathsf{Int}  \mathrel{\mid} \ottnt{e_{{\mathrm{12}}}} \}   \rangle^{ \ottnt{l_{{\mathrm{1}}}} } ~  {} \\  &  \qquad  \quad   (  \langle   \{ \mathit{x} \mathord{:}  \mathsf{Int}  \mathrel{\mid} \ottnt{e_{{\mathrm{13}}}} \}   \mathord{ \overset{\bullet}{\Rightarrow} }   \{ \mathit{x} \mathord{:}  \mathsf{Int}  \mathrel{\mid} \ottnt{e_{{\mathrm{12}}}} \}   \rangle^{ \ottnt{l_{{\mathrm{2}}}} } ~  \mathit{v'}  )   )   )   )  
\end{array} \]
The key observation for eliminating redundant checks is that only the check run first
can fail---there's no point in checking a predicate contract twice on
the same value. So eidetic \lambdah merges like so:
\[ \begin{array}{r@{~}l}
  \ottnt{e} \,  \longrightarrow ^{*}_{  \mathsf{E}  }  \,    &  \langle  \ottnt{T_{{\mathrm{2}}}}  \mathord{ \overset{  \{ \mathit{x} \mathord{:}  \mathsf{Int}  \mathrel{\mid} \ottnt{e_{{\mathrm{12}}}} \}^{ \ottnt{l_{{\mathrm{2}}}} }   \mapsto   \{ \mathit{x} \mathord{:}  \mathsf{Int}  \mathrel{\mid} \ottnt{e_{{\mathrm{22}}}} \}^{ \ottnt{l_{{\mathrm{2}}}} }  }{\Rightarrow} }  \ottnt{T_{{\mathrm{3}}}}  \rangle^{\bullet} ~  {} \\  &  \quad   (  \langle  \ottnt{T_{{\mathrm{1}}}}  \mathord{ \overset{  \{ \mathit{x} \mathord{:}  \mathsf{Int}  \mathrel{\mid} \ottnt{e_{{\mathrm{11}}}} \}^{ \ottnt{l_{{\mathrm{1}}}} }   \mapsto   \{ \mathit{x} \mathord{:}  \mathsf{Int}  \mathrel{\mid} \ottnt{e_{{\mathrm{22}}}} \}^{ \ottnt{l_{{\mathrm{1}}}} }  }{\Rightarrow} }  \ottnt{T_{{\mathrm{2}}}}  \rangle^{\bullet} ~  \mathit{v}  )   \\
   \longrightarrow _{  \mathsf{E}  }  &  \langle  \ottnt{T_{{\mathrm{1}}}}  \mathord{ \overset{ \ottnt{c} }{\Rightarrow} }  \ottnt{T_{{\mathrm{3}}}}  \rangle^{\bullet} ~  \mathit{v}  \\
\end{array} \]
where
\[ \begin{array}{r@{~}l}
  \ottnt{c} = &  \mathsf{join} (  \{ \mathit{x} \mathord{:}  \mathsf{Int}  \mathrel{\mid} \ottnt{e_{{\mathrm{12}}}} \}^{ \ottnt{l_{{\mathrm{2}}}} }  ,  \{ \mathit{x} \mathord{:}  \mathsf{Int}  \mathrel{\mid} \ottnt{e_{{\mathrm{11}}}} \}^{ \ottnt{l_{{\mathrm{1}}}} }  )   \mapsto  {} \\  &   \mathsf{join} (  \{ \mathit{x} \mathord{:}  \mathsf{Int}  \mathrel{\mid} \ottnt{e_{{\mathrm{22}}}} \}^{ \ottnt{l_{{\mathrm{1}}}} }  ,  \{ \mathit{x} \mathord{:}  \mathsf{Int}  \mathrel{\mid} \ottnt{e_{{\mathrm{22}}}} \}^{ \ottnt{l_{{\mathrm{2}}}} }  )  \\
  = &   \{ \mathit{x} \mathord{:}  \mathsf{Int}  \mathrel{\mid} \ottnt{e_{{\mathrm{12}}}} \}^{ \ottnt{l_{{\mathrm{2}}}} }  ,  \{ \mathit{x} \mathord{:}  \mathsf{Int}  \mathrel{\mid} \ottnt{e_{{\mathrm{11}}}} \}^{ \ottnt{l_{{\mathrm{1}}}} }    \mapsto   \{ \mathit{x} \mathord{:}  \mathsf{Int}  \mathrel{\mid} \ottnt{e_{{\mathrm{22}}}} \}^{ \ottnt{l_{{\mathrm{1}}}} } 
\end{array} \]
The coercion merge operator eliminates the redundant codomain check,
choosing to keep the one with blame label $\ottnt{l_{{\mathrm{1}}}}$. Choosing $\ottnt{l_{{\mathrm{1}}}}$
makes sense here because the codomain is a positive position and
$\ottnt{l_{{\mathrm{1}}}}$ is the older, innermost cast.
We construct a similar example for merges in negative positions.
\[ \begin{array}{r@{~}l}
   \ottnt{T_{{\mathrm{1}}}} = &   \{ \mathit{x} \mathord{:}  \mathsf{Int}  \mathrel{\mid} \ottnt{e_{{\mathrm{11}}}} \}  \mathord{ \rightarrow }  \{ \mathit{x} \mathord{:}  \mathsf{Int}  \mathrel{\mid} \ottnt{e_{{\mathrm{21}}}} \}   \\
  \ottnt{T'_{{\mathrm{2}}}} = &   \{ \mathit{x} \mathord{:}  \mathsf{Int}  \mathrel{\mid} \ottnt{e_{{\mathrm{11}}}} \}  \mathord{ \rightarrow }  \{ \mathit{x} \mathord{:}  \mathsf{Int}  \mathrel{\mid} \ottnt{e_{{\mathrm{22}}}} \}   \\
  \ottnt{T'_{{\mathrm{3}}}} = &   \{ \mathit{x} \mathord{:}  \mathsf{Int}  \mathrel{\mid} \ottnt{e_{{\mathrm{13}}}} \}  \mathord{ \rightarrow }  \{ \mathit{x} \mathord{:}  \mathsf{Int}  \mathrel{\mid} \ottnt{e_{{\mathrm{23}}}} \}   \\
   \ottnt{e'} = &  \langle  \ottnt{T'_{{\mathrm{2}}}}  \mathord{ \overset{\bullet}{\Rightarrow} }  \ottnt{T'_{{\mathrm{3}}}}  \rangle^{ \ottnt{l_{{\mathrm{2}}}} } ~   (  \langle  \ottnt{T_{{\mathrm{1}}}}  \mathord{ \overset{\bullet}{\Rightarrow} }  \ottnt{T'_{{\mathrm{2}}}}  \rangle^{ \ottnt{l_{{\mathrm{1}}}} } ~  \mathit{v}  )   \\
\end{array} \]
Again, the unfolding runs the positive parts inside-out and the negative
parts outside-in when applied to a value $\mathit{v'}$:
\[ \begin{array}{l}
   \langle   \{ \mathit{x} \mathord{:}  \mathsf{Int}  \mathrel{\mid} \ottnt{e_{{\mathrm{22}}}} \}   \mathord{ \overset{\bullet}{\Rightarrow} }   \{ \mathit{x} \mathord{:}  \mathsf{Int}  \mathrel{\mid} \ottnt{e_{{\mathrm{23}}}} \}   \rangle^{ \ottnt{l_{{\mathrm{2}}}} } ~  {} \\  \quad   (  \langle   \{ \mathit{x} \mathord{:}  \mathsf{Int}  \mathrel{\mid} \ottnt{e_{{\mathrm{21}}}} \}   \mathord{ \overset{\bullet}{\Rightarrow} }   \{ \mathit{x} \mathord{:}  \mathsf{Int}  \mathrel{\mid} \ottnt{e_{{\mathrm{22}}}} \}   \rangle^{ \ottnt{l_{{\mathrm{1}}}} } ~  {} \\  \qquad   (  \mathit{v} ~  (  \langle   \{ \mathit{x} \mathord{:}  \mathsf{Int}  \mathrel{\mid} \ottnt{e_{{\mathrm{11}}}} \}   \mathord{ \overset{\bullet}{\Rightarrow} }   \{ \mathit{x} \mathord{:}  \mathsf{Int}  \mathrel{\mid} \ottnt{e_{{\mathrm{11}}}} \}   \rangle^{ \ottnt{l_{{\mathrm{1}}}} } ~  {} \\  \qquad  \quad   (  \langle   \{ \mathit{x} \mathord{:}  \mathsf{Int}  \mathrel{\mid} \ottnt{e_{{\mathrm{13}}}} \}   \mathord{ \overset{\bullet}{\Rightarrow} }   \{ \mathit{x} \mathord{:}  \mathsf{Int}  \mathrel{\mid} \ottnt{e_{{\mathrm{11}}}} \}   \rangle^{ \ottnt{l_{{\mathrm{2}}}} } ~  \mathit{v'}  )   )   )   )  
\end{array} \]
Running the example in eidetic \lambdah, we reduce the redundant
checks in the domain:
\[ \begin{array}{r@{~}l}
  \ottnt{e'} \,  \longrightarrow ^{*}_{  \mathsf{E}  }  \,    &  \langle  \ottnt{T'_{{\mathrm{2}}}}  \mathord{ \overset{  \{ \mathit{x} \mathord{:}  \mathsf{Int}  \mathrel{\mid} \ottnt{e_{{\mathrm{11}}}} \}^{ \ottnt{l_{{\mathrm{2}}}} }   \mapsto   \{ \mathit{x} \mathord{:}  \mathsf{Int}  \mathrel{\mid} \ottnt{e_{{\mathrm{23}}}} \}^{ \ottnt{l_{{\mathrm{2}}}} }  }{\Rightarrow} }  \ottnt{T'_{{\mathrm{3}}}}  \rangle^{\bullet} ~  {} \\  &  \quad   (  \langle  \ottnt{T_{{\mathrm{1}}}}  \mathord{ \overset{  \{ \mathit{x} \mathord{:}  \mathsf{Int}  \mathrel{\mid} \ottnt{e_{{\mathrm{11}}}} \}^{ \ottnt{l_{{\mathrm{1}}}} }   \mapsto   \{ \mathit{x} \mathord{:}  \mathsf{Int}  \mathrel{\mid} \ottnt{e_{{\mathrm{22}}}} \}^{ \ottnt{l_{{\mathrm{1}}}} }  }{\Rightarrow} }  \ottnt{T'_{{\mathrm{2}}}}  \rangle^{\bullet} ~  \mathit{v}  )   \\
   \longrightarrow _{  \mathsf{E}  }  &  \langle  \ottnt{T_{{\mathrm{1}}}}  \mathord{ \overset{ \ottnt{c} }{\Rightarrow} }  \ottnt{T'_{{\mathrm{3}}}}  \rangle^{\bullet} ~  \mathit{v}  \\
  \multicolumn{2}{l}{\text{where }} \\
  \ottnt{c} = &  \mathsf{join} (  \{ \mathit{x} \mathord{:}  \mathsf{Int}  \mathrel{\mid} \ottnt{e_{{\mathrm{11}}}} \}^{ \ottnt{l_{{\mathrm{2}}}} }  ,  \{ \mathit{x} \mathord{:}  \mathsf{Int}  \mathrel{\mid} \ottnt{e_{{\mathrm{11}}}} \}^{ \ottnt{l_{{\mathrm{1}}}} }  )   \mapsto  {} \\  &   \mathsf{join} (  \{ \mathit{x} \mathord{:}  \mathsf{Int}  \mathrel{\mid} \ottnt{e_{{\mathrm{22}}}} \}^{ \ottnt{l_{{\mathrm{1}}}} }  ,  \{ \mathit{x} \mathord{:}  \mathsf{Int}  \mathrel{\mid} \ottnt{e_{{\mathrm{23}}}} \}^{ \ottnt{l_{{\mathrm{2}}}} }  )  \\
        = &  \{ \mathit{x} \mathord{:}  \mathsf{Int}  \mathrel{\mid} \ottnt{e_{{\mathrm{12}}}} \}^{ \ottnt{l_{{\mathrm{2}}}} }   \mapsto    \{ \mathit{x} \mathord{:}  \mathsf{Int}  \mathrel{\mid} \ottnt{e_{{\mathrm{22}}}} \}^{ \ottnt{l_{{\mathrm{1}}}} }  ,  \{ \mathit{x} \mathord{:}  \mathsf{Int}  \mathrel{\mid} \ottnt{e_{{\mathrm{23}}}} \}^{ \ottnt{l_{{\mathrm{2}}}} }  
\end{array} \]
Following the outside-in rule for negative positions, we keep the
blame label $\ottnt{l_{{\mathrm{2}}}}$ from the newer, outermost cast.

{\ifpopl 

\subsection{Metatheory}
\label{sec:eideticmetatheory}

The proof of type soundness is a standard syntactic proof, relying on
a few small lemmas concerning refinement list well formedness and the
generic metatheory described in Section~\ref{sec:metatheory}. The full
proofs are in Appendix~\ref{app:eideticsoundness}.

\begin{lemma}[Eidetic canonical forms (\ref{lem:eideticcanonicalforms})]
  If $ \emptyset   \vdash _{  \mathsf{E}  }  \ottnt{e}  :  \ottnt{T} $ and $ \mathsf{val} _{  \mathsf{E}  }~ \ottnt{e} $ then:
  \begin{itemize}
  \item If $\ottnt{T}  \ottsym{=}   \{ \mathit{x} \mathord{:} \ottnt{B} \mathrel{\mid} \ottnt{e'} \} $, then $\ottnt{e}  \ottsym{=}  \ottnt{k}$ and $ \mathsf{ty} ( \ottnt{k} )   \ottsym{=}  \ottnt{B}$
    and $ \ottnt{e'}  [  \ottnt{e} / \mathit{x}  ]  \,  \longrightarrow ^{*}_{  \mathsf{E}  }  \,  \mathsf{true} $.
  \item If $\ottnt{T}  \ottsym{=}   \ottnt{T_{{\mathrm{21}}}} \mathord{ \rightarrow } \ottnt{T_{{\mathrm{22}}}} $, then either $\ottnt{e}  \ottsym{=}   \lambda \mathit{x} \mathord{:} \ottnt{T} .~  \ottnt{e'} $ or $\ottnt{e}  \ottsym{=}   \langle   \ottnt{T_{{\mathrm{11}}}} \mathord{ \rightarrow } \ottnt{T_{{\mathrm{12}}}}   \mathord{ \overset{ \ottnt{c_{{\mathrm{1}}}}  \mapsto  \ottnt{c_{{\mathrm{2}}}} }{\Rightarrow} }   \ottnt{T_{{\mathrm{21}}}} \mathord{ \rightarrow } \ottnt{T_{{\mathrm{22}}}}   \rangle^{\bullet} ~   \lambda \mathit{x} \mathord{:} \ottnt{T_{{\mathrm{11}}}} .~  \ottnt{e'}  $.
  \end{itemize}
\end{lemma}

\begin{lemma}[Eidetic progress (\ref{lem:eideticprogress})]
  If $ \emptyset   \vdash _{  \mathsf{E}  }  \ottnt{e}  :  \ottnt{T} $, then either:
  \begin{enumerate}
  \item $ \mathsf{result} _{  \mathsf{E}  }~ \ottnt{e} $, i.e., $\ottnt{e}  \ottsym{=}   \mathord{\Uparrow}  \ottnt{l} $ or $ \mathsf{val} _{  \mathsf{E}  }~ \ottnt{e} $;
    or
  \item there exists an $\ottnt{e'}$ such that $\ottnt{e} \,  \longrightarrow _{  \mathsf{E}  }  \, \ottnt{e'}$.
  \end{enumerate}
\end{lemma}

\begin{lemma}[Eidetic preservation (\ref{lem:eideticpreservation})]
  If $ \emptyset   \vdash _{  \mathsf{E}  }  \ottnt{e}  :  \ottnt{T} $ and $\ottnt{e} \,  \longrightarrow _{  \mathsf{E}  }  \, \ottnt{e'}$ then $ \emptyset   \vdash _{  \mathsf{E}  }  \ottnt{e'}  :  \ottnt{T} $.
\end{lemma}

\else

As we did for the other calculi, we present the
routine syntactic proof of type soundness in 
(Appendix~\ref{app:eideticsoundness}). 

\fi}
\ifpopl Eidetic \else Like forgetful and heedful \lambdah before,
eidetic \fi \lambdah shares source programs
(Definition~\ref{def:sourceprogram}) with classic
\lambdah. {\ifpopl
  We can therefore say that classic and eidetic \lambdah are really
  just \textit{modes} of a single language.
\begin{lemma}[Source program typing for eidetic \lambdah (\ref{lem:eideticsource})] ~ \\
  \noindent
  Source programs are well typed in $ \mathsf{C} $ iff they are well typed in
  $ \mathsf{E} $, i.e.:
  \begin{itemize}
  \item $ \Gamma   \vdash _{  \mathsf{C}  }  \ottnt{e}  :  \ottnt{T} $ as a source program iff $ \Gamma   \vdash _{  \mathsf{E}  }  \ottnt{e}  :  \ottnt{T} $ as a source program.
  \item $ \mathord{  \vdash _{  \mathsf{C}  } }~ \ottnt{T} $ as a source program iff $ \mathord{  \vdash _{  \mathsf{E}  } }~ \ottnt{T} $ as a source program.
  \item $ \mathord{  \vdash _{  \mathsf{C}  } }~ \Gamma $ as a source program iff $ \mathord{  \vdash _{  \mathsf{E}  } }~ \Gamma $ as a source program.
  \end{itemize}
\end{lemma}

\else 
With this final lemma, we know that all modes share the same well
typed source programs. \fi}

{}

\section{Soundness for space efficiency}
\label{sec:soundness}

{\iffull\begin{figure*}
    % \forgetfulcongruence
    \begin{tikzpicture}[description/.style={fill=white,inner sep=2pt},align at top]
    \matrix (m) [matrix of math nodes, row sep=4pt, nodes in empty cells,
                 text height=1.5ex, text depth=0.25ex]
    {    
                       & \textbf{Forgetful \lambdah} & \\
      \ottnt{e_{{\mathrm{1}}}}           & & \ottnt{e_{{\mathrm{2}}}} \\
                       & \Downarrow & \\
       \langle  \ottnt{T_{{\mathrm{1}}}}  \mathord{ \overset{\bullet}{\Rightarrow} }  \ottnt{T_{{\mathrm{2}}}}  \rangle^{ \ottnt{l} } ~  \ottnt{e_{{\mathrm{1}}}}  & &  \langle  \ottnt{T_{{\mathrm{1}}}}  \mathord{ \overset{\bullet}{\Rightarrow} }  \ottnt{T_{{\mathrm{2}}}}  \rangle^{ \ottnt{l} } ~  \ottnt{e_{{\mathrm{2}}}}  \\[20pt]
                       & &  \mathsf{val} _{  \mathsf{F}  }~ \ottnt{e}  \\
    };

    \path[->] (m-2-1) edge[F] (m-2-3);
    \path[dashed,->] (m-4-1.south) edge[F*] (m-5-3);
    \path[->] (m-4-3) edge[F*] (m-5-3);
  \end{tikzpicture}
  \hspace{1em}
  % \heedfulcongruence
  \begin{tikzpicture}[description/.style={fill=white,inner sep=2pt},align at top]
  \tikzset{align at top/.style={baseline=(current bounding box.north)}}

    \matrix (m) [matrix of math nodes, row sep=4pt, nodes in empty cells,
                 text height=1.5ex, text depth=0.25ex]
    { 
                       & \textbf{Heedful \lambdah} & \\
      \ottnt{e_{{\mathrm{1}}}}           & & \ottnt{e_{{\mathrm{2}}}} \\
                       & \Downarrow & \\
       \langle  \ottnt{T_{{\mathrm{1}}}}  \mathord{ \overset{\bullet}{\Rightarrow} }  \ottnt{T_{{\mathrm{2}}}}  \rangle^{ \ottnt{l} } ~  \ottnt{e_{{\mathrm{1}}}}  & &  \langle  \ottnt{T_{{\mathrm{1}}}}  \mathord{ \overset{\bullet}{\Rightarrow} }  \ottnt{T_{{\mathrm{2}}}}  \rangle^{ \ottnt{l} } ~  \ottnt{e_{{\mathrm{2}}}}  \\[20pt]
       \mathsf{result} _{  \mathsf{H}  }~ \ottnt{e'_{{\mathrm{1}}}}  &  \sim  &  \mathsf{result} _{  \mathsf{H}  }~ \ottnt{e'_{{\mathrm{2}}}}  \\
      &  \mathsf{val} _{  \mathsf{H}  }~ \ottnt{e}  & \\
    };

    \path[->] (m-2-1) edge[H] (m-2-3);
    \path[dashed,->] (m-4-1) edge[H*] (m-5-1);
    \path[->] (m-4-3) edge[H*] (m-5-3);
    \path[color=white] (m-5-1) edge node[sloped,color=black] {$=$} (m-6-2);
    \path[color=white] (m-5-3) edge node[sloped,color=black] {$=$} (m-6-2);
  \end{tikzpicture}
  \hspace{1em}
  % \eideticcongruence
  \begin{tikzpicture}[description/.style={fill=white,inner sep=2pt},align at top]
    \matrix (m) [matrix of math nodes, row sep=4pt, nodes in empty cells,
                 text height=1.5ex, text depth=0.25ex]
    { 
      \ottnt{e_{{\mathrm{1}}}}           & & \ottnt{e_{{\mathrm{2}}}} \\
                       & \Downarrow & \\
       \langle  \ottnt{T_{{\mathrm{1}}}}  \mathord{ \overset{\bullet}{\Rightarrow} }  \ottnt{T_{{\mathrm{2}}}}  \rangle^{ \ottnt{l} } ~  \ottnt{e_{{\mathrm{1}}}}  & &  \langle  \ottnt{T_{{\mathrm{1}}}}  \mathord{ \overset{\bullet}{\Rightarrow} }  \ottnt{T_{{\mathrm{2}}}}  \rangle^{ \ottnt{l} } ~  \ottnt{e_{{\mathrm{2}}}}  \\[20pt]
      &  \mathsf{result} _{  \mathsf{E}  }~ \ottnt{e}  & \\
    };

    \path[->] (m-1-1) edge[E] (m-1-3);
    \path[->] (m-3-1) edge[E*] (m-4-2.north west);
    \path[->] (m-3-3) edge[E*] (m-4-2.north east);
  \end{tikzpicture}
  \vspace*{-10pt}
  \caption{Cast congruence lemmas as commutative diagrams}
  \label{fig:castcongruence}
\end{figure*}
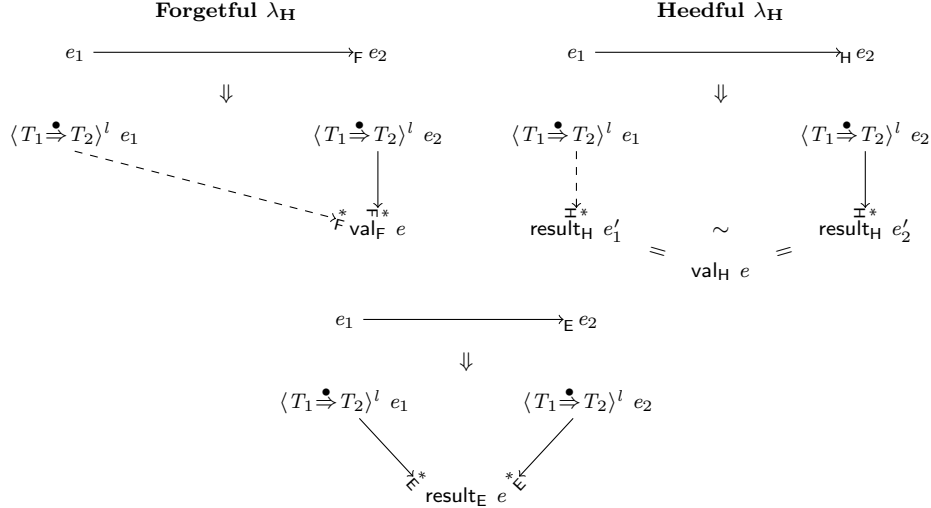\fi}

We want space efficiency to be \textit{sound}: it would be
space efficient to never check anything.
Classic \lambdah is normative: the more a mode behaves like classic
\lambdah, the ``sounder'' it is.

A single property summarizes how a space-efficient calculus behaves
with respect to classic \lambdah: cast congruence.
In classic \lambdah, if $\ottnt{e_{{\mathrm{1}}}} \,  \longrightarrow _{  \mathsf{C}  }  \, \ottnt{e_{{\mathrm{2}}}}$ then $ \langle  \ottnt{T_{{\mathrm{1}}}}  \mathord{ \overset{\bullet}{\Rightarrow} }  \ottnt{T_{{\mathrm{2}}}}  \rangle^{ \ottnt{l} } ~  \ottnt{e_{{\mathrm{1}}}} $ and
$ \langle  \ottnt{T_{{\mathrm{1}}}}  \mathord{ \overset{\bullet}{\Rightarrow} }  \ottnt{T_{{\mathrm{2}}}}  \rangle^{ \ottnt{l} } ~  \ottnt{e_{{\mathrm{2}}}} $ behave identically. This cast congruence principle
is easy to see, because \E{CastInnerC} applies freely.
In \ifpopl eidetic \lambdah\else the space-efficient modes\fi, however, \E{CastInner} can only apply
when \ECastMerge\ doesn't. Merged casts may not behave the same as
running the two casts separately.
{\ifpopl
  Eidetic \lambdah recovers a complete cast congruence, just like
  classic \lambdah has. Diagrammatically: 
\vspace*{-1em}
\begin{center}
  % \eideticcongruence
  \begin{tikzpicture}[description/.style={fill=white,inner sep=2pt},align at top]
    \matrix (m) [matrix of math nodes, row sep=4pt, nodes in empty cells,
                 text height=1.5ex, text depth=0.25ex]
    { 
      \ottnt{e_{{\mathrm{1}}}}           & & \ottnt{e_{{\mathrm{2}}}} \\
                       & \Downarrow & \\
       \langle  \ottnt{T_{{\mathrm{1}}}}  \mathord{ \overset{\bullet}{\Rightarrow} }  \ottnt{T_{{\mathrm{2}}}}  \rangle^{ \ottnt{l} } ~  \ottnt{e_{{\mathrm{1}}}}  & &  \langle  \ottnt{T_{{\mathrm{1}}}}  \mathord{ \overset{\bullet}{\Rightarrow} }  \ottnt{T_{{\mathrm{2}}}}  \rangle^{ \ottnt{l} } ~  \ottnt{e_{{\mathrm{2}}}}  \\[20pt]
      &  \mathsf{result} _{  \mathsf{E}  }~ \ottnt{e}  & \\
    };

    \path[->] (m-1-1) edge[E] (m-1-3);
    \path[->] (m-3-1) edge[E*] (m-4-2.north west);
    \path[->] (m-3-3) edge[E*] (m-4-2.north east);
  \end{tikzpicture}
\end{center}
\else
We summarize the results in commutative diagrams in
Figure~\ref{fig:castcongruence}.
Forgetful \lambdah has the property that if the unmerged casts reduce
to a value, then so do the merged ones. But the merged casts may
reduce to a value when the unmerged ones reduce to blame, because
forgetful merging skips checks.
Heedful \lambdah has a stronger property: the merged and unmerged
casts \textit{coterminate} at results, if the merged term reduces to
blame or a value, so does the unmerged term. If they both go to
values, they go to the exact same value; but if they both go to blame,
they may blame different labels. This is a direct result of
\ECastMerge\ saving only one label from casts.
Finally, eidetic \lambdah has a property as strong as heedful
\lambdah: the merged and unmerged casts coterminate exactly.

It is particularly nice that the key property for relating modes can
be proved entirely within each mode, i.e., the cast congruence lemma
for forgetful \lambdah is proved \textit{independently} of classic
\lambdah.

\fi}
The \iffull proofs are \else proof is \fi{} in Appendix~\ref{app:sesoundness}, but
\ifpopl it is worth observing here that eidetic \lambdah needs a
  proof of idempotency to justify the way it uses reflexivity to
  eliminate redundant coercions\refasm{implrefl}: \else there are
  two points worth observing here.
First, we need strong normalization to prove cast congruence for
heedful \lambdah: if we reorder checks, we need to know that
reordering checks doesn't change the observable behavior.
Second, both heedful and eidetic \lambdah eliminate redundant checks
when merging casts, the former by using sets and the latter by means
of the $ \mathsf{join} $ operator and the reflexivity of
$ \supset $\refasm{implrefl}. These two calculi require proofs that checking is
idempotent: \fi checking a property once is as good as checking it
twice. Naturally, this property only holds without state.

{\iffull
\begin{table}
\begin{tabular}{|r|c|c|}
  \hline
  \multicolumn{1}{|c|}{\textbf{Mode}} & \multicolumn{2}{|c|}{\textbf{Reduction behavior}} \\ 
  \hline
  Classic ($\ottnt{m}  \ottsym{=}   \mathsf{C} $)   & \multicolumn{2}{|c|}{Normative} \\ \hline
  Forgetful ($\ottnt{m}  \ottsym{=}   \mathsf{F} $) & \multicolumn{2}{|c|}{$\mathord{ \longrightarrow ^{*}_{  \mathsf{C}  } }~ \mathsf{val}  \Rightarrow \mathord{ \longrightarrow ^{*}_{  \mathsf{F}  } }~ \mathsf{val} $ (Lemma~\ref{lem:forgetfullr})} \\ \hline
  Heedful ($\ottnt{m}  \ottsym{=}   \mathsf{H} $)   & \multicolumn{2}{|c|}{$\mathord{ \longrightarrow ^{*}_{  \mathsf{C}  } }~ \mathsf{result}  \Leftrightarrow \mathord{ \longrightarrow ^{*}_{  \mathsf{H}  } }~ \mathsf{result} $ (Lemma~\ref{lem:heedfullr})} \\ \hline
  Eidetic ($\ottnt{m}  \ottsym{=}   \mathsf{E} $)   & \multicolumn{2}{|c|}{$\mathord{ \longrightarrow ^{*}_{  \mathsf{C}  } }~ \mathsf{result}  = \mathord{ \longrightarrow ^{*}_{  \mathsf{E}  } }~ \mathsf{result} $ (Lemma~\ref{lem:eideticlr})} \\ \hline 
\end{tabular}
\caption{Soundness results for \lambdah}
\label{tab:soundness}
\end{table}
\fi}

{\ifpopl
\begin{figure}[t]
  \hdr{Value rules}{\qquad \fbox{$ \ottnt{e_{{\mathrm{1}}}}   \sim _{  \mathsf{E}  }  \ottnt{e_{{\mathrm{2}}}}  :  \ottnt{T} $}}
  \[ \begin{array}{rcl}
   \ottnt{k}   \sim _{  \mathsf{E}  }  \ottnt{k}  :   \{ \mathit{x} \mathord{:} \ottnt{B} \mathrel{\mid} \ottnt{e} \}   &\iff&  \mathsf{ty} ( \ottnt{k} )   \ottsym{=}  \ottnt{B} \wedge  \ottnt{e}  [  \ottnt{k} / \mathit{x}  ]  \,  \longrightarrow ^{*}_{  \mathsf{E}  }  \,  \mathsf{true}  \\
   \ottnt{e_{{\mathrm{11}}}}   \sim _{  \mathsf{E}  }  \ottnt{e_{{\mathrm{21}}}}  :   \ottnt{T_{{\mathrm{1}}}} \mathord{ \rightarrow } \ottnt{T_{{\mathrm{2}}}}   &\iff&  \mathsf{val} _{  \mathsf{C}  }~ \ottnt{e_{{\mathrm{1}}}}  \wedge  \mathsf{val} _{  \mathsf{E}  }~ \ottnt{e_{{\mathrm{2}}}}  \wedge {} \\
  \multicolumn{3}{r}{ \forall   \ottnt{e_{{\mathrm{12}}}}   \sim _{  \mathsf{E}  }  \ottnt{e_{{\mathrm{22}}}}  :  \ottnt{T_{{\mathrm{1}}}}   . ~    \ottnt{e_{{\mathrm{11}}}} ~ \ottnt{e_{{\mathrm{12}}}}    \simeq _{  \mathsf{E}  }   \ottnt{e_{{\mathrm{21}}}} ~ \ottnt{e_{{\mathrm{22}}}}   :  \ottnt{T_{{\mathrm{2}}}}  }
  \end{array} \]

  \hdr{Term rules}{\qquad \fbox{$ \ottnt{e_{{\mathrm{1}}}}   \simeq _{  \mathsf{E}  }  \ottnt{e_{{\mathrm{2}}}}  :  \ottnt{T} $}}
  \[ \begin{array}{c}
     \ottnt{e_{{\mathrm{1}}}}   \simeq _{  \mathsf{E}  }  \ottnt{e_{{\mathrm{2}}}}  :  \ottnt{T}  \\ \iff \\
  \left( \begin{array}{@{}l@{}}
      \ottnt{e_{{\mathrm{1}}}} \,  \longrightarrow ^{*}_{  \mathsf{C}  }  \,  \mathord{\Uparrow}  \ottnt{l}  \wedge {}\\
      \ottnt{e_{{\mathrm{2}}}} \,  \longrightarrow ^{*}_{  \mathsf{E}  }  \,  \mathord{\Uparrow}  \ottnt{l} 
    \end{array} \right) \vee \left( \begin{array}{@{}l@{}}
      \ottnt{e_{{\mathrm{1}}}} \,  \longrightarrow ^{*}_{  \mathsf{C}  }  \, \ottnt{e'_{{\mathrm{1}}}} \wedge  \mathsf{val} _{  \mathsf{C}  }~ \ottnt{e'_{{\mathrm{1}}}}  \wedge{} \\
      \ottnt{e_{{\mathrm{2}}}} \,  \longrightarrow ^{*}_{  \mathsf{E}  }  \, \ottnt{e'_{{\mathrm{2}}}} \wedge  \mathsf{val} _{  \mathsf{E}  }~ \ottnt{e'_{{\mathrm{2}}}}  \wedge{} \\
       \ottnt{e'_{{\mathrm{1}}}}   \sim _{  \mathsf{E}  }  \ottnt{e'_{{\mathrm{2}}}}  :  \ottnt{T} 
      \end{array} \right) 
  \end{array} \]
  
  \hdr{Type rules}{\qquad \fbox{$ \ottnt{T_{{\mathrm{1}}}}   \sim _{  \mathsf{E}  }  \ottnt{T_{{\mathrm{2}}}} $}}
  \[ \begin{array}{rcl}
      \{ \mathit{x} \mathord{:} \ottnt{B} \mathrel{\mid} \ottnt{e_{{\mathrm{1}}}} \}    \sim _{  \mathsf{E}  }   \{ \mathit{x} \mathord{:} \ottnt{B} \mathrel{\mid} \ottnt{e_{{\mathrm{2}}}} \}   &\iff& \\
    \multicolumn{3}{r}{ \forall   \ottnt{e'_{{\mathrm{1}}}}   \sim _{  \mathsf{E}  }  \ottnt{e'_{{\mathrm{2}}}}  :   \{ \mathit{x} \mathord{:} \ottnt{B} \mathrel{\mid}  \mathsf{true}  \}    . ~    \ottnt{e_{{\mathrm{1}}}}  [  \ottnt{e'_{{\mathrm{1}}}} / \mathit{x}  ]    \simeq _{  \mathsf{E}  }   \ottnt{e_{{\mathrm{2}}}}  [  \ottnt{e'_{{\mathrm{2}}}} / \mathit{x}  ]   :   \{ \mathit{x} \mathord{:}  \mathsf{Bool}  \mathrel{\mid}  \mathsf{true}  \}   } \\
      \ottnt{T_{{\mathrm{11}}}} \mathord{ \rightarrow } \ottnt{T_{{\mathrm{12}}}}    \sim _{  \mathsf{E}  }   \ottnt{T_{{\mathrm{21}}}} \mathord{ \rightarrow } \ottnt{T_{{\mathrm{22}}}}   &\iff&  \ottnt{T_{{\mathrm{11}}}}   \sim _{  \mathsf{E}  }  \ottnt{T_{{\mathrm{21}}}}  \wedge  \ottnt{T_{{\mathrm{12}}}}   \sim _{  \mathsf{E}  }  \ottnt{T_{{\mathrm{22}}}} 
  \end{array}\]

  \hdr{Closing substitutions and open terms}{\qquad \fbox{$ \Gamma   \models _{  \mathsf{E}  }  \delta $} \qquad \fbox{$ \Gamma   \vdash   \ottnt{e_{{\mathrm{1}}}}   \simeq _{  \mathsf{E}  }  \ottnt{e_{{\mathrm{2}}}}  :  \ottnt{T} $}}

  \[ \begin{array}{r@{~}c@{~}l}
     \Gamma   \models _{  \mathsf{E}  }  \delta  &\iff&  \forall  \mathit{x}  \in  \mathsf{dom} ( \Gamma ) . ~   \delta_{{\mathrm{1}}}  \ottsym{(}  \mathit{x}  \ottsym{)}   \sim _{  \mathsf{E}  }  \delta_{{\mathrm{2}}}  \ottsym{(}  \mathit{x}  \ottsym{)}  :   \Gamma ( \mathit{x} )    \\
     \Gamma   \vdash   \ottnt{e_{{\mathrm{1}}}}   \simeq _{  \mathsf{E}  }  \ottnt{e_{{\mathrm{2}}}}  :  \ottnt{T}  &\iff&  \forall   \Gamma   \models _{  \mathsf{E}  }  \delta   . ~   \delta_{{\mathrm{1}}}  \ottsym{(}  \ottnt{e_{{\mathrm{1}}}}  \ottsym{)}   \simeq _{  \mathsf{E}  }  \delta_{{\mathrm{2}}}  \ottsym{(}  \ottnt{e_{{\mathrm{2}}}}  \ottsym{)}  :  \ottnt{T}  
  \end{array} \]

  \caption{Blame-exact, symmetric logical relation between classic \lambdah and eidetic \lambdah}
  \label{fig:eideticlr}
\end{figure}
\else
\begin{figure}
    \hdr{Value rules}{\qquad \fbox{$ \ottnt{e_{{\mathrm{1}}}}   \sim _{ \ottnt{m} }  \ottnt{e_{{\mathrm{2}}}}  :  \ottnt{T} $}}
  \[ \begin{array}{r@{~}c@{~}l}
   \ottnt{k}   \sim _{ \ottnt{m} }  \ottnt{k}  :   \{ \mathit{x} \mathord{:} \ottnt{B} \mathrel{\mid} \ottnt{e} \}   &\iff&  \mathsf{ty} ( \ottnt{k} )   \ottsym{=}  \ottnt{B} \wedge  \ottnt{e}  [  \ottnt{k} / \mathit{x}  ]  \,  \longrightarrow ^{*}_{ \ottnt{m} }  \,  \mathsf{true}  \\
   \ottnt{e_{{\mathrm{11}}}}   \sim _{ \ottnt{m} }  \ottnt{e_{{\mathrm{21}}}}  :   \ottnt{T_{{\mathrm{1}}}} \mathord{ \rightarrow } \ottnt{T_{{\mathrm{2}}}}   &\iff&  \mathsf{val} _{  \mathsf{C}  }~ \ottnt{e_{{\mathrm{11}}}}  \wedge  \mathsf{val} _{ \ottnt{m} }~ \ottnt{e_{{\mathrm{12}}}}  \wedge {} \\
  \multicolumn{3}{r}{ \forall   \ottnt{e_{{\mathrm{12}}}}   \sim _{ \ottnt{m} }  \ottnt{e_{{\mathrm{22}}}}  :  \ottnt{T_{{\mathrm{1}}}}   . ~    \ottnt{e_{{\mathrm{11}}}} ~ \ottnt{e_{{\mathrm{12}}}}    \simeq _{ \ottnt{m} }   \ottnt{e_{{\mathrm{21}}}} ~ \ottnt{e_{{\mathrm{22}}}}   :  \ottnt{T_{{\mathrm{2}}}}  }
  \end{array} \]

  \hdr{Term rules}{\qquad \fbox{$ \ottnt{e_{{\mathrm{1}}}}  \Downarrow_{ \ottnt{m} }  \ottnt{e_{{\mathrm{2}}}}  :  \ottnt{T} $} \qquad \fbox{$ \ottnt{e_{{\mathrm{1}}}}   \simeq _{ \ottnt{m} }  \ottnt{e_{{\mathrm{2}}}}  :  \ottnt{T} $}}
  \[ \begin{array}{r@{~}c@{~}l}
  % coterm
     \ottnt{e_{{\mathrm{1}}}}  \Downarrow_{ \ottnt{m} }  \ottnt{e_{{\mathrm{2}}}}  :  \ottnt{T}  &\iff& \ottnt{e_{{\mathrm{1}}}} \,  \longrightarrow ^{*}_{  \mathsf{C}  }  \, \ottnt{e'_{{\mathrm{1}}}} \wedge  \mathsf{val} _{  \mathsf{C}  }~ \ottnt{e'_{{\mathrm{1}}}}  \wedge{} \\
    && \ottnt{e_{{\mathrm{2}}}} \,  \longrightarrow ^{*}_{ \ottnt{m} }  \, \ottnt{e'_{{\mathrm{2}}}} \wedge  \mathsf{val} _{ \ottnt{m} }~ \ottnt{e'_{{\mathrm{2}}}}  \wedge{} \\
    &&  \ottnt{e'_{{\mathrm{1}}}}   \sim _{ \ottnt{m} }  \ottnt{e'_{{\mathrm{2}}}}  :  \ottnt{T}  \\[2em]
  %
  % forgetful
\iffull
     \ottnt{e_{{\mathrm{1}}}}   \simeq _{  \mathsf{F}  }  \ottnt{e_{{\mathrm{2}}}}  :  \ottnt{T}  &\iff& \ottnt{e_{{\mathrm{1}}}} \,  \longrightarrow ^{*}_{  \mathsf{C}  }  \,  \mathord{\Uparrow}  \ottnt{l}  \vee  \ottnt{e_{{\mathrm{1}}}}  \Downarrow_{  \mathsf{F}  }  \ottnt{e_{{\mathrm{2}}}}  :  \ottnt{T}  \\
  % heedful
     \ottnt{e_{{\mathrm{1}}}}   \simeq _{  \mathsf{H}  }  \ottnt{e_{{\mathrm{2}}}}  :  \ottnt{T}  &\iff& (\ottnt{e_{{\mathrm{1}}}} \,  \longrightarrow ^{*}_{  \mathsf{C}  }  \,  \mathord{\Uparrow}  \ottnt{l}  \wedge \ottnt{e_{{\mathrm{2}}}} \,  \longrightarrow ^{*}_{  \mathsf{H}  }  \,  \mathord{\Uparrow}  \ottnt{l'} ) \vee  \ottnt{e_{{\mathrm{1}}}}  \Downarrow_{  \mathsf{H}  }  \ottnt{e_{{\mathrm{2}}}}  :  \ottnt{T}  \\ \fi
  % eidetic
     \ottnt{e_{{\mathrm{1}}}}   \simeq _{  \mathsf{E}  }  \ottnt{e_{{\mathrm{2}}}}  :  \ottnt{T}  &\iff& (\ottnt{e_{{\mathrm{1}}}} \,  \longrightarrow ^{*}_{  \mathsf{C}  }  \,  \mathord{\Uparrow}  \ottnt{l}  \wedge \ottnt{e_{{\mathrm{2}}}} \,  \longrightarrow ^{*}_{  \mathsf{E}  }  \,  \mathord{\Uparrow}  \ottnt{l} ) \vee \ottnt{e_{{\mathrm{1}}}}  \Downarrow_{  \mathsf{E}  }  \ottnt{e_{{\mathrm{2}}}}  :  \ottnt{T} 
  \end{array} \]
    
  \hdr{Type rules}{\qquad \fbox{$ \ottnt{T_{{\mathrm{1}}}}   \sim _{ \ottnt{m} }  \ottnt{T_{{\mathrm{2}}}} $}}
  \[ \begin{array}{r@{~}c@{~}l}
      \{ \mathit{x} \mathord{:} \ottnt{B} \mathrel{\mid} \ottnt{e_{{\mathrm{1}}}} \}    \sim _{ \ottnt{m} }   \{ \mathit{x} \mathord{:} \ottnt{B} \mathrel{\mid} \ottnt{e_{{\mathrm{2}}}} \}   &\iff& \\
    \multicolumn{3}{r}{\begin{array}{@{}l@{}}
         \forall   \ottnt{e'_{{\mathrm{1}}}}   \sim _{ \ottnt{m} }  \ottnt{e'_{{\mathrm{2}}}}  :   \{ \mathit{x} \mathord{:} \ottnt{B} \mathrel{\mid}  \mathsf{true}  \}    . ~    {} \\  \quad  \ottnt{e_{{\mathrm{1}}}}  [  \ottnt{e'_{{\mathrm{1}}}} / \mathit{x}  ]    \simeq _{ \ottnt{m} }   \ottnt{e_{{\mathrm{2}}}}  [  \ottnt{e'_{{\mathrm{2}}}} / \mathit{x}  ]   :   \{ \mathit{x} \mathord{:}  \mathsf{Bool}  \mathrel{\mid}  \mathsf{true}  \}   
      \end{array}} \\
      \ottnt{T_{{\mathrm{11}}}} \mathord{ \rightarrow } \ottnt{T_{{\mathrm{12}}}}    \sim _{ \ottnt{m} }   \ottnt{T_{{\mathrm{21}}}} \mathord{ \rightarrow } \ottnt{T_{{\mathrm{22}}}}   &\iff&  \ottnt{T_{{\mathrm{11}}}}   \sim _{ \ottnt{m} }  \ottnt{T_{{\mathrm{21}}}}  \wedge  \ottnt{T_{{\mathrm{12}}}}   \sim _{ \ottnt{m} }  \ottnt{T_{{\mathrm{22}}}} 
  \end{array}\]

  \hdr{Closing substitutions and open terms}{\qquad \fbox{$ \Gamma   \models _{ \ottnt{m} }  \delta $} \qquad \fbox{$ \Gamma   \vdash   \ottnt{e_{{\mathrm{1}}}}   \simeq _{ \ottnt{m} }  \ottnt{e_{{\mathrm{2}}}}  :  \ottnt{T} $}}

  \[ \begin{array}{r@{~}c@{~}l}
     \Gamma   \models _{ \ottnt{m} }  \delta  &\iff&  \forall  \mathit{x}  \in  \mathsf{dom} ( \Gamma ) . ~   \delta_{{\mathrm{1}}}  \ottsym{(}  \mathit{x}  \ottsym{)}   \sim _{ \ottnt{m} }  \delta_{{\mathrm{2}}}  \ottsym{(}  \mathit{x}  \ottsym{)}  :   \Gamma ( \mathit{x} )    \\
     \Gamma   \vdash   \ottnt{e_{{\mathrm{1}}}}   \simeq _{ \ottnt{m} }  \ottnt{e_{{\mathrm{2}}}}  :  \ottnt{T}  &\iff&  \forall   \Gamma   \models _{ \ottnt{m} }  \delta   . ~   \delta_{{\mathrm{1}}}  \ottsym{(}  \ottnt{e_{{\mathrm{1}}}}  \ottsym{)}   \simeq _{ \ottnt{m} }  \delta_{{\mathrm{2}}}  \ottsym{(}  \ottnt{e_{{\mathrm{2}}}}  \ottsym{)}  :  \ottnt{T}  
  \end{array} \]

  \caption{Modal logical relations relating classic \lambdah to space-efficient modes}
  \label{fig:lr}
\end{figure}
\fi}

{\iffull
We summarize the relationships between each mode in Table~\ref{tab:soundness}.
\fi}
Our proofs relating classic \lambdah and \ifpopl eidetic \lambdah \else the space-efficient modes \fi are
by \iffull(mode-indexed)\fi logical relations, found in
Figure~\reflr.
{\ifpopl
  In the extended version, the soundness proofs for all three different
  space-efficient modes use a single mode-indexed logical relation. Here we give its restriction to eidetic \lambdah.
\else
The relation is modal: in $ \ottnt{e_{{\mathrm{1}}}}   \sim _{ \ottnt{m} }  \ottnt{e_{{\mathrm{2}}}}  :  \ottnt{T} $, the term $\ottnt{e_{{\mathrm{1}}}}$ is a
classic \lambdah term, while $\ottnt{e_{{\mathrm{2}}}}$ and $\ottnt{T}$ are in mode $\ottnt{m}$.
Each mode's logical relation matches its cast
congruence lemma: the forgetful logical relation allows more blame on
the classic side (not unlike the asymmetric logical relations of
\citet{Greenberg12contracts}); the heedful logical relation is
blame-inexact, allowing classic and heedful \lambdah to raise
different labels; the eidetic logical relation is exact.
The proofs can be found in Appendix~\ref{app:sesoundness}. They follow
a fairly standard pattern in each mode $\ottnt{m}$: we show that applying
$ \mathsf{C} $-casts and $\ottnt{m}$-casts between similar and related types to
related values yields related values (i.e., casts are applicative); we
then show that well typed $ \mathsf{C} $-source programs are related to
$\ottnt{m}$-source programs.
\fi}
As far as alternative techniques go, an induction over evaluation
derivations wouldn't give us enough information about evaluations that
return lambda abstractions. Other contextual equivalence
techniques (e.g., bisimulation) would probably work, too.

{\iffull
Our equivalence results for forgetful and heedful \lambdah are subtle:
they would break down if we had effects other than blame. Forgetful
\lambdah changes which contracts are checked, and so which code is
run. Heedful \lambdah can reorder when code is run. Well typed
\lambdah programs in this paper are strongly normalizing. If we
allowed nontermination, for example, then we could construct source
programs that diverge in classic \lambdah and converge in forgetful
\lambdah, or source programs that diverge in one of classic and
heedful \lambdah and converge in the other.
Similarly, if blame were a \textit{catchable} exception, we would have
no relation for these two modes at all: since they can raise different
blame labels, different exception handlers could have entirely
different behavior.
Eidetic \lambdah doesn't reorder checks, though, so its result is more
durable. As long as checks are pure---they return the same result
every time---eidetic and classic \lambdah coincide.
\fi}

{\iffull
Why bother proving strong results for
forgetful and heedful if they only adhere in such a restricted
setting?
First, we wish to explore the design space. Moreover, forgetful and heedful
offer insights into the semantics and structure of casts.
Second, we want to show soundness of space efficiency in
isolation. Implementations always differ from the theory. {\iffull
  Once this soundness has been established, whether an inefficient
  classic implementation would differ for a given program is less
  relevant when the reference implementation has a heedful
  semantics. \fi}
Analagously, languages with first-class stack traces make tail-call
optimization observable, but this change in semantics is typically
considered worthwhile---space efficiency is more important.

\else

\begin{lemma}[Similar casts are logically related (\ref{lem:eideticlrcast})]
  If $ \ottnt{T_{{\mathrm{1}}}}   \sim _{  \mathsf{E}  }  \ottnt{T'_{{\mathrm{1}}}} $ and $ \ottnt{T_{{\mathrm{2}}}}   \sim _{  \mathsf{E}  }  \ottnt{T'_{{\mathrm{2}}}} $ and $ \ottnt{e_{{\mathrm{1}}}}   \sim _{  \mathsf{E}  }  \ottnt{e_{{\mathrm{2}}}}  :  \ottnt{T_{{\mathrm{1}}}} $, then
  $  \langle  \ottnt{T_{{\mathrm{1}}}}  \mathord{ \overset{\bullet}{\Rightarrow} }  \ottnt{T_{{\mathrm{2}}}}  \rangle^{ \ottnt{l} } ~  \ottnt{e_{{\mathrm{1}}}}    \simeq _{  \mathsf{E}  }   \langle  \ottnt{T'_{{\mathrm{1}}}}  \mathord{ \overset{\bullet}{\Rightarrow} }  \ottnt{T'_{{\mathrm{2}}}}  \rangle^{ \ottnt{l} } ~  \ottnt{e_{{\mathrm{2}}}}   :  \ottnt{T_{{\mathrm{2}}}} $.
\end{lemma}

\begin{lemma}[Relating classic and eidetic source programs (\ref{lem:eideticlr})]
  ~

  \noindent
  \begin{enumerate}
  \item If $ \Gamma   \vdash _{  \mathsf{C}  }  \ottnt{e}  :  \ottnt{T} $ as a source program then $ \Gamma   \vdash   \ottnt{e}   \simeq _{  \mathsf{E}  }  \ottnt{e}  :  \ottnt{T} $.
  \item If $ \mathord{  \vdash _{  \mathsf{C}  } }~ \ottnt{T} $ as a source program then $ \ottnt{T}   \sim _{  \mathsf{E}  }  \ottnt{T} $.
  \end{enumerate}
\end{lemma}

\fi}

\section{Bounds for space efficiency}
\label{sec:bounds}

\begin{table}
\begin{tabular}{|r|c|c|}
  \hline
  \multicolumn{1}{|c|}{\textbf{Mode}} & \textbf{Cast size} & \textbf{Pending casts} \\
  \hline 
  Classic ($\ottnt{m}  \ottsym{=}   \mathsf{C} $)   & $2W_h + L$ & $\infty$ \\ \hline
\iffull  Forgetful ($\ottnt{m}  \ottsym{=}   \mathsf{F} $) & $2W_h + L$ & $\abs{\ottnt{e}}$ \\ \hline
  Heedful ($\ottnt{m}  \ottsym{=}   \mathsf{H} $)   & $2W_h + 2^{W_h} + L$ & $\abs{\ottnt{e}}$ \\ \hline \fi
  Eidetic ($\ottnt{m}  \ottsym{=}   \mathsf{E} $)   & $s 2^{L + W_{\ottnt{B}}}$ & $\abs{\ottnt{e}}$ \\ \hline
\end{tabular}
\caption{Space efficiency of \lambdah }
\label{tab:space}
\end{table}

We have claimed that {\ifpopl eidetic \lambdah is \else forgetful,
heedful, and eidetic \lambdah are \fi} space efficient: what do we mean?
What sort of space efficiency have we achieved\iffull in our various calculi\fi?
We summarize the results in Table~\ref{tab:space}; proofs are in
Appendix~\ref{app:bounds}.
From a high level, there are only a finite number of types that
appear in our programs, and this set of types can only reduce as
the program runs. We can effectively code each type in the program as
an integer, allowing us to efficiently run the $ \supset $ predicate.

Suppose that a type of height $h$ can be represented in $W_h$ bits and
a label in $L$ bits. (Type heights are defined in
Figure~\ref{fig:types} in Appendix~\ref{app:bounds}.) Casts in classic
\iffull and forgetful\fi \lambdah each take up $2W_h + L$ bits: two
types and a blame label.
{\iffull
Casts in heedful \lambdah take up more space---$2W_h + 2^{W_h} + L$
bits---because they need to keep track of the type set. \fi}
Coercions in eidetic \lambdah have a different form: the only types
recorded are those of height $1$, i.e., refinements of base
types. Pessimistically, each of these may appear at every position in
a function coercion $\ottnt{c_{{\mathrm{1}}}}  \mapsto  \ottnt{c_{{\mathrm{2}}}}$. We use $s$ to indicate the ``size''
of a function type, i.e., the number of positions it has. As a first
pass, a set of refinements and blame labels take up $2^{L + W_1}$
space. But in fact these coercions must all be between refinements
\textit{of the same base type}, leading to $2^{L + W_{\ottnt{B}}}$ space
per coercion, where $W_{\ottnt{B}}$ is the highest number of refinements
of any single base type.
We now have our worst-case space complexity: $s 2^{L + W_{\ottnt{B}}}$.
A more precise bound might track which refinements appear in which
parts of a function type, but in the worst case---each
refinement appears in every position---it degenerates to
the bound we give here.
Classic \lambdah can have an infinite number of ``pending
casts''---casts and function proxies---in a program. \ifpopl Eidetic
\else Forgetful, heedful and eidetic \fi \lambdah can have no more
than one pending cast per term node. Abstractions are limited to a
single function proxy, and \ECastMerge\ merges adjacent pending casts.

The text of a program $\ottnt{e}$ is finite, so the set of types appearing
in the program, $ \mathsf{types} ( \ottnt{e} ) $, is also finite. Since reduction
doesn't introduce types, we can bound the number of types in a program
(and therefore the size of casts).
We can therefore fix a numerical coding for types at runtime, where we
can encode a type in $W = \log_2(\abs{ \mathsf{types} ( \ottnt{e} ) })$ bits.
In a given cast, $W$ over-approximates how many types can appear: the
source, target, and annotation must all be compatible, which means
they must also be of the same height.
We can therefore represent the types in casts with fewer bits: $W_h =
\log_2(\abs{\set{\ottnt{T} \mid  \ottnt{T}  \in   \mathsf{types} ( \ottnt{e} )   \wedge  \mathsf{height} ( \ottnt{T} )  =
    h}})$.
In the worst case, we revert to the original bound: all types in the
program are of height 1. Even so, there are never casts between
different base types $\ottnt{B}$ and $\ottnt{B'}$, so $W_{\ottnt{B}} = \max_{\ottnt{B}}
\log_2(\abs{\set{  \{ \mathit{x} \mathord{:} \ottnt{B} \mathrel{\mid} \ottnt{e} \}   \in   \mathsf{types} ( \ottnt{e} )  }})$.
Eidetic \lambdah's coercions never hold types greater than height 1.
The types on its casts are erasable once the coercions are
generated, because coercions drive the checking.

\subsection{Representation choices}
\label{sec:representation}

{\iffull
We have shown that sound space efficiency is possible, and it is future
work to produce a feasible implementation.
\fi}
The bounds we find here are \textit{galactic}. Having established that
contracts are theoretically space efficient, making an implementation
practically space efficient is a different endeavor, involving careful
choices of representations and calling conventions.

Eidetic \lambdah's space bounds rely only on the reflexivity of the
$ \supset $ predicate, since we leave it abstract. We have identified
one situation where the relation allows us to find better space
bounds: mutual implication.

If $ \{ \mathit{x} \mathord{:} \ottnt{B} \mathrel{\mid} \ottnt{e_{{\mathrm{1}}}} \}  \, \supset \,  \{ \mathit{x} \mathord{:} \ottnt{B} \mathrel{\mid} \ottnt{e_{{\mathrm{2}}}} \} $ and $ \{ \mathit{x} \mathord{:} \ottnt{B} \mathrel{\mid} \ottnt{e_{{\mathrm{2}}}} \}  \, \supset \,  \{ \mathit{x} \mathord{:} \ottnt{B} \mathrel{\mid} \ottnt{e_{{\mathrm{1}}}} \} $, then
these two types are equivalent, and only one ever need be
checked. Which to check could be determined by a compiler with a
suitably clever cost model.
Note that our proofs don't entirely justify this optimization. By
default, our $ \mathsf{join} $ operator will take whichever of $ \{ \mathit{x} \mathord{:} \ottnt{B} \mathrel{\mid} \ottnt{e_{{\mathrm{1}}}} \} $
and $ \{ \mathit{x} \mathord{:} \ottnt{B} \mathrel{\mid} \ottnt{e_{{\mathrm{2}}}} \} $ was meant to be checked first. Adapting $ \mathsf{join} $
to always choose one based on some preference relation would not be
particularly hard, and we believe the proofs adapt easily.

Other analyses of the relation seem promising at first, but in fact do
not allow more compact representations.
Suppose we have a program where $ \{ \mathit{x} \mathord{:} \ottnt{B} \mathrel{\mid} \ottnt{e_{{\mathrm{1}}}} \}  \, \supset \,  \{ \mathit{x} \mathord{:} \ottnt{B} \mathrel{\mid} \ottnt{e_{{\mathrm{2}}}} \} $ but not
vice versa, and that $\ottnt{B}$ is our worst case type. That is, $W_{\ottnt{B}} =
2$, because there are 2 different refinements of $\ottnt{B}$ and fewer
refinements of other base types.
The worst-case representation for a refinement list is 2 bits, with
bit $b_i$ indicating whether $\ottnt{e_{\ottmv{i}}}$ is present in the list. Can we
do any better than 2 bits, since $\ottnt{e_{{\mathrm{1}}}}$ can stand in for $\ottnt{e_{{\mathrm{2}}}}$?
Could we represent the two types as just 1 bit?
We cannot when (a) there are constants that pass one type but not the
other and (b) when refinement lists are in the reverse order of
implication.
Suppose there is some $\ottnt{k}$ such that $ \ottnt{e_{{\mathrm{2}}}}  [  \ottnt{k} / \mathit{x}  ]  \,  \longrightarrow ^{*}_{  \mathsf{E}  }  \,  \mathsf{true} $ and
$ \ottnt{e_{{\mathrm{1}}}}  [  \ottnt{k} / \mathit{x}  ]  \,  \longrightarrow ^{*}_{  \mathsf{E}  }  \,  \mathsf{false} $. Now we consider a concatenation of
refinement lists in the reverse ordering: $ \mathsf{join} (  \{ \mathit{x} \mathord{:} \ottnt{B} \mathrel{\mid} \ottnt{e_{{\mathrm{2}}}} \}^{ \ottnt{l} }  ,  \{ \mathit{x} \mathord{:} \ottnt{B} \mathrel{\mid} \ottnt{e_{{\mathrm{1}}}} \}^{ \ottnt{l'} }  ) $. We must retain both checks, since different failures
lead to different blame. The $\ottnt{k}$ that passes $\ottnt{e_{{\mathrm{2}}}}$ but not
$\ottnt{e_{{\mathrm{1}}}}$ should raise $ \mathord{\Uparrow}  \ottnt{l'} $, but other $\ottnt{k'}$ that fail
for both types should raise $ \mathord{\Uparrow}  \ottnt{l} $. One bit isn't enough to
capture the situation of having the coercion
$  \{ \mathit{x} \mathord{:} \ottnt{B} \mathrel{\mid} \ottnt{e_{{\mathrm{2}}}} \}^{ \ottnt{l} }  ,  \{ \mathit{x} \mathord{:} \ottnt{B} \mathrel{\mid} \ottnt{e_{{\mathrm{1}}}} \}^{ \ottnt{l'} }  $.

Finally, what is the right representation for a function? When calling
a function, do we need to run coercions or not?
Jeremy Siek suggested a ``smart closure'' which holds the logic for
branching inside its own code; this may support better branch
prediction than an indirect jump or branching at call sites.

\section{Related work}
\label{sec:related}

Some earlier work uses first-class casts, whereas our casts are always
applied to a term~\cite{Belo11fh,Knowles10hybrid}. It is of course
possible to $\eta$-expand a cast with an abstraction, so no
expressiveness is lost. Leaving casts fully applied saves us from the
puzzling rules managing how casts work on other casts in
space-efficient semantics, like:
$ \langle   \ottnt{T_{{\mathrm{11}}}} \mathord{ \rightarrow } \ottnt{T_{{\mathrm{12}}}}   \mathord{ \overset{\bullet}{\Rightarrow} }   \ottnt{T_{{\mathrm{21}}}} \mathord{ \rightarrow } \ottnt{T_{{\mathrm{22}}}}   \rangle^{ \ottnt{l} } ~   \langle  \ottnt{T_{{\mathrm{11}}}}  \mathord{ \overset{\bullet}{\Rightarrow} }  \ottnt{T_{{\mathrm{12}}}}  \rangle^{ \ottnt{l'} } ~      \!\!  \longrightarrow _{  \mathsf{F}  }   \langle  \ottnt{T_{{\mathrm{21}}}}  \mathord{ \overset{\bullet}{\Rightarrow} }  \ottnt{T_{{\mathrm{22}}}}  \rangle^{ \ottnt{l} } ~     $\!\!.

Previous approaches to space-efficiency have focused on gradual
typing~\cite{Siek06gradual}. This work uses coercions~\cite{Henglein94dynamic},
casts, casts annotated with intermediate types a/k/a
\textit{threesomes}, or some combination of all
three~\cite{Siek09casts, Siek10threesomes, Herman10space,
  Siek12interpretations, Garcia13threesomes}. Recent work relates all
three frameworks, making particular use of
coercions~\cite{Siek15coercions}.
Our type structure differs from that of gradual types, so our space
bounds come in a somewhat novel form.
Gradual types, without the more complicated checking that comes with
predicate contracts, allow for simpler
designs. \Citet{Siek10threesomes} can define a simple recursive
operator on labeled types with a strong relationship to subtyping, the
fundamental property of casts. We haven't been able to discover a
connection in our setting. Instead, we ignore the type structure of
functions and focus our attention on managing labels in lists
of first-order predicate contracts.
In the gradual world, only \Citet{Rastogi12inference} take a similar
approach, ``recursively deconstruct[ing] higher-order types down to
their first-order parts'' when they compute the closure of flows into
and out of type variables.
Gradual types occasionally have simpler proofs, too, e.g., by
induction on evaluation~\cite{Siek09casts}; even when strong reasoning
principles are needed, the presence of dynamic types leads them to use
bisimulation~\cite{Siek10threesomes,Garcia13threesomes,Siek15coercions}.
We use logical relations because \lambdah's type structure is readily
available, and because they allow us to easily reason about how checks
evaluate.

Our coercions are inspired by Henglein's coercions for modeling
injection to and projection from the dynamic
type~\cite{Henglein94dynamic}. Henglein's primitive coercions tag and
untag values, while ours represent checks to be performed on base
types; both our formulation and Henglein's use structural function
coercions.

\citet{Greenberg13thesis}, the most closely related work, offers a
coercion language combining the dynamic types of Henglein's original
work with predicate contracts; his \Efficient language does not quite
achieve ``sound'' space efficeincy. Rather, it is forgetful,
occasionally dropping casts. He omits blame, though he conjectures
that blame for coercions reads left to right (as it does in
\citet{Siek12interpretations}); our eidetic \lambdah verifies this
conjecture.
While Greenberg's languages offer dynamic, simple, and refined types,
our types here are entirely refined. His coercions use Henglein's $!$
and $?$ syntax for injection and projection, while our coercions lack
such a distinction. In our refinement lists, each coercion
simultaneously projects from one refinement type and injects into
another (possibly producing blame). We reduce notation by omitting the
interrobang `\interrobang'.
\TODO{one reviewer didn't like this :(}

\Citet{Dimoulas13options} introduce \textit{option contracts}, which
offer a programmatic way of turning off contract checking, as well as
a controlled way to ``pass the buck'', handing off contracts from
component to component. Option contracts address time efficiency, not space efficiency.
\Citet{Findler08immutable} studied space and time efficiency for
datatype contracts, as did \citet{Koukoutos14data}.

Racket contracts have a mild form of space efficiency: the
\texttt{tail-marks-match?} predicate\footnote{From
  \texttt{racket/collects/racket/contract/private/arrow.rkt}.} checks
for exact duplicate contracts and blame at tail positions. The redundancy it
detects seems to rely on pointer equality. Since Racket contracts
are (a) module-oriented ``macro'' contracts and (b) first class, this
optimization is somewhat unpredictable---and limited compared with our
\ifpopl eidetic calculus\else heedful and eidetic calculi\fi, which can
handle differing contracts and blame labels.

{\iffull
\finishlater{
From Robby:
FWIW, the main reason we avoid this problem is that contracts are not
generally enforced on recursive calls, due to the way we set up
boundaries, it is possible to put a contract inside a recursive knot
using higher-order functions or by using a low-level contract
operator.

Here's some examples, one where the library detects the redundancy and
one where it doesn't.}

\begin{verbatim}
#lang racket

(define (count-em-integer? x)
  (printf "checking ~s\n" x)
  (integer? x))

;; yes:
(letrec ([f
          (contract (-> any/c count-em-integer?)
                    (lambda (x)
                      (printf "x: ~s\n" x)
                      (if (zero? x) x (f (- x 1))))
                    'pos
                    'neg)])
  (f 3))

;; no:
(letrec ([f
          (contract (-> any/c count-em-integer?)
                    (contract (-> any/c count-em-integer?)
                              (lambda (x)
                                (printf "x: ~s\n" x)
                                (if (zero? x) x (f (- x 1))))
                              'pos
                              'neg)
                    'pos
                    'neg)])
  (f 3))
\end{verbatim}
\fi}

\section{Conclusion and future work}
\label{sec:conclusion}

Semantics-preserving space efficiency for manifest contracts is
possible---leaving the admissibility of state as the final barrier to
practical utility. We established that eidetic \lambdah behaves
exactly like its classic counterpart without compromising space usage.
{\iffull Forgetful \lambdah is an interesting middle ground: if contracts exist
to make partial operations safe (and not abstraction or information
hiding), forgetfulness may be a good strategy. \fi}

We believe it would be easy to design a latent version of eidetic
\lambdah, following the translations in \citet{Greenberg10contracts}.

In our simple (i.e., not dependent) case, our refinement types close
over a single variable of base type.
Space efficiency for a dependent calculus remains open.
The first step towards dependent types would be extending $ \supset $
with a context (and a source of closing substitutions, a serious
issue~\cite{Belo11fh}).
In a dependent setting the definition of what it means to compare
closures isn't at all clear. Closures' environments may contain
functions, and closures over extensionally equivalent functions may
not be intensionally equal.
A more nominal approach to contract comparison may resolve some of the
issues here. Comparisons might be more straightforward when contracts
are explicitly declared and referenced by name. Similarly, a dependent
$ \supset $ predicate might be more easily defined over some explicit
structured family of types, like a lattice. \Citet{Findler08immutable}
has made some progress in this direction.

Finally, a host of practical issues remain. Beyond representation
choices, having expensive checks makes it important to predict
\textit{when} checks happen. The $ \supset $ predicate compares closures
and will surely have delicate interactions with optimizations.

%\vspace*{-1em}
{\iffull
\subsection*{Acknowledgments}
\else
\acks
\fi}
Comments from Rajeev Alur, Ron Garcia, Fritz Henglein, Greg Morrisett,
Stephanie Weirich, Phil Wadler, and Steve Zdancewic improved a
previous version of this work done at the University of
Pennsylvania. Some comments from Benjamin Pierce led me to realize
that a cast formulation was straightforward.
Discussions with Atsushi Igarashi, Robby Findler, and Sam
Tobin-Hochstadt greatly improved the quality of the exposition. Phil
Wadler encouraged me to return to coercions to understand the eidetic
formulation. The POPL reviewers had many excellent suggestions, and
Robby Findler helped once more with angelic guidance. Hannah de
Keijzer proofread the paper.

This work was supported in part by the NSF under grants TC 0915671 and
SHF 1016937 and by the DARPA CRASH program through the United States
Air Force Research Laboratory (AFRL) under Contract
No. FA8650-10-C-7090.  The views expressed are the author's and do not
reflect the official policy or position of the Department of Defense
or the U.S. Government.

% We recommend abbrvnat bibliography style.

\vspace*{1.0em}
\bibliographystyle{abbrvnat}
\bibliography{../../mgree}

% The bibliography should be embedded for final submission.

\pagebreak
\appendix
\numbertheoremstrue

\section{Proofs of type soundness}
\label{app:typesoundness}

This appendix includes the proofs of type soundness for all four modes
of \lambdah; we first prove some universally applicable
metatheoretical properties.

\subsection{Generic metatheory}
\label{app:genericmetatheory}

\begin{lemma}[Weakening]
  \label{lem:weakening}
  If $ \Gamma_{{\mathrm{1}}}  \ottsym{,}  \Gamma_{{\mathrm{2}}}   \vdash _{ \ottnt{m} }  \ottnt{e}  :  \ottnt{T} $ and $ \mathord{  \vdash _{ \ottnt{m} } }~ \ottnt{T'} $ and $\mathit{x}$ is fresh, then
  $ \mathord{  \vdash _{ \ottnt{m} } }~  \Gamma_{{\mathrm{1}}} , \mathit{x} \mathord{:} \ottnt{T'}   \ottsym{,}  \Gamma_{{\mathrm{2}}} $ and $  \Gamma_{{\mathrm{1}}} , \mathit{x} \mathord{:} \ottnt{T'}   \ottsym{,}  \Gamma_{{\mathrm{2}}}   \vdash _{ \ottnt{m} }  \ottnt{e}  :  \ottnt{T} $.
  {\iffull
  \begin{proof}
    The context well formedness is by induction on $\Gamma_{{\mathrm{2}}}$.
    \item[(\WF{Empty})] By \WF{Extend}.
    \item[(\WF{Extend})] By \WF{Extend} and the IH.

    By induction on $\ottnt{e}$, leaving $\Gamma_{{\mathrm{2}}}$ generalized.
    \item[($\ottnt{e}  \ottsym{=}  \mathit{y}$)] If $ \mathit{y}  \mathord{:}  \ottnt{T}  \in  \Gamma_{{\mathrm{1}}}  \ottsym{,}  \Gamma_{{\mathrm{2}}} $, then $ \mathit{y}  \mathord{:}  \ottnt{T}  \in   \Gamma_{{\mathrm{1}}} , \mathit{x} \mathord{:} \ottnt{T'}   \ottsym{,}  \Gamma_{{\mathrm{2}}} $.
    \item[($\ottnt{e}  \ottsym{=}  \ottnt{k}$)] By \T{Const} and the first part of the IH.
    \item[($\ottnt{e}  \ottsym{=}   \lambda \mathit{y} \mathord{:} \ottnt{T_{{\mathrm{1}}}} .~  \ottnt{e} $)] By \T{Abs} and the IH on $\ottnt{e}$, using
      $ \Gamma_{{\mathrm{2}}} , \mathit{y} \mathord{:} \ottnt{T_{{\mathrm{1}}}} $.
    \item[($\ottnt{e}  \ottsym{=}   \langle  \ottnt{T_{{\mathrm{1}}}}  \mathord{ \overset{\bullet}{\Rightarrow} }  \ottnt{T_{{\mathrm{2}}}}  \rangle^{ \ottnt{l} } ~  \ottnt{e'} $)] By \T{Cast} and the IHs.
    \item[($\ottnt{e}  \ottsym{=}   \langle  \ottnt{T_{{\mathrm{1}}}}  \mathord{ \overset{ \ottnt{c} }{\Rightarrow} }  \ottnt{T_{{\mathrm{2}}}}  \rangle^{\bullet} ~  \ottnt{e'} $)] By \T{Coerce} and the IHs.
    \item[($\ottnt{e}  \ottsym{=}   \mathord{\Uparrow}  \ottnt{l} $)] By \T{Blame} and the IH.
    \item[($\ottnt{e}  \ottsym{=}   \ottnt{e_{{\mathrm{1}}}} ~ \ottnt{e_{{\mathrm{2}}}} $)] By \T{App} and the IHs.
    \item[($\ottnt{e}  \ottsym{=}  \ottnt{op}  \ottsym{(}  \ottnt{e_{{\mathrm{1}}}}  \ottsym{,}  \dots  \ottsym{,}  \ottnt{e_{\ottmv{n}}}  \ottsym{)}$)] By \T{Op} and the IHs.
    \item[($\ottnt{e}  \ottsym{=}   \langle   \{ \mathit{x} \mathord{:} \ottnt{B} \mathrel{\mid} \ottnt{e_{{\mathrm{1}}}} \}  ,  \ottnt{e_{{\mathrm{2}}}} ,  \ottnt{k}  \rangle^{ \ottnt{l} } $)] By \T{Check} and the IH,
      observing that $\ottnt{e_{{\mathrm{2}}}}$ and $\ottnt{k}$ are closed.
  \end{proof}
  \fi}
\end{lemma}

\begin{lemma}[Substitution]
  \label{lem:substitution}
  If $  \Gamma_{{\mathrm{1}}} , \mathit{x} \mathord{:} \ottnt{T'}   \ottsym{,}  \Gamma_{{\mathrm{2}}}   \vdash _{ \ottnt{m} }  \ottnt{e}  :  \ottnt{T} $ and $ \emptyset   \vdash _{ \ottnt{m} }  \ottnt{e'}  :  \ottnt{T'} $, then
  $ \Gamma_{{\mathrm{1}}}  \ottsym{,}  \Gamma_{{\mathrm{2}}}   \vdash _{ \ottnt{m} }   \ottnt{e}  [  \ottnt{e'} / \mathit{x}  ]   :  \ottnt{T} $ and $ \mathord{  \vdash _{ \ottnt{m} } }~ \Gamma_{{\mathrm{1}}}  \ottsym{,}  \Gamma_{{\mathrm{2}}} $.
  {\iffull
  \begin{proof}
    By induction on the typing derivation.
    \begin{itemize}
    \item[(\T{Var})] By the assumption of typing and weakening,
      Lemma~\ref{lem:weakening}.

    \item[(\T{Const})] By \T{Const} and the IH (for $ \mathord{  \vdash _{ \ottnt{m} } }~ \Gamma_{{\mathrm{1}}}  \ottsym{,}  \Gamma_{{\mathrm{2}}} $).

    \item[(\T{Abs})] By \T{Abs} and the IH, renaming the variable.

    \item[(\T{Op})] By \T{Op} and the IHs.

    \item[(\T{App})] By \T{App} and the IHs.

    \item[(\T{Cast})] By \T{Cast} and the IHs, observing that all
      types are closed, though the term may not be.

    \item[(\T{Coerce})] By \T{Coerce} and the IHs, observing that all
      types are closed, though the term may not be.

    \item[(\T{Blame})] By \T{Blame} and the IH (for $ \mathord{  \vdash _{ \ottnt{m} } }~ \Gamma_{{\mathrm{1}}}  \ottsym{,}  \Gamma_{{\mathrm{2}}} $).

    \item[(\T{Check})] By \T{Check} and the IH (for $ \mathord{  \vdash _{ \ottnt{m} } }~ \Gamma_{{\mathrm{1}}}  \ottsym{,}  \Gamma_{{\mathrm{2}}} $). Note that the subterms are all closed, so the
      substitution actually has no effect other than strengthening the
      context.

    \end{itemize}    
  \end{proof}
  \fi}
\end{lemma}

{\iffull
\begin{lemma}[Coercion typing regularity]
  \label{lem:coercionregularity}
  If $ \mathord{  \vdash _{  \mathsf{E}  } }~ \ottnt{c}   \mathrel{\parallel}   \ottnt{T_{{\mathrm{1}}}}  \Rightarrow  \ottnt{T_{{\mathrm{2}}}} $ then $ \mathord{  \vdash _{  \mathsf{E}  } }~ \ottnt{T_{{\mathrm{1}}}} $ and $ \mathord{  \vdash _{  \mathsf{E}  } }~ \ottnt{T_{{\mathrm{2}}}} $.
  \begin{proof}
    By induction on the coercion well formedness derivation.
    \begin{itemize}
      \item[(\A{Refine})] By assumption.
      \item[(\A{Fun})] By \WF{Fun} on the IHs.
    \end{itemize}
  \end{proof}
\end{lemma}
\fi}

\begin{lemma}[Regularity]
  \label{lem:regularity}
  If $ \Gamma   \vdash _{ \ottnt{m} }  \ottnt{e}  :  \ottnt{T} $, then $ \mathord{  \vdash _{ \ottnt{m} } }~ \Gamma $ and $ \mathord{  \vdash _{ \ottnt{m} } }~ \ottnt{T} $.
  {\iffull
  \begin{proof}
    By induction on the typing relation.
    \begin{itemize}
    \item[(\T{Var})] We have $ \mathord{  \vdash _{ \ottnt{m} } }~ \Gamma $ by assumption, and $ \mathord{  \vdash _{ \ottnt{m} } }~ \ottnt{T} $
      by induction on the derivation of $ \mathord{  \vdash _{ \ottnt{m} } }~ \Gamma $.

    \item[(\T{Const})] We have $ \mathord{  \vdash _{ \ottnt{m} } }~ \Gamma $ and $ \mathord{  \vdash _{ \ottnt{m} } }~ \ottnt{T} $ by assumption.

    \item[(\T{Abs})] By the IH, we get $ \mathord{  \vdash _{ \ottnt{m} } }~  \Gamma , \mathit{x} \mathord{:} \ottnt{T_{{\mathrm{1}}}}  $ and $ \mathord{  \vdash _{ \ottnt{m} } }~ \ottnt{T_{{\mathrm{2}}}} $. By inversion we find $ \mathord{  \vdash _{ \ottnt{m} } }~ \Gamma $; by \WF{Fun} and the
      assumption that $ \mathord{  \vdash _{ \ottnt{m} } }~ \ottnt{T_{{\mathrm{1}}}} $, we find $ \mathord{  \vdash _{ \ottnt{m} } }~  \ottnt{T_{{\mathrm{1}}}} \mathord{ \rightarrow } \ottnt{T_{{\mathrm{2}}}}  $.

    \item[(\T{Op})] We have $ \mathord{  \vdash _{ \ottnt{m} } }~ \Gamma $ and $ \mathord{  \vdash _{ \ottnt{m} } }~ \ottnt{T} $ by assumption.

    \item[(\T{App})] By the IH, $ \mathord{  \vdash _{ \ottnt{m} } }~  \ottnt{T_{{\mathrm{1}}}} \mathord{ \rightarrow } \ottnt{T_{{\mathrm{2}}}}  $ and $ \mathord{  \vdash _{ \ottnt{m} } }~ \Gamma $. By
      inversion, $ \mathord{  \vdash _{ \ottnt{m} } }~ \ottnt{T_{{\mathrm{2}}}} $.

    \item[(\T{Cast})] By \WF{Fun} and inversion of the type set well
      formedness; $ \mathord{  \vdash _{ \ottnt{m} } }~ \Gamma $ is by the IH on $ \Gamma   \vdash _{ \ottnt{m} }  \ottnt{e}  :  \ottnt{T_{{\mathrm{1}}}} $.

    \item[(\T{Coerce})] It must be that $\ottnt{m}  \ottsym{=}   \mathsf{E} $; by coercion
      regularity (Lemma~\ref{lem:coercionregularity}).

    \item[(\T{Blame})] By assumption.

    \item[(\T{Check})] By assumption.

    \end{itemize}
  \end{proof}
  \fi}
\end{lemma}

\begin{lemma}[Similarity is reflexive]
  \label{lem:similarityreflexive}
  If $\vdash  \ottnt{T}  \mathrel{\parallel}  \ottnt{T}$.
  \begin{proof}
    By induction on $\ottnt{T}$.
    \begin{itemize}
    \item[($\ottnt{T}  \ottsym{=}   \{ \mathit{x} \mathord{:} \ottnt{B} \mathrel{\mid} \ottnt{e} \} $)] By \S{Refine}.
    \item[($\ottnt{T}  \ottsym{=}   \ottnt{T_{{\mathrm{1}}}} \mathord{ \rightarrow } \ottnt{T_{{\mathrm{2}}}} $)] By \S{Fun} and the IHs.
    \end{itemize}
  \end{proof}
\end{lemma}

\begin{lemma}[Similarity is symmetric]
  \label{lem:similaritysymmetric}
  If $\vdash  \ottnt{T_{{\mathrm{1}}}}  \mathrel{\parallel}  \ottnt{T_{{\mathrm{2}}}}$, then $\vdash  \ottnt{T_{{\mathrm{2}}}}  \mathrel{\parallel}  \ottnt{T_{{\mathrm{1}}}}$.
  \begin{proof}
    By induction on the similarity derivation.
    \begin{itemize}
    \item[(\S{Refine})] By \S{Refine}.
    \item[(\S{Fun})] By \S{Fun} and the IHs.
    \end{itemize}
  \end{proof}
\end{lemma}

\begin{lemma}[Similarity is transitive]
  \label{lem:similaritytransitive}
  If $\vdash  \ottnt{T_{{\mathrm{1}}}}  \mathrel{\parallel}  \ottnt{T_{{\mathrm{2}}}}$ and $\vdash  \ottnt{T_{{\mathrm{2}}}}  \mathrel{\parallel}  \ottnt{T_{{\mathrm{3}}}}$, then $\vdash  \ottnt{T_{{\mathrm{1}}}}  \mathrel{\parallel}  \ottnt{T_{{\mathrm{3}}}}$.
  \begin{proof}
    By induction on the derivation of $\vdash  \ottnt{T_{{\mathrm{1}}}}  \mathrel{\parallel}  \ottnt{T_{{\mathrm{2}}}}$.
    \begin{itemize}
    \item[(\S{Refine})] The other derivation must also be by \S{Refine}; by \S{Refine}.
    \item[(\S{Fun})] The other derivation must also be by \S{Fun}; by
      \S{Fun} and the IHs.
    \end{itemize}
  \end{proof}
\end{lemma}

\begin{lemma}[Well formed type sets have similar indices]
  \label{lem:typesetsimilar}
  ~

  \noindent
  If $ \mathord{  \vdash _{ \ottnt{m} } }~ \mathcal{S}   \mathrel{\parallel}   \ottnt{T_{{\mathrm{1}}}}  \Rightarrow  \ottnt{T_{{\mathrm{2}}}} $ then $\vdash  \ottnt{T_{{\mathrm{1}}}}  \mathrel{\parallel}  \ottnt{T_{{\mathrm{2}}}}$.
  \begin{proof}
    Immediate, by inversion.
  \end{proof}  
\end{lemma}

\begin{lemma}[Type set well formedness is symmetric]
  \label{lem:typesetsymmetric}
  $ \mathord{  \vdash _{ \ottnt{m} } }~ \ottnt{a}   \mathrel{\parallel}   \ottnt{T_{{\mathrm{1}}}}  \Rightarrow  \ottnt{T_{{\mathrm{2}}}} $ iff $ \mathord{  \vdash _{ \ottnt{m} } }~ \ottnt{a}   \mathrel{\parallel}   \ottnt{T_{{\mathrm{2}}}}  \Rightarrow  \ottnt{T_{{\mathrm{1}}}} $ for all $\ottnt{m}  \neq   \mathsf{E} $.
  \begin{proof}
    We immediately have $ \mathord{  \vdash _{ \ottnt{m} } }~ \ottnt{T_{{\mathrm{1}}}} $ and $ \mathord{  \vdash _{ \ottnt{m} } }~ \ottnt{T_{{\mathrm{2}}}} $, and $\vdash  \ottnt{T_{{\mathrm{1}}}}  \mathrel{\parallel}  \ottnt{T_{{\mathrm{2}}}}$ iff $\vdash  \ottnt{T_{{\mathrm{2}}}}  \mathrel{\parallel}  \ottnt{T_{{\mathrm{1}}}}$ by
    Lemma~\ref{lem:similaritysymmetric}.

    If $\ottnt{m}  \ottsym{=}   \mathsf{C} $ or $\ottnt{m}  \ottsym{=}   \mathsf{F} $, then by \A{None} and symmetry of
    similarity (Lemma~\ref{lem:similaritysymmetric}.

    If $\ottnt{m}  \ottsym{=}   \mathsf{H} $, then let $ \ottnt{T}  \in  \mathcal{S} $ be given. The $ \mathord{  \vdash _{  \mathsf{H}  } }~ \ottnt{T} $
    premises hold immediately; we are then done by transitivity
    (Lemma~\ref{lem:similaritytransitive}) and symmetry
    (Lemma~\ref{lem:similaritysymmetric}) of similarity ($\vdash  \ottnt{T}  \mathrel{\parallel}  \ottnt{T_{{\mathrm{1}}}}$ iff $\vdash  \ottnt{T}  \mathrel{\parallel}  \ottnt{T_{{\mathrm{2}}}}$ when $\vdash  \ottnt{T_{{\mathrm{1}}}}  \mathrel{\parallel}  \ottnt{T_{{\mathrm{2}}}}$).
  \end{proof}  
\end{lemma}

\begin{lemma}[Type set well formedness is transitive]
  \label{lem:typesettransitive}
  If $\vdash  \ottnt{T_{{\mathrm{1}}}}  \mathrel{\parallel}  \ottnt{T_{{\mathrm{2}}}}$ and $ \mathord{  \vdash _{ \ottnt{m} } }~ \ottnt{a}   \mathrel{\parallel}   \ottnt{T_{{\mathrm{2}}}}  \Rightarrow  \ottnt{T_{{\mathrm{3}}}} $ and $ \mathord{  \vdash _{ \ottnt{m} } }~ \ottnt{T_{{\mathrm{1}}}} $
  and $\ottnt{m}  \neq   \mathsf{E} $ then $ \mathord{  \vdash _{ \ottnt{m} } }~ \ottnt{a}   \mathrel{\parallel}   \ottnt{T_{{\mathrm{1}}}}  \Rightarrow  \ottnt{T_{{\mathrm{3}}}} $.
  \begin{proof}
    We immediately have $ \mathord{  \vdash _{ \ottnt{m} } }~ \ottnt{T_{{\mathrm{1}}}} $ and $ \mathord{  \vdash _{ \ottnt{m} } }~ \ottnt{T_{{\mathrm{3}}}} $; we have $\vdash  \ottnt{T_{{\mathrm{1}}}}  \mathrel{\parallel}  \ottnt{T_{{\mathrm{3}}}}$ by transitivity of similarity
    (Lemma~\ref{lem:similaritytransitive}).

    If $\ottnt{m}  \ottsym{=}   \mathsf{C} $ or $\ottnt{m}  \ottsym{=}   \mathsf{F} $, we are done immediately by
    \A{None}.

    If, on the other hand, $\ottnt{m}  \ottsym{=}   \mathsf{H} $, let $ \ottnt{T}  \in  \mathcal{S} $ be given. We
    know that $ \mathord{  \vdash _{  \mathsf{H}  } }~ \ottnt{T} $ and $\vdash  \ottnt{T}  \mathrel{\parallel}  \ottnt{T_{{\mathrm{2}}}}$; by symmetry
    (Lemma~\ref{lem:similaritysymmetric}) and transitivity
    (Lemma~\ref{lem:similaritytransitive}) of similarity, we are done
    by \A{TypeSet}.
  \end{proof}
\end{lemma}

\subsection{Classic type soundness}
\label{app:classicsoundness}

\begin{lemma}[Classic determinism]
  \label{lem:classicdeterminism}
  If $\ottnt{e} \,  \longrightarrow _{  \mathsf{C}  }  \, \ottnt{e_{{\mathrm{1}}}}$ and $\ottnt{e} \,  \longrightarrow _{  \mathsf{C}  }  \, \ottnt{e_{{\mathrm{2}}}}$ then $\ottnt{e_{{\mathrm{1}}}}  \ottsym{=}  \ottnt{e_{{\mathrm{2}}}}$.
  \begin{proof}
    By induction on the first evaluation derivation.
  \end{proof}  
\end{lemma}

\begin{lemma}[Classic canonical forms]
  \label{lem:classiccanonicalforms}
  If $ \emptyset   \vdash _{  \mathsf{C}  }  \ottnt{e}  :  \ottnt{T} $ and $ \mathsf{val} _{  \mathsf{C}  }~ \ottnt{e} $ then:
  \begin{itemize}
  \item If $\ottnt{T}  \ottsym{=}   \{ \mathit{x} \mathord{:} \ottnt{B} \mathrel{\mid} \ottnt{e'} \} $, then $\ottnt{e}  \ottsym{=}  \ottnt{k}$ and $ \mathsf{ty} ( \ottnt{k} )   \ottsym{=}  \ottnt{B}$
    and $ \ottnt{e'}  [  \ottnt{e} / \mathit{x}  ]  \,  \longrightarrow ^{*}_{  \mathsf{C}  }  \,  \mathsf{true} $.
  \item If $\ottnt{T}  \ottsym{=}   \ottnt{T_{{\mathrm{1}}}} \mathord{ \rightarrow } \ottnt{T_{{\mathrm{2}}}} $, then either $\ottnt{e}  \ottsym{=}   \lambda \mathit{x} \mathord{:} \ottnt{T} .~  \ottnt{e'} $ or $\ottnt{e}  \ottsym{=}   \langle   \ottnt{T_{{\mathrm{11}}}} \mathord{ \rightarrow } \ottnt{T_{{\mathrm{12}}}}   \mathord{ \overset{\bullet}{\Rightarrow} }   \ottnt{T_{{\mathrm{21}}}} \mathord{ \rightarrow } \ottnt{T_{{\mathrm{22}}}}   \rangle^{ \ottnt{l} } ~  \ottnt{e'} $.
  \end{itemize}
  {\iffull
  \begin{proof}
    By inversion of $ \mathsf{val} _{  \mathsf{C}  }~ \ottnt{e} $ and inspection of the typing
    rules. \V{Const}/\T{Const} are the only rules that types values at
    base types; \V{Abs}/\T{Abs} and \V{Proxy}/\T{Cast} are the only
    rules that type values at function types.
  \end{proof}
  \fi}
\end{lemma}

\begin{lemma}[Classic progress]
  \label{lem:classicprogress}
  If $ \emptyset   \vdash _{  \mathsf{C}  }  \ottnt{e}  :  \ottnt{T} $, then either:
  \begin{enumerate}
  \item $ \mathsf{result} _{  \mathsf{C}  }~ \ottnt{e} $, i.e., $\ottnt{e}  \ottsym{=}   \mathord{\Uparrow}  \ottnt{l} $ or $ \mathsf{val} _{  \mathsf{C}  }~ \ottnt{e} $; or
  \item there exists an $\ottnt{e'}$ such that $\ottnt{e} \,  \longrightarrow _{  \mathsf{C}  }  \, \ottnt{e'}$.
  \end{enumerate}
  \begin{proof}
    By induction on the typing derivation.
    {\iffull
    \begin{itemize}
    \item[(\T{Var})] A contradiction---$\mathit{x}$ isn't well typed in the
      empty environment.

    \item[(\T{Const})] $\ottnt{e}  \ottsym{=}  \ottnt{k}$ is a result by \V{Const} and \R{Val}.

    \item[(\T{Abs})] $\ottnt{e}  \ottsym{=}   \lambda \mathit{x} \mathord{:} \ottnt{T} .~  \ottnt{e'} $ is a result by \V{Abs} and \R{Val}.

    \item[(\T{Op})] We know that $ \mathsf{ty} (\mathord{ \ottnt{op} }) $ is a first order type
      $ {}   \{ \mathit{x} \mathord{:} \ottnt{B_{{\mathrm{1}}}} \mathrel{\mid} \ottnt{e'_{{\mathrm{1}}}} \}   \rightarrow \, ... \, \rightarrow   \{ \mathit{x} \mathord{:} \ottnt{B_{\ottmv{n}}} \mathrel{\mid} \ottnt{e'_{\ottmv{n}}} \}   {} \mathord{ \rightarrow } \ottnt{T} $.
      From left to right, we apply the IH on $\ottnt{e_{\ottmv{i}}}$. If $\ottnt{e_{\ottmv{i}}}$ is
      a result, there are two cases: either $\ottnt{e_{\ottmv{i}}}  \ottsym{=}   \mathord{\Uparrow}  \ottnt{l} $, and
      we step by \E{OpRaise}; or $\ottnt{e_{\ottmv{i}}}  \ottsym{=}  \ottnt{k_{\ottmv{i}}}$, since constants are
      the only values at base types by canonical forms
      (Lemma~\ref{lem:classiccanonicalforms}), and we continue on to the next
      $\ottnt{e_{\ottmv{i}}}$. If any of the $\ottnt{e_{\ottmv{i}}}$ step, we know that all of terms
      before them are values, so we can step by \E{OpInner}. If all of
      the $\ottnt{e_{\ottmv{i}}}$ are constants $\ottnt{k_{\ottmv{i}}}$, then $ \emptyset   \vdash _{  \mathsf{C}  }  \ottnt{k_{\ottmv{i}}}  :   \{ \mathit{x} \mathord{:} \ottnt{B_{\ottmv{i}}} \mathrel{\mid} \ottnt{e'_{\ottmv{i}}} \}  $, and so $ \ottnt{e'_{\ottmv{i}}}  [  \ottnt{k_{\ottmv{i}}} / \mathit{x}  ]  \,  \longrightarrow ^{*}_{  \mathsf{C}  }  \,  \mathsf{true} $. Therefore
      $\denot{ op } \, \ottsym{(}  \ottnt{k_{{\mathrm{1}}}}  \ottsym{,} \, ... \, \ottsym{,}  \ottnt{k_{\ottmv{n}}}  \ottsym{)}$ is defined, and we can step by \E{Op}.

    \item[(\T{App})] By the IH on $\ottnt{e_{{\mathrm{1}}}}$, we know that $\ottnt{e_{{\mathrm{1}}}}$: is
      blame, is a value, or steps to some $\ottnt{e'_{{\mathrm{1}}}}$. In the first
      case, we take a step by \E{AppRaiseL}. In the latter, we take a
      step by \E{AppL}.

      If $ \mathsf{val} _{  \mathsf{C}  }~ \ottnt{e_{{\mathrm{1}}}} $, then we can apply the IH on $\ottnt{e_{{\mathrm{2}}}}$, which
      is blame, is a value, or steps to some $\ottnt{e'_{{\mathrm{2}}}}$. The first and
      last cases are as before, using \E{AppRaiseR} and \E{AppR}.

      If $ \mathsf{val} _{  \mathsf{C}  }~ \ottnt{e_{{\mathrm{2}}}} $, we must go by cases on the shape of
      $\ottnt{e_{{\mathrm{1}}}}$. Since $ \emptyset   \vdash _{  \mathsf{C}  }  \ottnt{e_{{\mathrm{1}}}}  :   \ottnt{T_{{\mathrm{1}}}} \mathord{ \rightarrow } \ottnt{T_{{\mathrm{2}}}}  $, by canonical forms
      (Lemma~\ref{lem:classiccanonicalforms}) we know that $\ottnt{e_{{\mathrm{1}}}}$ can only
      be an abstraction, a wrapped abstraction, or a cast.
      \begin{itemize}
      \item[($\ottnt{e_{{\mathrm{1}}}}  \ottsym{=}   \lambda \mathit{x} \mathord{:} \ottnt{T_{{\mathrm{1}}}} .~  \ottnt{e'_{{\mathrm{1}}}} $)] We step to $ \ottnt{e'_{{\mathrm{1}}}}  [  \ottnt{e_{{\mathrm{2}}}} / \mathit{x}  ] $ by
        \E{Beta}.

      \item[($\ottnt{e_{{\mathrm{1}}}}  \ottsym{=}   \langle   \ottnt{T_{{\mathrm{11}}}} \mathord{ \rightarrow } \ottnt{T_{{\mathrm{12}}}}   \mathord{ \overset{\bullet}{\Rightarrow} }   \ottnt{T_{{\mathrm{21}}}} \mathord{ \rightarrow } \ottnt{T_{{\mathrm{22}}}}   \rangle^{ \ottnt{l} } ~  \ottnt{e'_{{\mathrm{1}}}} $)] We step to
        $ \langle  \ottnt{T_{{\mathrm{12}}}}  \mathord{ \overset{\bullet}{\Rightarrow} }  \ottnt{T_{{\mathrm{22}}}}  \rangle^{ \ottnt{l} } ~   (  \ottnt{e'_{{\mathrm{1}}}} ~  (  \langle  \ottnt{T_{{\mathrm{21}}}}  \mathord{ \overset{\bullet}{\Rightarrow} }  \ottnt{T_{{\mathrm{11}}}}  \rangle^{ \ottnt{l} } ~  \ottnt{e_{{\mathrm{2}}}}  )   )  $ by \E{Unwrap}, noting
        that all of the type sets are empty.
      \end{itemize}

    \item[(\T{Cast})] If $\ottnt{e'} \,  \longrightarrow _{  \mathsf{C}  }  \, \ottnt{e''}$, then by \E{CastInnerC}. If
      $\ottnt{e'}$ is blame, then by \E{CastRaise}. If $ \mathsf{val} _{  \mathsf{C}  }~ \ottnt{e'} $,
      then we invert $\vdash  \ottnt{T_{{\mathrm{1}}}}  \mathrel{\parallel}  \ottnt{T_{{\mathrm{2}}}}$, finding that either:
      \begin{itemize}
      \item[($\ottnt{T_{\ottmv{i}}}  \ottsym{=}   \{ \mathit{x} \mathord{:} \ottnt{B} \mathrel{\mid} \ottnt{e_{\ottmv{i}}} \} $)] By canonical forms
        (Lemma~\ref{lem:classiccanonicalforms}), $\ottnt{e_{{\mathrm{2}}}}  \ottsym{=}  \ottnt{k}$. We can step to
        $ \langle   \{ \mathit{x} \mathord{:} \ottnt{B} \mathrel{\mid} \ottnt{e_{{\mathrm{2}}}} \}  ,   \ottnt{e_{{\mathrm{2}}}}  [  \ottnt{k} / \mathit{x}  ]  ,  \ottnt{k}  \rangle^{ \ottnt{l} } $ by \ECheckNone.

      \item[($\ottnt{T_{\ottmv{i}}}  \ottsym{=}   \ottnt{T_{\ottmv{i}\,{\mathrm{1}}}} \mathord{ \rightarrow } \ottnt{T_{\ottmv{i}\,{\mathrm{2}}}} $)] This is a value, \V{ProxyC}.
      \end{itemize}

    \item[(\T{Blame})] $ \mathord{\Uparrow}  \ottnt{l} $ is a result by \R{Blame}.

    \item[(\T{Check})] By the IH on $\ottnt{e_{{\mathrm{2}}}}$, we know that $\ottnt{e_{{\mathrm{2}}}}$:
      is $ \mathord{\Uparrow}  \ottnt{l'} $, is a value by $ \mathsf{val} _{  \mathsf{C}  }~ \ottnt{e_{{\mathrm{2}}}} $, or takes a
      step to some $\ottnt{e'_{{\mathrm{2}}}}$. In the first case, we step to
      $ \mathord{\Uparrow}  \ottnt{l'} $ by \E{CheckRaise}. In the last case, we step by
      \E{CheckInner}. If $ \mathsf{val} _{  \mathsf{C}  }~ \ottnt{e_{{\mathrm{2}}}} $, by canonical forms
      (Lemma~\ref{lem:classiccanonicalforms}) we know that $\ottnt{e_{{\mathrm{2}}}}$ is a
      $\ottnt{k}$ such that $ \mathsf{ty} ( \ottnt{k} )   \ottsym{=}   \mathsf{Bool} $, i.e., $\ottnt{e_{{\mathrm{2}}}}$ is either
      $ \mathsf{true} $ or $ \mathsf{false} $. In the former case, we step to
      $\ottnt{k}$ by \E{CheckOK}; in the latter case, we step to
      $ \mathord{\Uparrow}  \ottnt{l} $ by \E{CheckFail}.

    \end{itemize}    
    \fi}
  \end{proof}
\end{lemma}

\begin{lemma}[Classic preservation]
  \label{lem:classicpreservation}
  If $ \emptyset   \vdash _{  \mathsf{C}  }  \ottnt{e}  :  \ottnt{T} $ and $\ottnt{e} \,  \longrightarrow _{  \mathsf{C}  }  \, \ottnt{e'}$, then $ \emptyset   \vdash _{  \mathsf{C}  }  \ottnt{e'}  :  \ottnt{T} $.
  \begin{proof}
    By induction on the typing derivation.
    {\iffull
    \begin{itemize}
    \item[(\T{Var})] Contradictory---$\mathit{x}$ isn't well typed in an
      empty context.

    \item[(\T{Const})] Contradictory---constants don't step.

    \item[(\T{Abs})] Contradictory---lambdas don't step.

    \item[(\T{Op})] By cases on the step taken.
      \begin{itemize}
      \item[(\E{Op})] $\denot{ op } \, \ottsym{(}  \ottnt{k_{{\mathrm{1}}}}  \ottsym{,} \, ... \, \ottsym{,}  \ottnt{k_{\ottmv{n}}}  \ottsym{)}  \ottsym{=}  \ottnt{k}$; we assume that
        $ \mathsf{ty} (\mathord{ \ottnt{op} }) $ correctly assigns types, i.e., if $ \mathsf{ty} (\mathord{ \ottnt{op} })   \ottsym{=}   {}   \{ \mathit{x} \mathord{:} \ottnt{B_{{\mathrm{1}}}} \mathrel{\mid} \ottnt{e'_{{\mathrm{1}}}} \}   \rightarrow \, ... \, \rightarrow   \{ \mathit{x} \mathord{:} \ottnt{B_{\ottmv{n}}} \mathrel{\mid} \ottnt{e'_{\ottmv{n}}} \}   {} \mathord{ \rightarrow }  \{ \mathit{x} \mathord{:} \ottnt{B} \mathrel{\mid} \ottnt{e} \}  $, then $ \ottnt{e}  [  \ottnt{k} / \mathit{x}  ]  \,  \longrightarrow ^{*}_{  \mathsf{C}  }  \,  \mathsf{true} $ and $ \mathord{  \vdash _{  \mathsf{C}  } }~  \{ \mathit{x} \mathord{:} \ottnt{B} \mathrel{\mid} \ottnt{e} \}  $. We can therefore conclude
        that $ \emptyset   \vdash _{  \mathsf{C}  }  \ottnt{k}  :   \{ \mathit{x} \mathord{:} \ottnt{B} \mathrel{\mid} \ottnt{e} \}  $ by \T{Const}.
      \item[(\E{OpInner})] By the IH and \T{Op}.
      \item[(\E{OpRaise})] We assume that $ \mathsf{ty} (\mathord{ \ottnt{op} }) $ only assigns
        well formed types, so $ \emptyset   \vdash _{  \mathsf{C}  }   \mathord{\Uparrow}  \ottnt{l}   :  \ottnt{T} $.
      \end{itemize}

    \item[(\T{App})] By cases on the step taken.
      \begin{itemize}
      \item[(\E{Beta})] We know that $  \mathit{x} \mathord{:} \ottnt{T_{{\mathrm{1}}}}    \vdash _{  \mathsf{C}  }  \ottnt{e_{{\mathrm{1}}}}  :  \ottnt{T_{{\mathrm{2}}}} $ and
        $ \emptyset   \vdash _{  \mathsf{C}  }  \ottnt{e_{{\mathrm{2}}}}  :  \ottnt{T_{{\mathrm{1}}}} $; we are done by substitution
        (Lemma~\ref{lem:substitution}).
      \item[(\E{Unwrap})] By inversion of $ \mathord{  \vdash _{  \mathsf{C}  } }~ \emptyset   \mathrel{\parallel}    \ottnt{T_{{\mathrm{11}}}} \mathord{ \rightarrow } \ottnt{T_{{\mathrm{12}}}}   \Rightarrow   \ottnt{T_{{\mathrm{21}}}} \mathord{ \rightarrow } \ottnt{T_{{\mathrm{22}}}}  $, we find well formedness judgments $ \mathord{  \vdash _{  \mathsf{C}  } }~ \ottnt{T_{{\mathrm{11}}}} $
        and $ \mathord{  \vdash _{  \mathsf{C}  } }~ \ottnt{T_{{\mathrm{12}}}} $ and $ \mathord{  \vdash _{  \mathsf{C}  } }~ \ottnt{T_{{\mathrm{21}}}} $ and $ \mathord{  \vdash _{  \mathsf{C}  } }~ \ottnt{T_{{\mathrm{22}}}} $ and
        similarity judgments $\vdash  \ottnt{T_{{\mathrm{11}}}}  \mathrel{\parallel}  \ottnt{T_{{\mathrm{21}}}}$ and $\vdash  \ottnt{T_{{\mathrm{12}}}}  \mathrel{\parallel}  \ottnt{T_{{\mathrm{22}}}}$; by symmetry (Lemma~\ref{lem:similaritysymmetric}),
        $\vdash  \ottnt{T_{{\mathrm{21}}}}  \mathrel{\parallel}  \ottnt{T_{{\mathrm{11}}}}$. Noting that $ \mathsf{dom} ( \bullet )  =  \mathsf{cod} ( \bullet )  =
         \bullet $, we can apply \A{None}, finding $ \mathord{  \vdash _{  \mathsf{C}  } }~ \bullet   \mathrel{\parallel}   \ottnt{T_{{\mathrm{21}}}}  \Rightarrow  \ottnt{T_{{\mathrm{11}}}} $ and $ \mathord{  \vdash _{  \mathsf{C}  } }~ \bullet   \mathrel{\parallel}   \ottnt{T_{{\mathrm{12}}}}  \Rightarrow  \ottnt{T_{{\mathrm{22}}}} $. We can then
        apply \T{Cast}, \T{App}, and assumptions to find $ \emptyset   \vdash _{  \mathsf{C}  }   \langle  \ottnt{T_{{\mathrm{12}}}}  \mathord{ \overset{\bullet}{\Rightarrow} }  \ottnt{T_{{\mathrm{22}}}}  \rangle^{ \ottnt{l} } ~   (  \ottnt{e_{{\mathrm{1}}}} ~  (  \langle  \ottnt{T_{{\mathrm{21}}}}  \mathord{ \overset{\bullet}{\Rightarrow} }  \ottnt{T_{{\mathrm{11}}}}  \rangle^{ \ottnt{l} } ~  \ottnt{e_{{\mathrm{2}}}}  )   )    :  \ottnt{T_{{\mathrm{22}}}} $.
      \item[(\E{AppL})] By \T{App} and the IH.
      \item[(\E{AppR})] By \T{App} and the IH.
      \item[(\E{AppRaiseL})] By regularity, $ \mathord{  \vdash _{  \mathsf{C}  } }~ \ottnt{T_{{\mathrm{2}}}} $, so we are
        done by \T{Blame}.
      \item[(\E{AppRaiseR})] By regularity, $ \mathord{  \vdash _{  \mathsf{C}  } }~ \ottnt{T_{{\mathrm{2}}}} $, so we are
        done by \T{Blame}.
      \end{itemize}

    \item[(\T{Cast})] By cases on the step taken.
      \begin{itemize}
      \item[(\ECheckNone)] We have $ \mathord{  \vdash _{  \mathsf{C}  } }~ \Gamma $ and $ \mathord{  \vdash _{  \mathsf{C}  } }~  \{ \mathit{x} \mathord{:} \ottnt{B} \mathrel{\mid} \ottnt{e_{{\mathrm{2}}}} \}  $ and
        $ \mathsf{ty} ( \ottnt{k} )   \ottsym{=}  \ottnt{B}$ by inversion and $ \ottnt{e_{{\mathrm{2}}}}  [  \ottnt{k} / \mathit{x}  ]  \,  \longrightarrow ^{*}_{  \mathsf{C}  }  \,  \ottnt{e_{{\mathrm{2}}}}  [  \ottnt{k} / \mathit{x}  ] $ by
        reflexivity. By substitution (and \T{Const}, to find $ \emptyset   \vdash _{  \mathsf{C}  }  \ottnt{k}  :   \{ \mathit{x} \mathord{:} \ottnt{B} \mathrel{\mid}  \mathsf{true}  \}  $), we find $ \emptyset   \vdash _{  \mathsf{C}  }   \ottnt{e_{{\mathrm{2}}}}  [  \ottnt{k} / \mathit{x}  ]   :   \{ \mathit{x} \mathord{:}  \mathsf{Bool}  \mathrel{\mid}  \mathsf{true}  \}  $. We can now apply \T{Check}, and are done.
      \item[(\E{CastInnerC})] By \T{Cast} and the IH.
      \item[(\E{CastRaise})] We have by assumption that $ \mathord{  \vdash _{  \mathsf{C}  } }~ \ottnt{T_{{\mathrm{2}}}} $,
        so we are done by \T{Blame}.
      \end{itemize}

    \item[(\T{Blame})] Contradictory---blame doesn't step.

    \item[(\T{Check})] By cases on the step taken.
      \begin{itemize}
      \item[(\E{CheckOK})] Since $ \mathord{  \vdash _{  \mathsf{C}  } }~ \emptyset $ and $ \mathsf{ty} ( \ottnt{k} )   \ottsym{=}  \ottnt{B}$
        and $ \mathord{  \vdash _{  \mathsf{C}  } }~  \{ \mathit{x} \mathord{:} \ottnt{B} \mathrel{\mid} \ottnt{e_{{\mathrm{1}}}} \}  $ and $ \ottnt{e_{{\mathrm{1}}}}  [  \ottnt{k} / \mathit{x}  ]  \,  \longrightarrow ^{*}_{  \mathsf{C}  }  \,  \mathsf{true} $, we can
        apply \T{Const} to find $ \emptyset   \vdash _{  \mathsf{C}  }  \ottnt{k}  :   \{ \mathit{x} \mathord{:} \ottnt{B} \mathrel{\mid} \ottnt{e_{{\mathrm{1}}}} \}  $.
      \item[(\E{CheckFail})] Since $ \mathord{  \vdash _{  \mathsf{C}  } }~ \emptyset $ and $ \mathord{  \vdash _{  \mathsf{C}  } }~  \{ \mathit{x} \mathord{:} \ottnt{B} \mathrel{\mid} \ottnt{e_{{\mathrm{1}}}} \}  $, \T{Blame} shows $ \emptyset   \vdash _{  \mathsf{C}  }   \mathord{\Uparrow}  \ottnt{l}   :   \{ \mathit{x} \mathord{:} \ottnt{B} \mathrel{\mid} \ottnt{e_{{\mathrm{1}}}} \}  $.
      \item[(\E{CheckInner})] By \T{Check} and the IH.
      \item[(\E{CheckRaise})] As for \E{CheckFail}---the differing
        label doesn't matter.
      \end{itemize}

    \end{itemize}
    \fi}
  \end{proof}
\end{lemma}

{\iffull
\subsection{Forgetful type soundness}
\label{app:forgetfulsoundness}

Just as we did for classic \lambdah in
Appendix~\ref{app:classicsoundness}, we reuse the theorems from
Appendix~\ref{app:genericmetatheory}.
Note that if $\ottnt{e}$ is a value in forgetful \lambdah, it's also a
value in classic \lambdah, i.e., $ \mathsf{val} _{  \mathsf{F}  }~ \ottnt{e} $ implies $ \mathsf{val} _{  \mathsf{C}  }~ \ottnt{e} $.

\begin{lemma}[Forgetful determinism]
  \label{lem:forgetfuldeterminism}
  ~

  \noindent
  If $\ottnt{e} \,  \longrightarrow _{  \mathsf{F}  }  \, \ottnt{e_{{\mathrm{1}}}}$ and $\ottnt{e} \,  \longrightarrow _{  \mathsf{F}  }  \, \ottnt{e_{{\mathrm{2}}}}$ then $\ottnt{e_{{\mathrm{1}}}}  \ottsym{=}  \ottnt{e_{{\mathrm{2}}}}$.
  \begin{proof}
    By induction on the first evaluation derivation.
  \end{proof}  
\end{lemma}

\begin{lemma}[Forgetful canonical forms]
  \label{lem:forgetfulcanonicalforms}
  If $ \emptyset   \vdash _{  \mathsf{F}  }  \ottnt{e}  :  \ottnt{T} $ and $ \mathsf{val} _{  \mathsf{F}  }~ \ottnt{e} $ then:
  \begin{itemize}
  \item If $\ottnt{T}  \ottsym{=}   \{ \mathit{x} \mathord{:} \ottnt{B} \mathrel{\mid} \ottnt{e'} \} $, then $\ottnt{e}  \ottsym{=}  \ottnt{k}$ and $ \mathsf{ty} ( \ottnt{k} )   \ottsym{=}  \ottnt{B}$ and
    $ \ottnt{e'}  [  \ottnt{e} / \mathit{x}  ]  \,  \longrightarrow ^{*}_{  \mathsf{F}  }  \,  \mathsf{true} $.
  \item If $\ottnt{T}  \ottsym{=}   \ottnt{T_{{\mathrm{1}}}} \mathord{ \rightarrow } \ottnt{T_{{\mathrm{2}}}} $, then either $\ottnt{e}  \ottsym{=}   \lambda \mathit{x} \mathord{:} \ottnt{T} .~  \ottnt{e'} $ or $\ottnt{e}  \ottsym{=}   \langle   \ottnt{T_{{\mathrm{11}}}} \mathord{ \rightarrow } \ottnt{T_{{\mathrm{12}}}}   \mathord{ \overset{\bullet}{\Rightarrow} }   \ottnt{T_{{\mathrm{21}}}} \mathord{ \rightarrow } \ottnt{T_{{\mathrm{22}}}}   \rangle^{ \ottnt{l} } ~   \lambda \mathit{x} \mathord{:} \ottnt{T_{{\mathrm{11}}}} .~  \ottnt{e'}  $.
  \end{itemize}
  {\iffull
  \begin{proof}
    By inspection of the rules: \T{Const} is the only rule that types
    values at base types; \T{Abs} and \T{Cast} are the only rules
    that type values at function types.
  \end{proof}
  \fi}
\end{lemma}

\begin{lemma}[Forgetful progress]
  \label{lem:forgetfulprogress}
  If $ \emptyset   \vdash _{  \mathsf{F}  }  \ottnt{e}  :  \ottnt{T} $, then either:
  \begin{enumerate}
  \item $ \mathsf{result} _{  \mathsf{F}  }~ \ottnt{e} $ is a result, i.e., $\ottnt{e}  \ottsym{=}   \mathord{\Uparrow}  \ottnt{l} $ or $ \mathsf{val} _{  \mathsf{F}  }~ \ottnt{e} $; or
  \item there exists an $\ottnt{e'}$ such that $\ottnt{e} \,  \longrightarrow _{  \mathsf{F}  }  \, \ottnt{e'}$.
  \end{enumerate}
  \begin{proof}
    By induction on the typing derivation.
    {\iffull
    \begin{itemize}
    \item[(\T{Var})] A contradiction---$\mathit{x}$ isn't well typed in the
      empty environment.

    \item[(\T{Const})] $\ottnt{e}  \ottsym{=}  \ottnt{k}$ is a result.

    \item[(\T{Abs})] $\ottnt{e}  \ottsym{=}   \lambda \mathit{x} \mathord{:} \ottnt{T} .~  \ottnt{e'} $ is a result.

    \item[(\T{Op})] We know that $ \mathsf{ty} (\mathord{ \ottnt{op} }) $ is a first order type
      $ {}   \{ \mathit{x} \mathord{:} \ottnt{B_{{\mathrm{1}}}} \mathrel{\mid} \ottnt{e'_{{\mathrm{1}}}} \}   \rightarrow \, ... \, \rightarrow   \{ \mathit{x} \mathord{:} \ottnt{B_{\ottmv{n}}} \mathrel{\mid} \ottnt{e'_{\ottmv{n}}} \}   {} \mathord{ \rightarrow } \ottnt{T} $.
      From left to right, we apply the IH on $\ottnt{e_{\ottmv{i}}}$. If $\ottnt{e_{\ottmv{i}}}$ is
      a result, there are two cases: either $\ottnt{e_{\ottmv{i}}}  \ottsym{=}   \mathord{\Uparrow}  \ottnt{l} $, and
      we step by \E{OpRaise}; or $\ottnt{e_{\ottmv{i}}}  \ottsym{=}  \ottnt{k_{\ottmv{i}}}$, since constants are
      the only values at base types by canonical forms
      (Lemma~\ref{lem:forgetfulcanonicalforms}), and we continue on to
      the next $\ottnt{e_{\ottmv{i}}}$. If any of the $\ottnt{e_{\ottmv{i}}}$ step, we know that all
      of terms before them are values, so we can step by
      \E{OpInner}. If all of the $\ottnt{e_{\ottmv{i}}}$ are constants $\ottnt{k_{\ottmv{i}}}$, then
      $ \emptyset   \vdash _{  \mathsf{F}  }  \ottnt{k_{\ottmv{i}}}  :   \{ \mathit{x} \mathord{:} \ottnt{B_{\ottmv{i}}} \mathrel{\mid} \ottnt{e'_{\ottmv{i}}} \}  $, and so $ \ottnt{e'_{\ottmv{i}}}  [  \ottnt{k_{\ottmv{i}}} / \mathit{x}  ]  \,  \longrightarrow ^{*}_{  \mathsf{F}  }  \,  \mathsf{true} $. Therefore $\denot{ op } \, \ottsym{(}  \ottnt{k_{{\mathrm{1}}}}  \ottsym{,} \, ... \, \ottsym{,}  \ottnt{k_{\ottmv{n}}}  \ottsym{)}$ is defined, and we
      can step by \E{Op}.

    \item[(\T{App})] By the IH on $\ottnt{e_{{\mathrm{1}}}}$, we know that $\ottnt{e_{{\mathrm{1}}}}$: is
      blame, is a value, or steps to some $\ottnt{e'_{{\mathrm{1}}}}$. In the first
      case, we take a step by \E{AppRaiseL}. In the latter, we take a
      step by \E{AppL}.

      If $ \mathsf{val} _{  \mathsf{C}  }~ \ottnt{e_{{\mathrm{1}}}} $, then we can apply the IH on $\ottnt{e_{{\mathrm{2}}}}$, which is
      blame, is a value, or steps to some $\ottnt{e'_{{\mathrm{2}}}}$. 

      If $\ottnt{e_{{\mathrm{2}}}}$ is blame, we step by \E{AppRaiseR}. Otherwise, we
      must go by cases. Since $ \emptyset   \vdash _{  \mathsf{F}  }  \ottnt{e_{{\mathrm{1}}}}  :   \ottnt{T_{{\mathrm{1}}}} \mathord{ \rightarrow } \ottnt{T_{{\mathrm{2}}}}  $ and $ \mathsf{val} _{  \mathsf{F}  }~ \ottnt{e_{{\mathrm{1}}}} $, by canonical forms
      (Lemma~\ref{lem:forgetfulcanonicalforms}) we know that $\ottnt{e_{{\mathrm{1}}}}$
      can only be an abstraction, a wrapped abstraction, or a cast.
      \begin{itemize}
      \item[($\ottnt{e_{{\mathrm{1}}}}  \ottsym{=}   \lambda \mathit{x} \mathord{:} \ottnt{T_{{\mathrm{1}}}} .~  \ottnt{e'_{{\mathrm{1}}}} $)] If $ \mathsf{val} _{  \mathsf{F}  }~ \ottnt{e_{{\mathrm{2}}}} $, we step to
        $ \ottnt{e'_{{\mathrm{1}}}}  [  \ottnt{e_{{\mathrm{2}}}} / \mathit{x}  ] $ by \E{Beta}. If $\ottnt{e_{{\mathrm{2}}}} \,  \longrightarrow _{  \mathsf{F}  }  \, \ottnt{e'_{{\mathrm{2}}}}$, we step by
        \E{AppR} (since $\ottnt{e_{{\mathrm{1}}}}$ isn't a cast).

      \item[($\ottnt{e_{{\mathrm{1}}}}  \ottsym{=}   \langle   \ottnt{T_{{\mathrm{11}}}} \mathord{ \rightarrow } \ottnt{T_{{\mathrm{12}}}}   \mathord{ \overset{\bullet}{\Rightarrow} }   \ottnt{T_{{\mathrm{21}}}} \mathord{ \rightarrow } \ottnt{T_{{\mathrm{22}}}}   \rangle^{ \ottnt{l} } ~  \ottnt{e'_{{\mathrm{1}}}} $)] If $ \mathsf{val} _{  \mathsf{F}  }~ \ottnt{e_{{\mathrm{2}}}} $, we step to $ \langle  \ottnt{T_{{\mathrm{12}}}}  \mathord{ \overset{\bullet}{\Rightarrow} }  \ottnt{T_{{\mathrm{22}}}}  \rangle^{ \ottnt{l} } ~   (  \ottnt{e'_{{\mathrm{1}}}} ~  (  \langle  \ottnt{T_{{\mathrm{21}}}}  \mathord{ \overset{\bullet}{\Rightarrow} }  \ottnt{T_{{\mathrm{11}}}}  \rangle^{ \ottnt{l} } ~  \ottnt{e_{{\mathrm{2}}}}  )   )  $ by
        \E{Unwrap} (noting that the annotation is $ \bullet $). If $\ottnt{e_{{\mathrm{2}}}} \,  \longrightarrow _{  \mathsf{F}  }  \, \ottnt{e'_{{\mathrm{2}}}}$, we step by \E{AppR} (since $\ottnt{e_{{\mathrm{1}}}}$ isn't a
        cast).
      \end{itemize}

    \item[(\T{Cast})] If $\ottnt{e'}  \ottsym{=}   \mathord{\Uparrow}  \ottnt{l'} $, then we step by
      \E{CastRaise}. If $ \mathsf{val} _{  \mathsf{F}  }~ \ottnt{e_{{\mathrm{2}}}} $, then we have $ \mathord{  \vdash _{  \mathsf{F}  } }~ \bullet   \mathrel{\parallel}   \ottnt{T_{{\mathrm{1}}}}  \Rightarrow  \ottnt{T_{{\mathrm{2}}}} $; by inversion $\vdash  \ottnt{T_{{\mathrm{1}}}}  \mathrel{\parallel}  \ottnt{T_{{\mathrm{2}}}}$. By inversion, one of the following cases adheres:
        \begin{itemize}
        \item[($\ottnt{T_{\ottmv{i}}}  \ottsym{=}   \{ \mathit{x} \mathord{:} \ottnt{B} \mathrel{\mid} \ottnt{e_{{\mathrm{1}}\,\ottmv{i}}} \} $)] By canonical forms
          (Lemma~\ref{lem:forgetfulcanonicalforms}), $\ottnt{e_{{\mathrm{2}}}}  \ottsym{=}  \ottnt{k}$. We
          can step to $ \langle   \{ \mathit{x} \mathord{:} \ottnt{B} \mathrel{\mid} \ottnt{e_{{\mathrm{12}}}} \}  ,   \ottnt{e_{{\mathrm{12}}}}  [  \ottnt{k} / \mathit{x}  ]  ,  \ottnt{k}  \rangle^{ \ottnt{l} } $ by \E{CheckNone}.

        \item[($\ottnt{T_{\ottmv{i}}}  \ottsym{=}   \ottnt{T_{\ottmv{i}\,{\mathrm{1}}}} \mathord{ \rightarrow } \ottnt{T_{\ottmv{i}\,{\mathrm{2}}}} $)] By canonical forms
          (Lemma~\ref{lem:forgetfulcanonicalforms}), $\ottnt{e_{{\mathrm{2}}}}$ is a
          lambda or a function proxy. In the latter case, we step by
          \E{CastMerge}; either way, we have a function proxy, which
          is a value.
        \end{itemize}
        If $\ottnt{e_{{\mathrm{2}}}}$ isn't a value or blame, then $\ottnt{e_{{\mathrm{2}}}} \,  \longrightarrow _{  \mathsf{F}  }  \, \ottnt{e'_{{\mathrm{2}}}}$. If
        $\ottnt{e_{{\mathrm{2}}}}  \neq   \langle   \ottnt{T_{{\mathrm{31}}}} \mathord{ \rightarrow } \ottnt{T_{{\mathrm{32}}}}   \mathord{ \overset{\bullet}{\Rightarrow} }   \ottnt{T_{{\mathrm{11}}}} \mathord{ \rightarrow } \ottnt{T_{{\mathrm{12}}}}   \rangle^{ \ottnt{l} } ~  \ottnt{e''_{{\mathrm{2}}}} $, then we can step by
        \E{CastInner}. If $\ottnt{e_{{\mathrm{2}}}}$ is in fact an application of a
        cast, we step by \E{CastMerge}.

    \item[(\T{Blame})] $ \mathord{\Uparrow}  \ottnt{l} $ is a result.

    \item[(\T{Check})] By the IH on $\ottnt{e_{{\mathrm{2}}}}$, we know that $\ottnt{e_{{\mathrm{2}}}}$:
      is $ \mathord{\Uparrow}  \ottnt{l'} $, is a value, or takes a step to some
      $\ottnt{e'_{{\mathrm{2}}}}$. In the first case, we step to $ \mathord{\Uparrow}  \ottnt{l'} $ by
      \E{CheckRaise}. In the last case, we step by \E{CheckInner}. If
      $ \mathsf{val} _{  \mathsf{F}  }~ \ottnt{e_{{\mathrm{2}}}} $, by canonical forms
      (Lemma~\ref{lem:forgetfulcanonicalforms}) we know that $\ottnt{e_{{\mathrm{2}}}}$
      is a $\ottnt{k}$ such that $ \mathsf{ty} ( \ottnt{k} )   \ottsym{=}   \mathsf{Bool} $, i.e., $\ottnt{e_{{\mathrm{2}}}}$ is
      either $ \mathsf{true} $ or $ \mathsf{false} $. In the former case, we step to
      $\ottnt{k}$ by \E{CheckOK}; in the latter case, we step to
      $ \mathord{\Uparrow}  \ottnt{l} $ by \E{CheckFail}.

    \end{itemize}  
    \fi}
  \end{proof}  
\end{lemma}

\begin{lemma}[Forgetful preservation]
  \label{lem:forgetfulpreservation}
  If $ \emptyset   \vdash _{  \mathsf{F}  }  \ottnt{e}  :  \ottnt{T} $ and $\ottnt{e} \,  \longrightarrow _{  \mathsf{F}  }  \, \ottnt{e'}$ then $ \emptyset   \vdash _{  \mathsf{F}  }  \ottnt{e'}  :  \ottnt{T} $.
  \begin{proof}
    By induction on the typing derivation.
    {\iffull
    \begin{itemize}
    \item[(\T{Var})] Contradictory---we assumed $\ottnt{e}$ was well typed
      in an empty context.
    \item[(\T{Const})] Contradictory---$\ottnt{k}$ is a value and doesn't step.
    \item[(\T{Abs})] Contradictory---$ \lambda \mathit{x} \mathord{:} \ottnt{T_{{\mathrm{1}}}} .~  \ottnt{e'} $ is a value and
      doesn't step.
    \item[(\T{Op})] By cases on the step taken.
      \begin{itemize}
      \item[(\E{Op})] $\denot{ op } \, \ottsym{(}  \ottnt{k_{{\mathrm{1}}}}  \ottsym{,} \, ... \, \ottsym{,}  \ottnt{k_{\ottmv{n}}}  \ottsym{)}  \ottsym{=}  \ottnt{k}$; we assume that
        $ \mathsf{ty} (\mathord{ \ottnt{op} }) $ correctly assigns types, i.e., if $ \mathsf{ty} (\mathord{ \ottnt{op} })   \ottsym{=}   {}   \{ \mathit{x} \mathord{:} \ottnt{B_{{\mathrm{1}}}} \mathrel{\mid} \ottnt{e'_{{\mathrm{1}}}} \}   \rightarrow \, ... \, \rightarrow   \{ \mathit{x} \mathord{:} \ottnt{B_{\ottmv{n}}} \mathrel{\mid} \ottnt{e'_{\ottmv{n}}} \}   {} \mathord{ \rightarrow }  \{ \mathit{x} \mathord{:} \ottnt{B} \mathrel{\mid} \ottnt{e} \}  $, then $ \ottnt{e}  [  \ottnt{k} / \mathit{x}  ]  \,  \longrightarrow ^{*}_{  \mathsf{F}  }  \,  \mathsf{true} $ and $ \mathord{  \vdash _{  \mathsf{F}  } }~  \{ \mathit{x} \mathord{:} \ottnt{B} \mathrel{\mid} \ottnt{e} \}  $. We can therefore conclude
        that $ \emptyset   \vdash _{  \mathsf{F}  }  \ottnt{k}  :   \{ \mathit{x} \mathord{:} \ottnt{B} \mathrel{\mid} \ottnt{e} \}  $ by \T{Const}.
      \item[(\E{OpInner})] By the IH and \T{Op}.
      \item[(\E{OpRaise})] We assume that $ \mathsf{ty} (\mathord{ \ottnt{op} }) $ only assigns
        well formed types, so $ \emptyset   \vdash _{  \mathsf{F}  }   \mathord{\Uparrow}  \ottnt{l}   :  \ottnt{T} $.
      \end{itemize}
    \item[(\T{App})] By cases on the step taken.
      \begin{itemize}
      \item[(\E{Beta})] We know that $  \mathit{x} \mathord{:} \ottnt{T_{{\mathrm{1}}}}    \vdash _{  \mathsf{F}  }  \ottnt{e_{{\mathrm{1}}}}  :  \ottnt{T_{{\mathrm{2}}}} $ and
        $ \emptyset   \vdash _{  \mathsf{F}  }  \ottnt{e_{{\mathrm{2}}}}  :  \ottnt{T_{{\mathrm{1}}}} $; we are done by substitution
        (Lemma~\ref{lem:substitution}).
      \item[(\E{Unwrap})] By inversion of $ \mathord{  \vdash _{  \mathsf{F}  } }~ \emptyset   \mathrel{\parallel}    \ottnt{T_{{\mathrm{11}}}} \mathord{ \rightarrow } \ottnt{T_{{\mathrm{12}}}}   \Rightarrow   \ottnt{T_{{\mathrm{21}}}} \mathord{ \rightarrow } \ottnt{T_{{\mathrm{22}}}}  $, we find well formedness judgments $ \mathord{  \vdash _{  \mathsf{F}  } }~ \ottnt{T_{{\mathrm{11}}}} $
        and $ \mathord{  \vdash _{  \mathsf{F}  } }~ \ottnt{T_{{\mathrm{12}}}} $ and $ \mathord{  \vdash _{  \mathsf{F}  } }~ \ottnt{T_{{\mathrm{21}}}} $ and $ \mathord{  \vdash _{  \mathsf{F}  } }~ \ottnt{T_{{\mathrm{22}}}} $ and
        similarity judgments $\vdash  \ottnt{T_{{\mathrm{11}}}}  \mathrel{\parallel}  \ottnt{T_{{\mathrm{21}}}}$ and $\vdash  \ottnt{T_{{\mathrm{12}}}}  \mathrel{\parallel}  \ottnt{T_{{\mathrm{22}}}}$; by symmetry (Lemma~\ref{lem:similaritysymmetric}),
        $\vdash  \ottnt{T_{{\mathrm{21}}}}  \mathrel{\parallel}  \ottnt{T_{{\mathrm{11}}}}$. Noting that $ \mathsf{dom} ( \bullet )  =  \mathsf{cod} ( \bullet )  =
         \bullet $, we can apply \A{None} to find derivations
        $ \mathord{  \vdash _{  \mathsf{F}  } }~ \bullet   \mathrel{\parallel}   \ottnt{T_{{\mathrm{21}}}}  \Rightarrow  \ottnt{T_{{\mathrm{11}}}} $ and $ \mathord{  \vdash _{  \mathsf{F}  } }~ \bullet   \mathrel{\parallel}   \ottnt{T_{{\mathrm{12}}}}  \Rightarrow  \ottnt{T_{{\mathrm{22}}}} $. We can then apply \T{Cast}, \T{App}, and
        assumptions to find $ \emptyset   \vdash _{  \mathsf{F}  }   \langle  \ottnt{T_{{\mathrm{12}}}}  \mathord{ \overset{\bullet}{\Rightarrow} }  \ottnt{T_{{\mathrm{22}}}}  \rangle^{ \ottnt{l} } ~   (  \ottnt{e_{{\mathrm{1}}}} ~  (  \langle  \ottnt{T_{{\mathrm{21}}}}  \mathord{ \overset{\bullet}{\Rightarrow} }  \ottnt{T_{{\mathrm{11}}}}  \rangle^{ \ottnt{l} } ~  \ottnt{e_{{\mathrm{2}}}}  )   )    :  \ottnt{T_{{\mathrm{22}}}} $.
      \item[(\E{AppL})] By \T{App} and the IH.
      \item[(\E{AppR})] By \T{App} and the IH.
      \item[(\E{AppRaiseL})] By regularity, $ \mathord{  \vdash _{  \mathsf{F}  } }~ \ottnt{T_{{\mathrm{2}}}} $, so we are done by \T{Blame}.
      \item[(\E{AppRaiseR})] By regularity, $ \mathord{  \vdash _{  \mathsf{F}  } }~ \ottnt{T_{{\mathrm{2}}}} $, so we are done by \T{Blame}.
      \end{itemize}
    \item[(\T{Cast})] By cases on the step taken.
      \begin{itemize}
      \item[(\E{CheckNone})] We have $ \mathord{  \vdash _{  \mathsf{F}  } }~ \Gamma $ and $ \mathord{  \vdash _{  \mathsf{F}  } }~  \{ \mathit{x} \mathord{:} \ottnt{B} \mathrel{\mid} \ottnt{e_{{\mathrm{2}}}} \}  $ and
        $ \mathsf{ty} ( \ottnt{k} )   \ottsym{=}  \ottnt{B}$ by inversion and $ \ottnt{e_{{\mathrm{2}}}}  [  \ottnt{k} / \mathit{x}  ]  \,  \longrightarrow ^{*}_{  \mathsf{F}  }  \,  \ottnt{e_{{\mathrm{2}}}}  [  \ottnt{k} / \mathit{x}  ] $ by
        reflexivity. By substitution (and \T{Const}, to find $ \emptyset   \vdash _{  \mathsf{F}  }  \ottnt{k}  :   \{ \mathit{x} \mathord{:} \ottnt{B} \mathrel{\mid}  \mathsf{true}  \}  $), we find $ \emptyset   \vdash _{  \mathsf{F}  }   \ottnt{e_{{\mathrm{2}}}}  [  \ottnt{k} / \mathit{x}  ]   :   \{ \mathit{x} \mathord{:}  \mathsf{Bool}  \mathrel{\mid}  \mathsf{true}  \}  $. We can now apply \T{Check}, and are done.
      \item[(\E{CastInner})] By \T{Cast} and the IH.
      \item[(\E{CastMerge})] We can combine the
        two \T{Cast} derivations (using
        Lemma~\ref{lem:typesettransitive} and the fact that
        $\ottnt{a}  \ottsym{=}  \bullet$ to find similarity). Then we are done by \T{Cast}.
      \item[(\E{CastRaise})] We have $ \mathord{  \vdash _{  \mathsf{F}  } }~ \ottnt{T_{{\mathrm{2}}}} $ by assumption, so we
        are done by \T{Blame}.
      \end{itemize}
    \item[(\T{Blame})] Contradictory---$ \mathord{\Uparrow}  \ottnt{l} $ is a result and
      doesn't step.
    \item[(\T{Check})]  By cases on the step taken.
      \begin{itemize}
      \item[(\E{CheckOK})] Since $ \mathord{  \vdash _{  \mathsf{F}  } }~ \emptyset $ and $ \mathsf{ty} ( \ottnt{k} )   \ottsym{=}  \ottnt{B}$
        and $ \mathord{  \vdash _{  \mathsf{F}  } }~  \{ \mathit{x} \mathord{:} \ottnt{B} \mathrel{\mid} \ottnt{e_{{\mathrm{1}}}} \}  $ and $ \ottnt{e_{{\mathrm{1}}}}  [  \ottnt{k} / \mathit{x}  ]  \,  \longrightarrow ^{*}_{  \mathsf{F}  }  \,  \mathsf{true} $, we can
        apply \T{Const} to find $ \emptyset   \vdash _{  \mathsf{F}  }  \ottnt{k}  :   \{ \mathit{x} \mathord{:} \ottnt{B} \mathrel{\mid} \ottnt{e_{{\mathrm{1}}}} \}  $.
      \item[(\E{CheckFail})] Since $ \mathord{  \vdash _{  \mathsf{F}  } }~ \emptyset $ and $ \mathord{  \vdash _{  \mathsf{F}  } }~  \{ \mathit{x} \mathord{:} \ottnt{B} \mathrel{\mid} \ottnt{e_{{\mathrm{1}}}} \}  $, \T{Blame} shows $ \emptyset   \vdash _{  \mathsf{F}  }   \mathord{\Uparrow}  \ottnt{l}   :   \{ \mathit{x} \mathord{:} \ottnt{B} \mathrel{\mid} \ottnt{e_{{\mathrm{1}}}} \}  $.
      \item[(\E{CheckInner})] By \T{Check} and the IH.
      \item[(\E{CheckRaise})] As for \E{CheckFail}---the differing
        label doesn't matter.
      \end{itemize}
    \end{itemize}
    \fi}
  \end{proof}
\end{lemma}

In addition to showing type soundness, we prove that a source program
(Definition~\ref{def:sourceprogram}) is well typed with $\ottnt{m}  \ottsym{=}   \mathsf{F} $ iff
it is well typed with $\ottnt{m}  \ottsym{=}   \mathsf{C} $.

\begin{lemma}[Source program typing for forgetful \lambdah]
  \label{lem:forgetfulsource}
  ~

  \noindent
  Source programs are well typed in $ \mathsf{C} $ iff they are well typed in
  $ \mathsf{F} $, i.e.:
  \begin{itemize}
  \item $ \Gamma   \vdash _{  \mathsf{C}  }  \ottnt{e}  :  \ottnt{T} $ as a source program iff $ \Gamma   \vdash _{  \mathsf{F}  }  \ottnt{e}  :  \ottnt{T} $ as a source program.
  \item $ \mathord{  \vdash _{  \mathsf{C}  } }~ \ottnt{T} $ as a source program iff $ \mathord{  \vdash _{  \mathsf{F}  } }~ \ottnt{T} $ as a source program.
  \item $ \mathord{  \vdash _{  \mathsf{C}  } }~ \Gamma $ as a source program iff $ \mathord{  \vdash _{  \mathsf{F}  } }~ \Gamma $ as a source program.
  \end{itemize}
  \begin{proof}
    By mutual induction on $\ottnt{e}$, $\ottnt{T}$, and $\Gamma$.
    {\iffull
    Since all of the rules are syntax directed, we use the rule names
    for cases (but prove both directions at once).
    \paragraph{Expressions $\ottnt{e}$}
    \begin{itemize}
    \item[(\T{Var})] By the IH on $\Gamma$ and \T{Var}.
    \item[(\T{Const})] By the IH on $\Gamma$ and $\T{Const}$, noting
      that $ \mathsf{true}  \,  \longrightarrow ^{*}_{ \ottnt{m} }  \,  \mathsf{true} $ in every mode $\ottnt{m}$.
    \item[(\T{Abs})] By the IH on $\ottnt{T_{{\mathrm{1}}}}$ and $\ottnt{e_{{\mathrm{12}}}}$ and \T{Abs}.
    \item[(\T{Op})] By the IHs on the arguments $\ottnt{e_{\ottmv{i}}}$ and \T{Op}.
    \item[(\T{App})] By the IHs on $\ottnt{e_{{\mathrm{1}}}}$ and $\ottnt{e_{{\mathrm{2}}}}$ and \T{App}.
    \item[(\T{Cast})] By the IHs on $\ottnt{T_{{\mathrm{1}}}}$ and $\ottnt{T_{{\mathrm{2}}}}$ and
      \T{Cast}, noting that similarity holds irrespective of modes and
      that the annotation is $ \bullet $.
    \item[(\T{Blame})] Contradictory---doesn't occur in source programs.
    \item[(\T{Check})] Contradictory---doesn't occur in source programs.
    \end{itemize}
    
    \paragraph{Types $\ottnt{T}$}
    \begin{itemize}
    \item[(\WF{Base})] Immediately true---\WF{Base} is an axiom.
    \item[(\WF{Refine})] By the IH on $\ottnt{e}$ and \WF{Refine}.
    \item[(\WF{Fun})] By the IHs on $\ottnt{T_{{\mathrm{1}}}}$ and $\ottnt{T_{{\mathrm{2}}}}$ and \WF{Fun}.
    \end{itemize}

    \paragraph{Contexts $\Gamma$}
    \begin{itemize}
    \item[(\WF{Empty})] Immediately true---\WF{Empty} is an axiom.
    \item[(\WF{Extend})] By the IHs on $\Gamma$ and $\ottnt{T}$ and
      \WF{Extend}.
    \end{itemize}
    \fi}
  \end{proof}
\end{lemma}

\subsection{Heedful type soundness}
\label{app:heedfulsoundness}

\begin{lemma}[Heedful canonical forms]
  \label{lem:heedfulcanonicalforms}
  If $ \emptyset   \vdash _{  \mathsf{H}  }  \ottnt{e}  :  \ottnt{T} $ and $ \mathsf{val} _{  \mathsf{H}  }~ \ottnt{e} $ then:
  \begin{itemize}
  \item If $\ottnt{T}  \ottsym{=}   \{ \mathit{x} \mathord{:} \ottnt{B} \mathrel{\mid} \ottnt{e'} \} $, then $\ottnt{e}  \ottsym{=}  \ottnt{k}$ and $ \mathsf{ty} ( \ottnt{k} )   \ottsym{=}  \ottnt{B}$
    and $ \ottnt{e'}  [  \ottnt{e} / \mathit{x}  ]  \,  \longrightarrow ^{*}_{  \mathsf{H}  }  \,  \mathsf{true} $.
  \item If $\ottnt{T}  \ottsym{=}   \ottnt{T_{{\mathrm{1}}}} \mathord{ \rightarrow } \ottnt{T_{{\mathrm{2}}}} $, then either $\ottnt{e}  \ottsym{=}   \lambda \mathit{x} \mathord{:} \ottnt{T} .~  \ottnt{e'} $ or $\ottnt{e}  \ottsym{=}   \langle   \ottnt{T_{{\mathrm{11}}}} \mathord{ \rightarrow } \ottnt{T_{{\mathrm{12}}}}   \mathord{ \overset{ \mathcal{S} }{\Rightarrow} }   \ottnt{T_{{\mathrm{21}}}} \mathord{ \rightarrow } \ottnt{T_{{\mathrm{22}}}}   \rangle^{ \ottnt{l} } ~   \lambda \mathit{x} \mathord{:} \ottnt{T_{{\mathrm{11}}}} .~  \ottnt{e'}  $.
  \end{itemize}
  {\iffull
  \begin{proof}
    By inspection of the rules: \T{Const} is the only rule that types
    values at base types; \T{Abs} and \T{Cast} are the only rules
    that type values at function types. Note that now our proxies may
    have type sets in them.
  \end{proof}
  \fi}
\end{lemma}

\begin{lemma}[Heedful progress]
  If $ \emptyset   \vdash _{  \mathsf{H}  }  \ottnt{e}  :  \ottnt{T} $, then either:
  \begin{enumerate}
  \item $ \mathsf{result} _{  \mathsf{H}  }~ \ottnt{e} $, i.e., $\ottnt{e}  \ottsym{=}   \mathord{\Uparrow}  \ottnt{l} $ or $ \mathsf{val} _{  \mathsf{H}  }~ \ottnt{e} $;
    or
  \item there exists an $\ottnt{e'}$ such that $\ottnt{e} \,  \longrightarrow _{  \mathsf{H}  }  \, \ottnt{e'}$.
  \end{enumerate}
  \begin{proof}
    By induction on the typing derivation.
    {\iffull
    \begin{itemize}
    \item[(\T{Var})] A contradiction---$\mathit{x}$ isn't well typed in the
      empty environment.

    \item[(\T{Const})] $\ottnt{e}  \ottsym{=}  \ottnt{k}$ is a result by \V{Const} and \R{Val}.

    \item[(\T{Abs})] $\ottnt{e}  \ottsym{=}   \lambda \mathit{x} \mathord{:} \ottnt{T} .~  \ottnt{e'} $ is a result by \V{Abs} and \R{Val}.

    \item[(\T{Op})] We know that $ \mathsf{ty} (\mathord{ \ottnt{op} }) $ is a first order type
      $ {}   \{ \mathit{x} \mathord{:} \ottnt{B_{{\mathrm{1}}}} \mathrel{\mid} \ottnt{e'_{{\mathrm{1}}}} \}   \rightarrow \, ... \, \rightarrow   \{ \mathit{x} \mathord{:} \ottnt{B_{\ottmv{n}}} \mathrel{\mid} \ottnt{e'_{\ottmv{n}}} \}   {} \mathord{ \rightarrow } \ottnt{T} $.
      From left to right, we apply the IH on $\ottnt{e_{\ottmv{i}}}$. If $\ottnt{e_{\ottmv{i}}}$ is
      a result, there are two cases: either $\ottnt{e_{\ottmv{i}}}  \ottsym{=}   \mathord{\Uparrow}  \ottnt{l} $, and
      we step by \E{OpRaise}; or $\ottnt{e_{\ottmv{i}}}  \ottsym{=}  \ottnt{k_{\ottmv{i}}}$, since constants are
      the only values at base types by canonical forms
      (Lemma~\ref{lem:heedfulcanonicalforms}), and we continue on to
      the next $\ottnt{e_{\ottmv{i}}}$. If any of the $\ottnt{e_{\ottmv{i}}}$ step, we know that all
      of terms before them are values, so we can step by
      \E{OpInner}. If all of the $\ottnt{e_{\ottmv{i}}}$ are constants $\ottnt{k_{\ottmv{i}}}$, then
      $ \emptyset   \vdash _{  \mathsf{E}  }  \ottnt{k_{\ottmv{i}}}  :   \{ \mathit{x} \mathord{:} \ottnt{B_{\ottmv{i}}} \mathrel{\mid} \ottnt{e'_{\ottmv{i}}} \}  $, and so $ \ottnt{e'_{\ottmv{i}}}  [  \ottnt{k_{\ottmv{i}}} / \mathit{x}  ]  \,  \longrightarrow ^{*}_{  \mathsf{H}  }  \,  \mathsf{true} $. Therefore $\denot{ op } \, \ottsym{(}  \ottnt{k_{{\mathrm{1}}}}  \ottsym{,} \, ... \, \ottsym{,}  \ottnt{k_{\ottmv{n}}}  \ottsym{)}$ is defined, and we
      can step by \E{Op}.

    \item[(\T{App})] By the IH on $\ottnt{e_{{\mathrm{1}}}}$, we know that $\ottnt{e_{{\mathrm{1}}}}$: is
      blame, is a value, or steps to some $\ottnt{e'_{{\mathrm{1}}}}$. In the first
      case, we take a step by \E{AppRaiseL}. In the latter, we take a
      step by \E{AppL}.

      If $ \mathsf{val} _{  \mathsf{H}  }~ \ottnt{e_{{\mathrm{1}}}} $, then we can apply the IH on $\ottnt{e_{{\mathrm{2}}}}$, which
      is blame, is a value, or steps to some $\ottnt{e'_{{\mathrm{2}}}}$. The first and
      last cases are as before, using \E{AppRaiseR} and \E{AppR}.

      If $ \mathsf{val} _{  \mathsf{H}  }~ \ottnt{e_{{\mathrm{2}}}} $, we must go by cases on the shape of
      $\ottnt{e_{{\mathrm{1}}}}$. Since $ \emptyset   \vdash _{  \mathsf{H}  }  \ottnt{e_{{\mathrm{1}}}}  :   \ottnt{T_{{\mathrm{1}}}} \mathord{ \rightarrow } \ottnt{T_{{\mathrm{2}}}}  $, by canonical forms
      (Lemma~\ref{lem:heedfulcanonicalforms}) we know that $\ottnt{e_{{\mathrm{1}}}}$ can only
      be an abstraction, a wrapped abstraction, or a cast.
      \begin{itemize}
      \item[($\ottnt{e_{{\mathrm{1}}}}  \ottsym{=}   \lambda \mathit{x} \mathord{:} \ottnt{T_{{\mathrm{1}}}} .~  \ottnt{e'_{{\mathrm{1}}}} $)] We step to $ \ottnt{e'_{{\mathrm{1}}}}  [  \ottnt{e_{{\mathrm{2}}}} / \mathit{x}  ] $ by
        \E{Beta}.

      \item[($\ottnt{e_{{\mathrm{1}}}}  \ottsym{=}   \langle   \ottnt{T_{{\mathrm{11}}}} \mathord{ \rightarrow } \ottnt{T_{{\mathrm{12}}}}   \mathord{ \overset{ \mathcal{S} }{\Rightarrow} }   \ottnt{T_{{\mathrm{21}}}} \mathord{ \rightarrow } \ottnt{T_{{\mathrm{22}}}}   \rangle^{ \ottnt{l} } ~  \ottnt{e'_{{\mathrm{1}}}} $)] We step to
        $ \langle  \ottnt{T_{{\mathrm{12}}}}  \mathord{ \overset{  \mathsf{cod} ( \mathcal{S} )  }{\Rightarrow} }  \ottnt{T_{{\mathrm{22}}}}  \rangle^{ \ottnt{l} } ~   (  \ottnt{e'_{{\mathrm{1}}}} ~  (  \langle  \ottnt{T_{{\mathrm{21}}}}  \mathord{ \overset{  \mathsf{dom} ( \mathcal{S} )  }{\Rightarrow} }  \ottnt{T_{{\mathrm{11}}}}  \rangle^{ \ottnt{l} } ~  \ottnt{e_{{\mathrm{2}}}}  )   )  $ by
        \E{Unwrap}.
      \end{itemize}

    \item[(\T{Cast})] If the annotation is $ \bullet $, we step by
      \E{TypeSet}. We have $\ottnt{e}  \ottsym{=}   \langle  \ottnt{T_{{\mathrm{1}}}}  \mathord{ \overset{ \mathcal{S} }{\Rightarrow} }  \ottnt{T_{{\mathrm{2}}}}  \rangle^{ \ottnt{l} } ~  \ottnt{e_{{\mathrm{1}}}} $. If $\ottnt{e_{{\mathrm{1}}}}$ is
      blame, then we step by \E{CastRaise}.
      If $\ottnt{e_{{\mathrm{1}}}}$ is a value, then we go by cases on $ \mathsf{val} _{  \mathsf{H}  }~ \ottnt{e_{{\mathrm{1}}}} $:
      \begin{itemize}
      \item[(\V{Const})] The case must be between refinements, and we
        step by \E{CheckSet} or \E{CheckEmpty}.
      \item[(\V{Abs})] The cast must be between function types, and we have a value.
      \item[(\V{ProxyH})] The cast must be between function types, and
        we step by \E{CastMerge}.
      \end{itemize}
      Finally, it may be the case that $\ottnt{e_{{\mathrm{1}}}} \,  \longrightarrow _{  \mathsf{H}  }  \, \ottnt{e'_{{\mathrm{1}}}}$.
      If $\ottnt{e_{{\mathrm{1}}}}  \neq   \langle  \ottnt{T_{{\mathrm{3}}}}  \mathord{ \overset{ \mathcal{S}_{{\mathrm{2}}} }{\Rightarrow} }  \ottnt{T_{{\mathrm{1}}}}  \rangle^{ \ottnt{l'} } ~  \ottnt{e''_{{\mathrm{1}}}} $, then we step by \E{CastInner}.
      On the other hand, if $\ottnt{e_{{\mathrm{1}}}}$ is a cast term, we step by
      \E{CastMerge}.

    \item[(\T{Blame})] $ \mathord{\Uparrow}  \ottnt{l} $ is a result by \R{Blame}.

    \item[(\T{Check})] By the IH on $\ottnt{e_{{\mathrm{2}}}}$, we know that $\ottnt{e_{{\mathrm{2}}}}$:
      is $ \mathord{\Uparrow}  \ottnt{l'} $, is a value by $ \mathsf{val} _{  \mathsf{H}  }~ \ottnt{e_{{\mathrm{2}}}} $, or takes a
      step to some $\ottnt{e'_{{\mathrm{2}}}}$. In the first case, we step to
      $ \mathord{\Uparrow}  \ottnt{l'} $ by \E{CheckRaise}. In the last case, we step by
      \E{CheckInner}. If $ \mathsf{val} _{  \mathsf{E}  }~ \ottnt{e_{{\mathrm{2}}}} $, by canonical forms
      (Lemma~\ref{lem:heedfulcanonicalforms}) we know that $\ottnt{e_{{\mathrm{2}}}}$ is
      a $\ottnt{k}$ such that $ \mathsf{ty} ( \ottnt{k} )   \ottsym{=}   \mathsf{Bool} $, i.e., $\ottnt{e_{{\mathrm{2}}}}$ is either
      $ \mathsf{true} $ or $ \mathsf{false} $. In the former case, we step to
      $\ottnt{k}$ by \E{CheckOK}; in the latter case, we step to
      $ \mathord{\Uparrow}  \ottnt{l} $ by \E{CheckFail}.
    \end{itemize}
    \fi}
  \end{proof}
\end{lemma}

Before proving preservation, we must establish some properties about
type sets: type sets as merged by \E{CastMerge} are well
formed; the $ \mathsf{dom} $ and $ \mathsf{cod} $ operators take type sets of
function types and produce well formed type sets.

\begin{lemma}[Merged type sets are well formed]
  \label{lem:heedfulmergewf}
  If $ \mathord{  \vdash _{  \mathsf{H}  } }~ \mathcal{S}_{{\mathrm{1}}}   \mathrel{\parallel}   \ottnt{T_{{\mathrm{1}}}}  \Rightarrow  \ottnt{T_{{\mathrm{2}}}} $ and $ \mathord{  \vdash _{  \mathsf{H}  } }~ \mathcal{S}_{{\mathrm{2}}}   \mathrel{\parallel}   \ottnt{T_{{\mathrm{2}}}}  \Rightarrow  \ottnt{T_{{\mathrm{3}}}} $ then $ \mathord{  \vdash _{  \mathsf{H}  } }~ \ottsym{(}    \mathcal{S}_{{\mathrm{1}}}  \cup  \mathcal{S}_{{\mathrm{2}}}   \cup   \set{  \ottnt{T_{{\mathrm{2}}}}  }    \ottsym{)}   \mathrel{\parallel}   \ottnt{T_{{\mathrm{1}}}}  \Rightarrow  \ottnt{T_{{\mathrm{3}}}} $.
  \begin{proof}
    By transitivity of similarity, we have $\vdash  \ottnt{T_{{\mathrm{1}}}}  \mathrel{\parallel}  \ottnt{T_{{\mathrm{3}}}}$. We have
    $ \mathord{  \vdash _{  \mathsf{H}  } }~ \ottnt{T_{{\mathrm{1}}}} $ and $ \mathord{  \vdash _{  \mathsf{H}  } }~ \ottnt{T_{{\mathrm{3}}}} $ from each of the \A{TypeSet}
    derivations, so it remains to show the premises for each $ \ottnt{T}  \in  \mathcal{S} $.

    Let $ \ottnt{T}  \in  \ottsym{(}    \mathcal{S}_{{\mathrm{1}}}  \cup  \mathcal{S}_{{\mathrm{2}}}   \cup   \set{  \ottnt{T_{{\mathrm{2}}}}  }    \ottsym{)} $. We have $\vdash  \ottnt{T}  \mathrel{\parallel}  \ottnt{T_{{\mathrm{1}}}}$ and
    $ \mathord{  \vdash _{  \mathsf{H}  } }~ \ottnt{T} $ (a) by assumption and symmetry
    (Lemma~\ref{lem:similaritysymmetric}) if $\ottnt{T}  \ottsym{=}  \ottnt{T_{{\mathrm{2}}}}$; and (b) by
    \A{TypeSet} and symmetry and transitivity
    (Lemma~\ref{lem:similaritytransitive}) if $ \ottnt{T}  \in   \mathcal{S}_{{\mathrm{1}}}  \cup  \mathcal{S}_{{\mathrm{2}}}  $. We
    can therefore apply \A{TypeSet}, and we are done.
  \end{proof}
\end{lemma}

\begin{lemma}[Domain type set well formedness]
  \label{lem:heedfuldomwf}
  If $ \mathord{  \vdash _{  \mathsf{H}  } }~ \mathcal{S}   \mathrel{\parallel}    \ottnt{T_{{\mathrm{11}}}} \mathord{ \rightarrow } \ottnt{T_{{\mathrm{12}}}}   \Rightarrow   \ottnt{T_{{\mathrm{21}}}} \mathord{ \rightarrow } \ottnt{T_{{\mathrm{22}}}}  $ then $ \mathord{  \vdash _{  \mathsf{H}  } }~  \mathsf{dom} ( \mathcal{S} )    \mathrel{\parallel}   \ottnt{T_{{\mathrm{21}}}}  \Rightarrow  \ottnt{T_{{\mathrm{11}}}} $.
  \begin{proof}
    First, observe that for every $ \ottnt{T}  \in  \mathcal{S} $, we know that $\vdash  \ottnt{T}  \mathrel{\parallel}   \ottnt{T_{{\mathrm{11}}}} \mathord{ \rightarrow } \ottnt{T_{{\mathrm{12}}}} $, so each $\ottnt{T_{\ottmv{i}}}  \ottsym{=}   \ottnt{T_{\ottmv{i}\,{\mathrm{1}}}} \mathord{ \rightarrow } \ottnt{T_{\ottmv{i}\,{\mathrm{2}}}} $ by inversion. This
    means that $ \mathsf{dom} ( \mathcal{S} ) $ is well defined.

    By inversion of similarity and type well formedness, we have $\vdash  \ottnt{T_{{\mathrm{11}}}}  \mathrel{\parallel}  \ottnt{T_{{\mathrm{21}}}}$ and $ \mathord{  \vdash _{  \mathsf{H}  } }~ \ottnt{T_{{\mathrm{11}}}} $ and $ \mathord{  \vdash _{  \mathsf{H}  } }~ \ottnt{T_{{\mathrm{21}}}} $. By symmetry of
    similarity, we have $\vdash  \ottnt{T_{{\mathrm{21}}}}  \mathrel{\parallel}  \ottnt{T_{{\mathrm{11}}}}$
    (Lemma~\ref{lem:similaritysymmetric}).

    Let $ \ottnt{T_{\ottmv{i}\,{\mathrm{1}}}}  \in   \mathsf{dom} ( \mathcal{S} )  $ by given. We know that there exists some
    $\ottnt{T_{\ottmv{i}\,{\mathrm{2}}}}$ such that $  \ottnt{T_{\ottmv{i}\,{\mathrm{1}}}} \mathord{ \rightarrow } \ottnt{T_{\ottmv{i}\,{\mathrm{2}}}}   \in  \mathcal{S} $ and $\vdash   \ottnt{T_{\ottmv{i}\,{\mathrm{1}}}} \mathord{ \rightarrow } \ottnt{T_{\ottmv{i}\,{\mathrm{2}}}}   \mathrel{\parallel}   \ottnt{T_{{\mathrm{11}}}} \mathord{ \rightarrow } \ottnt{T_{{\mathrm{12}}}} $ and $ \mathord{  \vdash _{  \mathsf{H}  } }~  \ottnt{T_{\ottmv{i}\,{\mathrm{1}}}} \mathord{ \rightarrow } \ottnt{T_{\ottmv{i}\,{\mathrm{2}}}}  $. By inversion, we find $\vdash  \ottnt{T_{\ottmv{i}\,{\mathrm{1}}}}  \mathrel{\parallel}  \ottnt{T_{{\mathrm{11}}}}$ and $ \mathord{  \vdash _{  \mathsf{H}  } }~ \ottnt{T_{\ottmv{i}\,{\mathrm{1}}}} $. By transitivity of similarity
    (Lemma~\ref{lem:similaritytransitive}), we have $\vdash  \ottnt{T_{\ottmv{i}\,{\mathrm{1}}}}  \mathrel{\parallel}  \ottnt{T_{{\mathrm{21}}}}$, and we are done by \A{TypeSet}.
  \end{proof}
\end{lemma}

\begin{lemma}[Codomain type set well formedness]
  \label{lem:heedfulcodwf}
  If $ \mathord{  \vdash _{  \mathsf{H}  } }~ \mathcal{S}   \mathrel{\parallel}    \ottnt{T_{{\mathrm{11}}}} \mathord{ \rightarrow } \ottnt{T_{{\mathrm{12}}}}   \Rightarrow   \ottnt{T_{{\mathrm{21}}}} \mathord{ \rightarrow } \ottnt{T_{{\mathrm{22}}}}  $ then $ \mathord{  \vdash _{  \mathsf{H}  } }~  \mathsf{cod} ( \mathcal{S} )    \mathrel{\parallel}   \ottnt{T_{{\mathrm{12}}}}  \Rightarrow  \ottnt{T_{{\mathrm{22}}}} $.
  \begin{proof}
    First, observe that for every $ \ottnt{T}  \in  \mathcal{S} $, we know that $\vdash  \ottnt{T}  \mathrel{\parallel}   \ottnt{T_{{\mathrm{11}}}} \mathord{ \rightarrow } \ottnt{T_{{\mathrm{12}}}} $, so each $\ottnt{T_{\ottmv{i}}}  \ottsym{=}   \ottnt{T_{\ottmv{i}\,{\mathrm{1}}}} \mathord{ \rightarrow } \ottnt{T_{\ottmv{i}\,{\mathrm{2}}}} $ by inversion. This
    means that $ \mathsf{dom} ( \mathcal{S} ) $ is well defined.

    By inversion of similarity and type well formedness, we have $\vdash  \ottnt{T_{{\mathrm{12}}}}  \mathrel{\parallel}  \ottnt{T_{{\mathrm{22}}}}$ and $ \mathord{  \vdash _{  \mathsf{H}  } }~ \ottnt{T_{{\mathrm{12}}}} $ and $ \mathord{  \vdash _{  \mathsf{H}  } }~ \ottnt{T_{{\mathrm{22}}}} $.

    Let $ \ottnt{T_{\ottmv{i}\,{\mathrm{2}}}}  \in   \mathsf{dom} ( \mathcal{S} )  $ by given. We know that there exists some
    $\ottnt{T_{\ottmv{i}\,{\mathrm{2}}}}$ such that $  \ottnt{T_{\ottmv{i}\,{\mathrm{1}}}} \mathord{ \rightarrow } \ottnt{T_{\ottmv{i}\,{\mathrm{2}}}}   \in  \mathcal{S} $ and $\vdash   \ottnt{T_{\ottmv{i}\,{\mathrm{1}}}} \mathord{ \rightarrow } \ottnt{T_{\ottmv{i}\,{\mathrm{2}}}}   \mathrel{\parallel}   \ottnt{T_{{\mathrm{12}}}} \mathord{ \rightarrow } \ottnt{T_{{\mathrm{12}}}} $ and $ \mathord{  \vdash _{  \mathsf{H}  } }~  \ottnt{T_{\ottmv{i}\,{\mathrm{1}}}} \mathord{ \rightarrow } \ottnt{T_{\ottmv{i}\,{\mathrm{2}}}}  $. By inversion, we find $\vdash  \ottnt{T_{\ottmv{i}\,{\mathrm{2}}}}  \mathrel{\parallel}  \ottnt{T_{{\mathrm{12}}}}$ and $ \mathord{  \vdash _{  \mathsf{H}  } }~ \ottnt{T_{\ottmv{i}\,{\mathrm{2}}}} $. We are done by \A{TypeSet}.
  \end{proof} 
\end{lemma}

\begin{lemma}[Reducing type sets]
  \label{lem:typesetreduce}
  If $ \mathord{  \vdash _{  \mathsf{H}  } }~ \mathcal{S}   \mathrel{\parallel}   \ottnt{T_{{\mathrm{1}}}}  \Rightarrow  \ottnt{T_{{\mathrm{3}}}} $ then $ \mathord{  \vdash _{  \mathsf{H}  } }~ \ottsym{(}   \mathcal{S}  \setminus  \ottnt{T_{{\mathrm{2}}}}   \ottsym{)}   \mathrel{\parallel}   \ottnt{T_{{\mathrm{1}}}}  \Rightarrow  \ottnt{T_{{\mathrm{3}}}} $.
  \begin{proof}
    All of the $ \ottnt{T}  \in  \mathcal{S} $ remain well formed and similar to $\ottnt{T_{{\mathrm{1}}}}$
    and $\ottnt{T_{{\mathrm{3}}}}$, as do the well formedness and similarity relations
    for $\ottnt{T_{{\mathrm{1}}}}$ and $\ottnt{T_{{\mathrm{3}}}}$.
  \end{proof}
\end{lemma}

\begin{lemma}[Heedful preservation]
  \label{lem:heedfulpreservation}
  If $ \emptyset   \vdash _{  \mathsf{H}  }  \ottnt{e}  :  \ottnt{T} $ and $\ottnt{e} \,  \longrightarrow _{  \mathsf{H}  }  \, \ottnt{e'}$ then $ \emptyset   \vdash _{  \mathsf{H}  }  \ottnt{e'}  :  \ottnt{T} $.
  \begin{proof}
    By induction on the typing derivation.
    {\iffull
    \begin{itemize}
    \item[(\T{Var})] Contradictory---we assumed $\ottnt{e}$ was well typed
      in an empty context.
    \item[(\T{Const})] Contradictory---$\ottnt{k}$ is a value and doesn't step.
    \item[(\T{Abs})] Contradictory---$ \lambda \mathit{x} \mathord{:} \ottnt{T_{{\mathrm{1}}}} .~  \ottnt{e'} $ is a value and
      doesn't step.
    \item[(\T{Op})] By cases on the step taken.
      \begin{itemize}
      \item[(\E{Op})] $\denot{ op } \, \ottsym{(}  \ottnt{k_{{\mathrm{1}}}}  \ottsym{,} \, ... \, \ottsym{,}  \ottnt{k_{\ottmv{n}}}  \ottsym{)}  \ottsym{=}  \ottnt{k}$; we assume that
        $ \mathsf{ty} (\mathord{ \ottnt{op} }) $ correctly assigns types, i.e., if $ \mathsf{ty} (\mathord{ \ottnt{op} })   \ottsym{=}   {}   \{ \mathit{x} \mathord{:} \ottnt{B_{{\mathrm{1}}}} \mathrel{\mid} \ottnt{e'_{{\mathrm{1}}}} \}   \rightarrow \, ... \, \rightarrow   \{ \mathit{x} \mathord{:} \ottnt{B_{\ottmv{n}}} \mathrel{\mid} \ottnt{e'_{\ottmv{n}}} \}   {} \mathord{ \rightarrow }  \{ \mathit{x} \mathord{:} \ottnt{B} \mathrel{\mid} \ottnt{e} \}  $, then $ \ottnt{e}  [  \ottnt{k} / \mathit{x}  ]  \,  \longrightarrow ^{*}_{  \mathsf{H}  }  \,  \mathsf{true} $ and $ \mathord{  \vdash _{  \mathsf{H}  } }~  \{ \mathit{x} \mathord{:} \ottnt{B} \mathrel{\mid} \ottnt{e} \}  $. We can therefore conclude
        that $ \emptyset   \vdash _{  \mathsf{H}  }  \ottnt{k}  :   \{ \mathit{x} \mathord{:} \ottnt{B} \mathrel{\mid} \ottnt{e} \}  $ by \T{Const}.
      \item[(\E{OpInner})] By the IH and \T{Op}.
      \item[(\E{OpRaise})] We assume that $ \mathsf{ty} (\mathord{ \ottnt{op} }) $ only assigns
        well formed types, so $ \emptyset   \vdash _{  \mathsf{H}  }   \mathord{\Uparrow}  \ottnt{l}   :  \ottnt{T} $.
      \end{itemize}
    \item[(\T{App})] By cases on the step taken.
      \begin{itemize}
      \item[(\E{Beta})] We know that $  \mathit{x} \mathord{:} \ottnt{T_{{\mathrm{1}}}}    \vdash _{  \mathsf{H}  }  \ottnt{e_{{\mathrm{1}}}}  :  \ottnt{T_{{\mathrm{2}}}} $ and
        $ \emptyset   \vdash _{  \mathsf{H}  }  \ottnt{e_{{\mathrm{2}}}}  :  \ottnt{T_{{\mathrm{1}}}} $; we are done by substitution
        (Lemma~\ref{lem:substitution}).
      \item[(\E{Unwrap})] We have $\ottnt{e} =
          (  \langle   \ottnt{T_{{\mathrm{11}}}} \mathord{ \rightarrow } \ottnt{T_{{\mathrm{12}}}}   \mathord{ \overset{ \mathcal{S} }{\Rightarrow} }   \ottnt{T_{{\mathrm{21}}}} \mathord{ \rightarrow } \ottnt{T_{{\mathrm{22}}}}   \rangle^{ \ottnt{l} } ~  \ottnt{e_{{\mathrm{1}}}}  )  ~ \ottnt{e_{{\mathrm{2}}}}  \,  \longrightarrow _{  \mathsf{H}  }  \,  \langle  \ottnt{T_{{\mathrm{12}}}}  \mathord{ \overset{  \mathsf{cod} ( \mathcal{S} )  }{\Rightarrow} }  \ottnt{T_{{\mathrm{22}}}}  \rangle^{ \ottnt{l} } ~   (  \ottnt{e_{{\mathrm{1}}}} ~  (  \langle  \ottnt{T_{{\mathrm{21}}}}  \mathord{ \overset{  \mathsf{dom} ( \mathcal{S} )  }{\Rightarrow} }  \ottnt{T_{{\mathrm{11}}}}  \rangle^{ \ottnt{l} } ~  \ottnt{e_{{\mathrm{2}}}}  )   )   = \ottnt{e'}$.

        By inversion, $\vdash   \ottnt{T_{{\mathrm{11}}}} \mathord{ \rightarrow } \ottnt{T_{{\mathrm{12}}}}   \mathrel{\parallel}   \ottnt{T_{{\mathrm{21}}}} \mathord{ \rightarrow } \ottnt{T_{{\mathrm{22}}}} $ and $ \mathord{  \vdash _{  \mathsf{H}  } }~ \mathcal{S}   \mathrel{\parallel}    \ottnt{T_{{\mathrm{11}}}} \mathord{ \rightarrow } \ottnt{T_{{\mathrm{12}}}}   \Rightarrow   \ottnt{T_{{\mathrm{21}}}} \mathord{ \rightarrow } \ottnt{T_{{\mathrm{22}}}}  $. By Lemma~\ref{lem:heedfuldomwf},
        $ \mathord{  \vdash _{  \mathsf{H}  } }~  \mathsf{dom} ( \mathcal{S} )    \mathrel{\parallel}   \ottnt{T_{{\mathrm{21}}}}  \Rightarrow  \ottnt{T_{{\mathrm{11}}}} $; by
        Lemma~\ref{lem:heedfulcodwf}, $ \mathord{  \vdash _{  \mathsf{H}  } }~  \mathsf{cod} ( \mathcal{S} )    \mathrel{\parallel}   \ottnt{T_{{\mathrm{12}}}}  \Rightarrow  \ottnt{T_{{\mathrm{22}}}} $.
        
        By inversion, $ \emptyset   \vdash _{  \mathsf{H}  }  \ottnt{e_{{\mathrm{2}}}}  :  \ottnt{T_{{\mathrm{21}}}} $; so $ \emptyset   \vdash _{  \mathsf{H}  }   \langle  \ottnt{T_{{\mathrm{21}}}}  \mathord{ \overset{  \mathsf{dom} ( \mathcal{S} )  }{\Rightarrow} }  \ottnt{T_{{\mathrm{11}}}}  \rangle^{ \ottnt{l} } ~  \ottnt{e_{{\mathrm{2}}}}   :  \ottnt{T_{{\mathrm{11}}}} $ by \T{Cast}. By
        inversion, $ \emptyset   \vdash _{  \mathsf{H}  }  \ottnt{e_{{\mathrm{1}}}}  :   \ottnt{T_{{\mathrm{11}}}} \mathord{ \rightarrow } \ottnt{T_{{\mathrm{12}}}}  $, so $ \emptyset   \vdash _{  \mathsf{H}  }   \ottnt{e_{{\mathrm{1}}}} ~  (  \langle  \ottnt{T_{{\mathrm{21}}}}  \mathord{ \overset{  \mathsf{dom} ( \mathcal{S} )  }{\Rightarrow} }  \ottnt{T_{{\mathrm{11}}}}  \rangle^{ \ottnt{l} } ~  \ottnt{e_{{\mathrm{2}}}}  )    :  \ottnt{T_{{\mathrm{12}}}} $ by \T{App}.
        By the type set typing for $ \mathsf{cod} ( \mathcal{S} ) $, we can apply \T{Cast}
        to finish the case, typing the whole term at $\ottnt{T_{{\mathrm{22}}}}$.

      \item[(\E{AppL})] By \T{App} and the IH.
      \item[(\E{AppR})] By \T{App} and the IH.
      \item[(\E{AppRaiseL})] By regularity, $ \mathord{  \vdash _{  \mathsf{H}  } }~ \ottnt{T_{{\mathrm{2}}}} $, so we are
        done by \T{Blame}.
      \item[(\E{AppRaiseR})] By regularity, $ \mathord{  \vdash _{  \mathsf{H}  } }~ \ottnt{T_{{\mathrm{2}}}} $, so we are
        done by \T{Blame}.
      \end{itemize}
    \item[(\T{Cast})] By cases on the step taken.
      \begin{itemize}
      \item[(\E{TypeSet})] By the assumptions, using \A{None} to
        derive \A{TypeSet}---which holds vacuously, since the set is
        empty.
      \item[(\E{CheckEmpty})] We have $ \mathord{  \vdash _{  \mathsf{H}  } }~ \Gamma $ and $ \mathord{  \vdash _{  \mathsf{H}  } }~  \{ \mathit{x} \mathord{:} \ottnt{B} \mathrel{\mid} \ottnt{e_{{\mathrm{2}}}} \}  $ and $ \mathsf{ty} ( \ottnt{k} )   \ottsym{=}  \ottnt{B}$ by inversion and $ \ottnt{e_{{\mathrm{2}}}}  [  \ottnt{k} / \mathit{x}  ]  \,  \longrightarrow ^{*}_{  \mathsf{H}  }  \,  \ottnt{e_{{\mathrm{2}}}}  [  \ottnt{k} / \mathit{x}  ] $ by reflexivity. By substitution (and
        \T{Const}, to find $ \emptyset   \vdash _{  \mathsf{H}  }  \ottnt{k}  :   \{ \mathit{x} \mathord{:} \ottnt{B} \mathrel{\mid}  \mathsf{true}  \}  $), we find
        $ \emptyset   \vdash _{  \mathsf{H}  }   \ottnt{e_{{\mathrm{2}}}}  [  \ottnt{k} / \mathit{x}  ]   :   \{ \mathit{x} \mathord{:}  \mathsf{Bool}  \mathrel{\mid}  \mathsf{true}  \}  $. We can now apply
        \T{Check}, and are done.
      \item[(\E{CheckSet})] We have \[  \langle   \{ \mathit{x} \mathord{:} \ottnt{B} \mathrel{\mid} \ottnt{e_{{\mathrm{1}}}} \}   \mathord{ \overset{ \mathcal{S} }{\Rightarrow} }   \{ \mathit{x} \mathord{:} \ottnt{B} \mathrel{\mid} \ottnt{e_{{\mathrm{3}}}} \}   \rangle^{ \ottnt{l} } ~  \ottnt{k}  \,  \longrightarrow _{  \mathsf{H}  }  \,  \langle   \{ \mathit{x} \mathord{:} \ottnt{B} \mathrel{\mid} \ottnt{e_{{\mathrm{2}}}} \}   \mathord{ \overset{  \mathcal{S}  \setminus   \{ \mathit{x} \mathord{:} \ottnt{B} \mathrel{\mid} \ottnt{e_{{\mathrm{2}}}} \}   }{\Rightarrow} }   \{ \mathit{x} \mathord{:} \ottnt{B} \mathrel{\mid} \ottnt{e_{{\mathrm{3}}}} \}   \rangle^{ \ottnt{l} } ~   \langle   \{ \mathit{x} \mathord{:} \ottnt{B} \mathrel{\mid} \ottnt{e_{{\mathrm{2}}}} \}  ,   \ottnt{e_{{\mathrm{2}}}}  [  \ottnt{k} / \mathit{x}  ]  ,  \ottnt{k}  \rangle^{ \ottnt{l} }   \] where $ \mathsf{choose} ( \mathcal{S} )   \ottsym{=}   \{ \mathit{x} \mathord{:} \ottnt{B} \mathrel{\mid} \ottnt{e_{{\mathrm{2}}}} \} $, so $  \{ \mathit{x} \mathord{:} \ottnt{B} \mathrel{\mid} \ottnt{e_{{\mathrm{2}}}} \}   \in  \mathcal{S} $.
        We have $ \mathord{  \vdash _{  \mathsf{H}  } }~ \Gamma $ and $ \mathord{  \vdash _{  \mathsf{H}  } }~  \{ \mathit{x} \mathord{:} \ottnt{B} \mathrel{\mid} \ottnt{e_{{\mathrm{2}}}} \}  $ and $ \mathsf{ty} ( \ottnt{k} )   \ottsym{=}  \ottnt{B}$
        by inversion and $ \ottnt{e_{{\mathrm{2}}}}  [  \ottnt{k} / \mathit{x}  ]  \,  \longrightarrow ^{*}_{  \mathsf{H}  }  \,  \ottnt{e_{{\mathrm{2}}}}  [  \ottnt{k} / \mathit{x}  ] $ by
        reflexivity. By substitution (and \T{Const}, to find $ \emptyset   \vdash _{  \mathsf{H}  }  \ottnt{k}  :   \{ \mathit{x} \mathord{:} \ottnt{B} \mathrel{\mid}  \mathsf{true}  \}  $), we find $ \emptyset   \vdash _{  \mathsf{H}  }   \ottnt{e_{{\mathrm{2}}}}  [  \ottnt{k} / \mathit{x}  ]   :   \{ \mathit{x} \mathord{:}  \mathsf{Bool}  \mathrel{\mid}  \mathsf{true}  \}  $. We can now apply \T{Check} to type the
        active check at $ \{ \mathit{x} \mathord{:} \ottnt{B} \mathrel{\mid} \ottnt{e_{{\mathrm{2}}}} \} $.

        By inversion of the original type set well formedness
        derivation, $\vdash   \{ \mathit{x} \mathord{:} \ottnt{B} \mathrel{\mid} \ottnt{e_{{\mathrm{2}}}} \}   \mathrel{\parallel}   \{ \mathit{x} \mathord{:} \ottnt{B} \mathrel{\mid} \ottnt{e_{{\mathrm{1}}}} \} $ and $ \mathord{  \vdash _{  \mathsf{H}  } }~   \set{   \{ \mathit{x} \mathord{:} \ottnt{B} \mathrel{\mid} \ottnt{e_{{\mathrm{2}}}} \}   }   \cup  \mathcal{S}    \mathrel{\parallel}    \{ \mathit{x} \mathord{:} \ottnt{B} \mathrel{\mid} \ottnt{e_{{\mathrm{1}}}} \}   \Rightarrow   \{ \mathit{x} \mathord{:} \ottnt{B} \mathrel{\mid} \ottnt{e_{{\mathrm{3}}}} \}  $; by
        Lemma~\ref{lem:typesetreduce}, $ \mathord{  \vdash _{  \mathsf{H}  } }~  \mathcal{S}  \setminus   \{ \mathit{x} \mathord{:} \ottnt{B} \mathrel{\mid} \ottnt{e_{{\mathrm{2}}}} \}     \mathrel{\parallel}    \{ \mathit{x} \mathord{:} \ottnt{B} \mathrel{\mid} \ottnt{e_{{\mathrm{2}}}} \}   \Rightarrow   \{ \mathit{x} \mathord{:} \ottnt{B} \mathrel{\mid} \ottnt{e_{{\mathrm{3}}}} \}  $. We have $\vdash   \{ \mathit{x} \mathord{:} \ottnt{B} \mathrel{\mid} \ottnt{e_{{\mathrm{2}}}} \}   \mathrel{\parallel}   \{ \mathit{x} \mathord{:} \ottnt{B} \mathrel{\mid} \ottnt{e_{{\mathrm{3}}}} \} $ transitivity of similarity ($\vdash   \{ \mathit{x} \mathord{:} \ottnt{B} \mathrel{\mid} \ottnt{e_{{\mathrm{1}}}} \}   \mathrel{\parallel}   \{ \mathit{x} \mathord{:} \ottnt{B} \mathrel{\mid} \ottnt{e_{{\mathrm{3}}}} \} $ and Lemma~\ref{lem:similaritytransitive}).
        So now we can apply \T{Cast} to our \T{Check} derivation, and
        we are done.
      \item[(\E{CastInner})] By \T{Cast} and the IH.
      \item[(\E{CastMerge})] We have \[ \begin{array}{l}
           \langle  \ottnt{T_{{\mathrm{2}}}}  \mathord{ \overset{ \mathcal{S}_{{\mathrm{2}}} }{\Rightarrow} }  \ottnt{T_{{\mathrm{3}}}}  \rangle^{ \ottnt{l_{{\mathrm{2}}}} } ~   (  \langle  \ottnt{T_{{\mathrm{1}}}}  \mathord{ \overset{ \mathcal{S}_{{\mathrm{1}}} }{\Rightarrow} }  \ottnt{T_{{\mathrm{2}}}}  \rangle^{ \ottnt{l_{{\mathrm{1}}}} } ~  \ottnt{e}  )    \longrightarrow _{  \mathsf{H}  }  {} \\
           \langle  \ottnt{T_{{\mathrm{1}}}}  \mathord{ \overset{ \ottsym{(}    \mathcal{S}_{{\mathrm{1}}}  \cup  \mathcal{S}_{{\mathrm{2}}}   \cup   \set{  \ottnt{T_{{\mathrm{2}}}}  }    \ottsym{)} }{\Rightarrow} }  \ottnt{T_{{\mathrm{3}}}}  \rangle^{ \ottnt{l_{{\mathrm{2}}}} } ~  \ottnt{e} . \end{array} \]
        By Lemma~\ref{lem:heedfulmergewf}, $ \mathord{  \vdash _{  \mathsf{H}  } }~ \ottsym{(}    \mathcal{S}_{{\mathrm{1}}}  \cup  \mathcal{S}_{{\mathrm{2}}}   \cup   \set{  \ottnt{T_{{\mathrm{2}}}}  }    \ottsym{)}   \mathrel{\parallel}   \ottnt{T_{{\mathrm{1}}}}  \Rightarrow  \ottnt{T_{{\mathrm{3}}}} $. We already know that $ \emptyset   \vdash _{  \mathsf{H}  }  \ottnt{e_{{\mathrm{1}}}}  :  \ottnt{T_{{\mathrm{1}}}} $ (by
        assumption) and $ \mathord{  \vdash _{  \mathsf{H}  } }~ \ottnt{T_{{\mathrm{3}}}} $ (by inversion of the outer cast's
        typing derivation), so we can apply \T{Cast} to type the
        resulting merged cast.
      \item[(\E{CastRaise})] We have $ \mathord{  \vdash _{  \mathsf{H}  } }~ \ottnt{T_{{\mathrm{2}}}} $ by assumption, so we
        are done by \T{Blame}.
      \end{itemize}
    \item[(\T{Blame})] Contradictory---$ \mathord{\Uparrow}  \ottnt{l} $ is a result and
      doesn't step.
    \item[(\T{Check})]  By cases on the step taken.
      \begin{itemize}
      \item[(\E{CheckOK})] Since $ \mathord{  \vdash _{  \mathsf{H}  } }~ \emptyset $ and $ \mathsf{ty} ( \ottnt{k} )   \ottsym{=}  \ottnt{B}$
        and $ \mathord{  \vdash _{  \mathsf{H}  } }~  \{ \mathit{x} \mathord{:} \ottnt{B} \mathrel{\mid} \ottnt{e_{{\mathrm{1}}}} \}  $ and $ \ottnt{e_{{\mathrm{1}}}}  [  \ottnt{k} / \mathit{x}  ]  \,  \longrightarrow ^{*}_{  \mathsf{H}  }  \,  \mathsf{true} $, we can
        apply \T{Const} to find $ \emptyset   \vdash _{  \mathsf{H}  }  \ottnt{k}  :   \{ \mathit{x} \mathord{:} \ottnt{B} \mathrel{\mid} \ottnt{e_{{\mathrm{1}}}} \}  $.
      \item[(\E{CheckFail})] Since $ \mathord{  \vdash _{  \mathsf{H}  } }~ \emptyset $ and $ \mathord{  \vdash _{  \mathsf{H}  } }~  \{ \mathit{x} \mathord{:} \ottnt{B} \mathrel{\mid} \ottnt{e_{{\mathrm{1}}}} \}  $, \T{Blame} shows $ \emptyset   \vdash _{  \mathsf{H}  }   \mathord{\Uparrow}  \ottnt{l}   :   \{ \mathit{x} \mathord{:} \ottnt{B} \mathrel{\mid} \ottnt{e_{{\mathrm{1}}}} \}  $.
      \item[(\E{CheckInner})] By \T{Check} and the IH.
      \item[(\E{CheckRaise})] As for \E{CheckFail}---the differing
        label doesn't matter.
      \end{itemize}
    \end{itemize}
    \fi}
  \end{proof}
\end{lemma}

Just as we did for forgetful \lambdah in
(Appendix~\ref{app:forgetfulsoundness}), we show that source programs
are well typed heedfully iff they are well typed classically---iff
they are well typed forgetfull (Lemma~\ref{lem:forgetfulsource}). that
is, source programs are valid staring points in any mode.

\begin{lemma}[Source program typing for heedful \lambdah]
  \label{lem:heedfulsource}
  ~

  \noindent
  Source programs are well typed in $ \mathsf{C} $ iff they are well typed in
  $ \mathsf{H} $, i.e.:
  \begin{itemize}
  \item $ \Gamma   \vdash _{  \mathsf{C}  }  \ottnt{e}  :  \ottnt{T} $ as a source program iff $ \Gamma   \vdash _{  \mathsf{H}  }  \ottnt{e}  :  \ottnt{T} $ as
    a source program.
  \item $ \mathord{  \vdash _{  \mathsf{C}  } }~ \ottnt{T} $ as a source program iff $ \mathord{  \vdash _{  \mathsf{H}  } }~ \ottnt{T} $ as a source program.
  \item $ \mathord{  \vdash _{  \mathsf{C}  } }~ \Gamma $ as a source program iff $ \mathord{  \vdash _{  \mathsf{H}  } }~ \Gamma $ as a source program.
  \end{itemize}
  \begin{proof}
    By mutual induction on $\ottnt{e}$, $\ottnt{T}$, and $\Gamma$.
    {\iffull
    Since all of the rules are syntax directed, we use the rule names
    for cases (but prove both directions at once).
    \paragraph{Expressions $\ottnt{e}$}
    \begin{itemize}
    \item[(\T{Var})] By the IH on $\Gamma$ and \T{Var}.
    \item[(\T{Const})] By the IH on $\Gamma$ and $\T{Const}$, noting
      that $ \mathsf{true}  \,  \longrightarrow ^{*}_{ \ottnt{m} }  \,  \mathsf{true} $ in every mode $\ottnt{m}$.
    \item[(\T{Abs})] By the IH on $\ottnt{T_{{\mathrm{1}}}}$ and $\ottnt{e_{{\mathrm{12}}}}$ and \T{Abs}.
    \item[(\T{Op})] By the IHs on the arguments $\ottnt{e_{\ottmv{i}}}$ and \T{Op}.
    \item[(\T{App})] By the IHs on $\ottnt{e_{{\mathrm{1}}}}$ and $\ottnt{e_{{\mathrm{2}}}}$ and \T{App}.
    \item[(\T{Cast})] By the IHs on $\ottnt{T_{{\mathrm{1}}}}$ and $\ottnt{T_{{\mathrm{2}}}}$ and
      \T{Cast}, using \A{None} for the source program.
    \item[(\T{Blame})] Contradictory---doesn't occur in source programs.
    \item[(\T{Check})] Contradictory---doesn't occur in source programs.
    \end{itemize}
    
    \paragraph{Types $\ottnt{T}$}
    \begin{itemize}
    \item[(\WF{Base})] Immediately true---\WF{Base} is an axiom.
    \item[(\WF{Refine})] By the IH on $\ottnt{e}$ and \WF{Refine}.
    \item[(\WF{Fun})] By the IHs on $\ottnt{T_{{\mathrm{1}}}}$ and $\ottnt{T_{{\mathrm{2}}}}$ and \WF{Fun}.
    \end{itemize}

    \paragraph{Contexts $\Gamma$}
    \begin{itemize}
    \item[(\WF{Empty})] Immediately true---\WF{Empty} is an axiom.
    \item[(\WF{Extend})] By the IHs on $\Gamma$ and $\ottnt{T}$ and
      \WF{Extend}.
    \end{itemize}
    \fi}
  \end{proof}
\end{lemma}
\fi}

\subsection{Eidetic type soundness}
\label{app:eideticsoundness}

\begin{lemma}[Determinism of eidetic \lambdah]
  \label{lem:eideticdeterminism}
  If $\ottnt{e} \,  \longrightarrow _{  \mathsf{E}  }  \, \ottnt{e_{{\mathrm{1}}}}$ and $\ottnt{e} \,  \longrightarrow _{  \mathsf{E}  }  \, \ottnt{e_{{\mathrm{2}}}}$ then $\ottnt{e_{{\mathrm{1}}}}  \ottsym{=}  \ottnt{e_{{\mathrm{2}}}}$.
  \begin{proof}
    By induction on the first evaluation derivation. In every case,
    only a single step can be taken.
  \end{proof}
\end{lemma}

\begin{lemma}[Eidetic canonical forms]
  \label{lem:eideticcanonicalforms}
  If $ \emptyset   \vdash _{  \mathsf{E}  }  \ottnt{e}  :  \ottnt{T} $ and $ \mathsf{val} _{  \mathsf{E}  }~ \ottnt{e} $ then:
  \begin{itemize}
  \item If $\ottnt{T}  \ottsym{=}   \{ \mathit{x} \mathord{:} \ottnt{B} \mathrel{\mid} \ottnt{e'} \} $, then $\ottnt{e}  \ottsym{=}  \ottnt{k}$ and $ \mathsf{ty} ( \ottnt{k} )   \ottsym{=}  \ottnt{B}$
    and $ \ottnt{e'}  [  \ottnt{e} / \mathit{x}  ]  \,  \longrightarrow ^{*}_{  \mathsf{E}  }  \,  \mathsf{true} $.
  \item If $\ottnt{T}  \ottsym{=}   \ottnt{T_{{\mathrm{21}}}} \mathord{ \rightarrow } \ottnt{T_{{\mathrm{22}}}} $, then either $\ottnt{e}  \ottsym{=}   \lambda \mathit{x} \mathord{:} \ottnt{T} .~  \ottnt{e'} $ or $\ottnt{e}  \ottsym{=}   \langle   \ottnt{T_{{\mathrm{11}}}} \mathord{ \rightarrow } \ottnt{T_{{\mathrm{12}}}}   \mathord{ \overset{ \ottnt{c_{{\mathrm{1}}}}  \mapsto  \ottnt{c_{{\mathrm{2}}}} }{\Rightarrow} }   \ottnt{T_{{\mathrm{21}}}} \mathord{ \rightarrow } \ottnt{T_{{\mathrm{22}}}}   \rangle^{\bullet} ~   \lambda \mathit{x} \mathord{:} \ottnt{T_{{\mathrm{11}}}} .~  \ottnt{e'}  $.
  \end{itemize}
  {\iffull
  \begin{proof}
    By inspection of the rules: \T{Const} is the only rule that types
    values at base types; \T{Abs} and \T{Cast} are the only rules
    that type values at function types. Note that now our proxies use
    coercions and not types and type sets.
  \end{proof}
  \fi}
\end{lemma}

\begin{lemma}[Eidetic progress]
  \label{lem:eideticprogress}
  If $ \emptyset   \vdash _{  \mathsf{E}  }  \ottnt{e}  :  \ottnt{T} $, then either:
  \begin{enumerate}
  \item $ \mathsf{result} _{  \mathsf{E}  }~ \ottnt{e} $, i.e., $\ottnt{e}  \ottsym{=}   \mathord{\Uparrow}  \ottnt{l} $ or $ \mathsf{val} _{  \mathsf{E}  }~ \ottnt{e} $;
    or
  \item there exists an $\ottnt{e'}$ such that $\ottnt{e} \,  \longrightarrow _{  \mathsf{E}  }  \, \ottnt{e'}$.
  \end{enumerate}
  \begin{proof}
    By induction on the typing derivation.
    {\iffull
    \begin{itemize}
    \item[(\T{Var})] A contradiction---$\mathit{x}$ isn't well typed in the
      empty environment.

    \item[(\T{Const})] $\ottnt{e}  \ottsym{=}  \ottnt{k}$ is a result by \V{Const} and \R{Val}.

    \item[(\T{Abs})] $\ottnt{e}  \ottsym{=}   \lambda \mathit{x} \mathord{:} \ottnt{T} .~  \ottnt{e'} $ is a result by \V{Abs} and \R{Val}.

    \item[(\T{Op})] We know that $ \mathsf{ty} (\mathord{ \ottnt{op} }) $ is a first order type
      $ {}   \{ \mathit{x} \mathord{:} \ottnt{B_{{\mathrm{1}}}} \mathrel{\mid} \ottnt{e'_{{\mathrm{1}}}} \}   \rightarrow \, ... \, \rightarrow   \{ \mathit{x} \mathord{:} \ottnt{B_{\ottmv{n}}} \mathrel{\mid} \ottnt{e'_{\ottmv{n}}} \}   {} \mathord{ \rightarrow } \ottnt{T} $.
      From left to right, we apply the IH on $\ottnt{e_{\ottmv{i}}}$. If $\ottnt{e_{\ottmv{i}}}$ is
      a result, there are two cases: either $\ottnt{e_{\ottmv{i}}}  \ottsym{=}   \mathord{\Uparrow}  \ottnt{l} $, and
      we step by \E{OpRaise}; or $\ottnt{e_{\ottmv{i}}}  \ottsym{=}  \ottnt{k_{\ottmv{i}}}$, since constants are
      the only values at base types by canonical forms
      (Lemma~\ref{lem:eideticcanonicalforms}), and we continue on to
      the next $\ottnt{e_{\ottmv{i}}}$. If any of the $\ottnt{e_{\ottmv{i}}}$ step, we know that all
      of terms before them are values, so we can step by
      \E{OpInner}. If all of the $\ottnt{e_{\ottmv{i}}}$ are constants $\ottnt{k_{\ottmv{i}}}$, then
      $ \emptyset   \vdash _{  \mathsf{E}  }  \ottnt{k_{\ottmv{i}}}  :   \{ \mathit{x} \mathord{:} \ottnt{B_{\ottmv{i}}} \mathrel{\mid} \ottnt{e'_{\ottmv{i}}} \}  $, and so $ \ottnt{e'_{\ottmv{i}}}  [  \ottnt{k_{\ottmv{i}}} / \mathit{x}  ]  \,  \longrightarrow ^{*}_{  \mathsf{H}  }  \,  \mathsf{true} $. Therefore $\denot{ op } \, \ottsym{(}  \ottnt{k_{{\mathrm{1}}}}  \ottsym{,} \, ... \, \ottsym{,}  \ottnt{k_{\ottmv{n}}}  \ottsym{)}$ is defined, and we
      can step by \E{Op}.

    \item[(\T{App})] By the IH on $\ottnt{e_{{\mathrm{1}}}}$, we know that $\ottnt{e_{{\mathrm{1}}}}$: is
      blame, is a value, or steps to some $\ottnt{e'_{{\mathrm{1}}}}$. In the first
      case, we take a step by \E{AppRaiseL}. In the latter, we take a
      step by \E{AppL}.

      If $ \mathsf{val} _{  \mathsf{E}  }~ \ottnt{e_{{\mathrm{1}}}} $, then we can apply the IH on $\ottnt{e_{{\mathrm{2}}}}$, which
      is blame, is a value, or steps to some $\ottnt{e'_{{\mathrm{2}}}}$. The first and
      last cases are as before, using \E{AppRaiseR} and \E{AppR}.

      If $ \mathsf{val} _{  \mathsf{E}  }~ \ottnt{e_{{\mathrm{2}}}} $, we must go by cases on the shape of
      $\ottnt{e_{{\mathrm{1}}}}$. Since $ \emptyset   \vdash _{  \mathsf{E}  }  \ottnt{e_{{\mathrm{1}}}}  :   \ottnt{T_{{\mathrm{1}}}} \mathord{ \rightarrow } \ottnt{T_{{\mathrm{2}}}}  $, by canonical forms
      (Lemma~\ref{lem:eideticcanonicalforms}) we know that $\ottnt{e_{{\mathrm{1}}}}$ can only
      be an abstraction, a wrapped abstraction, or a cast.
      \begin{itemize}
      \item[($\ottnt{e_{{\mathrm{1}}}}  \ottsym{=}   \lambda \mathit{x} \mathord{:} \ottnt{T_{{\mathrm{1}}}} .~  \ottnt{e'_{{\mathrm{1}}}} $)] We step to $ \ottnt{e'_{{\mathrm{1}}}}  [  \ottnt{e_{{\mathrm{2}}}} / \mathit{x}  ] $ by
        \E{Beta}.

      \item[($\ottnt{e_{{\mathrm{1}}}}  \ottsym{=}   \langle   \ottnt{T_{{\mathrm{11}}}} \mathord{ \rightarrow } \ottnt{T_{{\mathrm{12}}}}   \mathord{ \overset{ \ottnt{c_{{\mathrm{1}}}}  \mapsto  \ottnt{c_{{\mathrm{2}}}} }{\Rightarrow} }   \ottnt{T_{{\mathrm{21}}}} \mathord{ \rightarrow } \ottnt{T_{{\mathrm{22}}}}   \rangle^{\bullet} ~   \lambda \mathit{x} \mathord{:} \ottnt{T_{{\mathrm{11}}}} .~  \ottnt{e'_{{\mathrm{1}}}}  $)] We
        step to $ \langle  \ottnt{T_{{\mathrm{12}}}}  \mathord{ \overset{ \ottnt{c_{{\mathrm{2}}}} }{\Rightarrow} }  \ottnt{T_{{\mathrm{22}}}}  \rangle^{\bullet} ~   (   (  \lambda \mathit{x} \mathord{:} \ottnt{T_{{\mathrm{11}}}} .~  \ottnt{e'_{{\mathrm{1}}}}  )  ~  (  \langle  \ottnt{T_{{\mathrm{21}}}}  \mathord{ \overset{ \ottnt{c_{{\mathrm{1}}}} }{\Rightarrow} }  \ottnt{T_{{\mathrm{11}}}}  \rangle^{\bullet} ~  \ottnt{e_{{\mathrm{2}}}}  )   )  $ by \E{Unwrap}.
      \end{itemize}

    \item[(\T{Cast})] We step by \E{Coerce} to $ \langle  \ottnt{T_{{\mathrm{1}}}}  \mathord{ \overset{  \mathsf{coerce} ( \ottnt{T_{{\mathrm{1}}}} , \ottnt{T_{{\mathrm{2}}}} , \ottnt{l} )  }{\Rightarrow} }  \ottnt{T_{{\mathrm{2}}}}  \rangle^{\bullet} ~  \ottnt{e} $.

    \item[(\T{Coerce})] We have $\ottnt{e}  \ottsym{=}   \langle  \ottnt{T_{{\mathrm{1}}}}  \mathord{ \overset{ \ottnt{c} }{\Rightarrow} }  \ottnt{T_{{\mathrm{2}}}}  \rangle^{\bullet} ~  \ottnt{e_{{\mathrm{1}}}} $. If $\ottnt{e_{{\mathrm{1}}}}$ is
      blame, then we step by \E{CoerceRaise}.
      If $\ottnt{e_{{\mathrm{1}}}}$ is a value, then we go by cases on $ \mathsf{val} _{  \mathsf{E}  }~ \ottnt{e_{{\mathrm{1}}}} $:
      \begin{itemize}
      \item[(\V{Const})] The case must be between refinements, and we
        step by \E{CoerceStack}.
      \item[(\V{Abs})] The cast must be between function types, and we
        have a value by \V{ProxyE}.
      \item[(\V{ProxyE})] We step by \ECastMerge.
      \end{itemize}
      Finally, it may be the case that $\ottnt{e_{{\mathrm{1}}}} \,  \longrightarrow _{  \mathsf{E}  }  \, \ottnt{e'_{{\mathrm{1}}}}$.
      If $\ottnt{e_{{\mathrm{1}}}}  \neq   \langle  \ottnt{T_{{\mathrm{3}}}}  \mathord{ \overset{ \ottnt{c'} }{\Rightarrow} }  \ottnt{T_{{\mathrm{1}}}}  \rangle^{\bullet} ~  \ottnt{e''_{{\mathrm{1}}}} $, then we step by
      \ECastInner.
      On the other hand, if $\ottnt{e_{{\mathrm{1}}}}$ is a coercion term, we step by
      \ECastMerge.

    \item[(\T{Blame})] $ \mathord{\Uparrow}  \ottnt{l} $ is a result by \R{Blame}.

    \item[(\T{Stack})] By cases on the shape of $\ottnt{r}$: if $\ottnt{r}  \ottsym{=}  \mathsf{nil}$, we know that $\ottnt{s}  \ottsym{=}   \mathord{\checkmark} $. We step by \E{StackDone} if
      $\ottnt{e}  \ottsym{=}  \ottnt{k}$; otherwise, we step by \E{StackInner} or
      \E{StackRaise}. If not, we go by cases on the shape of
      $\ottnt{e}$. If $\ottnt{e}  \ottsym{=}  \ottnt{k}$, then we step by \E{StackPop}. Otherwise,
      we step by \E{StackInner} or \E{StackRaise}.

    \item[(\T{Check})] By the IH on $\ottnt{e_{{\mathrm{2}}}}$, we know that $\ottnt{e_{{\mathrm{2}}}}$:
      is $ \mathord{\Uparrow}  \ottnt{l'} $, is a value by $ \mathsf{val} _{  \mathsf{E}  }~ \ottnt{e_{{\mathrm{2}}}} $, or takes a
      step to some $\ottnt{e'_{{\mathrm{2}}}}$. In the first case, we step to
      $ \mathord{\Uparrow}  \ottnt{l'} $ by \E{CheckRaise}. In the last case, we step by
      \E{CheckInner}. If $ \mathsf{val} _{  \mathsf{E}  }~ \ottnt{e_{{\mathrm{2}}}} $, by canonical forms
      (Lemma~\ref{lem:eideticcanonicalforms}) we know that $\ottnt{e_{{\mathrm{2}}}}$ is
      a $\ottnt{k}$ such that $ \mathsf{ty} ( \ottnt{k} )   \ottsym{=}   \mathsf{Bool} $, i.e., $\ottnt{e_{{\mathrm{2}}}}$ is either
      $ \mathsf{true} $ or $ \mathsf{false} $. In the former case, we step to
      $\ottnt{k}$ by \E{CheckOK}; in the latter case, we step to
      $ \mathord{\Uparrow}  \ottnt{l} $ by \E{CheckFail}.

    \end{itemize}
    \fi}
  \end{proof}
\end{lemma}

\begin{lemma}[Extended refinement lists are well formed]
  \label{lem:eideticextendrefinements}
  ~

  \noindent
  If $ \mathord{  \vdash _{  \mathsf{E}  } }~  \{ \mathit{x} \mathord{:} \ottnt{B} \mathrel{\mid} \ottnt{e} \}  $ and $ \mathord{  \vdash _{  \mathsf{E}  } }~ \ottnt{r}   \mathrel{\parallel}    \{ \mathit{x} \mathord{:} \ottnt{B} \mathrel{\mid} \ottnt{e_{{\mathrm{1}}}} \}   \Rightarrow   \{ \mathit{x} \mathord{:} \ottnt{B} \mathrel{\mid} \ottnt{e_{{\mathrm{2}}}} \}  $ then $ \mathord{  \vdash _{  \mathsf{E}  } }~  \mathsf{join} (  \{ \mathit{x} \mathord{:} \ottnt{B} \mathrel{\mid} \ottnt{e} \}^{ \ottnt{l} }  , \ottnt{r} )    \mathrel{\parallel}    \{ \mathit{x} \mathord{:} \ottnt{B} \mathrel{\mid} \ottnt{e_{{\mathrm{1}}}} \}   \Rightarrow   \{ \mathit{x} \mathord{:} \ottnt{B} \mathrel{\mid} \ottnt{e_{{\mathrm{2}}}} \}  $.
  \begin{proof}
    By cases on the rule used.
    \begin{itemize}
    \item[(\A{Refine})] All of the premises are immediately restored
      except in one tricky case. When $ \{ \mathit{x} \mathord{:} \ottnt{B} \mathrel{\mid} \ottnt{e} \}  \, \supset \,  \{ \mathit{x} \mathord{:} \ottnt{B} \mathrel{\mid} \ottnt{e'} \} $
      where $ \{ \mathit{x} \mathord{:} \ottnt{B} \mathrel{\mid} \ottnt{e'} \} \in  \ottnt{r} $ is the only type implying
      $ \{ \mathit{x} \mathord{:} \ottnt{B} \mathrel{\mid} \ottnt{e_{{\mathrm{2}}}} \} $. Then $\mathsf{drop} \, \ottsym{(}  \ottnt{r}  \ottsym{,}   \{ \mathit{x} \mathord{:} \ottnt{B} \mathrel{\mid} \ottnt{e} \}   \ottsym{)}$ isn't well formed on its
      own, but adding $ \{ \mathit{x} \mathord{:} \ottnt{B} \mathrel{\mid} \ottnt{e} \}^{ \ottnt{l} } $ makes it so by
      transitivity\refasm{impltrans}. If not, then we know that $\mathsf{drop} \, \ottsym{(}  \ottnt{r}  \ottsym{,}   \{ \mathit{x} \mathord{:} \ottnt{B} \mathrel{\mid} \ottnt{e} \}   \ottsym{)}$ is well formed, and so is its extensions by
      assumption.

      We know that there are no duplicates by reflexivity of
      $ \supset $.\refasm{implrefl}
    \item[(\A{Fun})] Contradictory.
    \end{itemize}
  \end{proof}
\end{lemma}

\begin{lemma}[Merged coercions are well formed]
  \label{lem:eideticmergewf}
  If $ \mathord{  \vdash _{  \mathsf{E}  } }~ \ottnt{c_{{\mathrm{1}}}}   \mathrel{\parallel}   \ottnt{T_{{\mathrm{1}}}}  \Rightarrow  \ottnt{T_{{\mathrm{2}}}} $ and $ \mathord{  \vdash _{  \mathsf{E}  } }~ \ottnt{c_{{\mathrm{2}}}}   \mathrel{\parallel}   \ottnt{T_{{\mathrm{2}}}}  \Rightarrow  \ottnt{T_{{\mathrm{3}}}} $ then $ \mathord{  \vdash _{  \mathsf{E}  } }~  \mathsf{join} ( \ottnt{c_{{\mathrm{1}}}} , \ottnt{c_{{\mathrm{2}}}} )    \mathrel{\parallel}   \ottnt{T_{{\mathrm{1}}}}  \Rightarrow  \ottnt{T_{{\mathrm{3}}}} $.
  \begin{proof}
    By induction on $\ottnt{c_{{\mathrm{1}}}}$'s typing derivation.
    \begin{itemize}
    \item[(\A{Refine})] By the IH,
      Lemma~\ref{lem:eideticextendrefinements}, and \A{Refine}.
    \item[(\A{Fun})] By the IHs and \A{Fun}.
    \end{itemize}
  \end{proof}
\end{lemma}

\begin{lemma}[$ \mathsf{coerce} $ generates well formed coercions]
  \label{lem:eideticcoercewf}
  ~

  \noindent
  If $\vdash  \ottnt{T_{{\mathrm{1}}}}  \mathrel{\parallel}  \ottnt{T_{{\mathrm{2}}}}$ then $ \mathord{  \vdash _{  \mathsf{E}  } }~  \mathsf{coerce} ( \ottnt{T_{{\mathrm{1}}}} , \ottnt{T_{{\mathrm{2}}}} , \ottnt{l} )    \mathrel{\parallel}   \ottnt{T_{{\mathrm{1}}}}  \Rightarrow  \ottnt{T_{{\mathrm{2}}}} $.
  \begin{proof}
    By induction on the similarity derivation.
    \begin{itemize}
    \item[(\S{Refine})] By \A{Refine}, with
      $ \mathsf{coerce} (  \{ \mathit{x} \mathord{:} \ottnt{B} \mathrel{\mid} \ottnt{e_{{\mathrm{1}}}} \}  ,  \{ \mathit{x} \mathord{:} \ottnt{B} \mathrel{\mid} \ottnt{e_{{\mathrm{2}}}} \}  , \ottnt{l} )   \ottsym{=}   \{ \mathit{x} \mathord{:} \ottnt{B} \mathrel{\mid} \ottnt{e_{{\mathrm{2}}}} \}^{ \ottnt{l} } $.
    \item[(\S{Fun})] By \A{Fun} and the IHs.
    \end{itemize}
  \end{proof}
\end{lemma}

\begin{lemma}[Eidetic preservation]
  \label{lem:eideticpreservation}
  If $ \emptyset   \vdash _{  \mathsf{E}  }  \ottnt{e}  :  \ottnt{T} $ and $\ottnt{e} \,  \longrightarrow _{  \mathsf{E}  }  \, \ottnt{e'}$ then $ \emptyset   \vdash _{  \mathsf{E}  }  \ottnt{e'}  :  \ottnt{T} $.
  \begin{proof}
    By induction on the typing derivation.
    {\iffull
    \begin{itemize}
    \item[(\T{Var})] Contradictory---we assumed $\ottnt{e}$ was well typed
      in an empty context.
    \item[(\T{Const})] Contradictory---$\ottnt{k}$ is a value and doesn't step.
    \item[(\T{Abs})] Contradictory---$ \lambda \mathit{x} \mathord{:} \ottnt{T_{{\mathrm{1}}}} .~  \ottnt{e'} $ is a value and
      doesn't step.
    \item[(\T{Op})] By cases on the step taken.
      \begin{itemize}
      \item[(\E{Op})] $\denot{ op } \, \ottsym{(}  \ottnt{k_{{\mathrm{1}}}}  \ottsym{,} \, ... \, \ottsym{,}  \ottnt{k_{\ottmv{n}}}  \ottsym{)}  \ottsym{=}  \ottnt{k}$; we assume that
        $ \mathsf{ty} (\mathord{ \ottnt{op} }) $ correctly assigns types, i.e., if $ \mathsf{ty} (\mathord{ \ottnt{op} })   \ottsym{=}   {}   \{ \mathit{x} \mathord{:} \ottnt{B_{{\mathrm{1}}}} \mathrel{\mid} \ottnt{e'_{{\mathrm{1}}}} \}   \rightarrow \, ... \, \rightarrow   \{ \mathit{x} \mathord{:} \ottnt{B_{\ottmv{n}}} \mathrel{\mid} \ottnt{e'_{\ottmv{n}}} \}   {} \mathord{ \rightarrow }  \{ \mathit{x} \mathord{:} \ottnt{B} \mathrel{\mid} \ottnt{e} \}  $, then $ \ottnt{e}  [  \ottnt{k} / \mathit{x}  ]  \,  \longrightarrow ^{*}_{  \mathsf{E}  }  \,  \mathsf{true} $ and $ \mathord{  \vdash _{  \mathsf{E}  } }~  \{ \mathit{x} \mathord{:} \ottnt{B} \mathrel{\mid} \ottnt{e} \}  $. We can therefore conclude
        that $ \emptyset   \vdash _{  \mathsf{E}  }  \ottnt{k}  :   \{ \mathit{x} \mathord{:} \ottnt{B} \mathrel{\mid} \ottnt{e} \}  $ by \T{Const}.
      \item[(\E{OpInner})] By the IH and \T{Op}.
      \item[(\E{OpRaise})] We assume that $ \mathsf{ty} (\mathord{ \ottnt{op} }) $ only assigns
        well formed types, so $ \emptyset   \vdash _{  \mathsf{E}  }   \mathord{\Uparrow}  \ottnt{l}   :  \ottnt{T} $.
      \end{itemize}
    \item[(\T{App})] By cases on the step taken.
      \begin{itemize}
      \item[(\E{Beta})] We know that $  \mathit{x} \mathord{:} \ottnt{T_{{\mathrm{1}}}}    \vdash _{  \mathsf{E}  }  \ottnt{e_{{\mathrm{1}}}}  :  \ottnt{T_{{\mathrm{2}}}} $ and
        $ \emptyset   \vdash _{  \mathsf{E}  }  \ottnt{e_{{\mathrm{2}}}}  :  \ottnt{T_{{\mathrm{1}}}} $; we are done by substitution
        (Lemma~\ref{lem:substitution}).
      \item[(\E{Unwrap})] We have $\ottnt{e} =
          (  \langle   \ottnt{T_{{\mathrm{11}}}} \mathord{ \rightarrow } \ottnt{T_{{\mathrm{12}}}}   \mathord{ \overset{ \ottnt{c_{{\mathrm{1}}}}  \mapsto  \ottnt{c_{{\mathrm{2}}}} }{\Rightarrow} }   \ottnt{T_{{\mathrm{21}}}} \mathord{ \rightarrow } \ottnt{T_{{\mathrm{22}}}}   \rangle^{\bullet} ~  \ottnt{e_{{\mathrm{1}}}}  )  ~ \ottnt{e_{{\mathrm{2}}}}  \,  \longrightarrow _{  \mathsf{E}  }  \,  \langle  \ottnt{T_{{\mathrm{12}}}}  \mathord{ \overset{ \ottnt{c_{{\mathrm{2}}}} }{\Rightarrow} }  \ottnt{T_{{\mathrm{22}}}}  \rangle^{\bullet} ~   (  \ottnt{e_{{\mathrm{1}}}} ~  (  \langle  \ottnt{T_{{\mathrm{21}}}}  \mathord{ \overset{ \ottnt{c_{{\mathrm{1}}}} }{\Rightarrow} }  \ottnt{T_{{\mathrm{11}}}}  \rangle^{\bullet} ~  \ottnt{e_{{\mathrm{2}}}}  )   )   = \ottnt{e'}$.

        By inversion, $ \mathord{  \vdash _{  \mathsf{E}  } }~ \ottnt{c_{{\mathrm{1}}}}   \mathrel{\parallel}   \ottnt{T_{{\mathrm{21}}}}  \Rightarrow  \ottnt{T_{{\mathrm{11}}}} $ and $ \mathord{  \vdash _{  \mathsf{E}  } }~ \ottnt{c_{{\mathrm{2}}}}   \mathrel{\parallel}   \ottnt{T_{{\mathrm{12}}}}  \Rightarrow  \ottnt{T_{{\mathrm{22}}}} $. We are done by \T{Coerce} and \T{App}.
      \item[(\E{AppL})] By \T{App} and the IH.
      \item[(\E{AppR})] By \T{App} and the IH.
      \item[(\E{AppRaiseL})] By regularity, $ \mathord{  \vdash _{  \mathsf{E}  } }~ \ottnt{T_{{\mathrm{2}}}} $, so we are
        done by \T{Blame}.
      \item[(\E{AppRaiseR})] By regularity, $ \mathord{  \vdash _{  \mathsf{E}  } }~ \ottnt{T_{{\mathrm{2}}}} $, so we are
        done by \T{Blame}.
      \end{itemize}
    \item[(\T{Cast})] If the annotation is $ \bullet $, we step by
      \E{Coerce}, which is well typed by \T{Coerce} (using
      Lemma~\ref{lem:eideticcoercewf}).

      Otherwise, by cases on the step taken.
      \begin{itemize}
      \item[(\E{CoerceStack}] We have $ \mathord{  \vdash _{  \mathsf{E}  } }~ \Gamma $ and $ \mathord{  \vdash _{  \mathsf{E}  } }~  \{ \mathit{x} \mathord{:} \ottnt{B} \mathrel{\mid} \ottnt{e_{{\mathrm{1}}}} \}  $ and $ \mathsf{ty} ( \ottnt{k} )   \ottsym{=}  \ottnt{B}$ by inversion. The
        quantification over $\ottnt{r}$ is also by inversion, of coercion
        well formedness. Since $\ottnt{s}  \ottsym{=}   \mathord{?} $, we can find $ \{ \mathit{x} \mathord{:} \ottnt{B} \mathrel{\mid} \ottnt{e} \} \in  \ottnt{r} $ such that $ \{ \mathit{x} \mathord{:} \ottnt{B} \mathrel{\mid} \ottnt{e} \}  \, \supset \,  \{ \mathit{x} \mathord{:} \ottnt{B} \mathrel{\mid} \ottnt{e_{{\mathrm{1}}}} \} $ by that same well
        formedness derivation. So: by \T{Stack}.
      \item[(\E{CoerceInner})] By \T{Coerce} and the IH.
      \item[(\E{CastMerge})] We have \[ \begin{array}{l}
           \langle  \ottnt{T_{{\mathrm{2}}}}  \mathord{ \overset{ \ottnt{c_{{\mathrm{2}}}} }{\Rightarrow} }  \ottnt{T_{{\mathrm{3}}}}  \rangle^{\bullet} ~   (  \langle  \ottnt{T_{{\mathrm{1}}}}  \mathord{ \overset{ \ottnt{c_{{\mathrm{1}}}} }{\Rightarrow} }  \ottnt{T_{{\mathrm{2}}}}  \rangle^{\bullet} ~  \ottnt{e}  )    \longrightarrow _{  \mathsf{E}  }  {} \\
           \langle  \ottnt{T_{{\mathrm{1}}}}  \mathord{ \overset{  \mathsf{join} ( \ottnt{c_{{\mathrm{1}}}} , \ottnt{c_{{\mathrm{2}}}} )  }{\Rightarrow} }  \ottnt{T_{{\mathrm{3}}}}  \rangle^{\bullet} ~  \ottnt{e} . \end{array} \]
        By Lemma~\ref{lem:eideticmergewf}, $ \mathord{  \vdash _{  \mathsf{E}  } }~  \mathsf{join} ( \ottnt{c_{{\mathrm{1}}}} , \ottnt{c_{{\mathrm{2}}}} )    \mathrel{\parallel}   \ottnt{T_{{\mathrm{1}}}}  \Rightarrow  \ottnt{T_{{\mathrm{3}}}} $. We already know that $ \emptyset   \vdash _{  \mathsf{E}  }  \ottnt{e_{{\mathrm{1}}}}  :  \ottnt{T_{{\mathrm{1}}}} $ (by
        assumption) and $ \mathord{  \vdash _{  \mathsf{E}  } }~ \ottnt{T_{{\mathrm{3}}}} $ (by inversion of the outer cast's
        typing derivation), so we can apply \T{Coerce} to type the
        resulting merged coercion.
      \item[(\E{CoerceRaise})] We have $ \mathord{  \vdash _{  \mathsf{E}  } }~ \ottnt{T_{{\mathrm{2}}}} $ by assumption, so we
        are done by \T{Blame}.
      \end{itemize}

    \item[(\T{Stack})] By cases on the step taken.
      \begin{itemize}
      \item[(\E{StackDone})] We know by assumption that $ \mathord{  \vdash _{  \mathsf{E}  } }~  \{ \mathit{x} \mathord{:} \ottnt{B} \mathrel{\mid} \ottnt{e} \}  $ and $ \ottnt{e}  [  \ottnt{k} / \mathit{x}  ]  \,  \longrightarrow ^{*}_{  \mathsf{E}  }  \,  \mathsf{true} $, so by \T{Const}.

      \item[(\E{StackPop})] We have $ \mathord{  \vdash _{  \mathsf{E}  } }~ \Gamma $ and $ \mathord{  \vdash _{  \mathsf{E}  } }~  \{ \mathit{x} \mathord{:} \ottnt{B} \mathrel{\mid} \ottnt{e_{{\mathrm{2}}}} \}  $ and $ \mathsf{ty} ( \ottnt{k} )   \ottsym{=}  \ottnt{B}$ by inversion. The
        quantification over $\ottnt{r}$ is also by inversion, of coercion
        well formedness.

        If $ \{ \mathit{x} \mathord{:} \ottnt{B} \mathrel{\mid} \ottnt{e'} \}  \, \supset \,  \{ \mathit{x} \mathord{:} \ottnt{B} \mathrel{\mid} \ottnt{e} \} $, then our new status is
        $ \mathord{\checkmark} $ and we enter a checking form---so the reduction
        $ \langle   \{ \mathit{x} \mathord{:} \ottnt{B} \mathrel{\mid} \ottnt{e'} \}  ,   \ottnt{e'}  [  \ottnt{k} / \mathit{x}  ]  ,  \ottnt{k}  \rangle^{ \ottnt{l} }  \,  \longrightarrow ^{*}_{  \mathsf{E}  }  \, \ottnt{e_{{\mathrm{2}}}}$ holds by reflexivity.

        If not, then our status is whatever it was before. If it was
        $ \mathord{\checkmark} $, then that is because we either (a) already knew that
        $ \ottnt{e_{{\mathrm{1}}}}  [  \ottnt{k} / \mathit{x}  ]  \,  \longrightarrow ^{*}_{  \mathsf{E}  }  \,  \mathsf{true} $ or because $ \langle   \{ \mathit{x} \mathord{:} \ottnt{B} \mathrel{\mid} \ottnt{e'} \}  ,   \ottnt{e'}  [  \ottnt{k} / \mathit{x}  ]  ,  \ottnt{k}  \rangle^{ \ottnt{l} }  \,  \longrightarrow ^{*}_{  \mathsf{E}  }  \, \ottnt{k}$ for some $ \{ \mathit{x} \mathord{:} \ottnt{B} \mathrel{\mid} \ottnt{e'} \}  \, \supset \,  \{ \mathit{x} \mathord{:} \ottnt{B} \mathrel{\mid} \ottnt{e_{{\mathrm{1}}}} \} $---which
        implies that $ \ottnt{e_{{\mathrm{1}}}}  [  \ottnt{k} / \mathit{x}  ]  \,  \longrightarrow ^{*}_{  \mathsf{E}  }  \,  \mathsf{true} $ by adequacy of
        $ \supset $\refasm{impldec}. So if $\ottnt{s}  \ottsym{=}   \mathord{\checkmark} $, our side
        condition is covered. If $\ottnt{s}  \ottsym{=}   \mathord{?} $, then we know that
        $ \{ \mathit{x} \mathord{:} \ottnt{B} \mathrel{\mid} \ottnt{e_{{\mathrm{1}}}} \} $ remains to be checked, and some $ \{ \mathit{x} \mathord{:} \ottnt{B} \mathrel{\mid} \ottnt{e'} \}  \, \supset \,  \{ \mathit{x} \mathord{:} \ottnt{B} \mathrel{\mid} \ottnt{e_{{\mathrm{1}}}} \} $ is in $\ottnt{r}$.
        
        We can type the active check using our assumptions, where
        $ \ottnt{e_{{\mathrm{2}}}}  [  \ottnt{k} / \mathit{x}  ]  \,  \longrightarrow ^{*}_{  \mathsf{E}  }  \,  \ottnt{e_{{\mathrm{2}}}}  [  \ottnt{k} / \mathit{x}  ] $ by reflexivity. By substitution
        (and \T{Const}, to find $ \emptyset   \vdash _{  \mathsf{E}  }  \ottnt{k}  :   \{ \mathit{x} \mathord{:} \ottnt{B} \mathrel{\mid}  \mathsf{true}  \}  $), we
        find $ \emptyset   \vdash _{  \mathsf{E}  }   \ottnt{e_{{\mathrm{2}}}}  [  \ottnt{k} / \mathit{x}  ]   :   \{ \mathit{x} \mathord{:}  \mathsf{Bool}  \mathrel{\mid}  \mathsf{true}  \}  $. We can now apply
        \T{Check}, and can then apply an outer \T{Stack}.
      \item[\E{StackInner}] By \T{Stack} and the IH. If $\ottnt{s}  \ottsym{=}   \mathord{\checkmark} $, we
        need to extend the evaluation derivation by one step.
      \item[\E{StackRaise}] We have $ \mathord{  \vdash _{  \mathsf{E}  } }~  \{ \mathit{x} \mathord{:} \ottnt{B} \mathrel{\mid} \ottnt{e} \}  $ and $ \mathord{  \vdash _{  \mathsf{E}  } }~ \Gamma $
        already, so by \T{Blame}.
      \end{itemize}
    \item[(\T{Blame})] Contradictory---$ \mathord{\Uparrow}  \ottnt{l} $ is a result and
      doesn't step.
    \item[(\T{Check})]  By cases on the step taken.
      \begin{itemize}
      \item[(\E{CheckOK})] Since $ \mathord{  \vdash _{  \mathsf{E}  } }~ \emptyset $ and $ \mathsf{ty} ( \ottnt{k} )   \ottsym{=}  \ottnt{B}$
        and $ \mathord{  \vdash _{  \mathsf{E}  } }~  \{ \mathit{x} \mathord{:} \ottnt{B} \mathrel{\mid} \ottnt{e_{{\mathrm{1}}}} \}  $ and $ \ottnt{e_{{\mathrm{1}}}}  [  \ottnt{k} / \mathit{x}  ]  \,  \longrightarrow ^{*}_{  \mathsf{E}  }  \,  \mathsf{true} $, we can
        apply \T{Const} to find $ \emptyset   \vdash _{  \mathsf{E}  }  \ottnt{k}  :   \{ \mathit{x} \mathord{:} \ottnt{B} \mathrel{\mid} \ottnt{e_{{\mathrm{1}}}} \}  $.
      \item[(\E{CheckFail})] Since $ \mathord{  \vdash _{  \mathsf{E}  } }~ \emptyset $ and $ \mathord{  \vdash _{  \mathsf{E}  } }~  \{ \mathit{x} \mathord{:} \ottnt{B} \mathrel{\mid} \ottnt{e_{{\mathrm{1}}}} \}  $, \T{Blame} shows $ \emptyset   \vdash _{  \mathsf{E}  }   \mathord{\Uparrow}  \ottnt{l}   :   \{ \mathit{x} \mathord{:} \ottnt{B} \mathrel{\mid} \ottnt{e_{{\mathrm{1}}}} \}  $.
      \item[(\E{CheckInner})] By \T{Check} and the IH.
      \item[(\E{CheckRaise})] As for \E{CheckFail}---the differing
        label doesn't matter.
      \end{itemize}
    \end{itemize}
    \fi}
  \end{proof}
\end{lemma}

\begin{lemma}[Source program typing for eidetic \lambdah]
  \label{lem:eideticsource}
  Source programs are well typed in $ \mathsf{C} $ iff they are well typed in
  $ \mathsf{E} $, i.e.:
  \begin{itemize}
  \item $ \Gamma   \vdash _{  \mathsf{C}  }  \ottnt{e}  :  \ottnt{T} $ as a source program iff $ \Gamma   \vdash _{  \mathsf{E}  }  \ottnt{e}  :  \ottnt{T} $ as a source program.
  \item $ \mathord{  \vdash _{  \mathsf{C}  } }~ \ottnt{T} $ as a source program iff $ \mathord{  \vdash _{  \mathsf{E}  } }~ \ottnt{T} $ as a source program.
  \item $ \mathord{  \vdash _{  \mathsf{C}  } }~ \Gamma $ as a source program iff $ \mathord{  \vdash _{  \mathsf{E}  } }~ \Gamma $ as a source program.
  \end{itemize}
  \begin{proof}
    By mutual induction on $\ottnt{e}$, $\ottnt{T}$, and $\Gamma$.
    {\iffull
    Since all of the rules are syntax directed, we use the rule names
    for cases (but prove both directions at once).
    \paragraph{Expressions $\ottnt{e}$}
    \begin{itemize}
    \item[(\T{Var})] By the IH on $\Gamma$ and \T{Var}.
    \item[(\T{Const})] By the IH on $\Gamma$ and $\T{Const}$, noting
      that $ \mathsf{true}  \,  \longrightarrow ^{*}_{ \ottnt{m} }  \,  \mathsf{true} $ in every mode $\ottnt{m}$.
    \item[(\T{Abs})] By the IH on $\ottnt{T_{{\mathrm{1}}}}$ and $\ottnt{e_{{\mathrm{12}}}}$ and \T{Abs}.
    \item[(\T{Op})] By the IHs on the arguments $\ottnt{e_{\ottmv{i}}}$ and \T{Op}.
    \item[(\T{App})] By the IHs on $\ottnt{e_{{\mathrm{1}}}}$ and $\ottnt{e_{{\mathrm{2}}}}$ and \T{App}.
    \item[(\T{Cast})] By the IHs on $\ottnt{T_{{\mathrm{1}}}}$ and $\ottnt{T_{{\mathrm{2}}}}$ and
      \T{Cast}, noting that similarity holds irrespective of modes and
      that the annotation is $ \bullet $.
    \item[(\T{Blame})] Contradictory---doesn't occur in source programs.
    \item[(\T{Check})] Contradictory---doesn't occur in source programs.
    \item[(\T{Stack})] Contradictory---doesn't occur in source programs.
    \end{itemize}
    
    \paragraph{Types $\ottnt{T}$}
    \begin{itemize}
    \item[(\WF{Base})] Immediately true---\WF{Base} is an axiom.
    \item[(\WF{Refine})] By the IH on $\ottnt{e}$ and \WF{Refine}.
    \item[(\WF{Fun})] By the IHs on $\ottnt{T_{{\mathrm{1}}}}$ and $\ottnt{T_{{\mathrm{2}}}}$ and \WF{Fun}.
    \end{itemize}

    \paragraph{Contexts $\Gamma$}
    \begin{itemize}
    \item[(\WF{Empty})] Immediately true---\WF{Empty} is an axiom.
    \item[(\WF{Extend})] By the IHs on $\Gamma$ and $\ottnt{T}$ and
      \WF{Extend}.
    \end{itemize}
    \fi}
  \end{proof}
\end{lemma}

\section{Proofs of space-efficiency soundness}
\label{app:sesoundness}

{\iffull

This appendix contains the proofs relating classic \lambdah to each
other mode: forgetful, heedful, and eidetic.

\subsection{Relating classic and forgetful manifest contracts}
\label{sec:forgetfullr}

If we evaluate a \lambdah term with the classic semantics and find a
value, then the forgetful semantics finds a similar
value---identical if they're constants. Since forgetful \lambdah drops
some casts, some terms reduce to blame in classic \lambdah while they
reduce to values in forgetful \lambdah.

The relationship between classic and forgetful \lambdah is
\textit{blame-inexact}, to borrow the terminology of
\citet{Greenberg12contracts}: we define an asymmetric logical relation
in Figure~\reflr, relating classic values to forgetful
values---and everything to classic blame.
The proof proceeds largely like that of \citet{Greenberg12contracts}:
we define a logical relation on terms and an inductive invariant
relation on types, prove that casts between related types are
logically related, and then show that well typed source programs are
logically related.

Before we explain the logical relation proof itself, there is one new
feature of the proof that merits discussion: we need to derive a
congruence principle for casts forgetful \lambdah.
When proving that casts between related types are related
(Lemma~\ref{lem:forgetfullrcast}), we want to be able to reason with
the logical relation---which involves reducing the cast's argument to
a value. But if $\ottnt{e} \,  \longrightarrow ^{*}_{  \mathsf{F}  }  \, \ottnt{e'}$ such that $ \mathsf{result} _{  \mathsf{F}  }~ \ottnt{e'} $, how to
$ \langle  \ottnt{T_{{\mathrm{1}}}}  \mathord{ \overset{\bullet}{\Rightarrow} }  \ottnt{T_{{\mathrm{2}}}}  \rangle^{ \ottnt{l} } ~  \ottnt{e} $ and $ \langle  \ottnt{T_{{\mathrm{1}}}}  \mathord{ \overset{\bullet}{\Rightarrow} }  \ottnt{T_{{\mathrm{2}}}}  \rangle^{ \ottnt{l} } ~  \ottnt{e'} $ relate?
If $\ottnt{e'}  \ottsym{=}   \mathord{\Uparrow}  \ottnt{l'} $ is blame, then it may be that $ \langle  \ottnt{T_{{\mathrm{1}}}}  \mathord{ \overset{\bullet}{\Rightarrow} }  \ottnt{T_{{\mathrm{2}}}}  \rangle^{ \ottnt{l} } ~  \ottnt{e} $
reduces to a value while $ \langle  \ottnt{T_{{\mathrm{1}}}}  \mathord{ \overset{\bullet}{\Rightarrow} }  \ottnt{T_{{\mathrm{2}}}}  \rangle^{ \ottnt{l} } ~   \mathord{\Uparrow}  \ottnt{l'}  $ propagates the
blame. But if $\ottnt{e'}$ is a value, then both casts reduce to the same
value.
We show this property first for a single step $\ottnt{e} \,  \longrightarrow _{  \mathsf{F}  }  \, \ottnt{e'}$, and
then lift it to many steps.

\begin{lemma}[Cast congruence (single step)]
  \label{lem:forgetfulcastcongruencesinglestep}
  If
  \begin{itemize}
  \item $ \emptyset   \vdash _{  \mathsf{F}  }  \ottnt{e}  :  \ottnt{T_{{\mathrm{1}}}} $ and and $ \mathord{  \vdash _{  \mathsf{F}  } }~ \emptyset   \mathrel{\parallel}   \ottnt{T_{{\mathrm{1}}}}  \Rightarrow  \ottnt{T_{{\mathrm{2}}}} $
    (and so $ \emptyset   \vdash _{  \mathsf{F}  }   \langle  \ottnt{T_{{\mathrm{1}}}}  \mathord{ \overset{\bullet}{\Rightarrow} }  \ottnt{T_{{\mathrm{2}}}}  \rangle^{ \ottnt{l} } ~  \ottnt{e}   :  \ottnt{T_{{\mathrm{2}}}} $),
  \item $\ottnt{e} \,  \longrightarrow _{  \mathsf{F}  }  \, \ottnt{e_{{\mathrm{1}}}}$ (and so $ \emptyset   \vdash _{  \mathsf{F}  }  \ottnt{e_{{\mathrm{1}}}}  :  \ottnt{T_{{\mathrm{1}}}} $),
  \item $ \langle  \ottnt{T_{{\mathrm{1}}}}  \mathord{ \overset{\bullet}{\Rightarrow} }  \ottnt{T_{{\mathrm{2}}}}  \rangle^{ \ottnt{l} } ~  \ottnt{e_{{\mathrm{1}}}}  \,  \longrightarrow ^{*}_{  \mathsf{F}  }  \, \ottnt{e_{{\mathrm{2}}}}$, and
  \item $ \mathsf{val} _{  \mathsf{F}  }~ \ottnt{e_{{\mathrm{2}}}} $
  \end{itemize}
  then $ \langle  \ottnt{T_{{\mathrm{1}}}}  \mathord{ \overset{\bullet}{\Rightarrow} }  \ottnt{T_{{\mathrm{2}}}}  \rangle^{ \ottnt{l} } ~  \ottnt{e}  \,  \longrightarrow ^{*}_{  \mathsf{F}  }  \, \ottnt{e_{{\mathrm{2}}}}$.
  \begin{proof}
    By cases on the step $\ottnt{e} \,  \longrightarrow _{  \mathsf{F}  }  \, \ottnt{e_{{\mathrm{1}}}}$. There are three groups of
    reductions: straightforward merge-free reductions, merging
    reductions (the interesting cases, where a reduction step taken in
    $\ottnt{e}$ has an exposed cast), and (contradictory) reductions where
    blame is raised.
    {\iffull
    \paragraph{Merge-free reductions}
    \begin{itemize}
    \item[(\E{Beta})] By \E{CastInner} and \E{Beta}.
    \item[(\E{Op})] By \E{CastInner} and \E{Op}.
    \item[(\E{Unwrap})] By \E{CastInner} and \E{Unwrap}.
    \item[(\E{AppL})] By \E{CastInner} with \E{AppL}.
    \item[(\E{AppR})] By \E{CastInner} with \E{AppR}.
    \item[(\E{CheckOK})] By \E{CastInner} and \E{CheckOK}.
    \item[(\E{OpInner})] By \E{CastInner} and \E{OpInner}.
    \end{itemize}

    \paragraph{Merging reductions}
    The interesting case of the proof occurs when the reduction step
    taken in $\ottnt{e}$ has an exposed cast: \E{CheckNone} or a
    congruence/merge rule (\E{CastInner} or \E{CastMerge}).
    Applying a cast to $\ottnt{e}$ and $\ottnt{e_{{\mathrm{1}}}}$ leads to slightly
    different reductions, because we merge the cast in $\ottnt{e}$
    and not in $\ottnt{e_{{\mathrm{1}}}}$. \textit{If no blame is raised}, then the
    reductions join back up.
    \begin{itemize}
    \item[(\E{CheckNone})] We have $\ottnt{e}  \ottsym{=}   (  \langle  \ottnt{T_{{\mathrm{3}}}}  \mathord{ \overset{\bullet}{\Rightarrow} }   \{ \mathit{x} \mathord{:} \ottnt{B} \mathrel{\mid} \ottnt{e_{{\mathrm{11}}}} \}   \rangle^{ \ottnt{l'} } ~  \ottnt{k}  ) $ where
      $\ottnt{T_{{\mathrm{1}}}}  \ottsym{=}   \{ \mathit{x} \mathord{:} \ottnt{B} \mathrel{\mid} \ottnt{e_{{\mathrm{11}}}} \} $ and $\ottnt{e_{{\mathrm{1}}}}  \ottsym{=}   \langle   \{ \mathit{x} \mathord{:} \ottnt{B} \mathrel{\mid} \ottnt{e_{{\mathrm{11}}}} \}  ,   \ottnt{e_{{\mathrm{11}}}}  [  \ottnt{k} / \mathit{x}  ]  ,  \ottnt{k}  \rangle^{ \ottnt{l'} } $ and
      $ \langle   \{ \mathit{x} \mathord{:} \ottnt{B} \mathrel{\mid} \ottnt{e_{{\mathrm{11}}}} \}   \mathord{ \overset{\bullet}{\Rightarrow} }  \ottnt{T_{{\mathrm{2}}}}  \rangle^{ \ottnt{l} } ~  \ottnt{e_{{\mathrm{1}}}}  \,  \longrightarrow ^{*}_{  \mathsf{F}  }  \, \ottnt{e_{{\mathrm{2}}}}$ such that $ \mathsf{val} _{  \mathsf{F}  }~ \ottnt{e_{{\mathrm{2}}}} $. We
      must show that $ \langle   \{ \mathit{x} \mathord{:} \ottnt{B} \mathrel{\mid} \ottnt{e_{{\mathrm{11}}}} \}   \mathord{ \overset{\bullet}{\Rightarrow} }  \ottnt{T_{{\mathrm{2}}}}  \rangle^{ \ottnt{l} } ~  \ottnt{e}  \,  \longrightarrow ^{*}_{  \mathsf{F}  }  \, \ottnt{e_{{\mathrm{2}}}}$.

      By inversion of the similarity relation $\vdash   \{ \mathit{x} \mathord{:} \ottnt{B} \mathrel{\mid} \ottnt{e_{{\mathrm{11}}}} \}   \mathrel{\parallel}  \ottnt{T_{{\mathrm{2}}}}$, we know that $\ottnt{T_{{\mathrm{2}}}}  \ottsym{=}   \{ \mathit{x} \mathord{:} \ottnt{B} \mathrel{\mid} \ottnt{e_{{\mathrm{12}}}} \} $.
      If $ \langle   \{ \mathit{x} \mathord{:} \ottnt{B} \mathrel{\mid} \ottnt{e_{{\mathrm{11}}}} \}   \mathord{ \overset{\bullet}{\Rightarrow} }   \{ \mathit{x} \mathord{:} \ottnt{B} \mathrel{\mid} \ottnt{e_{{\mathrm{12}}}} \}   \rangle^{ \ottnt{l} } ~  \ottnt{e_{{\mathrm{1}}}} $ reduces to a value, then it
      must be the case that $ \ottnt{e_{{\mathrm{1}}}}  \mathrel{=}   \langle   \{ \mathit{x} \mathord{:} \ottnt{B} \mathrel{\mid} \ottnt{e_{{\mathrm{11}}}} \}  ,   \ottnt{e_{{\mathrm{11}}}}  [  \ottnt{k} / \mathit{x}  ]  ,  \ottnt{k}  \rangle^{ \ottnt{l'} }   \,  \longrightarrow ^{*}_{  \mathsf{F}  }  \, \ottnt{k}$ and that $ \ottnt{e_{{\mathrm{12}}}}  [  \ottnt{k} / \mathit{x}  ]  \,  \longrightarrow ^{*}_{  \mathsf{F}  }  \,  \mathsf{true} $ (and so the entire term
      reduces $ \langle   \{ \mathit{x} \mathord{:} \ottnt{B} \mathrel{\mid} \ottnt{e_{{\mathrm{11}}}} \}   \mathord{ \overset{\bullet}{\Rightarrow} }   \{ \mathit{x} \mathord{:} \ottnt{B} \mathrel{\mid} \ottnt{e_{{\mathrm{12}}}} \}   \rangle^{ \ottnt{l} } ~  \ottnt{e_{{\mathrm{1}}}}  \,  \longrightarrow ^{*}_{  \mathsf{F}  }  \, \ottnt{k} = \ottnt{e_{{\mathrm{2}}}}$). If not, we
      would have gotten $ \mathord{\Uparrow}  \ottnt{l'} $ or $ \mathord{\Uparrow}  \ottnt{l} $.
      
      Instead, we find that: 
      \[ \begin{array}{r@{~}l}
        &  \langle   \{ \mathit{x} \mathord{:} \ottnt{B} \mathrel{\mid} \ottnt{e_{{\mathrm{11}}}} \}   \mathord{ \overset{\bullet}{\Rightarrow} }   \{ \mathit{x} \mathord{:} \ottnt{B} \mathrel{\mid} \ottnt{e_{{\mathrm{12}}}} \}   \rangle^{ \ottnt{l} } ~   (  \langle  \ottnt{T_{{\mathrm{3}}}}  \mathord{ \overset{\bullet}{\Rightarrow} }   \{ \mathit{x} \mathord{:} \ottnt{B} \mathrel{\mid} \ottnt{e_{{\mathrm{11}}}} \}   \rangle^{ \ottnt{l'} } ~  \ottnt{k}  )    \\
         \longrightarrow _{  \mathsf{F}  } &  \langle  \ottnt{T_{{\mathrm{3}}}}  \mathord{ \overset{\bullet}{\Rightarrow} }   \{ \mathit{x} \mathord{:} \ottnt{B} \mathrel{\mid} \ottnt{e_{{\mathrm{12}}}} \}   \rangle^{ \ottnt{l} } ~  \ottnt{k}  \\
         \longrightarrow _{  \mathsf{F}  } &  \langle   \{ \mathit{x} \mathord{:} \ottnt{B} \mathrel{\mid} \ottnt{e_{{\mathrm{12}}}} \}  ,   \ottnt{e_{{\mathrm{12}}}}  [  \ottnt{k} / \mathit{x}  ]  ,  \ottnt{k}  \rangle^{ \ottnt{l} }  \\
         \longrightarrow ^{*}_{  \mathsf{F}  } &  \langle   \{ \mathit{x} \mathord{:} \ottnt{B} \mathrel{\mid} \ottnt{e_{{\mathrm{12}}}} \}  ,   \mathsf{true}  ,  \ottnt{k}  \rangle^{ \ottnt{l} }  \\
         \longrightarrow _{  \mathsf{F}  } & \ottnt{k}  \ottsym{=}  \ottnt{e_{{\mathrm{2}}}}
      \end{array} \]
      
    \item[(\E{CastInner})] We have:
      \[ \ottnt{e} =  \langle  \ottnt{T_{{\mathrm{3}}}}  \mathord{ \overset{\bullet}{\Rightarrow} }  \ottnt{T_{{\mathrm{1}}}}  \rangle^{ \ottnt{l'} } ~  \ottnt{e_{{\mathrm{11}}}}   \longrightarrow _{  \mathsf{F}  }   \langle  \ottnt{T_{{\mathrm{3}}}}  \mathord{ \overset{\bullet}{\Rightarrow} }  \ottnt{T_{{\mathrm{1}}}}  \rangle^{ \ottnt{l'} } ~  \ottnt{e_{{\mathrm{12}}}}  =
      \ottnt{e_{{\mathrm{1}}}} \] with $\ottnt{e_{{\mathrm{11}}}} \,  \longrightarrow _{  \mathsf{F}  }  \, \ottnt{e_{{\mathrm{12}}}}$ and $\ottnt{e_{{\mathrm{11}}}}  \neq   \langle  \ottnt{T_{{\mathrm{4}}}}  \mathord{ \overset{\bullet}{\Rightarrow} }  \ottnt{T_{{\mathrm{3}}}}  \rangle^{ \ottnt{l''} } ~  \ottnt{e''_{{\mathrm{2}}}} $.

        In the original derivation with $\ottnt{e_{{\mathrm{1}}}}$, we have \[ \begin{array}{r@{~}c@{~}l} 
           \langle  \ottnt{T_{{\mathrm{1}}}}  \mathord{ \overset{\bullet}{\Rightarrow} }  \ottnt{T_{{\mathrm{2}}}}  \rangle^{ \ottnt{l} } ~   (  \langle  \ottnt{T_{{\mathrm{3}}}}  \mathord{ \overset{\bullet}{\Rightarrow} }  \ottnt{T_{{\mathrm{1}}}}  \rangle^{ \ottnt{l'} } ~  \ottnt{e_{{\mathrm{12}}}}  )   & \longrightarrow _{  \mathsf{F}  } &  \langle  \ottnt{T_{{\mathrm{3}}}}  \mathord{ \overset{\bullet}{\Rightarrow} }  \ottnt{T_{{\mathrm{2}}}}  \rangle^{ \ottnt{l} } ~  \ottnt{e_{{\mathrm{12}}}}  \\
          & \longrightarrow ^{*}_{  \mathsf{F}  } & \ottnt{e_{{\mathrm{2}}}} \end{array} \] by
        \E{CastMerge} and then assumption. We find a new derivation with
        $\ottnt{e}$ as follows:
      \[ \begin{array}{rl}
        &  \langle  \ottnt{T_{{\mathrm{1}}}}  \mathord{ \overset{\bullet}{\Rightarrow} }  \ottnt{T_{{\mathrm{2}}}}  \rangle^{ \ottnt{l} } ~   (  \langle  \ottnt{T_{{\mathrm{3}}}}  \mathord{ \overset{\bullet}{\Rightarrow} }  \ottnt{T_{{\mathrm{1}}}}  \rangle^{ \ottnt{l'} } ~  \ottnt{e_{{\mathrm{11}}}}  )   \qquad (\E{CastMerge}) \\
         \longrightarrow _{  \mathsf{F}  } &  \langle  \ottnt{T_{{\mathrm{3}}}}  \mathord{ \overset{\bullet}{\Rightarrow} }  \ottnt{T_{{\mathrm{2}}}}  \rangle^{ \ottnt{l} } ~  \ottnt{e_{{\mathrm{11}}}}  \qquad\qquad\qquad\qquad (\E{CastInner}) \\
        \multicolumn{2}{r}{\text{~since $\ottnt{e_{{\mathrm{11}}}}  \neq   \langle  \ottnt{T_{{\mathrm{4}}}}  \mathord{ \overset{\bullet}{\Rightarrow} }  \ottnt{T_{{\mathrm{3}}}}  \rangle^{ \ottnt{l''} } ~  \ottnt{e''_{{\mathrm{2}}}} $}} \\
         \longrightarrow _{  \mathsf{F}  } &  \langle  \ottnt{T_{{\mathrm{3}}}}  \mathord{ \overset{\bullet}{\Rightarrow} }  \ottnt{T_{{\mathrm{2}}}}  \rangle^{ \ottnt{l} } ~  \ottnt{e_{{\mathrm{12}}}}  \qquad\qquad\qquad\qquad \text{(assumption)} \\
         \longrightarrow ^{*}_{  \mathsf{F}  } & \ottnt{e_{{\mathrm{2}}}} 
      \end{array} \]

    \item[(\E{CastMerge})] We have: \[ \ottnt{e} =  \langle  \ottnt{T_{{\mathrm{3}}}}  \mathord{ \overset{\bullet}{\Rightarrow} }  \ottnt{T_{{\mathrm{1}}}}  \rangle^{ \ottnt{l'} } ~   (  \langle  \ottnt{T_{{\mathrm{4}}}}  \mathord{ \overset{\bullet}{\Rightarrow} }  \ottnt{T_{{\mathrm{3}}}}  \rangle^{ \ottnt{l''} } ~  \ottnt{e_{{\mathrm{11}}}}  )   \,  \longrightarrow _{  \mathsf{F}  }  \,  \langle  \ottnt{T_{{\mathrm{4}}}}  \mathord{ \overset{\bullet}{\Rightarrow} }  \ottnt{T_{{\mathrm{1}}}}  \rangle^{ \ottnt{l'} } ~  \ottnt{e_{{\mathrm{11}}}}  = \ottnt{e_{{\mathrm{1}}}} \]
      In the original derivation with $\ottnt{e_{{\mathrm{1}}}}$, we have \[  \langle  \ottnt{T_{{\mathrm{1}}}}  \mathord{ \overset{\bullet}{\Rightarrow} }  \ottnt{T_{{\mathrm{2}}}}  \rangle^{ \ottnt{l} } ~   (  \langle  \ottnt{T_{{\mathrm{4}}}}  \mathord{ \overset{\bullet}{\Rightarrow} }  \ottnt{T_{{\mathrm{1}}}}  \rangle^{ \ottnt{l'} } ~  \ottnt{e_{{\mathrm{11}}}}  )   \,  \longrightarrow _{  \mathsf{F}  }  \,  \langle  \ottnt{T_{{\mathrm{4}}}}  \mathord{ \overset{\bullet}{\Rightarrow} }  \ottnt{T_{{\mathrm{2}}}}  \rangle^{ \ottnt{l} } ~  \ottnt{e_{{\mathrm{11}}}}   \longrightarrow ^{*}_{  \mathsf{F}  }  \ottnt{e_{{\mathrm{2}}}} \] We can
      build a new derivation with $\ottnt{e}$ as follows, stepping twice by \E{CastMerge}:
      \[ \begin{array}{rl}
        &  \langle  \ottnt{T_{{\mathrm{1}}}}  \mathord{ \overset{\bullet}{\Rightarrow} }  \ottnt{T_{{\mathrm{2}}}}  \rangle^{ \ottnt{l} } ~   (  \langle  \ottnt{T_{{\mathrm{3}}}}  \mathord{ \overset{\bullet}{\Rightarrow} }  \ottnt{T_{{\mathrm{1}}}}  \rangle^{ \ottnt{l'} } ~   (  \langle  \ottnt{T_{{\mathrm{4}}}}  \mathord{ \overset{\bullet}{\Rightarrow} }  \ottnt{T_{{\mathrm{3}}}}  \rangle^{ \ottnt{l''} } ~  \ottnt{e_{{\mathrm{11}}}}  )   )    \\
         \longrightarrow _{  \mathsf{F}  } &  \langle  \ottnt{T_{{\mathrm{3}}}}  \mathord{ \overset{\bullet}{\Rightarrow} }  \ottnt{T_{{\mathrm{2}}}}  \rangle^{ \ottnt{l} } ~   (  \langle  \ottnt{T_{{\mathrm{4}}}}  \mathord{ \overset{\bullet}{\Rightarrow} }  \ottnt{T_{{\mathrm{3}}}}  \rangle^{ \ottnt{l''} } ~  \ottnt{e_{{\mathrm{11}}}}  )    \\
         \longrightarrow _{  \mathsf{F}  } &  \langle  \ottnt{T_{{\mathrm{4}}}}  \mathord{ \overset{\bullet}{\Rightarrow} }  \ottnt{T_{{\mathrm{2}}}}  \rangle^{ \ottnt{l} } ~  \ottnt{e_{{\mathrm{11}}}}  \qquad\qquad\qquad\qquad \text{(assumption)} \\
         \longrightarrow ^{*}_{  \mathsf{F}  } & \ottnt{e_{{\mathrm{2}}}} 
      \end{array} \]
    \end{itemize}

    \paragraph{Contradictory blame-raising reductions}
    \begin{itemize}
    \item[(\E{AppRaiseL})] Contradiction---in this case, $\ottnt{e_{{\mathrm{1}}}}  \ottsym{=}   \mathord{\Uparrow}  \ottnt{l'} $, and $ \langle  \ottnt{T_{{\mathrm{1}}}}  \mathord{ \overset{\bullet}{\Rightarrow} }  \ottnt{T_{{\mathrm{2}}}}  \rangle^{ \ottnt{l} } ~  \ottnt{e_{{\mathrm{1}}}}  \,  \longrightarrow ^{*}_{  \mathsf{F}  }  \,  \mathord{\Uparrow}  \ottnt{l'} $, which isn't
      a value.
    \item[(\E{AppRaiseR})] Contradiction---in this case, $\ottnt{e_{{\mathrm{1}}}}  \ottsym{=}   \mathord{\Uparrow}  \ottnt{l'} $, and $ \langle  \ottnt{T_{{\mathrm{1}}}}  \mathord{ \overset{\bullet}{\Rightarrow} }  \ottnt{T_{{\mathrm{2}}}}  \rangle^{ \ottnt{l} } ~  \ottnt{e_{{\mathrm{1}}}}  \,  \longrightarrow ^{*}_{  \mathsf{F}  }  \,  \mathord{\Uparrow}  \ottnt{l'} $, which isn't
      a value.
    \item[(\E{CastRaise})]  Contradiction---in this case, $\ottnt{e_{{\mathrm{1}}}}  \ottsym{=}   \mathord{\Uparrow}  \ottnt{l'} $, and $ \langle  \ottnt{T_{{\mathrm{1}}}}  \mathord{ \overset{\bullet}{\Rightarrow} }  \ottnt{T_{{\mathrm{2}}}}  \rangle^{ \ottnt{l} } ~  \ottnt{e_{{\mathrm{1}}}}  \,  \longrightarrow ^{*}_{  \mathsf{F}  }  \,  \mathord{\Uparrow}  \ottnt{l'} $, which isn't
      a value.
    \item[(\E{CheckFail})] Contradiction---in this case, $\ottnt{e_{{\mathrm{1}}}}  \ottsym{=}   \mathord{\Uparrow}  \ottnt{l'} $, and $ \langle  \ottnt{T_{{\mathrm{1}}}}  \mathord{ \overset{\bullet}{\Rightarrow} }  \ottnt{T_{{\mathrm{2}}}}  \rangle^{ \ottnt{l} } ~  \ottnt{e_{{\mathrm{1}}}}  \,  \longrightarrow ^{*}_{  \mathsf{F}  }  \,  \mathord{\Uparrow}  \ottnt{l'} $, which isn't
      a value.
    \item[(\E{OpRaise})] Contradiction---in this case, $\ottnt{e_{{\mathrm{1}}}}  \ottsym{=}   \mathord{\Uparrow}  \ottnt{l'} $, and $ \langle  \ottnt{T_{{\mathrm{1}}}}  \mathord{ \overset{\bullet}{\Rightarrow} }  \ottnt{T_{{\mathrm{2}}}}  \rangle^{ \ottnt{l} } ~  \ottnt{e_{{\mathrm{1}}}}  \,  \longrightarrow ^{*}_{  \mathsf{F}  }  \,  \mathord{\Uparrow}  \ottnt{l'} $, which isn't
      a value.
    \item[(\E{CheckRaise})] Contradiction---in this case, $\ottnt{e_{{\mathrm{1}}}}  \ottsym{=}   \mathord{\Uparrow}  \ottnt{l'} $, and $ \langle  \ottnt{T_{{\mathrm{1}}}}  \mathord{ \overset{\bullet}{\Rightarrow} }  \ottnt{T_{{\mathrm{2}}}}  \rangle^{ \ottnt{l} } ~  \ottnt{e_{{\mathrm{1}}}}  \,  \longrightarrow ^{*}_{  \mathsf{F}  }  \,  \mathord{\Uparrow}  \ottnt{l'} $, which isn't
      a value.
    \end{itemize}
    \else

    None of the blame propagation rules (i.e., \E{*Raise*}) can occur:
    we would have $\ottnt{e_{{\mathrm{1}}}}  \ottsym{=}   \mathord{\Uparrow}  \ottnt{l'} $, and then $ \langle  \ottnt{T_{{\mathrm{1}}}}  \mathord{ \overset{\bullet}{\Rightarrow} }  \ottnt{T_{{\mathrm{2}}}}  \rangle^{ \ottnt{l} } ~   \mathord{\Uparrow}  \ottnt{l'}  $ doesn't reduce to a value, contradicting our
    assumption.

    \fi}
  \end{proof}
\end{lemma}

Once we have cast congruence for a single step, a straightforward
induction gives us reasoning principle applicable to many steps.

{\iffull
\begin{lemma}[Cast congruence]
  \label{lem:forgetfulcastcongruence}
  If
  \begin{itemize}
  \item $ \emptyset   \vdash _{  \mathsf{F}  }  \ottnt{e}  :  \ottnt{T_{{\mathrm{1}}}} $ and $ \mathord{  \vdash _{  \mathsf{F}  } }~ \emptyset   \mathrel{\parallel}   \ottnt{T_{{\mathrm{1}}}}  \Rightarrow  \ottnt{T_{{\mathrm{2}}}} $ (and so
    $ \emptyset   \vdash _{  \mathsf{F}  }   \langle  \ottnt{T_{{\mathrm{1}}}}  \mathord{ \overset{\bullet}{\Rightarrow} }  \ottnt{T_{{\mathrm{2}}}}  \rangle^{ \ottnt{l} } ~  \ottnt{e}   :  \ottnt{T_{{\mathrm{2}}}} $),
  \item $\ottnt{e} \,  \longrightarrow ^{*}_{  \mathsf{F}  }  \, \ottnt{e_{{\mathrm{1}}}}$ (and so $ \emptyset   \vdash _{  \mathsf{F}  }  \ottnt{e_{{\mathrm{1}}}}  :  \ottnt{T_{{\mathrm{1}}}} $),
  \item $ \langle  \ottnt{T_{{\mathrm{1}}}}  \mathord{ \overset{\bullet}{\Rightarrow} }  \ottnt{T_{{\mathrm{2}}}}  \rangle^{ \ottnt{l} } ~  \ottnt{e_{{\mathrm{1}}}}  \,  \longrightarrow ^{*}_{  \mathsf{F}  }  \, \ottnt{e_{{\mathrm{2}}}}$, and
  \item $ \mathsf{val} _{  \mathsf{F}  }~ \ottnt{e_{{\mathrm{2}}}} $
  \end{itemize}
  then $ \langle  \ottnt{T_{{\mathrm{1}}}}  \mathord{ \overset{\bullet}{\Rightarrow} }  \ottnt{T_{{\mathrm{2}}}}  \rangle^{ \ottnt{l} } ~  \ottnt{e}  \,  \longrightarrow ^{*}_{  \mathsf{F}  }  \, \ottnt{e_{{\mathrm{2}}}}$. Diagrammatically:
  %
    % \forgetfulcongruence
    
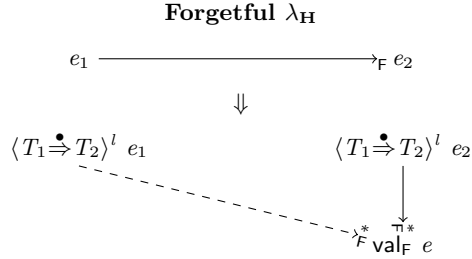
\begin{center}
    \begin{tikzpicture}[description/.style={fill=white,inner sep=2pt},align at top]
    \matrix (m) [matrix of math nodes, row sep=4pt, nodes in empty cells,
                 text height=1.5ex, text depth=0.25ex]
    {    
                       & \textbf{Forgetful \lambdah} & \\
      \ottnt{e_{{\mathrm{1}}}}           & & \ottnt{e_{{\mathrm{2}}}} \\
                       & \Downarrow & \\
       \langle  \ottnt{T_{{\mathrm{1}}}}  \mathord{ \overset{\bullet}{\Rightarrow} }  \ottnt{T_{{\mathrm{2}}}}  \rangle^{ \ottnt{l} } ~  \ottnt{e_{{\mathrm{1}}}}  & &  \langle  \ottnt{T_{{\mathrm{1}}}}  \mathord{ \overset{\bullet}{\Rightarrow} }  \ottnt{T_{{\mathrm{2}}}}  \rangle^{ \ottnt{l} } ~  \ottnt{e_{{\mathrm{2}}}}  \\[20pt]
                       & &  \mathsf{val} _{  \mathsf{F}  }~ \ottnt{e}  \\
    };

    \path[->] (m-2-1) edge[F] (m-2-3);
    \path[dashed,->] (m-4-1.south) edge[F*] (m-5-3);
    \path[->] (m-4-3) edge[F*] (m-5-3);
  \end{tikzpicture}
  \end{center}
  \begin{proof}
    By induction on the derivation $\ottnt{e} \,  \longrightarrow ^{*}_{  \mathsf{F}  }  \, \ottnt{e_{{\mathrm{1}}}}$, using the
    single-step cast congruence
    (Lemma~\ref{lem:forgetfulcastcongruencesinglestep}).
  \end{proof}
\end{lemma}
\fi}

We define the logical relation in Figure~\reflr. It is
defined in a split style, with separate definitions for values and
terms. Note that terms that classically reduce to blame are related to
all forgetful terms, but terms that classically reduce to values
reduce forgetfully to similar values. We lift these closed relations
on values and terms to open terms by means of dual closing
substitutions.
As in \citet{Greenberg12contracts}, we define an inductive invariant
to relate types, using it to show that casts between related types on
related values yield related values, i.e., casts are applicative
(Lemma~\ref{lem:forgetfullrcast}).
One important subtle technicality is that the type indices of this
logical relation are forgetful types---in the constant case of the
value relation, we evaluate the predicate in the forgetful
semantics. We believe the choice is arbitrary, but have not tried the
proof using classic type indices.

{\iffull
\begin{lemma}[Value relation relates only values]
  \label{lem:forgetfullrvalue}
  If $ \ottnt{e_{{\mathrm{1}}}}   \sim _{  \mathsf{F}  }  \ottnt{e_{{\mathrm{2}}}}  :  \ottnt{T} $ then $ \mathsf{val} _{  \mathsf{C}  }~ \ottnt{e_{{\mathrm{1}}}} $ and $ \mathsf{val} _{  \mathsf{F}  }~ \ottnt{e_{{\mathrm{2}}}} $.
  \begin{proof}
    By induction on $\ottnt{T}$. We have $\ottnt{e_{{\mathrm{1}}}}  \ottsym{=}  \ottnt{e_{{\mathrm{2}}}} = \ottnt{k}$ when $\ottnt{T}  \ottsym{=}   \{ \mathit{x} \mathord{:} \ottnt{B} \mathrel{\mid} \ottnt{e} \} $ (and so we are done by \V{Const}). When $\ottnt{T}  \ottsym{=}   \ottnt{T_{{\mathrm{1}}}} \mathord{ \rightarrow } \ottnt{T_{{\mathrm{2}}}} $, we have the value derivations as assumptions.
  \end{proof}  
\end{lemma}

\begin{lemma}[Relation implies similarity]
  \label{lem:forgetfullrsimilar}
  If $ \ottnt{T_{{\mathrm{1}}}}   \sim _{  \mathsf{F}  }  \ottnt{T_{{\mathrm{2}}}} $ then $\vdash  \ottnt{T_{{\mathrm{1}}}}  \mathrel{\parallel}  \ottnt{T_{{\mathrm{2}}}}$.
  \begin{proof}
    By induction on $\ottnt{T_{{\mathrm{1}}}}$, using \S{Refine} and \S{Fun}.
  \end{proof}
\end{lemma}
\fi}

\begin{lemma}[Relating classic and forgetful casts]
  \label{lem:forgetfullrcast}
  If $ \ottnt{T_{{\mathrm{11}}}}   \sim _{  \mathsf{F}  }  \ottnt{T_{{\mathrm{21}}}} $ and $ \ottnt{T_{{\mathrm{12}}}}   \sim _{  \mathsf{F}  }  \ottnt{T_{{\mathrm{22}}}} $ and $\vdash  \ottnt{T_{{\mathrm{11}}}}  \mathrel{\parallel}  \ottnt{T_{{\mathrm{12}}}}$,
  then forall $ \ottnt{e_{{\mathrm{1}}}}   \sim _{  \mathsf{F}  }  \ottnt{e_{{\mathrm{2}}}}  :  \ottnt{T_{{\mathrm{21}}}} $, we have $  \langle  \ottnt{T_{{\mathrm{11}}}}  \mathord{ \overset{\bullet}{\Rightarrow} }  \ottnt{T_{{\mathrm{12}}}}  \rangle^{ \ottnt{l} } ~  \ottnt{e_{{\mathrm{1}}}}    \simeq _{  \mathsf{F}  }   \langle  \ottnt{T_{{\mathrm{21}}}}  \mathord{ \overset{\bullet}{\Rightarrow} }  \ottnt{T_{{\mathrm{22}}}}  \rangle^{ \ottnt{l'} } ~  \ottnt{e_{{\mathrm{2}}}}   :  \ottnt{T_{{\mathrm{22}}}} $.
  \begin{proof}
    By induction on the sum of the heights of $\ottnt{T_{{\mathrm{21}}}}$ and
    $\ottnt{T_{{\mathrm{22}}}}$.
    {\iffull
    By Lemma~\ref{lem:forgetfullrsimilar}, we know that
    $\vdash  \ottnt{T_{{\mathrm{11}}}}  \mathrel{\parallel}  \ottnt{T_{{\mathrm{21}}}}$ and $\vdash  \ottnt{T_{{\mathrm{12}}}}  \mathrel{\parallel}  \ottnt{T_{{\mathrm{22}}}}$; by
    Lemma~\ref{lem:similaritytransitive}, we know that $\vdash  \ottnt{T_{{\mathrm{21}}}}  \mathrel{\parallel}  \ottnt{T_{{\mathrm{22}}}}$. We go by cases on $\ottnt{T_{{\mathrm{22}}}}$.
    \begin{itemize}
    \item[($\ottnt{T_{{\mathrm{22}}}}  \ottsym{=}   \{ \mathit{x} \mathord{:} \ottnt{B} \mathrel{\mid} \ottnt{e_{{\mathrm{22}}}} \} $)] It must be the case (by similarity)
      that all of the other types are also refinements. Moreover, it
      must be that case that $\ottnt{e_{{\mathrm{1}}}}  \ottsym{=}  \ottnt{e_{{\mathrm{2}}}} = \ottnt{k}$.

      Both sides step by \E{CheckNone}. Since $ \ottnt{e_{{\mathrm{1}}}}   \sim _{  \mathsf{F}  }  \ottnt{e_{{\mathrm{2}}}}  :  \ottnt{T_{{\mathrm{21}}}}  =
       \{ \mathit{x} \mathord{:} \ottnt{B} \mathrel{\mid} \ottnt{e_{{\mathrm{21}}}} \} $, we can find that $ \ottnt{e_{{\mathrm{1}}}}   \sim _{  \mathsf{F}  }  \ottnt{e_{{\mathrm{2}}}}  :   \{ \mathit{x} \mathord{:} \ottnt{B} \mathrel{\mid}  \mathsf{true}  \}  $
      trivially. Then, since $  \{ \mathit{x} \mathord{:} \ottnt{B} \mathrel{\mid} \ottnt{e_{{\mathrm{12}}}} \}    \sim _{  \mathsf{F}  }   \{ \mathit{x} \mathord{:} \ottnt{B} \mathrel{\mid} \ottnt{e_{{\mathrm{22}}}} \}  $, we know
      that $  \ottnt{e_{{\mathrm{12}}}}  [  \ottnt{k} / \mathit{x}  ]    \simeq _{  \mathsf{F}  }   \ottnt{e_{{\mathrm{22}}}}  [  \ottnt{k} / \mathit{x}  ]   :   \{ \mathit{x} \mathord{:}  \mathsf{Bool}  \mathrel{\mid}  \mathsf{true}  \}  $.

      If $ \ottnt{e_{{\mathrm{12}}}}  [  \ottnt{k} / \mathit{x}  ]  \,  \longrightarrow ^{*}_{  \mathsf{C}  }  \,  \mathord{\Uparrow}  \ottnt{l'} $, then the entire classic side
      steps to $ \mathord{\Uparrow}  \ottnt{l'} $ by \E{CheckInner} and \E{CheckRaise},
      and then we are done. If not, then both predicates reduce to a
      boolean together. If they reduce to $ \mathsf{false} $, then the
      classic side eventually reduces to $ \mathord{\Uparrow}  \ottnt{l} $ via
      \E{CheckInner} and \E{CheckFail}, and we are done. If they both
      go to $ \mathsf{true} $, then both sides step by \E{CheckInner} and
      \E{CheckOK} to yield $\ottnt{k}$, and we can find $ \ottnt{k}   \sim _{  \mathsf{F}  }  \ottnt{k}  :   \{ \mathit{x} \mathord{:} \ottnt{B} \mathrel{\mid} \ottnt{e_{{\mathrm{22}}}} \}  $ easily---we have a derivation for $ \ottnt{e_{{\mathrm{22}}}}  [  \ottnt{k} / \mathit{x}  ]  \,  \longrightarrow ^{*}_{  \mathsf{F}  }  \,  \mathsf{true} $ handy.

    \item[($\ottnt{T_{{\mathrm{22}}}}  \ottsym{=}   \ottnt{T_{{\mathrm{221}}}} \mathord{ \rightarrow } \ottnt{T_{{\mathrm{222}}}} $)] By
      Lemma~\ref{lem:forgetfullrvalue}, we know that $ \mathsf{val} _{  \mathsf{C}  }~ \ottnt{e_{{\mathrm{1}}}} $
      and $ \mathsf{val} _{  \mathsf{F}  }~ \ottnt{e_{{\mathrm{2}}}} $. The classic side is a value $\ottnt{e_{{\mathrm{11}}}}$ (by
      \V{ProxyC}), while the forgetful side steps by one of its
      \E{CastMerge} rules to some value $\ottnt{e_{{\mathrm{21}}}}$, depending on the
      shape of $\ottnt{e_{{\mathrm{2}}}}$: an abstraction yields a value by \V{ProxyF},
      while a function proxy yields another function proxy by
      \E{CastMerge}.

      We must now show that $ \ottnt{e_{{\mathrm{11}}}}   \sim _{  \mathsf{F}  }  \ottnt{e_{{\mathrm{21}}}}  :   \ottnt{T_{{\mathrm{221}}}} \mathord{ \rightarrow } \ottnt{T_{{\mathrm{222}}}}  $, knowing
      that $ \ottnt{e_{{\mathrm{1}}}}   \sim _{  \mathsf{F}  }  \ottnt{e_{{\mathrm{2}}}}  :   \ottnt{T_{{\mathrm{211}}}} \mathord{ \rightarrow } \ottnt{T_{{\mathrm{212}}}}  $. Let $ \ottnt{e_{{\mathrm{12}}}}   \sim _{  \mathsf{F}  }  \ottnt{e_{{\mathrm{22}}}}  :  \ottnt{T_{{\mathrm{221}}}} $ be
      given. On the classic side, we step by \E{Unwrap} to find
      $ \langle  \ottnt{T_{{\mathrm{112}}}}  \mathord{ \overset{\bullet}{\Rightarrow} }  \ottnt{T_{{\mathrm{122}}}}  \rangle^{ \ottnt{l} } ~   (  \ottnt{e_{{\mathrm{1}}}} ~  (  \langle  \ottnt{T_{{\mathrm{121}}}}  \mathord{ \overset{\bullet}{\Rightarrow} }  \ottnt{T_{{\mathrm{111}}}}  \rangle^{ \ottnt{l} } ~  \ottnt{e_{{\mathrm{12}}}}  )   )  $. (Recall that the
      annotations are all $ \bullet $.)

      We now go by cases on the step taken on the whether or not
      $\ottnt{e_{{\mathrm{2}}}}$ is a value or needed to merge:
      \begin{itemize}
      \item[(\V{ProxyF})] We have $\ottnt{e_{{\mathrm{21}}}}  \ottsym{=}   \langle   \ottnt{T_{{\mathrm{211}}}} \mathord{ \rightarrow } \ottnt{T_{{\mathrm{212}}}}   \mathord{ \overset{\bullet}{\Rightarrow} }   \ottnt{T_{{\mathrm{221}}}} \mathord{ \rightarrow } \ottnt{T_{{\mathrm{222}}}}   \rangle^{ \ottnt{l} } ~   \lambda \mathit{x} \mathord{:} \ottnt{T_{{\mathrm{211}}}} .~  \ottnt{e'_{{\mathrm{2}}}}  $ since $\ottnt{e_{{\mathrm{2}}}}  \ottsym{=}   \lambda \mathit{x} \mathord{:} \ottnt{T_{{\mathrm{211}}}} .~  \ottnt{e'_{{\mathrm{2}}}} $. We must show that:
        \[   \langle  \ottnt{T_{{\mathrm{112}}}}  \mathord{ \overset{\bullet}{\Rightarrow} }  \ottnt{T_{{\mathrm{122}}}}  \rangle^{ \ottnt{l} } ~   (  \ottnt{e_{{\mathrm{1}}}} ~  (  \langle  \ottnt{T_{{\mathrm{121}}}}  \mathord{ \overset{\bullet}{\Rightarrow} }  \ottnt{T_{{\mathrm{111}}}}  \rangle^{ \ottnt{l} } ~  \ottnt{e_{{\mathrm{12}}}}  )   )     \simeq _{  \mathsf{F}  }    (  \langle   \ottnt{T_{{\mathrm{211}}}} \mathord{ \rightarrow } \ottnt{T_{{\mathrm{212}}}}   \mathord{ \overset{\bullet}{\Rightarrow} }   \ottnt{T_{{\mathrm{221}}}} \mathord{ \rightarrow } \ottnt{T_{{\mathrm{222}}}}   \rangle^{ \ottnt{l} } ~   \lambda \mathit{x} \mathord{:} \ottnt{T_{{\mathrm{211}}}} .~  \ottnt{e'_{{\mathrm{2}}}}   )  ~ \ottnt{e_{{\mathrm{22}}}}   :  \ottnt{T_{{\mathrm{222}}}}  \]
        The forgetful side steps by \E{Unwrap},
        yielding $ \langle  \ottnt{T_{{\mathrm{212}}}}  \mathord{ \overset{\bullet}{\Rightarrow} }  \ottnt{T_{{\mathrm{222}}}}  \rangle^{ \ottnt{l} } ~   (   (  \lambda \mathit{x} \mathord{:} \ottnt{T_{{\mathrm{211}}}} .~  \ottnt{e'_{{\mathrm{2}}}}  )  ~  (  \langle  \ottnt{T_{{\mathrm{221}}}}  \mathord{ \overset{\bullet}{\Rightarrow} }  \ottnt{T_{{\mathrm{211}}}}  \rangle^{ \ottnt{l} } ~  \ottnt{e_{{\mathrm{22}}}}  )   )  $.
        By the IH, we know that $  \langle  \ottnt{T_{{\mathrm{121}}}}  \mathord{ \overset{\bullet}{\Rightarrow} }  \ottnt{T_{{\mathrm{111}}}}  \rangle^{ \ottnt{l} } ~  \ottnt{e_{{\mathrm{12}}}}    \simeq _{  \mathsf{F}  }   \langle  \ottnt{T_{{\mathrm{221}}}}  \mathord{ \overset{\bullet}{\Rightarrow} }  \ottnt{T_{{\mathrm{211}}}}  \rangle^{ \ottnt{l} } ~  \ottnt{e_{{\mathrm{22}}}}   :  \ottnt{T_{{\mathrm{211}}}} $. If we get blame on the classic side, we are
        done immediately.  Otherwise, each side reduces to values
        $ \ottnt{e'_{{\mathrm{12}}}}   \sim _{  \mathsf{F}  }  \ottnt{e'_{{\mathrm{22}}}}  :  \ottnt{T_{{\mathrm{211}}}} $. We know by assumption that $  \ottnt{e_{{\mathrm{1}}}} ~ \ottnt{e'_{{\mathrm{12}}}}    \simeq _{  \mathsf{F}  }   \ottnt{e_{{\mathrm{2}}}} ~ \ottnt{e'_{{\mathrm{22}}}}   :  \ottnt{T_{{\mathrm{212}}}} $; again, blame on the classic side
        finishes this case. So suppose both sides go to values
        $ \ottnt{e''_{{\mathrm{12}}}}   \sim _{  \mathsf{F}  }  \ottnt{e''_{{\mathrm{22}}}}  :  \ottnt{T_{{\mathrm{212}}}} $. By the IH, we know that
        $  \langle  \ottnt{T_{{\mathrm{112}}}}  \mathord{ \overset{\bullet}{\Rightarrow} }  \ottnt{T_{{\mathrm{122}}}}  \rangle^{ \ottnt{l} } ~  \ottnt{e''_{{\mathrm{12}}}}    \simeq _{  \mathsf{F}  }   \langle  \ottnt{T_{{\mathrm{212}}}}  \mathord{ \overset{\bullet}{\Rightarrow} }  \ottnt{T_{{\mathrm{222}}}}  \rangle^{ \ottnt{l} } ~  \ottnt{e''_{{\mathrm{22}}}}   :  \ottnt{T_{{\mathrm{222}}}} $, and
        we are done.

      \item[(\E{CastMerge})] We have $\ottnt{e_{{\mathrm{21}}}}  \ottsym{=}   \langle   \ottnt{T_{{\mathrm{31}}}} \mathord{ \rightarrow } \ottnt{T_{{\mathrm{32}}}}   \mathord{ \overset{\bullet}{\Rightarrow} }   \ottnt{T_{{\mathrm{221}}}} \mathord{ \rightarrow } \ottnt{T_{{\mathrm{222}}}}   \rangle^{ \ottnt{l} } ~   \lambda \mathit{x} \mathord{:} \ottnt{T_{{\mathrm{31}}}} .~  \ottnt{e'_{{\mathrm{2}}}}  $ since $\ottnt{e_{{\mathrm{2}}}}  \ottsym{=}   \langle   \ottnt{T_{{\mathrm{31}}}} \mathord{ \rightarrow } \ottnt{T_{{\mathrm{32}}}}   \mathord{ \overset{\bullet}{\Rightarrow} }   \ottnt{T_{{\mathrm{211}}}} \mathord{ \rightarrow } \ottnt{T_{{\mathrm{212}}}}   \rangle^{ \ottnt{l'} } ~   \lambda \mathit{x} \mathord{:} \ottnt{T_{{\mathrm{31}}}} .~  \ottnt{e'_{{\mathrm{2}}}}  $.
        We must show that:
        \[   \langle  \ottnt{T_{{\mathrm{112}}}}  \mathord{ \overset{\bullet}{\Rightarrow} }  \ottnt{T_{{\mathrm{122}}}}  \rangle^{ \ottnt{l} } ~   (  \ottnt{e_{{\mathrm{1}}}} ~  (  \langle  \ottnt{T_{{\mathrm{121}}}}  \mathord{ \overset{\bullet}{\Rightarrow} }  \ottnt{T_{{\mathrm{111}}}}  \rangle^{ \ottnt{l} } ~  \ottnt{e_{{\mathrm{12}}}}  )   )     \simeq _{  \mathsf{F}  }    (  \langle   \ottnt{T_{{\mathrm{31}}}} \mathord{ \rightarrow } \ottnt{T_{{\mathrm{32}}}}   \mathord{ \overset{\bullet}{\Rightarrow} }   \ottnt{T_{{\mathrm{221}}}} \mathord{ \rightarrow } \ottnt{T_{{\mathrm{222}}}}   \rangle^{ \ottnt{l} } ~   \lambda \mathit{x} \mathord{:} \ottnt{T_{{\mathrm{31}}}} .~  \ottnt{e'_{{\mathrm{2}}}}   )  ~ \ottnt{e_{{\mathrm{22}}}}   :  \ottnt{T_{{\mathrm{222}}}}  \]
        The right hand steps by \E{Unwrap},
        yielding $ \langle  \ottnt{T_{{\mathrm{32}}}}  \mathord{ \overset{\bullet}{\Rightarrow} }  \ottnt{T_{{\mathrm{222}}}}  \rangle^{ \ottnt{l} } ~   (   (  \lambda \mathit{x} \mathord{:} \ottnt{T_{{\mathrm{31}}}} .~  \ottnt{e'_{{\mathrm{2}}}}  )  ~  (  \langle  \ottnt{T_{{\mathrm{221}}}}  \mathord{ \overset{\bullet}{\Rightarrow} }  \ottnt{T_{{\mathrm{31}}}}  \rangle^{ \ottnt{l} } ~  \ottnt{e_{{\mathrm{22}}}}  )   )  $. We must show that this forgetful term is related to
        the classic term $ \langle  \ottnt{T_{{\mathrm{112}}}}  \mathord{ \overset{\bullet}{\Rightarrow} }  \ottnt{T_{{\mathrm{122}}}}  \rangle^{ \ottnt{l} } ~   (  \ottnt{e_{{\mathrm{1}}}} ~  (  \langle  \ottnt{T_{{\mathrm{121}}}}  \mathord{ \overset{\bullet}{\Rightarrow} }  \ottnt{T_{{\mathrm{111}}}}  \rangle^{ \ottnt{l} } ~  \ottnt{e_{{\mathrm{12}}}}  )   )  $.

        We must now make a brief digression to examine the behavior of
        the cast that was eliminated by \E{CastMerge}.
        We know by the IH that $  \langle  \ottnt{T_{{\mathrm{121}}}}  \mathord{ \overset{\bullet}{\Rightarrow} }  \ottnt{T_{{\mathrm{111}}}}  \rangle^{ \ottnt{l} } ~  \ottnt{e_{{\mathrm{12}}}}    \simeq _{  \mathsf{F}  }   \langle  \ottnt{T_{{\mathrm{221}}}}  \mathord{ \overset{\bullet}{\Rightarrow} }  \ottnt{T_{{\mathrm{211}}}}  \rangle^{ \ottnt{l} } ~  \ottnt{e_{{\mathrm{22}}}}   :  \ottnt{T_{{\mathrm{211}}}} $, so either the classic side goes to blame---and
        we are done---or both sides go to values $ \ottnt{e'_{{\mathrm{12}}}}   \sim _{  \mathsf{F}  }  \ottnt{e'_{{\mathrm{22}}}}  :  \ottnt{T_{{\mathrm{211}}}} $.  By Lemma~\ref{lem:forgetfulcastcongruence}, we can
        find that $ \langle  \ottnt{T_{{\mathrm{211}}}}  \mathord{ \overset{\bullet}{\Rightarrow} }  \ottnt{T_{{\mathrm{31}}}}  \rangle^{ \ottnt{l} } ~  \ottnt{e'_{{\mathrm{22}}}}  \,  \longrightarrow ^{*}_{  \mathsf{F}  }  \, \ottnt{e''_{{\mathrm{22}}}}$ implies
        $ \langle  \ottnt{T_{{\mathrm{211}}}}  \mathord{ \overset{\bullet}{\Rightarrow} }  \ottnt{T_{{\mathrm{31}}}}  \rangle^{ \ottnt{l} } ~   (  \langle  \ottnt{T_{{\mathrm{221}}}}  \mathord{ \overset{\bullet}{\Rightarrow} }  \ottnt{T_{{\mathrm{211}}}}  \rangle^{ \ottnt{l} } ~  \ottnt{e_{{\mathrm{22}}}}  )   \,  \longrightarrow ^{*}_{  \mathsf{F}  }  \, \ottnt{e''_{{\mathrm{22}}}}$.
        But then we have that $ \langle  \ottnt{T_{{\mathrm{211}}}}  \mathord{ \overset{\bullet}{\Rightarrow} }  \ottnt{T_{{\mathrm{31}}}}  \rangle^{ \ottnt{l} } ~   (  \langle  \ottnt{T_{{\mathrm{221}}}}  \mathord{ \overset{\bullet}{\Rightarrow} }  \ottnt{T_{{\mathrm{211}}}}  \rangle^{ \ottnt{l} } ~  \ottnt{e_{{\mathrm{22}}}}  )   \,  \longrightarrow _{  \mathsf{F}  }  \,  \langle  \ottnt{T_{{\mathrm{211}}}}  \mathord{ \overset{\bullet}{\Rightarrow} }  \ottnt{T_{{\mathrm{31}}}}  \rangle^{ \ottnt{l} } ~  \ottnt{e_{{\mathrm{22}}}} $, so we then know that $ \langle  \ottnt{T_{{\mathrm{221}}}}  \mathord{ \overset{\bullet}{\Rightarrow} }  \ottnt{T_{{\mathrm{31}}}}  \rangle^{ \ottnt{l} } ~  \ottnt{e_{{\mathrm{22}}}}  \,  \longrightarrow ^{*}_{  \mathsf{F}  }  \, \ottnt{e''_{{\mathrm{22}}}}$, just as if it were applied to $\ottnt{e'_{{\mathrm{22}}}}$.

        Now we can return to the meat of our proof. If
        $ \langle  \ottnt{T_{{\mathrm{121}}}}  \mathord{ \overset{\bullet}{\Rightarrow} }  \ottnt{T_{{\mathrm{111}}}}  \rangle^{ \ottnt{l} } ~  \ottnt{e_{{\mathrm{12}}}}  \,  \longrightarrow ^{*}_{  \mathsf{F}  }  \,  \mathord{\Uparrow}  \ottnt{l'} $, we are done. If it
        reduces to a value $\ottnt{e'_{{\mathrm{12}}}}$, then we are left considering
        the term $ \langle  \ottnt{T_{{\mathrm{112}}}}  \mathord{ \overset{\bullet}{\Rightarrow} }  \ottnt{T_{{\mathrm{122}}}}  \rangle^{ \ottnt{l} } ~   (  \ottnt{e_{{\mathrm{1}}}} ~ \ottnt{e'_{{\mathrm{12}}}}  )  $ on the classic side. We
        know that $ \ottnt{e_{{\mathrm{1}}}}   \sim _{  \mathsf{F}  }  \ottnt{e_{{\mathrm{2}}}}  :  \ottnt{T_{{\mathrm{21}}}} $. Unfolding the definition of
        $\ottnt{e_{{\mathrm{2}}}}$, this means that $  \ottnt{e_{{\mathrm{1}}}} ~ \ottnt{e'_{{\mathrm{12}}}}    \simeq _{  \mathsf{F}  }   \langle  \ottnt{T_{{\mathrm{32}}}}  \mathord{ \overset{\bullet}{\Rightarrow} }  \ottnt{T_{{\mathrm{212}}}}  \rangle^{ \ottnt{l} } ~   (   (  \lambda \mathit{x} \mathord{:} \ottnt{T_{{\mathrm{31}}}} .~  \ottnt{e'_{{\mathrm{2}}}}  )  ~  (  \langle  \ottnt{T_{{\mathrm{211}}}}  \mathord{ \overset{\bullet}{\Rightarrow} }  \ottnt{T_{{\mathrm{31}}}}  \rangle^{ \ottnt{l} } ~  \ottnt{e'_{{\mathrm{22}}}}  )   )    :  \ottnt{T_{{\mathrm{212}}}} $.
        If the classic side produces blame, we are done, as indicated
        in the digression above. If not, then both sides produce
        values. For these terms to produce values, it must be the case
        that (a) the domain cast on the forgetful side produces a
        value, (b) the forgetful function produces a value given that
        input, and (c) the forgetful codomain cast produces a
        value. Now, we know from our digression above that
        $ \langle  \ottnt{T_{{\mathrm{211}}}}  \mathord{ \overset{\bullet}{\Rightarrow} }  \ottnt{T_{{\mathrm{31}}}}  \rangle^{ \ottnt{l} } ~  \ottnt{e'_{{\mathrm{22}}}} $ and $ \langle  \ottnt{T_{{\mathrm{211}}}}  \mathord{ \overset{\bullet}{\Rightarrow} }  \ottnt{T_{{\mathrm{31}}}}  \rangle^{ \ottnt{l} } ~  \ottnt{e_{{\mathrm{22}}}} $ reduce to
        the exact same value, $\ottnt{e''_{{\mathrm{22}}}}$. So if $ \langle  \ottnt{T_{{\mathrm{32}}}}  \mathord{ \overset{\bullet}{\Rightarrow} }  \ottnt{T_{{\mathrm{212}}}}  \rangle^{ \ottnt{l} } ~   (   (  \lambda \mathit{x} \mathord{:} \ottnt{T_{{\mathrm{31}}}} .~  \ottnt{e'_{{\mathrm{2}}}}  )  ~  (  \langle  \ottnt{T_{{\mathrm{211}}}}  \mathord{ \overset{\bullet}{\Rightarrow} }  \ottnt{T_{{\mathrm{31}}}}  \rangle^{ \ottnt{l} } ~  \ottnt{e'_{{\mathrm{22}}}}  )   )   \,  \longrightarrow ^{*}_{  \mathsf{F}  }  \, \ottnt{e''_{{\mathrm{22}}}}$ then we
        can also see \[  \langle  \ottnt{T_{{\mathrm{32}}}}  \mathord{ \overset{\bullet}{\Rightarrow} }  \ottnt{T_{{\mathrm{212}}}}  \rangle^{ \ottnt{l} } ~   (   (  \lambda \mathit{x} \mathord{:} \ottnt{T_{{\mathrm{31}}}} .~  \ottnt{e'_{{\mathrm{2}}}}  )  ~  (  \langle  \ottnt{T_{{\mathrm{211}}}}  \mathord{ \overset{\bullet}{\Rightarrow} }  \ottnt{T_{{\mathrm{31}}}}  \rangle^{ \ottnt{l} } ~  \ottnt{e_{{\mathrm{22}}}}  )   )   \,  \longrightarrow ^{*}_{  \mathsf{F}  }  \,  \langle  \ottnt{T_{{\mathrm{32}}}}  \mathord{ \overset{\bullet}{\Rightarrow} }  \ottnt{T_{{\mathrm{212}}}}  \rangle^{ \ottnt{l} } ~  \ottnt{e_{{\mathrm{32}}}}   \longrightarrow ^{*}_{  \mathsf{F}  }  \ottnt{e''_{{\mathrm{22}}}}. \]

        We have shown that the domains and then the applied inner
        functions are equivalent.  It now remains to show that
        \[   \langle  \ottnt{T_{{\mathrm{112}}}}  \mathord{ \overset{\bullet}{\Rightarrow} }  \ottnt{T_{{\mathrm{122}}}}  \rangle^{ \ottnt{l} } ~  \ottnt{e''_{{\mathrm{11}}}}    \simeq _{  \mathsf{F}  }   \langle  \ottnt{T_{{\mathrm{32}}}}  \mathord{ \overset{\bullet}{\Rightarrow} }  \ottnt{T_{{\mathrm{222}}}}  \rangle^{ \ottnt{l} } ~   (   (  \lambda \mathit{x} \mathord{:} \ottnt{T_{{\mathrm{31}}}} .~  \ottnt{e'_{{\mathrm{2}}}}  )  ~  (  \langle  \ottnt{T_{{\mathrm{221}}}}  \mathord{ \overset{\bullet}{\Rightarrow} }  \ottnt{T_{{\mathrm{31}}}}  \rangle^{ \ottnt{l} } ~  \ottnt{e_{{\mathrm{22}}}}  )   )    :  \ottnt{T_{{\mathrm{222}}}}  \]
        We write the \textit{entire} forgetful term to highlight the
        fact that we cannot freely apply congruence, but must instead
        carefully apply cast congruence
        (Lemma~\ref{lem:forgetfulcastcongruence}) as we go.
        
        By the IH, we know that either $ \langle  \ottnt{T_{{\mathrm{112}}}}  \mathord{ \overset{\bullet}{\Rightarrow} }  \ottnt{T_{{\mathrm{122}}}}  \rangle^{ \ottnt{l} } ~  \ottnt{e''_{{\mathrm{11}}}} $ goes
        to blame or it goes to a value along with $ \langle  \ottnt{T_{{\mathrm{212}}}}  \mathord{ \overset{\bullet}{\Rightarrow} }  \ottnt{T_{{\mathrm{222}}}}  \rangle^{ \ottnt{l} } ~  \ottnt{e''_{{\mathrm{22}}}} $. In the former case we are done; in the latter case,
        we already know that $ \langle  \ottnt{T_{{\mathrm{32}}}}  \mathord{ \overset{\bullet}{\Rightarrow} }  \ottnt{T_{{\mathrm{212}}}}  \rangle^{ \ottnt{l} } ~  \ottnt{e_{{\mathrm{32}}}}  \,  \longrightarrow ^{*}_{  \mathsf{F}  }  \, \ottnt{e''_{{\mathrm{22}}}}$, so
        we can apply cast congruence
        (Lemma~\ref{lem:forgetfulcastcongruence}) to see that if
        $ \langle  \ottnt{T_{{\mathrm{212}}}}  \mathord{ \overset{\bullet}{\Rightarrow} }  \ottnt{T_{{\mathrm{222}}}}  \rangle^{ \ottnt{l} } ~  \ottnt{e''_{{\mathrm{22}}}}  \,  \longrightarrow ^{*}_{  \mathsf{F}  }  \, \ottnt{e'''_{{\mathrm{22}}}}$ then $ \langle  \ottnt{T_{{\mathrm{212}}}}  \mathord{ \overset{\bullet}{\Rightarrow} }  \ottnt{T_{{\mathrm{222}}}}  \rangle^{ \ottnt{l} } ~   (  \langle  \ottnt{T_{{\mathrm{32}}}}  \mathord{ \overset{\bullet}{\Rightarrow} }  \ottnt{T_{{\mathrm{212}}}}  \rangle^{ \ottnt{l} } ~  \ottnt{e_{{\mathrm{32}}}}  )   \,  \longrightarrow ^{*}_{  \mathsf{F}  }  \, \ottnt{e'''_{{\mathrm{22}}}}$.
        But we know that $ \langle  \ottnt{T_{{\mathrm{212}}}}  \mathord{ \overset{\bullet}{\Rightarrow} }  \ottnt{T_{{\mathrm{222}}}}  \rangle^{ \ottnt{l} } ~   (  \langle  \ottnt{T_{{\mathrm{32}}}}  \mathord{ \overset{\bullet}{\Rightarrow} }  \ottnt{T_{{\mathrm{212}}}}  \rangle^{ \ottnt{l} } ~  \ottnt{e_{{\mathrm{32}}}}  )   \,  \longrightarrow _{  \mathsf{F}  }  \,  \langle  \ottnt{T_{{\mathrm{32}}}}  \mathord{ \overset{\bullet}{\Rightarrow} }  \ottnt{T_{{\mathrm{222}}}}  \rangle^{ \ottnt{l} } ~  \ottnt{e_{{\mathrm{32}}}} $, so we then we know that $ \langle  \ottnt{T_{{\mathrm{32}}}}  \mathord{ \overset{\bullet}{\Rightarrow} }  \ottnt{T_{{\mathrm{222}}}}  \rangle^{ \ottnt{l} } ~  \ottnt{e_{{\mathrm{32}}}}  \,  \longrightarrow ^{*}_{  \mathsf{F}  }  \, \ottnt{e'''_{{\mathrm{22}}}}$. Since $ \langle  \ottnt{T_{{\mathrm{32}}}}  \mathord{ \overset{\bullet}{\Rightarrow} }  \ottnt{T_{{\mathrm{222}}}}  \rangle^{ \ottnt{l} } ~   (   (  \lambda \mathit{x} \mathord{:} \ottnt{T_{{\mathrm{31}}}} .~  \ottnt{e'_{{\mathrm{2}}}}  )  ~  (  \langle  \ottnt{T_{{\mathrm{221}}}}  \mathord{ \overset{\bullet}{\Rightarrow} }  \ottnt{T_{{\mathrm{31}}}}  \rangle^{ \ottnt{l} } ~  \ottnt{e_{{\mathrm{22}}}}  )   )   \,  \longrightarrow ^{*}_{  \mathsf{F}  }  \,  \langle  \ottnt{T_{{\mathrm{32}}}}  \mathord{ \overset{\bullet}{\Rightarrow} }  \ottnt{T_{{\mathrm{222}}}}  \rangle^{ \ottnt{l} } ~  \ottnt{e_{{\mathrm{32}}}} $, we have shown
        that the classic term and forgetful term reduce to values
        $ \ottnt{e'''_{{\mathrm{12}}}}   \sim _{  \mathsf{F}  }  \ottnt{e'''_{{\mathrm{22}}}}  :  \ottnt{T_{{\mathrm{222}}}} $, and we are done.
      \end{itemize}
    \end{itemize}
    \fi}
  \end{proof}
\end{lemma}

\begin{lemma}[Relating classic and forgetful source programs]
  \label{lem:forgetfullr}
  ~

  \noindent
  \begin{enumerate}
  \item \label{flr:term} If $ \Gamma   \vdash _{  \mathsf{C}  }  \ottnt{e}  :  \ottnt{T} $ as a source program then
    $ \Gamma   \vdash   \ottnt{e}   \simeq _{  \mathsf{F}  }  \ottnt{e}  :  \ottnt{T} $.
  \item \label{flr:type} If $ \mathord{  \vdash _{  \mathsf{C}  } }~ \ottnt{T} $ as a source program then $ \ottnt{T}   \sim _{  \mathsf{F}  }  \ottnt{T} $.
  \end{enumerate}
  \begin{proof}
    By mutual induction on the typing derivations.
    
    {\iffull

    \paragraph{Term typing \fbox{$ \Gamma   \vdash _{  \mathsf{C}  }  \ottnt{e}  :  \ottnt{T} $}}
    \begin{itemize}
    \item[\T{Var}] We know by assumption that $ \delta_{{\mathrm{1}}}  \ottsym{(}  \mathit{x}  \ottsym{)}   \sim _{  \mathsf{F}  }  \delta_{{\mathrm{2}}}  \ottsym{(}  \mathit{x}  \ottsym{)}  :  \ottnt{T} $.
    \item[\T{Const}] Since we are dealing with a source program, $\ottnt{T}  \ottsym{=}   \{ \mathit{x} \mathord{:} \ottnt{B} \mathrel{\mid}  \mathsf{true}  \} $. We have immediately that $ \mathsf{ty} ( \ottnt{k} )   \ottsym{=}  \ottnt{B}$ and
      $   \mathsf{true}   [  \ottnt{k} / \mathit{x}  ]    \simeq _{  \mathsf{F}  }    \mathsf{true}   [  \ottnt{k} / \mathit{x}  ]   :   \{ \mathit{x} \mathord{:}  \mathsf{Bool}  \mathrel{\mid}  \mathsf{true}  \}  $, so $ \ottnt{k}   \simeq _{  \mathsf{F}  }  \ottnt{k}  :   \{ \mathit{x} \mathord{:} \ottnt{B} \mathrel{\mid}  \mathsf{true}  \}  $.
    \item[\T{Abs}] Let $ \Gamma   \models _{  \mathsf{F}  }  \delta $. We must show that
      $  \lambda \mathit{x} \mathord{:} \ottnt{T_{{\mathrm{1}}}} .~  \delta_{{\mathrm{1}}}  \ottsym{(}  \ottnt{e_{{\mathrm{1}}}}  \ottsym{)}    \sim _{  \mathsf{F}  }   \lambda \mathit{x} \mathord{:} \ottnt{T_{{\mathrm{2}}}} .~  \delta_{{\mathrm{2}}}  \ottsym{(}  \ottnt{e_{{\mathrm{1}}}}  \ottsym{)}   :   \ottnt{T_{{\mathrm{1}}}} \mathord{ \rightarrow } \ottnt{T_{{\mathrm{2}}}}  $. Let $ \ottnt{e_{{\mathrm{2}}}}   \sim _{  \mathsf{F}  }  \ottnt{e'_{{\mathrm{2}}}}  :  \ottnt{T_{{\mathrm{1}}}} $. We must show that applying the abstractions to
      these values yields related values. Both sides step by \E{Beta},
      to $ \delta_{{\mathrm{1}}}  \ottsym{(}  \ottnt{e_{{\mathrm{1}}}}  \ottsym{)}  [  \ottnt{e_{{\mathrm{2}}}} / \mathit{x}  ] $ and $ \delta_{{\mathrm{2}}}  \ottsym{(}  \ottnt{e_{{\mathrm{1}}}}  \ottsym{)}  [  \ottnt{e'_{{\mathrm{2}}}} / \mathit{x}  ] $,
      respectively. But $  \Gamma , \mathit{x} \mathord{:} \ottnt{T_{{\mathrm{1}}}}    \models _{  \mathsf{F}  }   \delta  [  \ottnt{e_{{\mathrm{2}}}} , \ottnt{e'_{{\mathrm{2}}}} / \mathit{x}  ]  $, so we can
      apply IH (\ref{flr:term}) $\ottnt{e_{{\mathrm{1}}}}$, the two sides reduce to related
      values.
    \item[\T{Op}] By IH (\ref{flr:term}) on each arguments, either one
      of the arguments goes to blame in the classic evaluation, and we
      are done by \E{OpRaise}. Otherwise, all of the arguments reduce
      to related values. Since $ \mathsf{ty} (\mathord{ \ottnt{op} }) $ is first order, these
      values must be related at refined base types, which means that
      they are in fact all equal constants. We then reduce by \E{Op}
      on both sides to have $\denot{ op } \, \ottsym{(}  \ottnt{k_{{\mathrm{1}}}}  \ottsym{,}  \dots  \ottsym{,}  \ottnt{k_{\ottmv{n}}}  \ottsym{)}$. We have assumed
      that the denotations of operations agree with their typings in
      \textit{all} modes, so then $\denot{ op } \, \ottsym{(}  \ottnt{k_{{\mathrm{1}}}}  \ottsym{,}  \dots  \ottsym{,}  \ottnt{k_{\ottmv{n}}}  \ottsym{)}$ satisfies
      the refinement for $ \longrightarrow _{  \mathsf{F}  } $ in particular, and we are done.
    \item[\T{App}] Let $ \Gamma   \models _{  \mathsf{F}  }  \delta $. We must show that
      $  \delta_{{\mathrm{1}}}  \ottsym{(}  \ottnt{e_{{\mathrm{1}}}}  \ottsym{)} ~ \delta_{{\mathrm{1}}}  \ottsym{(}  \ottnt{e_{{\mathrm{2}}}}  \ottsym{)}    \simeq _{  \mathsf{F}  }   \delta_{{\mathrm{2}}}  \ottsym{(}  \ottnt{e_{{\mathrm{1}}}}  \ottsym{)} ~ \delta_{{\mathrm{2}}}  \ottsym{(}  \ottnt{e_{{\mathrm{2}}}}  \ottsym{)}   :  \ottnt{T_{{\mathrm{2}}}} $. But
      by IH (\ref{flr:term}) on $\ottnt{e_{{\mathrm{1}}}}$ and $\ottnt{e_{{\mathrm{2}}}}$, we are done
      directly.
    \item[\T{Cast}] Let $ \Gamma   \models _{  \mathsf{F}  }  \delta $. By IH (\ref{flr:term}) on
      $\ottnt{e'}$, $ \delta_{{\mathrm{1}}}  \ottsym{(}  \ottnt{e'}  \ottsym{)}   \simeq _{  \mathsf{F}  }  \delta_{{\mathrm{2}}}  \ottsym{(}  \ottnt{e'}  \ottsym{)}  :  \ottnt{T_{{\mathrm{1}}}} $, either
      $\delta_{{\mathrm{1}}}  \ottsym{(}  \ottnt{e'}  \ottsym{)} \,  \longrightarrow ^{*}_{  \mathsf{C}  }  \,  \mathord{\Uparrow}  \ottnt{l'} $ (and we are done) or
      $\delta_{{\mathrm{1}}}  \ottsym{(}  \ottnt{e'}  \ottsym{)}$ and $\delta_{{\mathrm{2}}}  \ottsym{(}  \ottnt{e'}  \ottsym{)}$ reduce to values $ \ottnt{e_{{\mathrm{1}}}}   \sim _{  \mathsf{F}  }  \ottnt{e_{{\mathrm{2}}}}  :  \ottnt{T_{{\mathrm{1}}}} $. By Lemma~\ref{lem:forgetfullrcast} (using IH
      (\ref{flr:type}) on the types), we know that $  \langle  \ottnt{T_{{\mathrm{1}}}}  \mathord{ \overset{\bullet}{\Rightarrow} }  \ottnt{T_{{\mathrm{2}}}}  \rangle^{ \ottnt{l} } ~  \ottnt{e_{{\mathrm{1}}}}    \simeq _{  \mathsf{F}  }   \langle  \ottnt{T_{{\mathrm{1}}}}  \mathord{ \overset{\bullet}{\Rightarrow} }  \ottnt{T_{{\mathrm{2}}}}  \rangle^{ \ottnt{l} } ~  \ottnt{e_{{\mathrm{2}}}}   :  \ottnt{T_{{\mathrm{2}}}} $.
      If $ \langle  \ottnt{T_{{\mathrm{1}}}}  \mathord{ \overset{\bullet}{\Rightarrow} }  \ottnt{T_{{\mathrm{2}}}}  \rangle^{ \ottnt{l} } ~  \ottnt{e_{{\mathrm{1}}}}  \,  \longrightarrow ^{*}_{  \mathsf{C}  }  \,  \mathord{\Uparrow}  \ottnt{l'} $, we are done. If not, then
      we know that the cast applied to both \textit{values} reduce to
      values $ \ottnt{e'_{{\mathrm{1}}}}   \sim _{  \mathsf{F}  }  \ottnt{e'_{{\mathrm{2}}}}  :  \ottnt{T_{{\mathrm{2}}}} $, but we must still show that
      $  \langle  \ottnt{T_{{\mathrm{1}}}}  \mathord{ \overset{\bullet}{\Rightarrow} }  \ottnt{T_{{\mathrm{2}}}}  \rangle^{ \ottnt{l} } ~  \delta_{{\mathrm{1}}}  \ottsym{(}  \ottnt{e'}  \ottsym{)}    \simeq _{  \mathsf{F}  }   \langle  \ottnt{T_{{\mathrm{1}}}}  \mathord{ \overset{\bullet}{\Rightarrow} }  \ottnt{T_{{\mathrm{2}}}}  \rangle^{ \ottnt{l} } ~  \delta_{{\mathrm{2}}}  \ottsym{(}  \ottnt{e'}  \ottsym{)}   :  \ottnt{T_{{\mathrm{2}}}} $ for the
      \textit{terms}. The classic side obviously goes to $\ottnt{e'_{{\mathrm{1}}}}$.
      On the forgetful side, we can see by
      Lemma~\ref{lem:forgetfulcastcongruence} that $ \langle  \ottnt{T_{{\mathrm{1}}}}  \mathord{ \overset{\bullet}{\Rightarrow} }  \ottnt{T_{{\mathrm{2}}}}  \rangle^{ \ottnt{l} } ~  \ottnt{e_{{\mathrm{2}}}}  \,  \longrightarrow ^{*}_{  \mathsf{F}  }  \, \ottnt{e'_{{\mathrm{2}}}}$ implies $ \langle  \ottnt{T_{{\mathrm{1}}}}  \mathord{ \overset{\bullet}{\Rightarrow} }  \ottnt{T_{{\mathrm{2}}}}  \rangle^{ \ottnt{l} } ~  \delta_{{\mathrm{2}}}  \ottsym{(}  \ottnt{e'}  \ottsym{)}  \,  \longrightarrow ^{*}_{  \mathsf{F}  }  \, \ottnt{e'_{{\mathrm{2}}}}$, since
      $\delta_{{\mathrm{2}}}  \ottsym{(}  \ottnt{e'}  \ottsym{)} \,  \longrightarrow ^{*}_{  \mathsf{F}  }  \, \ottnt{e_{{\mathrm{2}}}}$. Constructing this derivation
      completes the case.
    \item[\T{Blame}] Contradiction---doesn't appear in source
      programs. Though in fact it is in the relation, since
      $  \mathord{\Uparrow}  \ottnt{l}    \simeq _{  \mathsf{F}  }  \ottnt{e}  :  \ottnt{T} $ for any $\ottnt{e}$ and $\ottnt{T}$.
    \item[\T{Check}] Contradiction---doesn't appear in source
      programs.
    \end{itemize}
    
    \paragraph{Type well formedness \fbox{$ \mathord{  \vdash _{  \mathsf{C}  } }~ \ottnt{T} $}}
    \begin{itemize}
    \item[\WF{Base}] We can immediately see $   \mathsf{true}   [  \ottnt{k} / \mathit{x}  ]    \simeq _{  \mathsf{F}  }    \mathsf{true}   [  \ottnt{k} / \mathit{x}  ]   :   \{ \mathit{x} \mathord{:}  \mathsf{Bool}  \mathrel{\mid}  \mathsf{true}  \}  $ for any $ \ottnt{k}   \sim _{  \mathsf{F}  }  \ottnt{k}  :   \{ \mathit{x} \mathord{:} \ottnt{B} \mathrel{\mid}  \mathsf{true}  \}  $, i.e., any
      $\ottnt{k}$ such that $ \mathsf{ty} ( \ottnt{k} )   \ottsym{=}  \ottnt{B}$.
    \item[\WF{Refine}] By inversion, we know that $  \mathit{x} \mathord{:}  \{ \mathit{x} \mathord{:} \ottnt{B} \mathrel{\mid}  \mathsf{true}  \}     \vdash _{  \mathsf{C}  }  \ottnt{e}  :   \{ \mathit{x} \mathord{:}  \mathsf{Bool}  \mathrel{\mid}  \mathsf{true}  \}  $; by IH (\ref{flr:term}), we find that
      $ \delta_{{\mathrm{1}}}  \ottsym{(}  \ottnt{e}  \ottsym{)}   \simeq _{  \mathsf{F}  }  \delta_{{\mathrm{2}}}  \ottsym{(}  \ottnt{e}  \ottsym{)}  :   \{ \mathit{x} \mathord{:}  \mathsf{Bool}  \mathrel{\mid}  \mathsf{true}  \}  $, i.e., that
      $  \ottnt{e}  [  \ottnt{e_{{\mathrm{1}}}} / \mathit{x}  ]    \simeq _{  \mathsf{F}  }   \ottnt{e}  [  \ottnt{e_{{\mathrm{2}}}} / \mathit{x}  ]   :   \{ \mathit{x} \mathord{:}  \mathsf{Bool}  \mathrel{\mid}  \mathsf{true}  \}  $ for all $ \ottnt{e_{{\mathrm{1}}}}   \sim _{  \mathsf{F}  }  \ottnt{e_{{\mathrm{2}}}}  :   \{ \mathit{x} \mathord{:} \ottnt{B} \mathrel{\mid}  \mathsf{true}  \}  $---which is what we needed to know.
    \item[\WF{Fun}] By IH (\ref{flr:type}) on each of the types.
    \end{itemize}

    \fi}
  \end{proof}
\end{lemma}

\subsection{Relating classic and heedful manifest contracts}
\label{sec:heedfullr}

Heedful \lambdah reorders casts, so we won't necessarily get the same
blame as we do in classic \lambdah. We can show, however, that they
blame the same amount: heedful \lambdah raises blame if and only if
classic \lambdah does, too. We define a blame-inexact, symmetric
logical relation.

The proof follows the same scheme as the proof for forgetful \lambdah
in Section~\ref{sec:forgetfullr}: we first prove a cast congruence
principle; then we define a logical relation relating classic and
heedful \lambdah; we prove a lemma establishing a notion of
applicativity for casts using an inductive invariant grounded in the
logical relation, and then use that lemma to prove that well typed
source programs are logically related.

Cast congruence---that $ \langle  \ottnt{T_{{\mathrm{1}}}}  \mathord{ \overset{ \mathcal{S} }{\Rightarrow} }  \ottnt{T_{{\mathrm{2}}}}  \rangle^{ \ottnt{l} } ~  \ottnt{e} $ and $ \langle  \ottnt{T_{{\mathrm{1}}}}  \mathord{ \overset{ \mathcal{S} }{\Rightarrow} }  \ottnt{T_{{\mathrm{2}}}}  \rangle^{ \ottnt{l} } ~  \ottnt{e_{{\mathrm{1}}}} $
behave identically when $\ottnt{e} \,  \longrightarrow _{  \mathsf{H}  }  \, \ottnt{e_{{\mathrm{1}}}}$---holds almost exactly.  The
pre- and post-step terms may end blaming different labels, but
otherwise return identical values.
Note that this cast congruence lemma (a) has annotations other than
$ \bullet $, and (b) is stronger than
Lemma~\ref{lem:forgetfulcastcongruencesinglestep}, since we not only
get the same value out, but we also get blame when the inner reduction
yields blame---though the label may be different.
The potentially different blame labels in heedful \lambdah's cast
congruence principle arises because of how casts are merged: heedful
\lambdah is heedful of types, but forgets blame labels. 

\finishlater{do we need this lemma anywhere?}

\begin{lemma}[First-order casts don't change their arguments]
  \label{lem:heedfulfocast}
  If $ \langle   \{ \mathit{x} \mathord{:} \ottnt{B} \mathrel{\mid} \ottnt{e_{{\mathrm{1}}}} \}   \mathord{ \overset{ \mathcal{S} }{\Rightarrow} }   \{ \mathit{x} \mathord{:} \ottnt{B} \mathrel{\mid} \ottnt{e_{{\mathrm{2}}}} \}   \rangle^{ \ottnt{l} } ~  \ottnt{k}  \,  \longrightarrow ^{*}_{  \mathsf{H}  }  \, \ottnt{e}$ and $ \mathsf{val} _{  \mathsf{H}  }~ \ottnt{e} $ then
  $\ottnt{e}  \ottsym{=}  \ottnt{k}$.
  \begin{proof}
    By induction on the size of $\mathcal{S}$.
    {\iffull
      \begin{itemize}
      \item[($\mathcal{S}  \ottsym{=}  \emptyset$)] The only possible step is $ \langle   \{ \mathit{x} \mathord{:} \ottnt{B} \mathrel{\mid} \ottnt{e_{{\mathrm{1}}}} \}   \mathord{ \overset{ \emptyset }{\Rightarrow} }   \{ \mathit{x} \mathord{:} \ottnt{B} \mathrel{\mid} \ottnt{e_{{\mathrm{2}}}} \}   \rangle^{ \ottnt{l} } ~  \ottnt{k}  \,  \longrightarrow _{  \mathsf{H}  }  \,  \langle   \{ \mathit{x} \mathord{:} \ottnt{B} \mathrel{\mid} \ottnt{e_{{\mathrm{2}}}} \}  ,   \ottnt{e_{{\mathrm{2}}}}  [  \ottnt{k} / \mathit{x}  ]  ,  \ottnt{k}  \rangle^{ \ottnt{l} } $ by
        \E{CheckEmpty}. So if the original cast term reduces to a
        value, then so must this term. But the only step out of an
        active check that produces a value produces $\ottnt{k}$ by
        \E{CheckOK}.

      \item[($\mathcal{S}  \ottsym{=}    \set{   \{ \mathit{x} \mathord{:} \ottnt{B} \mathrel{\mid} \ottnt{e_{{\mathrm{3}}}} \}   }   \cup  \mathcal{S}' $)] If $\mathcal{S}$ is nonempty,
        then we must step by \E{CheckSet} for some $  \{ \mathit{x} \mathord{:} \ottnt{B} \mathrel{\mid} \ottnt{e_{{\mathrm{3}}}} \}   \in  \mathcal{S} $
        to $ \langle   \{ \mathit{x} \mathord{:} \ottnt{B} \mathrel{\mid} \ottnt{e_{{\mathrm{3}}}} \}   \mathord{ \overset{  \mathcal{S}  \setminus   \{ \mathit{x} \mathord{:} \ottnt{B} \mathrel{\mid} \ottnt{e_{{\mathrm{3}}}} \}   }{\Rightarrow} }   \{ \mathit{x} \mathord{:} \ottnt{B} \mathrel{\mid} \ottnt{e_{{\mathrm{2}}}} \}   \rangle^{ \ottnt{l} } ~   \langle   \{ \mathit{x} \mathord{:} \ottnt{B} \mathrel{\mid} \ottnt{e_{{\mathrm{3}}}} \}  ,   \ottnt{e_{{\mathrm{3}}}}  [  \ottnt{k} / \mathit{x}  ]  ,  \ottnt{k}  \rangle^{ \ottnt{l} }  $.
        For this entire term to reduce to a value, the active check
        must reduce to a value---if it goes to blame, so does the
        whole term. But the only value it can produce is $\ottnt{k}$
        itself, by \E{CheckOK}.
        By the IH, we know that $ \langle   \{ \mathit{x} \mathord{:} \ottnt{B} \mathrel{\mid} \ottnt{e_{{\mathrm{3}}}} \}   \mathord{ \overset{  \mathcal{S}  \setminus   \{ \mathit{x} \mathord{:} \ottnt{B} \mathrel{\mid} \ottnt{e_{{\mathrm{3}}}} \}   }{\Rightarrow} }   \{ \mathit{x} \mathord{:} \ottnt{B} \mathrel{\mid} \ottnt{e_{{\mathrm{2}}}} \}   \rangle^{ \ottnt{l} } ~  \ottnt{k} $ goes to $\ottnt{k}$ if it reduces to a value.
      \end{itemize}
    \fi}
  \end{proof}
\end{lemma}

\begin{lemma}[Determinism of heedful \lambdah]
  \label{lem:heedfuldeterminism}
  If $\ottnt{e} \,  \longrightarrow _{  \mathsf{H}  }  \, \ottnt{e_{{\mathrm{1}}}}$ and $\ottnt{e} \,  \longrightarrow _{  \mathsf{H}  }  \, \ottnt{e_{{\mathrm{2}}}}$ then $\ottnt{e_{{\mathrm{1}}}}  \ottsym{=}  \ottnt{e_{{\mathrm{2}}}}$.
  \begin{proof}
    By induction on the first evaluation derivation. In every case,
    only a single step can be taken. Critically, \E{CheckSet} uses the
    $ \mathsf{choose} $ function, which makes some deterministic choice.
  \end{proof}
\end{lemma}

Heedful \lambdah's cast congruence proof requires an extra
principle. We first show that casting is idempotent: we can safely
remove the source type from a type set.

\begin{lemma}[Idempotence of casts]
  \label{lem:heedfulcastidempotence}
  ~

  \noindent
  If $ \emptyset   \vdash _{  \mathsf{H}  }   \langle   \{ \mathit{x} \mathord{:} \ottnt{B} \mathrel{\mid} \ottnt{e_{{\mathrm{1}}}} \}   \mathord{ \overset{ \mathcal{S} }{\Rightarrow} }   \{ \mathit{x} \mathord{:} \ottnt{B} \mathrel{\mid} \ottnt{e_{{\mathrm{2}}}} \}   \rangle^{ \ottnt{l} } ~  \ottnt{k}   :   \{ \mathit{x} \mathord{:} \ottnt{B} \mathrel{\mid} \ottnt{e_{{\mathrm{2}}}} \}  $ and
  $ \emptyset   \vdash _{  \mathsf{H}  }  \ottnt{k}  :   \{ \mathit{x} \mathord{:} \ottnt{B} \mathrel{\mid} \ottnt{e_{{\mathrm{3}}}} \}  $ then for all $ \mathsf{result} _{  \mathsf{H}  }~ \ottnt{e} $, then: \\
  (a) $ \langle   \{ \mathit{x} \mathord{:} \ottnt{B} \mathrel{\mid} \ottnt{e_{{\mathrm{1}}}} \}   \mathord{ \overset{ \mathcal{S} }{\Rightarrow} }   \{ \mathit{x} \mathord{:} \ottnt{B} \mathrel{\mid} \ottnt{e_{{\mathrm{2}}}} \}   \rangle^{ \ottnt{l} } ~  \ottnt{k}  \,  \longrightarrow ^{*}_{  \mathsf{H}  }  \, \ottnt{e}$ iff \\
  (b) $ \langle   \{ \mathit{x} \mathord{:} \ottnt{B} \mathrel{\mid} \ottnt{e_{{\mathrm{1}}}} \}   \mathord{ \overset{  \mathcal{S}  \setminus   \{ \mathit{x} \mathord{:} \ottnt{B} \mathrel{\mid} \ottnt{e_{{\mathrm{3}}}} \}   }{\Rightarrow} }   \{ \mathit{x} \mathord{:} \ottnt{B} \mathrel{\mid} \ottnt{e_{{\mathrm{2}}}} \}   \rangle^{ \ottnt{l} } ~  \ottnt{k}  \,  \longrightarrow ^{*}_{  \mathsf{H}  }  \, \ottnt{e}$.
  \begin{proof}
    {\iffull
    If $  \{ \mathit{x} \mathord{:} \ottnt{B} \mathrel{\mid} \ottnt{e_{{\mathrm{1}}}} \}   \not\in  \mathcal{S} $, then the proof is trivial, since the
    (a) and (b) are the same. The rest of the proof assumes that
    $  \{ \mathit{x} \mathord{:} \ottnt{B} \mathrel{\mid} \ottnt{e_{{\mathrm{1}}}} \}   \in  \mathcal{S} $.

    We prove both directions by induction on $\mathcal{S}$.
    In both cases, $\mathcal{S}  \ottsym{=}  \emptyset$ is immediate---the two are the same!

    For the only if ($\Leftarrow$) direction, if $ \mathsf{choose} ( \mathcal{S} )   \ottsym{=}   \{ \mathit{x} \mathord{:} \ottnt{B} \mathrel{\mid} \ottnt{e_{{\mathrm{1}}}} \} $, then we step (a) by \E{CheckSet} choosing
    $ \{ \mathit{x} \mathord{:} \ottnt{B} \mathrel{\mid} \ottnt{e_{{\mathrm{1}}}} \} $, finding $ \langle   \{ \mathit{x} \mathord{:} \ottnt{B} \mathrel{\mid} \ottnt{e_{{\mathrm{1}}}} \}   \mathord{ \overset{  \mathcal{S}  \setminus   \{ \mathit{x} \mathord{:} \ottnt{B} \mathrel{\mid} \ottnt{e_{{\mathrm{1}}}} \}   }{\Rightarrow} }   \{ \mathit{x} \mathord{:} \ottnt{B} \mathrel{\mid} \ottnt{e_{{\mathrm{2}}}} \}   \rangle^{ \ottnt{l} } ~   \langle   \{ \mathit{x} \mathord{:} \ottnt{B} \mathrel{\mid} \ottnt{e_{{\mathrm{1}}}} \}  ,   \ottnt{e_{{\mathrm{1}}}}  [  \ottnt{k} / \mathit{x}  ]  ,  \ottnt{k}  \rangle^{ \ottnt{l} }  $. By inversion of the typing derivation,
    $ \ottnt{e_{{\mathrm{1}}}}  [  \ottnt{k} / \mathit{x}  ]  \,  \longrightarrow ^{*}_{  \mathsf{H}  }  \,  \mathsf{true} $, so we can step by \E{CheckInner} and
    \E{CheckOK} to $ \langle   \{ \mathit{x} \mathord{:} \ottnt{B} \mathrel{\mid} \ottnt{e_{{\mathrm{1}}}} \}   \mathord{ \overset{  \mathcal{S}  \setminus   \{ \mathit{x} \mathord{:} \ottnt{B} \mathrel{\mid} \ottnt{e_{{\mathrm{1}}}} \}   }{\Rightarrow} }   \{ \mathit{x} \mathord{:} \ottnt{B} \mathrel{\mid} \ottnt{e_{{\mathrm{2}}}} \}   \rangle^{ \ottnt{l} } ~  \ottnt{k} $, which
    is exactly (b), and we are done.
    If not, the two casts step to the same sub-checks and co-reduce
    until eventually $ \mathsf{choose} ( \mathcal{S} )   \ottsym{=}   \{ \mathit{x} \mathord{:} \ottnt{B} \mathrel{\mid} \ottnt{e_{{\mathrm{1}}}} \} $.

    For the if ($\Rightarrow$) direction, (a) chooses a type to check
    and steps by \E{CheckSet} to $ \langle   \{ \mathit{x} \mathord{:} \ottnt{B} \mathrel{\mid} \ottnt{e_{{\mathrm{3}}}} \}   \mathord{ \overset{  \mathcal{S}  \setminus   \{ \mathit{x} \mathord{:} \ottnt{B} \mathrel{\mid} \ottnt{e_{{\mathrm{3}}}} \}   }{\Rightarrow} }   \{ \mathit{x} \mathord{:} \ottnt{B} \mathrel{\mid} \ottnt{e_{{\mathrm{2}}}} \}   \rangle^{ \ottnt{l} } ~   \langle   \{ \mathit{x} \mathord{:} \ottnt{B} \mathrel{\mid} \ottnt{e_{{\mathrm{3}}}} \}  ,   \ottnt{e_{{\mathrm{3}}}}  [  \ottnt{k} / \mathit{x}  ]  ,  \ottnt{k}  \rangle^{ \ottnt{l} }  $ where $ \mathsf{choose} ( \mathcal{S} )   \ottsym{=}   \{ \mathit{x} \mathord{:} \ottnt{B} \mathrel{\mid} \ottnt{e_{{\mathrm{3}}}} \} $.
    If $\ottnt{e_{{\mathrm{3}}}}  \ottsym{=}  \ottnt{e_{{\mathrm{1}}}}$, then we know by the typing derivation that
    $ \ottnt{e_{{\mathrm{3}}}}  [  \ottnt{k} / \mathit{x}  ]  \,  \longrightarrow ^{*}_{  \mathsf{H}  }  \,  \mathsf{true} $, so the active check must succeed and (a)
    necessarily steps to $ \langle   \{ \mathit{x} \mathord{:} \ottnt{B} \mathrel{\mid} \ottnt{e_{{\mathrm{1}}}} \}   \mathord{ \overset{  \mathcal{S}  \setminus   \{ \mathit{x} \mathord{:} \ottnt{B} \mathrel{\mid} \ottnt{e_{{\mathrm{1}}}} \}   }{\Rightarrow} }   \{ \mathit{x} \mathord{:} \ottnt{B} \mathrel{\mid} \ottnt{e_{{\mathrm{2}}}} \}   \rangle^{ \ottnt{l} } ~  \ottnt{k} $
    by determinism\iffull (Lemma~\ref{lem:heedfuldeterminism})\fi---and we are
    done, since this is the term for which we needed to produce a
    derivation.

    If $\ottnt{e_{{\mathrm{3}}}}  \neq  \ottnt{e_{{\mathrm{1}}}}$, then we take a similar step in (b) to
    $ \langle   \{ \mathit{x} \mathord{:} \ottnt{B} \mathrel{\mid} \ottnt{e_{{\mathrm{3}}}} \}   \mathord{ \overset{   \mathcal{S}  \setminus   \{ \mathit{x} \mathord{:} \ottnt{B} \mathrel{\mid} \ottnt{e_{{\mathrm{1}}}} \}    \setminus   \{ \mathit{x} \mathord{:} \ottnt{B} \mathrel{\mid} \ottnt{e_{{\mathrm{3}}}} \}   }{\Rightarrow} }   \{ \mathit{x} \mathord{:} \ottnt{B} \mathrel{\mid} \ottnt{e_{{\mathrm{2}}}} \}   \rangle^{ \ottnt{l} } ~   \langle   \{ \mathit{x} \mathord{:} \ottnt{B} \mathrel{\mid} \ottnt{e_{{\mathrm{3}}}} \}  ,   \ottnt{e_{{\mathrm{3}}}}  [  \ottnt{k} / \mathit{x}  ]  ,  \ottnt{k}  \rangle^{ \ottnt{l} }  $. Now whatever the derivation for (a) does
    to the active check, we can recapitulate in (b).
    If (a) produces $ \mathord{\Uparrow}  \ottnt{l'} $ for some $\ottnt{l'}$, via either
    \E{CheckRaise} or \E{CheckFail}, then we are done with the whole
    proof. If (a) produces a value, it must produce $\ottnt{k}$ itself by
    \E{CheckOK}.
    But by the IH we know that $ \langle   \{ \mathit{x} \mathord{:} \ottnt{B} \mathrel{\mid} \ottnt{e_{{\mathrm{3}}}} \}   \mathord{ \overset{  \mathcal{S}  \setminus   \{ \mathit{x} \mathord{:} \ottnt{B} \mathrel{\mid} \ottnt{e_{{\mathrm{3}}}} \}   }{\Rightarrow} }   \{ \mathit{x} \mathord{:} \ottnt{B} \mathrel{\mid} \ottnt{e_{{\mathrm{2}}}} \}   \rangle^{ \ottnt{l} } ~  \ottnt{k}  \,  \longrightarrow ^{*}_{  \mathsf{H}  }  \, \ottnt{e}$ implies that $ \langle   \{ \mathit{x} \mathord{:} \ottnt{B} \mathrel{\mid} \ottnt{e_{{\mathrm{3}}}} \}   \mathord{ \overset{   \mathcal{S}  \setminus   \{ \mathit{x} \mathord{:} \ottnt{B} \mathrel{\mid} \ottnt{e_{{\mathrm{3}}}} \}    \setminus   \{ \mathit{x} \mathord{:} \ottnt{B} \mathrel{\mid} \ottnt{e_{{\mathrm{1}}}} \}   }{\Rightarrow} }   \{ \mathit{x} \mathord{:} \ottnt{B} \mathrel{\mid} \ottnt{e_{{\mathrm{2}}}} \}   \rangle^{ \ottnt{l} } ~  \ottnt{k}  \,  \longrightarrow ^{*}_{  \mathsf{H}  }  \, \ottnt{e}$, and we are done.

    \else
    
    By induction on the size of $\mathcal{S}$, with the terms in lock step
    until $ \mathsf{choose} $ produces $ \{ \mathit{x} \mathord{:} \ottnt{B} \mathrel{\mid} \ottnt{e_{{\mathrm{3}}}} \} $ and we can discharge
    its check with the fact that $ \ottnt{e_{{\mathrm{3}}}}  [  \ottnt{k} / \mathit{x}  ]  \,  \longrightarrow ^{*}_{  \mathsf{H}  }  \,  \mathsf{true} $.

    \fi}
  \end{proof}
\end{lemma}

\begin{figure}
  \hdr{Normalizing closed terms}{\qquad \fbox{$\ottnt{e} \, \in \,  \denot{ \ottnt{T} } $}}

  \[ \begin{array}{r@{~~}c@{~~}l}
    \ottnt{e} \, \in \,  \denot{  \{ \mathit{x} \mathord{:} \ottnt{B} \mathrel{\mid} \ottnt{e} \}  }  &\iff& \ottnt{e} \,  \longrightarrow ^{*}_{  \mathsf{H}  }  \,  \mathord{\Uparrow}  \ottnt{l}  \, \vee \, {} \\  &  &  \ottnt{e} \,  \longrightarrow ^{*}_{  \mathsf{H}  }  \, \ottnt{k} \, \wedge \,  \mathsf{ty} ( \ottnt{k} )   \ottsym{=}  \ottnt{B} \\
    \ottnt{e} \, \in \,  \denot{  \ottnt{T_{{\mathrm{1}}}} \mathord{ \rightarrow } \ottnt{T_{{\mathrm{2}}}}  }   &\iff&  \forall  \ottnt{e'} \, \in \,  \denot{ \ottnt{T_{{\mathrm{1}}}} }   . ~   \mathsf{result} _{  \mathsf{H}  }~ \ottnt{e'}    \Rightarrow   \ottnt{e} ~ \ottnt{e'}  \, \in \,  \denot{ \ottnt{T_{{\mathrm{2}}}} } 
  \end{array} \]

  \hdr{Normalizing open terms}{\qquad \fbox{$\Gamma  \models  \ottnt{e}  \ottsym{:}  \ottnt{T}$} \qquad \fbox{$\Gamma  \models  \sigma$}}

  \[ \begin{array}{r@{~~}c@{~~}l} 
    \Gamma  \models  \ottnt{e}  \ottsym{:}  \ottnt{T} &\iff&  \forall  \sigma  . ~  \Gamma  \models  \sigma   \Rightarrow  \sigma  \ottsym{(}  \ottnt{e}  \ottsym{)} \, \in \,  \denot{ \ottnt{T} }  \\
    \Gamma  \models  \sigma &\iff&  \forall   \mathit{x}  \mathord{:}  \ottnt{T}  \in  \Gamma   . ~  \sigma  \ottsym{(}  \mathit{x}  \ottsym{)} \, \in \,  \denot{ \ottnt{T} }  
  \end{array} \]

  \hdr{Normalizing types and type sets}{\qquad \fbox{$\models  \ottnt{T}$} \qquad \fbox{$\models  \mathcal{S}  \mathrel{\parallel}  \ottnt{T_{{\mathrm{1}}}}  \Rightarrow  \ottnt{T_{{\mathrm{2}}}}$}}

  %\sidebyside
    {\ottusedrule{\ottdruleSWFXXRefine{}}}
    {\ottusedrule{\ottdruleSWFXXFun{}}}
  \ottusedrule{\ottdruleSWFXXTypeSet{}}

  \caption{Strong normalization for heedful \lambdah}
  \label{fig:heedfulsn}
\end{figure}

We need strong normalization to prove cast congruence: if we reorder
checks, we need to know that reordering checks doesn't change the
observable behavior. We define a unary logical relation to show strong
normalization in Figure~\ref{fig:heedfulsn}.
We assume throughout at the terms are well
typed at their indices: $\ottnt{e} \, \in \,  \denot{ \ottnt{T} } $ implies $ \emptyset   \vdash _{  \mathsf{H}  }  \ottnt{e}  :  \ottnt{T} $
and $\models  \ottnt{T}$ implies $ \mathord{  \vdash _{  \mathsf{H}  } }~ \ottnt{T} $ and $\models  \mathcal{S}  \mathrel{\parallel}  \ottnt{T_{{\mathrm{1}}}}  \Rightarrow  \ottnt{T_{{\mathrm{2}}}}$ implies
$ \mathord{  \vdash _{  \mathsf{H}  } }~ \mathcal{S}   \mathrel{\parallel}   \ottnt{T_{{\mathrm{1}}}}  \Rightarrow  \ottnt{T_{{\mathrm{2}}}} $ and $\Gamma  \models  \ottnt{e}  \ottsym{:}  \ottnt{T}$ implies $ \Gamma   \vdash _{  \mathsf{H}  }  \ottnt{e}  :  \ottnt{T} $
\textit{by definition}. Making this assumption simplifies many of the
technicalities. First, typed terms stay well typed as they evaluate
(by preservation, Lemma~\ref{lem:heedfulpreservation}), so a well typed relation allows
us to reason exclusively over typed terms. Second, it allows us to
ignore the refinements in our relation, essentially using the simple
type structure.
After proving cast congruence, we show that all well typed terms are
in fact in the relation, i.e., that all heedful terms normalize.

\begin{lemma}[Expansion and contraction]
  \label{lem:heedfulsneval}
  If $\ottnt{e_{{\mathrm{1}}}} \,  \longrightarrow ^{*}_{  \mathsf{H}  }  \, \ottnt{e_{{\mathrm{2}}}}$ then $\ottnt{e_{{\mathrm{1}}}} \, \in \,  \denot{ \ottnt{T} } $ iff $\ottnt{e_{{\mathrm{2}}}} \, \in \,  \denot{ \ottnt{T} } $.
  \begin{proof}
    By induction on $\ottnt{T}$.
    \begin{itemize}
    \item[($\ottnt{T}  \ottsym{=}   \{ \mathit{x} \mathord{:} \ottnt{B} \mathrel{\mid} \ottnt{e''} \} $)] By determinism (Lemma~\ref{lem:heedfuldeterminism}).
    \item[($\ottnt{T}  \ottsym{=}   \ottnt{T_{{\mathrm{1}}}} \mathord{ \rightarrow } \ottnt{T_{{\mathrm{2}}}} $)] Given some $\ottnt{e'} \, \in \,  \denot{ \ottnt{T_{{\mathrm{1}}}} } $, we must show
      that $ \ottnt{e_{{\mathrm{1}}}} ~ \ottnt{e'}  \, \in \,  \denot{ \ottnt{T_{{\mathrm{2}}}} } $ iff $ \ottnt{e_{{\mathrm{2}}}} ~ \ottnt{e'}  \, \in \,  \denot{ \ottnt{T_{{\mathrm{2}}}} } $. We have
      $ \ottnt{e_{{\mathrm{1}}}} ~ \ottnt{e'}  \,  \longrightarrow ^{*}_{  \mathsf{H}  }  \,  \ottnt{e_{{\mathrm{2}}}} ~ \ottnt{e'} $ by induction on the length of the
      evaluation derivation and \E{AppL}, so we are done by the IH on
      $\ottnt{T_{{\mathrm{2}}}}$.
    \end{itemize}
  \end{proof}  
\end{lemma}

\begin{lemma}[Blame inhabits every type]
  \label{lem:heedfulsnblame}
  $ \mathord{\Uparrow}  \ottnt{l}  \, \in \,  \denot{ \ottnt{T} } $ for all $\ottnt{T}$.
  \begin{proof}
    By induction on $\ottnt{T}$.
    \begin{itemize}
    \item[($\ottnt{T}  \ottsym{=}   \{ \mathit{x} \mathord{:} \ottnt{B} \mathrel{\mid} \ottnt{e''} \} $)] By definition.
    \item[($\ottnt{T}  \ottsym{=}   \ottnt{T_{{\mathrm{1}}}} \mathord{ \rightarrow } \ottnt{T_{{\mathrm{2}}}} $)] By the IH, $ \mathord{\Uparrow}  \ottnt{l'}  \, \in \,  \denot{ \ottnt{T_{{\mathrm{1}}}} } $. We
      must show that $  \mathord{\Uparrow}  \ottnt{l}  ~  \mathord{\Uparrow}  \ottnt{l'}   \, \in \,  \denot{ \ottnt{T_{{\mathrm{2}}}} } $. This term
      steps to $ \mathord{\Uparrow}  \ottnt{l} $ by \E{AppRaiseL}, and then we are done
      by contraction (Lemma~\ref{lem:heedfulsneval}).
    \end{itemize}
  \end{proof}
\end{lemma}

\begin{lemma}[Strong normalization]
  \label{lem:heedfulsn}
  If $\ottnt{e} \, \in \,  \denot{ \ottnt{T} } $ then $\ottnt{e} \,  \longrightarrow ^{*}_{  \mathsf{H}  }  \, \ottnt{e'}$ uniquely such that
  $ \mathsf{result} _{  \mathsf{H}  }~ \ottnt{e'} $.
  \begin{proof}
    Uniqueness is immediate by determinism\iffull
    (Lemma~\ref{lem:heedfuldeterminism})\fi. We show normalization by
    induction on $\ottnt{T}$, observing that blame inhabits every type.
    {\iffull
    \begin{itemize}
    \item[($\ottnt{T}  \ottsym{=}   \{ \mathit{x} \mathord{:} \ottnt{B} \mathrel{\mid} \ottnt{e''} \} $)] By definition.
    \item[($\ottnt{T}  \ottsym{=}   \ottnt{T_{{\mathrm{1}}}} \mathord{ \rightarrow } \ottnt{T_{{\mathrm{2}}}} $)] By Lemma~\ref{lem:heedfulsnblame}, we know
      that at least one result is in the domain type: $ \mathord{\Uparrow}  \ottnt{l}  \, \in \,  \denot{ \ottnt{T_{{\mathrm{1}}}} } $. So by assumption, $ \ottnt{e} ~  \mathord{\Uparrow}  \ottnt{l}   \, \in \,  \denot{ \ottnt{T_{{\mathrm{2}}}} } $. By the
      IH, this term is strong normalizing---but that can only be so if
      $\ottnt{e}$ reduces to a result.
    \end{itemize}
    \fi}
  \end{proof}
\end{lemma}

\begin{lemma}[Cast congruence (single step)]
  \label{lem:heedfulcastcongruencesinglestep}
  If
  \begin{itemize}
  \item $\ottnt{e} \, \in \,  \denot{ \ottnt{T_{{\mathrm{1}}}} } $ and $\models  \mathcal{S}  \mathrel{\parallel}  \ottnt{T_{{\mathrm{1}}}}  \Rightarrow  \ottnt{T_{{\mathrm{2}}}}$ (and so
    $ \emptyset   \vdash _{  \mathsf{H}  }   \langle  \ottnt{T_{{\mathrm{1}}}}  \mathord{ \overset{\bullet}{\Rightarrow} }  \ottnt{T_{{\mathrm{2}}}}  \rangle^{ \ottnt{l} } ~  \ottnt{e}   :  \ottnt{T_{{\mathrm{2}}}} $),
  \item $\ottnt{e} \,  \longrightarrow _{  \mathsf{H}  }  \, \ottnt{e_{{\mathrm{1}}}}$ (and so $ \emptyset   \vdash _{  \mathsf{H}  }  \ottnt{e_{{\mathrm{1}}}}  :  \ottnt{T_{{\mathrm{1}}}} $),
  \item $ \langle  \ottnt{T_{{\mathrm{1}}}}  \mathord{ \overset{ \mathcal{S} }{\Rightarrow} }  \ottnt{T_{{\mathrm{2}}}}  \rangle^{ \ottnt{l} } ~  \ottnt{e_{{\mathrm{1}}}}  \,  \longrightarrow ^{*}_{  \mathsf{H}  }  \, \ottnt{e_{{\mathrm{2}}}}$, and
  \item $ \mathsf{result} _{  \mathsf{H}  }~ \ottnt{e_{{\mathrm{2}}}} $
  \end{itemize}
  then $ \langle  \ottnt{T_{{\mathrm{1}}}}  \mathord{ \overset{ \mathcal{S} }{\Rightarrow} }  \ottnt{T_{{\mathrm{2}}}}  \rangle^{ \ottnt{l} } ~  \ottnt{e}  \,  \longrightarrow ^{*}_{  \mathsf{H}  }  \,  \mathord{\Uparrow}  \ottnt{l'} $ if $\ottnt{e_{{\mathrm{2}}}}  \ottsym{=}   \mathord{\Uparrow}  \ottnt{l} $ or
  to $\ottnt{e_{{\mathrm{2}}}}$ itself if $ \mathsf{val} _{  \mathsf{H}  }~ \ottnt{e_{{\mathrm{2}}}} $.
  \begin{proof}
    By cases on the step taken; the proof is as for forgetful \lambdah
    (Lemma~\ref{lem:forgetfulcastcongruencesinglestep}), though we
    need to use strong normalization to handle the reorderings.
    {\iffull
    There are two groups of reductions: straightforward merge-free
    reductions and merging reductions.

    \paragraph{Merge-free reductions}
    In these cases, we apply \E{CastInner} and whatever rule
    derived $\ottnt{e} \,  \longrightarrow _{  \mathsf{H}  }  \, \ottnt{e_{{\mathrm{1}}}}$.
    \begin{itemize}
    \item[(\E{Beta})] By \E{CastInner} and \E{Beta}.
    \item[(\E{Op})] By \E{CastInner} and \E{Op}.
    \item[(\E{Unwrap})] By \E{CastInner} and \E{Unwrap}.
    \item[(\E{TypeSet})] By \E{CastInner} and \E{TypeSet}.
    \item[(\E{AppL})] By \E{CastInner} with \E{AppL}.
    \item[(\E{AppR})] By \E{CastInner} with \E{AppR}.
    \item[(\E{AppRaiseL})] By \E{CastInner} with \E{AppRaiseL}; then
      by \E{CastRaise} on both sides.
    \item[(\E{AppRaiseR})] By \E{CastInner} with \E{AppRaiseR}; then
      by \E{CastRaise} on both sides.
    \item[(\E{CheckOK})] By \E{CastInner} and \E{CheckOK}.
    \item[(\E{CheckFail})] By \E{CastInner} and \E{CheckRaise}.
    \item[(\E{CheckFail})] By \E{CastInner} and \E{CheckRaise}.
    \item[(\E{OpInner})] By \E{CastInner} and \E{OpInner}.
    \item[(\E{OpRaise})] By \E{CastInner} and \E{OpRaise}.

    \end{itemize}

    \paragraph{Merging reductions}
    In these cases, some cast in $\ottnt{e}$ reduces when we step $\ottnt{e} \,  \longrightarrow _{  \mathsf{H}  }  \, \ottnt{e_{{\mathrm{1}}}}$, but merges when we consider $ \langle  \ottnt{T_{{\mathrm{1}}}}  \mathord{ \overset{ \mathcal{S} }{\Rightarrow} }  \ottnt{T_{{\mathrm{2}}}}  \rangle^{ \ottnt{l} } ~  \ottnt{e} $. We
    must show that the merged term and $ \langle  \ottnt{T_{{\mathrm{1}}}}  \mathord{ \overset{ \mathcal{S} }{\Rightarrow} }  \ottnt{T_{{\mathrm{2}}}}  \rangle^{ \ottnt{l} } ~  \ottnt{e_{{\mathrm{1}}}} $
    eventually meet. After merging the cast in $\ottnt{e}$---and possibly
    some steps in $\ottnt{e_{{\mathrm{1}}}}$--- the $\ottnt{e}$ and $\ottnt{e_{{\mathrm{1}}}}$ terms 
    reduce to a common term, which immediately gives us the common
    reduction to results we need.
    \begin{itemize}
    \item[(\E{CheckEmpty})] We have $\ottnt{e}  \ottsym{=}   (  \langle  \ottnt{T_{{\mathrm{3}}}}  \mathord{ \overset{ \emptyset }{\Rightarrow} }   \{ \mathit{x} \mathord{:} \ottnt{B} \mathrel{\mid} \ottnt{e_{{\mathrm{11}}}} \}   \rangle^{ \ottnt{l'} } ~  \ottnt{k}  ) $ where $\ottnt{T_{{\mathrm{1}}}}  \ottsym{=}   \{ \mathit{x} \mathord{:} \ottnt{B} \mathrel{\mid} \ottnt{e_{{\mathrm{11}}}} \} $ and $\ottnt{e_{{\mathrm{1}}}}  \ottsym{=}   \langle   \{ \mathit{x} \mathord{:} \ottnt{B} \mathrel{\mid} \ottnt{e_{{\mathrm{11}}}} \}  ,   \ottnt{e_{{\mathrm{11}}}}  [  \ottnt{k} / \mathit{x}  ]  ,  \ottnt{k}  \rangle^{ \ottnt{l'} } $ and $ \langle   \{ \mathit{x} \mathord{:} \ottnt{B} \mathrel{\mid} \ottnt{e_{{\mathrm{11}}}} \}   \mathord{ \overset{ \mathcal{S} }{\Rightarrow} }  \ottnt{T_{{\mathrm{2}}}}  \rangle^{ \ottnt{l} } ~  \ottnt{e_{{\mathrm{1}}}}  \,  \longrightarrow ^{*}_{  \mathsf{H}  }  \, \ottnt{e_{{\mathrm{2}}}}$ such that $ \mathsf{result} _{  \mathsf{H}  }~ \ottnt{e_{{\mathrm{2}}}} $. We must show that
      $ \langle   \{ \mathit{x} \mathord{:} \ottnt{B} \mathrel{\mid} \ottnt{e_{{\mathrm{11}}}} \}   \mathord{ \overset{ \mathcal{S} }{\Rightarrow} }  \ottnt{T_{{\mathrm{2}}}}  \rangle^{ \ottnt{l} } ~  \ottnt{e}  \,  \longrightarrow ^{*}_{  \mathsf{H}  }  \, \ottnt{e'_{{\mathrm{2}}}}$ such that $\ottnt{e_{{\mathrm{2}}}}  \ottsym{=}   \mathord{\Uparrow}  \ottnt{l} $ and $\ottnt{e'_{{\mathrm{2}}}}  \ottsym{=}   \mathord{\Uparrow}  \ottnt{l'} $ or $ \mathsf{val} _{  \mathsf{H}  }~ \ottnt{e_{{\mathrm{2}}}} $ and
      $\ottnt{e_{{\mathrm{2}}}}  \ottsym{=}  \ottnt{e'_{{\mathrm{2}}}}$.
      We find that both terms go to blame (at possibly different
      labels), or reduce to $ \langle   \{ \mathit{x} \mathord{:} \ottnt{B} \mathrel{\mid} \ottnt{e_{{\mathrm{11}}}} \}   \mathord{ \overset{ \mathcal{S} }{\Rightarrow} }  \ottnt{T_{{\mathrm{2}}}}  \rangle^{ \ottnt{l} } ~  \ottnt{k} $.
      
      We step the $\ottnt{e}$ term:
      \[ \begin{array}{rlr}
        &  \langle   \{ \mathit{x} \mathord{:} \ottnt{B} \mathrel{\mid} \ottnt{e_{{\mathrm{11}}}} \}   \mathord{ \overset{ \mathcal{S} }{\Rightarrow} }  \ottnt{T_{{\mathrm{2}}}}  \rangle^{ \ottnt{l} } ~   (  \langle  \ottnt{T_{{\mathrm{3}}}}  \mathord{ \overset{\bullet}{\Rightarrow} }   \{ \mathit{x} \mathord{:} \ottnt{B} \mathrel{\mid} \ottnt{e_{{\mathrm{11}}}} \}   \rangle^{ \ottnt{l'} } ~  \ottnt{k}  )   & \E{CastMerge} \\
         \longrightarrow _{  \mathsf{H}  }  &  \langle  \ottnt{T_{{\mathrm{3}}}}  \mathord{ \overset{  \mathcal{S}  \cup   \set{   \{ \mathit{x} \mathord{:} \ottnt{B} \mathrel{\mid} \ottnt{e_{{\mathrm{11}}}} \}   }   }{\Rightarrow} }  \ottnt{T_{{\mathrm{2}}}}  \rangle^{ \ottnt{l} } ~  \ottnt{k}  & \E{CheckSet} \\
         \longrightarrow _{  \mathsf{H}  }  &  \langle   \{ \mathit{x} \mathord{:} \ottnt{B} \mathrel{\mid} \ottnt{e_{{\mathrm{11}}}} \}   \mathord{ \overset{  \mathcal{S}  \setminus   \{ \mathit{x} \mathord{:} \ottnt{B} \mathrel{\mid} \ottnt{e_{{\mathrm{11}}}} \}   }{\Rightarrow} }  \ottnt{T_{{\mathrm{2}}}}  \rangle^{ \ottnt{l} } ~   \langle   \{ \mathit{x} \mathord{:} \ottnt{B} \mathrel{\mid} \ottnt{e_{{\mathrm{11}}}} \}  ,   \ottnt{e_{{\mathrm{11}}}}  [  \ottnt{k} / \mathit{x}  ]  ,  \ottnt{k}  \rangle^{ \ottnt{l} }   &
      \end{array} \]
      Knowing that $ \langle   \{ \mathit{x} \mathord{:} \ottnt{B} \mathrel{\mid} \ottnt{e_{{\mathrm{11}}}} \}   \mathord{ \overset{ \mathcal{S} }{\Rightarrow} }  \ottnt{T_{{\mathrm{2}}}}  \rangle^{ \ottnt{l} } ~   \langle   \{ \mathit{x} \mathord{:} \ottnt{B} \mathrel{\mid} \ottnt{e_{{\mathrm{11}}}} \}  ,   \ottnt{e_{{\mathrm{11}}}}  [  \ottnt{k} / \mathit{x}  ]  ,  \ottnt{k}  \rangle^{ \ottnt{l'} }   \,  \longrightarrow ^{*}_{  \mathsf{H}  }  \, \ottnt{e_{{\mathrm{2}}}}$, we know that $ \ottnt{e_{{\mathrm{11}}}}  [  \ottnt{k} / \mathit{x}  ]  \,  \longrightarrow ^{*}_{  \mathsf{H}  }  \, \ottnt{e'_{{\mathrm{11}}}}$ such that
      $ \mathsf{result} _{  \mathsf{H}  }~ \ottnt{e'_{{\mathrm{11}}}} $.
      If it goes to $ \mathord{\Uparrow}  \ottnt{l''} $, so do both the $\ottnt{e_{{\mathrm{1}}}}$ and
      $\ottnt{e}$ terms by \E{CheckRaise}.  If it goes to $ \mathsf{false} $, the
      $\ottnt{e_{{\mathrm{1}}}}$ term goes to $ \mathord{\Uparrow}  \ottnt{l'} $ while the $\ottnt{e}$ term
      goes to $ \mathord{\Uparrow}  \ottnt{l} $, both by \E{CheckFail}.
      Finally, if it goes to $ \mathsf{true} $, then we know that $ \langle   \{ \mathit{x} \mathord{:} \ottnt{B} \mathrel{\mid} \ottnt{e_{{\mathrm{11}}}} \}   \mathord{ \overset{ \mathcal{S} }{\Rightarrow} }  \ottnt{T_{{\mathrm{2}}}}  \rangle^{ \ottnt{l} } ~  \ottnt{e_{{\mathrm{1}}}}  =  \langle   \{ \mathit{x} \mathord{:} \ottnt{B} \mathrel{\mid} \ottnt{e_{{\mathrm{11}}}} \}   \mathord{ \overset{ \mathcal{S} }{\Rightarrow} }  \ottnt{T_{{\mathrm{2}}}}  \rangle^{ \ottnt{l} } ~   \langle   \{ \mathit{x} \mathord{:} \ottnt{B} \mathrel{\mid} \ottnt{e_{{\mathrm{11}}}} \}  ,   \ottnt{e_{{\mathrm{11}}}}  [  \ottnt{k} / \mathit{x}  ]  ,  \ottnt{k}  \rangle^{ \ottnt{l'} }   \,  \longrightarrow ^{*}_{  \mathsf{H}  }  \,  \langle  \ottnt{T_{{\mathrm{1}}}}  \mathord{ \overset{ \mathcal{S} }{\Rightarrow} }  \ottnt{T_{{\mathrm{2}}}}  \rangle^{ \ottnt{l} } ~  \ottnt{k} $.
      If $  \{ \mathit{x} \mathord{:} \ottnt{B} \mathrel{\mid} \ottnt{e_{{\mathrm{11}}}} \}   \not\in  \mathcal{S} $, then we are already done---we say
      that the $\ottnt{e}$ term stepped to this.

      If, on the other hand, $  \{ \mathit{x} \mathord{:} \ottnt{B} \mathrel{\mid} \ottnt{e_{{\mathrm{11}}}} \}   \in  \mathcal{S} $, then the $\ottnt{e}$
      term stepped to $ \langle   \{ \mathit{x} \mathord{:} \ottnt{B} \mathrel{\mid} \ottnt{e_{{\mathrm{11}}}} \}   \mathord{ \overset{  \mathcal{S}  \setminus   \{ \mathit{x} \mathord{:} \ottnt{B} \mathrel{\mid} \ottnt{e_{{\mathrm{11}}}} \}   }{\Rightarrow} }  \ottnt{T_{{\mathrm{2}}}}  \rangle^{ \ottnt{l} } ~  \ottnt{k} $ while
      the $\ottnt{e_{{\mathrm{1}}}}$ term stepped to $ \langle   \{ \mathit{x} \mathord{:} \ottnt{B} \mathrel{\mid} \ottnt{e_{{\mathrm{11}}}} \}   \mathord{ \overset{ \mathcal{S} }{\Rightarrow} }  \ottnt{T_{{\mathrm{2}}}}  \rangle^{ \ottnt{l} } ~  \ottnt{k} $.
      We can apply reflexivity of casts
      (Lemma~\ref{lem:heedfulcastidempotence}) to see that these terms
      reduce to the same results.
      
    \item[(\E{CheckSet})] This case is quite similar to \E{CheckEmpty}.
      We have $\ottnt{e}  \ottsym{=}   (  \langle  \ottnt{T_{{\mathrm{3}}}}  \mathord{ \overset{ \mathcal{S}_{{\mathrm{2}}} }{\Rightarrow} }  \ottnt{T_{{\mathrm{1}}}}  \rangle^{ \ottnt{l'} } ~  \ottnt{k}  ) $ where $ \mathsf{choose} ( \mathcal{S}_{{\mathrm{2}}} )   \ottsym{=}   \{ \mathit{x} \mathord{:} \ottnt{B} \mathrel{\mid} \ottnt{e_{{\mathrm{11}}}} \} $ and $\ottnt{e_{{\mathrm{1}}}}  \ottsym{=}   \langle   \{ \mathit{x} \mathord{:} \ottnt{B} \mathrel{\mid} \ottnt{e_{{\mathrm{11}}}} \}   \mathord{ \overset{  \mathcal{S}_{{\mathrm{2}}}  \setminus   \{ \mathit{x} \mathord{:} \ottnt{B} \mathrel{\mid} \ottnt{e_{{\mathrm{11}}}} \}   }{\Rightarrow} }  \ottnt{T_{{\mathrm{1}}}}  \rangle^{ \ottnt{l'} } ~   \langle   \{ \mathit{x} \mathord{:} \ottnt{B} \mathrel{\mid} \ottnt{e_{{\mathrm{11}}}} \}  ,   \ottnt{e_{{\mathrm{11}}}}  [  \ottnt{k} / \mathit{x}  ]  ,  \ottnt{k}  \rangle^{ \ottnt{l'} }  $ and $ \langle   \{ \mathit{x} \mathord{:} \ottnt{B} \mathrel{\mid} \ottnt{e_{{\mathrm{11}}}} \}   \mathord{ \overset{ \mathcal{S} }{\Rightarrow} }  \ottnt{T_{{\mathrm{2}}}}  \rangle^{ \ottnt{l} } ~  \ottnt{e_{{\mathrm{1}}}}  \,  \longrightarrow ^{*}_{  \mathsf{H}  }  \, \ottnt{e_{{\mathrm{2}}}}$ such that $ \mathsf{result} _{  \mathsf{H}  }~ \ottnt{e_{{\mathrm{2}}}} $.
      We find that both sides reduce to blame (at possibly
      different labels) or the common term $ \langle   \{ \mathit{x} \mathord{:} \ottnt{B} \mathrel{\mid} \ottnt{e_{{\mathrm{11}}}} \}   \mathord{ \overset{  \ottsym{(}    \mathcal{S}_{{\mathrm{1}}}  \cup  \mathcal{S}_{{\mathrm{2}}}   \cup   \set{  \ottnt{T_{{\mathrm{1}}}}  }    \ottsym{)}  \setminus   \{ \mathit{x} \mathord{:} \ottnt{B} \mathrel{\mid} \ottnt{e_{{\mathrm{11}}}} \}   }{\Rightarrow} }  \ottnt{T_{{\mathrm{2}}}}  \rangle^{ \ottnt{l} } ~  \ottnt{k} $.
      
      We step the $\ottnt{e}$ term:
      \[ \begin{array}{rlr}
        &  \langle  \ottnt{T_{{\mathrm{1}}}}  \mathord{ \overset{ \mathcal{S}_{{\mathrm{1}}} }{\Rightarrow} }  \ottnt{T_{{\mathrm{2}}}}  \rangle^{ \ottnt{l} } ~   (  \langle  \ottnt{T_{{\mathrm{3}}}}  \mathord{ \overset{ \mathcal{S}_{{\mathrm{2}}} }{\Rightarrow} }  \ottnt{T_{{\mathrm{1}}}}  \rangle^{ \ottnt{l'} } ~  \ottnt{k}  )   & \E{CastMerge} \\
         \longrightarrow _{  \mathsf{H}  }  &  \langle  \ottnt{T_{{\mathrm{3}}}}  \mathord{ \overset{   \mathcal{S}_{{\mathrm{1}}}  \cup  \mathcal{S}_{{\mathrm{2}}}   \cup   \set{  \ottnt{T_{{\mathrm{1}}}}  }   }{\Rightarrow} }  \ottnt{T_{{\mathrm{2}}}}  \rangle^{ \ottnt{l} } ~  \ottnt{k}  & \E{CheckSet} %\\
%         \longrightarrow _{  \mathsf{H}  }  &  \langle   \{ \mathit{x} \mathord{:} \ottnt{B} \mathrel{\mid} \ottnt{e_{{\mathrm{11}}}} \}   \mathord{ \overset{  \ottsym{(}    \mathcal{S}_{{\mathrm{1}}}  \cup  \mathcal{S}_{{\mathrm{2}}}   \cup   \set{  \ottnt{T_{{\mathrm{1}}}}  }    \ottsym{)}  \setminus   \{ \mathit{x} \mathord{:} \ottnt{B} \mathrel{\mid} \ottnt{e_{{\mathrm{11}}}} \}   }{\Rightarrow} }  \ottnt{T_{{\mathrm{2}}}}  \rangle^{ \ottnt{l} } ~   \langle   \{ \mathit{x} \mathord{:} \ottnt{B} \mathrel{\mid} \ottnt{e_{{\mathrm{11}}}} \}  ,   \ottnt{e_{{\mathrm{11}}}}  [  \ottnt{k} / \mathit{x}  ]  ,  \ottnt{k}  \rangle^{ \ottnt{l} }   &
      \end{array} \]
      Similarly, we know that the $\ottnt{e_{{\mathrm{1}}}}$ term must step by
      \E{CastMerge} as well:
      \[ \begin{array}{l}
         \langle  \ottnt{T_{{\mathrm{1}}}}  \mathord{ \overset{ \mathcal{S}_{{\mathrm{1}}} }{\Rightarrow} }  \ottnt{T_{{\mathrm{2}}}}  \rangle^{ \ottnt{l} } ~   (  \langle   \{ \mathit{x} \mathord{:} \ottnt{B} \mathrel{\mid} \ottnt{e_{{\mathrm{11}}}} \}   \mathord{ \overset{  \mathcal{S}_{{\mathrm{2}}}  \setminus   \{ \mathit{x} \mathord{:} \ottnt{B} \mathrel{\mid} \ottnt{e_{{\mathrm{11}}}} \}   }{\Rightarrow} }  \ottnt{T_{{\mathrm{1}}}}  \rangle^{ \ottnt{l} } ~   \langle   \{ \mathit{x} \mathord{:} \ottnt{B} \mathrel{\mid} \ottnt{e_{{\mathrm{11}}}} \}  ,   \ottnt{e_{{\mathrm{11}}}}  [  \ottnt{k} / \mathit{x}  ]  ,  \ottnt{k}  \rangle^{ \ottnt{l'} }   )    \longrightarrow _{  \mathsf{H}  }  {} \\
         \langle   \{ \mathit{x} \mathord{:} \ottnt{B} \mathrel{\mid} \ottnt{e_{{\mathrm{11}}}} \}   \mathord{ \overset{  \ottsym{(}    \mathcal{S}_{{\mathrm{1}}}  \cup  \mathcal{S}_{{\mathrm{2}}}   \cup   \set{  \ottnt{T_{{\mathrm{1}}}}  }    \ottsym{)}  \setminus   \{ \mathit{x} \mathord{:} \ottnt{B} \mathrel{\mid} \ottnt{e_{{\mathrm{11}}}} \}   }{\Rightarrow} }  \ottnt{T_{{\mathrm{2}}}}  \rangle^{ \ottnt{l} } ~   \langle   \{ \mathit{x} \mathord{:} \ottnt{B} \mathrel{\mid} \ottnt{e_{{\mathrm{11}}}} \}  ,   \ottnt{e_{{\mathrm{11}}}}  [  \ottnt{k} / \mathit{x}  ]  ,  \ottnt{k}  \rangle^{ \ottnt{l'} }  
      \end{array} \]      
      Knowing that $ \langle   \{ \mathit{x} \mathord{:} \ottnt{B} \mathrel{\mid} \ottnt{e_{{\mathrm{11}}}} \}   \mathord{ \overset{  \mathcal{S}_{{\mathrm{2}}}  \setminus   \{ \mathit{x} \mathord{:} \ottnt{B} \mathrel{\mid} \ottnt{e_{{\mathrm{11}}}} \}   }{\Rightarrow} }  \ottnt{T_{{\mathrm{2}}}}  \rangle^{ \ottnt{l} } ~   \langle   \{ \mathit{x} \mathord{:} \ottnt{B} \mathrel{\mid} \ottnt{e_{{\mathrm{11}}}} \}  ,   \ottnt{e_{{\mathrm{11}}}}  [  \ottnt{k} / \mathit{x}  ]  ,  \ottnt{k}  \rangle^{ \ottnt{l'} }   \,  \longrightarrow ^{*}_{  \mathsf{H}  }  \, \ottnt{e_{{\mathrm{2}}}}$, we know that $ \ottnt{e_{{\mathrm{11}}}}  [  \ottnt{k} / \mathit{x}  ]  \,  \longrightarrow ^{*}_{  \mathsf{H}  }  \, \ottnt{e'_{{\mathrm{11}}}}$ such that
      $ \mathsf{result} _{  \mathsf{H}  }~ \ottnt{e'_{{\mathrm{11}}}} $.
      If it goes to $ \mathord{\Uparrow}  \ottnt{l''} $, so does the $\ottnt{e_{{\mathrm{1}}}}$ and
      by \E{CheckRaise} followed by \E{CastRaise}. 
      The $\ottnt{e}$ term, depending on what $ \mathsf{choose} $ selects, 
      produces either a different blame label (because the types it
      checks first fail) or it eventually chooses $ \{ \mathit{x} \mathord{:} \ottnt{B} \mathrel{\mid} \ottnt{e_{{\mathrm{11}}}} \} $
      and raises blame, too. Note that here we are relying critically
      on strong normalization, Lemma~\ref{lem:heedfulsn} and
      \SWF{TypeSet}, to see that all checks reduce to results on both
      sides.
      A similar case adheres when the check goes to $ \mathsf{false} $.
      Finally, if it goes to $ \mathsf{true} $, then we know that the $\ottnt{e_{{\mathrm{1}}}}$ term reduces
      $ \langle   \{ \mathit{x} \mathord{:} \ottnt{B} \mathrel{\mid} \ottnt{e_{{\mathrm{11}}}} \}   \mathord{ \overset{  \ottsym{(}    \mathcal{S}_{{\mathrm{1}}}  \cup  \mathcal{S}_{{\mathrm{2}}}   \cup   \set{  \ottnt{T_{{\mathrm{1}}}}  }    \ottsym{)}  \setminus   \{ \mathit{x} \mathord{:} \ottnt{B} \mathrel{\mid} \ottnt{e_{{\mathrm{11}}}} \}   }{\Rightarrow} }  \ottnt{T_{{\mathrm{2}}}}  \rangle^{ \ottnt{l} } ~   \langle   \{ \mathit{x} \mathord{:} \ottnt{B} \mathrel{\mid} \ottnt{e_{{\mathrm{11}}}} \}  ,   \ottnt{e_{{\mathrm{11}}}}  [  \ottnt{k} / \mathit{x}  ]  ,  \ottnt{k}  \rangle^{ \ottnt{l'} }   \,  \longrightarrow ^{*}_{  \mathsf{H}  }  \,  \langle   \{ \mathit{x} \mathord{:} \ottnt{B} \mathrel{\mid} \ottnt{e_{{\mathrm{11}}}} \}   \mathord{ \overset{  \ottsym{(}    \mathcal{S}_{{\mathrm{1}}}  \cup  \mathcal{S}_{{\mathrm{2}}}   \cup   \set{  \ottnt{T_{{\mathrm{1}}}}  }    \ottsym{)}  \setminus   \{ \mathit{x} \mathord{:} \ottnt{B} \mathrel{\mid} \ottnt{e_{{\mathrm{11}}}} \}   }{\Rightarrow} }  \ottnt{T_{{\mathrm{2}}}}  \rangle^{ \ottnt{l} } ~  \ottnt{k} $ on its way to the result $\ottnt{e_{{\mathrm{2}}}}$
      We can then reduce the two terms together as new types are
      chosen (from $   \mathcal{S}_{{\mathrm{1}}}  \cup  \mathcal{S}_{{\mathrm{2}}}   \cup   \set{  \ottnt{T_{{\mathrm{1}}}}  }    \setminus   \{ \mathit{x} \mathord{:} \ottnt{B} \mathrel{\mid} \ottnt{e_{{\mathrm{11}}}} \}  $ and from $  \mathcal{S}_{{\mathrm{1}}}  \cup  \mathcal{S}_{{\mathrm{2}}}   \cup   \set{  \ottnt{T_{{\mathrm{1}}}}  }  $) until $ \{ \mathit{x} \mathord{:} \ottnt{B} \mathrel{\mid} \ottnt{e_{{\mathrm{11}}}} \} $ is eliminated from the
      $\ottnt{e}$-term's type set, and the two terms are the same.

    \item[(\E{CastInner})] We have $\ottnt{e} =  \langle  \ottnt{T_{{\mathrm{3}}}}  \mathord{ \overset{ \mathcal{S}_{{\mathrm{2}}} }{\Rightarrow} }  \ottnt{T_{{\mathrm{1}}}}  \rangle^{ \ottnt{l'} } ~  \ottnt{e_{{\mathrm{11}}}}  \,  \longrightarrow _{  \mathsf{H}  }  \,  \langle  \ottnt{T_{{\mathrm{3}}}}  \mathord{ \overset{ \mathcal{S}_{{\mathrm{2}}} }{\Rightarrow} }  \ottnt{T_{{\mathrm{1}}}}  \rangle^{ \ottnt{l'} } ~  \ottnt{e_{{\mathrm{12}}}}  = \ottnt{e_{{\mathrm{1}}}}$, with $\ottnt{e_{{\mathrm{11}}}} \,  \longrightarrow _{  \mathsf{H}  }  \, \ottnt{e_{{\mathrm{12}}}}$ and
      $\ottnt{e_{{\mathrm{11}}}}  \neq   \langle  \ottnt{T_{{\mathrm{4}}}}  \mathord{ \overset{\bullet}{\Rightarrow} }  \ottnt{T_{{\mathrm{3}}}}  \rangle^{ \ottnt{l''} } ~  \ottnt{e''_{{\mathrm{2}}}} $. We reduce both to the common
      term $ \langle  \ottnt{T_{{\mathrm{3}}}}  \mathord{ \overset{   \mathcal{S}_{{\mathrm{1}}}  \cup  \mathcal{S}_{{\mathrm{2}}}   \cup   \set{  \ottnt{T_{{\mathrm{1}}}}  }   }{\Rightarrow} }  \ottnt{T_{{\mathrm{2}}}}  \rangle^{ \ottnt{l} } ~  \ottnt{e_{{\mathrm{12}}}} $.

      In the original derivation with $\ottnt{e_{{\mathrm{1}}}}$, the only step we can take is $ \langle  \ottnt{T_{{\mathrm{1}}}}  \mathord{ \overset{ \mathcal{S}_{{\mathrm{1}}} }{\Rightarrow} }  \ottnt{T_{{\mathrm{2}}}}  \rangle^{ \ottnt{l} } ~   (  \langle  \ottnt{T_{{\mathrm{3}}}}  \mathord{ \overset{ \mathcal{S}_{{\mathrm{2}}} }{\Rightarrow} }  \ottnt{T_{{\mathrm{1}}}}  \rangle^{ \ottnt{l'} } ~  \ottnt{e_{{\mathrm{12}}}}  )   \,  \longrightarrow _{  \mathsf{H}  }  \,  \langle  \ottnt{T_{{\mathrm{3}}}}  \mathord{ \overset{   \mathcal{S}_{{\mathrm{1}}}  \cup  \mathcal{S}_{{\mathrm{2}}}   \cup   \set{  \ottnt{T_{{\mathrm{1}}}}  }   }{\Rightarrow} }  \ottnt{T_{{\mathrm{2}}}}  \rangle^{ \ottnt{l} } ~  \ottnt{e_{{\mathrm{12}}}} $
      \E{CastMerge}. We find a new derivation with $\ottnt{e}$ as
      follows:
      \[ \begin{array}{rlr}
        &  \langle  \ottnt{T_{{\mathrm{1}}}}  \mathord{ \overset{ \mathcal{S}_{{\mathrm{1}}} }{\Rightarrow} }  \ottnt{T_{{\mathrm{2}}}}  \rangle^{ \ottnt{l} } ~   (  \langle  \ottnt{T_{{\mathrm{3}}}}  \mathord{ \overset{ \mathcal{S}_{{\mathrm{2}}} }{\Rightarrow} }  \ottnt{T_{{\mathrm{1}}}}  \rangle^{ \ottnt{l'} } ~  \ottnt{e_{{\mathrm{11}}}}  )   & \E{CastMerge} \\
         \longrightarrow _{  \mathsf{H}  } &  \langle  \ottnt{T_{{\mathrm{3}}}}  \mathord{ \overset{   \mathcal{S}_{{\mathrm{1}}}  \cup  \mathcal{S}_{{\mathrm{2}}}   \cup   \set{  \ottnt{T_{{\mathrm{1}}}}  }   }{\Rightarrow} }  \ottnt{T_{{\mathrm{2}}}}  \rangle^{ \ottnt{l} } ~  \ottnt{e_{{\mathrm{11}}}}  & \E{CastInner} \text{~since $\ottnt{e_{{\mathrm{11}}}}  \neq   \langle  \ottnt{T_{{\mathrm{4}}}}  \mathord{ \overset{\bullet}{\Rightarrow} }  \ottnt{T_{{\mathrm{3}}}}  \rangle^{ \ottnt{l''} } ~  \ottnt{e''_{{\mathrm{2}}}} $} \\
         \longrightarrow _{  \mathsf{H}  } &  \langle  \ottnt{T_{{\mathrm{3}}}}  \mathord{ \overset{   \mathcal{S}_{{\mathrm{1}}}  \cup  \mathcal{S}_{{\mathrm{2}}}   \cup   \set{  \ottnt{T_{{\mathrm{1}}}}  }   }{\Rightarrow} }  \ottnt{T_{{\mathrm{2}}}}  \rangle^{ \ottnt{l} } ~  \ottnt{e_{{\mathrm{12}}}}  & \text{(assumption)}
      \end{array} \]

    \item[(\E{CastMerge})] We have $\ottnt{e} =  \langle  \ottnt{T_{{\mathrm{3}}}}  \mathord{ \overset{ \mathcal{S}_{{\mathrm{2}}} }{\Rightarrow} }  \ottnt{T_{{\mathrm{1}}}}  \rangle^{ \ottnt{l'} } ~   (  \langle  \ottnt{T_{{\mathrm{4}}}}  \mathord{ \overset{ \mathcal{S}_{{\mathrm{3}}} }{\Rightarrow} }  \ottnt{T_{{\mathrm{3}}}}  \rangle^{ \ottnt{l''} } ~  \ottnt{e_{{\mathrm{11}}}}  )   \,  \longrightarrow _{  \mathsf{H}  }  \,  \langle  \ottnt{T_{{\mathrm{4}}}}  \mathord{ \overset{   \mathcal{S}_{{\mathrm{2}}}  \cup  \mathcal{S}_{{\mathrm{3}}}   \cup   \set{  \ottnt{T_{{\mathrm{3}}}}  }   }{\Rightarrow} }  \ottnt{T_{{\mathrm{1}}}}  \rangle^{ \ottnt{l'} } ~  \ottnt{e_{{\mathrm{11}}}}  = \ottnt{e_{{\mathrm{1}}}}$.

      In the original derivation with $\ottnt{e_{{\mathrm{1}}}}$, the only step we can
      take is $ \langle  \ottnt{T_{{\mathrm{1}}}}  \mathord{ \overset{ \mathcal{S}_{{\mathrm{1}}} }{\Rightarrow} }  \ottnt{T_{{\mathrm{2}}}}  \rangle^{ \ottnt{l} } ~   (  \langle  \ottnt{T_{{\mathrm{4}}}}  \mathord{ \overset{   \mathcal{S}_{{\mathrm{2}}}  \cup  \mathcal{S}_{{\mathrm{3}}}   \cup   \set{  \ottnt{T_{{\mathrm{3}}}}  }   }{\Rightarrow} }  \ottnt{T_{{\mathrm{1}}}}  \rangle^{ \ottnt{l'} } ~  \ottnt{e_{{\mathrm{11}}}}  )   \,  \longrightarrow _{  \mathsf{H}  }  \,  \langle  \ottnt{T_{{\mathrm{4}}}}  \mathord{ \overset{     \mathcal{S}_{{\mathrm{1}}}  \cup  \mathcal{S}_{{\mathrm{2}}}   \cup  \mathcal{S}_{{\mathrm{3}}}   \cup   \set{  \ottnt{T_{{\mathrm{1}}}}  }    \cup   \set{  \ottnt{T_{{\mathrm{3}}}}  }   }{\Rightarrow} }  \ottnt{T_{{\mathrm{2}}}}  \rangle^{ \ottnt{l} } ~  \ottnt{e_{{\mathrm{11}}}} $. We can build a new
      derivation with $\ottnt{e}$ as follows:
      \[ \begin{array}{rlr}
        &  \langle  \ottnt{T_{{\mathrm{1}}}}  \mathord{ \overset{ \mathcal{S}_{{\mathrm{1}}} }{\Rightarrow} }  \ottnt{T_{{\mathrm{2}}}}  \rangle^{ \ottnt{l} } ~   (  \langle  \ottnt{T_{{\mathrm{3}}}}  \mathord{ \overset{ \mathcal{S}_{{\mathrm{2}}} }{\Rightarrow} }  \ottnt{T_{{\mathrm{1}}}}  \rangle^{ \ottnt{l'} } ~   (  \langle  \ottnt{T_{{\mathrm{4}}}}  \mathord{ \overset{ \mathcal{S}_{{\mathrm{3}}} }{\Rightarrow} }  \ottnt{T_{{\mathrm{3}}}}  \rangle^{ \ottnt{l''} } ~  \ottnt{e_{{\mathrm{11}}}}  )   )   & \E{CastMerge} \\
         \longrightarrow _{  \mathsf{H}  } &  \langle  \ottnt{T_{{\mathrm{3}}}}  \mathord{ \overset{   \mathcal{S}_{{\mathrm{1}}}  \cup  \mathcal{S}_{{\mathrm{2}}}   \cup   \set{  \ottnt{T_{{\mathrm{1}}}}  }   }{\Rightarrow} }  \ottnt{T_{{\mathrm{2}}}}  \rangle^{ \ottnt{l} } ~   (  \langle  \ottnt{T_{{\mathrm{4}}}}  \mathord{ \overset{ \mathcal{S}_{{\mathrm{3}}} }{\Rightarrow} }  \ottnt{T_{{\mathrm{3}}}}  \rangle^{ \ottnt{l''} } ~  \ottnt{e_{{\mathrm{11}}}}  )   & \E{CastMerge} \\
         \longrightarrow _{  \mathsf{H}  } &  \langle  \ottnt{T_{{\mathrm{4}}}}  \mathord{ \overset{     \mathcal{S}_{{\mathrm{1}}}  \cup  \mathcal{S}_{{\mathrm{2}}}   \cup  \mathcal{S}_{{\mathrm{3}}}   \cup   \set{  \ottnt{T_{{\mathrm{1}}}}  }    \cup   \set{  \ottnt{T_{{\mathrm{3}}}}  }   }{\Rightarrow} }  \ottnt{T_{{\mathrm{2}}}}  \rangle^{ \ottnt{l} } ~  \ottnt{e_{{\mathrm{11}}}}  & \text{(assumption)} \\
      \end{array} \]

    \item[(\E{CastRaise})] By \E{CastMerge}, we can reduce $\ottnt{e}$ to
      $ \langle  \ottnt{T_{{\mathrm{1}}}}  \mathord{ \overset{ \mathcal{S} }{\Rightarrow} }  \ottnt{T_{{\mathrm{2}}}}  \rangle^{ \ottnt{l} } ~   \mathord{\Uparrow}  \ottnt{l'}  $, which is just the same term as
      $ \langle  \ottnt{T_{{\mathrm{1}}}}  \mathord{ \overset{ \mathcal{S} }{\Rightarrow} }  \ottnt{T_{{\mathrm{2}}}}  \rangle^{ \ottnt{l} } ~  \ottnt{e_{{\mathrm{1}}}} $.
    \end{itemize}
    \fi}
  \end{proof}
\end{lemma}

\begin{lemma}[Cast congruence]
  \label{lem:heedfulcastcongruence}
  If
  \begin{itemize}
  \item $\emptyset  \models  \ottnt{e}  \ottsym{:}  \ottnt{T_{{\mathrm{1}}}}$ and $\models  \mathcal{S}  \mathrel{\parallel}  \ottnt{T_{{\mathrm{1}}}}  \Rightarrow  \ottnt{T_{{\mathrm{2}}}}$ (and so
    $ \emptyset   \vdash _{  \mathsf{H}  }   \langle  \ottnt{T_{{\mathrm{1}}}}  \mathord{ \overset{ \mathcal{S} }{\Rightarrow} }  \ottnt{T_{{\mathrm{2}}}}  \rangle^{ \ottnt{l} } ~  \ottnt{e}   :  \ottnt{T_{{\mathrm{2}}}} $),
  \item $\ottnt{e} \,  \longrightarrow ^{*}_{  \mathsf{H}  }  \, \ottnt{e_{{\mathrm{1}}}}$ (and so $ \emptyset   \vdash _{  \mathsf{H}  }  \ottnt{e_{{\mathrm{1}}}}  :  \ottnt{T_{{\mathrm{1}}}} $),
  \item $ \langle  \ottnt{T_{{\mathrm{1}}}}  \mathord{ \overset{ \mathcal{S} }{\Rightarrow} }  \ottnt{T_{{\mathrm{2}}}}  \rangle^{ \ottnt{l} } ~  \ottnt{e_{{\mathrm{1}}}}  \,  \longrightarrow ^{*}_{  \mathsf{H}  }  \, \ottnt{e_{{\mathrm{2}}}}$, and
  \item $ \mathsf{result} _{  \mathsf{H}  }~ \ottnt{e_{{\mathrm{2}}}} $
  \end{itemize}
  then $ \langle  \ottnt{T_{{\mathrm{1}}}}  \mathord{ \overset{ \mathcal{S} }{\Rightarrow} }  \ottnt{T_{{\mathrm{2}}}}  \rangle^{ \ottnt{l} } ~  \ottnt{e}  \,  \longrightarrow ^{*}_{  \mathsf{H}  }  \,  \mathord{\Uparrow}  \ottnt{l'} $ if $\ottnt{e_{{\mathrm{2}}}}  \ottsym{=}   \mathord{\Uparrow}  \ottnt{l} $ or
  to $\ottnt{e_{{\mathrm{2}}}}$ itself if $ \mathsf{val} _{  \mathsf{H}  }~ \ottnt{e_{{\mathrm{2}}}} $. Diagrammatically:
  %
  % \heedfulcongruence
  \begin{center}
  \begin{tikzpicture}[description/.style={fill=white,inner sep=2pt},align at top]
  \tikzset{align at top/.style={baseline=(current bounding box.north)}}

    \matrix (m) [matrix of math nodes, row sep=4pt, nodes in empty cells,
                 text height=1.5ex, text depth=0.25ex]
    { 
                       & \textbf{Heedful \lambdah} & \\
      \ottnt{e_{{\mathrm{1}}}}           & & \ottnt{e_{{\mathrm{2}}}} \\
                       & \Downarrow & \\
       \langle  \ottnt{T_{{\mathrm{1}}}}  \mathord{ \overset{\bullet}{\Rightarrow} }  \ottnt{T_{{\mathrm{2}}}}  \rangle^{ \ottnt{l} } ~  \ottnt{e_{{\mathrm{1}}}}  & &  \langle  \ottnt{T_{{\mathrm{1}}}}  \mathord{ \overset{\bullet}{\Rightarrow} }  \ottnt{T_{{\mathrm{2}}}}  \rangle^{ \ottnt{l} } ~  \ottnt{e_{{\mathrm{2}}}}  \\[20pt]
       \mathsf{result} _{  \mathsf{H}  }~ \ottnt{e'_{{\mathrm{1}}}}  &  \sim  &  \mathsf{result} _{  \mathsf{H}  }~ \ottnt{e'_{{\mathrm{2}}}}  \\
      &  \mathsf{val} _{  \mathsf{H}  }~ \ottnt{e}  & \\
    };

    \path[->] (m-2-1) edge[H] (m-2-3);
    \path[dashed,->] (m-4-1) edge[H*] (m-5-1);
    \path[->] (m-4-3) edge[H*] (m-5-3);
    \path[color=white] (m-5-1) edge node[sloped,color=black] {$=$} (m-6-2);
    \path[color=white] (m-5-3) edge node[sloped,color=black] {$=$} (m-6-2);
  \end{tikzpicture}
  \end{center}
  \begin{proof}
    By induction on the derivation $\ottnt{e} \,  \longrightarrow ^{*}_{  \mathsf{H}  }  \, \ottnt{e_{{\mathrm{1}}}}$, using the
    single-step cast congruence
    (Lemma~\ref{lem:heedfulcastcongruencesinglestep}).
  \end{proof}
\end{lemma}

\begin{lemma}[Strong normalization of casts]
  \label{lem:heedfulsncast}
  If $\models  \mathcal{S}  \mathrel{\parallel}  \ottnt{T_{{\mathrm{1}}}}  \Rightarrow  \ottnt{T_{{\mathrm{2}}}}$ and $\ottnt{e} \, \in \,  \denot{ \ottnt{T_{{\mathrm{1}}}} } $ then $ \langle  \ottnt{T_{{\mathrm{1}}}}  \mathord{ \overset{ \mathcal{S} }{\Rightarrow} }  \ottnt{T_{{\mathrm{2}}}}  \rangle^{ \ottnt{l} } ~  \ottnt{e}  \, \in \,  \denot{ \ottnt{T_{{\mathrm{2}}}} } $.
  \begin{proof}
    By induction on the sum of the heights of $\ottnt{T_{{\mathrm{1}}}}$ and $\ottnt{T_{{\mathrm{2}}}}$.
    {\iffull
    We go by cases on the shape of the types.
    \begin{itemize}
    \item[($\ottnt{T_{\ottmv{i}}}  \ottsym{=}   \{ \mathit{x} \mathord{:} \ottnt{B} \mathrel{\mid} \ottnt{e_{\ottmv{i}}} \} $)] We know that $\ottnt{e}$ goes to blame or a
      constant. In the former case, the entire term goes to blame by
      \E{CastRaise}. Otherwise, we go by \E{CheckSet} and the
      normalization assumptions in \SWF{TypeSet} until we run out of
      types in $\mathcal{S}$, at which time we apply \E{CheckNone} and the
      normalization assumption in $\models  \ottnt{T_{{\mathrm{2}}}}$.
    \item[($\ottnt{T_{\ottmv{i}}}  \ottsym{=}   \ottnt{T_{\ottmv{i}\,{\mathrm{1}}}} \mathord{ \rightarrow } \ottnt{T_{\ottmv{i}\,{\mathrm{2}}}} $)] We know that $\ottnt{e} \, \in \,  \denot{  \ottnt{T_{{\mathrm{1}}}} \mathord{ \rightarrow } \ottnt{T_{{\mathrm{2}}}}  } $, so
      it normalizes to some $\ottnt{e'}$. By cast congruence
      (Lemma~\ref{lem:heedfulcastcongruence}), we know that $ \langle  \ottnt{T_{{\mathrm{1}}}}  \mathord{ \overset{ \mathcal{S} }{\Rightarrow} }  \ottnt{T_{{\mathrm{2}}}}  \rangle^{ \ottnt{l} } ~  \ottnt{e} $ and $ \langle  \ottnt{T_{{\mathrm{1}}}}  \mathord{ \overset{ \mathcal{S} }{\Rightarrow} }  \ottnt{T_{{\mathrm{2}}}}  \rangle^{ \ottnt{l} } ~  \ottnt{e'} $ terminate together.
      We go by cases on the shape of the result $\ottnt{e'} \, \in \,  \denot{ \ottnt{T_{{\mathrm{1}}}} } $.
      \begin{itemize}
      \item[($\ottnt{e'}  \ottsym{=}   \mathord{\Uparrow}  \ottnt{l'} $)] We are done by \E{CastRaise} and
        Lemma~\ref{lem:heedfulsnblame}.
      \item[($\ottnt{e'}  \ottsym{=}   \lambda \mathit{x} \mathord{:} \ottnt{T_{{\mathrm{11}}}} .~  \ottnt{e_{{\mathrm{1}}}} $)] We have a value. We must
        show that $ \langle   \ottnt{T_{{\mathrm{11}}}} \mathord{ \rightarrow } \ottnt{T_{{\mathrm{12}}}}   \mathord{ \overset{ \mathcal{S} }{\Rightarrow} }   \ottnt{T_{{\mathrm{21}}}} \mathord{ \rightarrow } \ottnt{T_{{\mathrm{22}}}}   \rangle^{ \ottnt{l} } ~   \lambda \mathit{x} \mathord{:} \ottnt{T_{{\mathrm{11}}}} .~  \ottnt{e_{{\mathrm{1}}}}   \, \in \,  \denot{  \ottnt{T_{{\mathrm{21}}}} \mathord{ \rightarrow } \ottnt{T_{{\mathrm{22}}}}  } $. Let $\ottnt{e_{{\mathrm{2}}}} \, \in \,  \denot{ \ottnt{T_{{\mathrm{21}}}} } $ be a heedful \lambdah result.
        We can step by \E{Unwrap} and then apply the IH on the smaller
        domain and codomain types.
      \item[($\ottnt{e'}  \ottsym{=}   \langle   \ottnt{T_{{\mathrm{31}}}} \mathord{ \rightarrow } \ottnt{T_{{\mathrm{32}}}}   \mathord{ \overset{ \mathcal{S}' }{\Rightarrow} }   \ottnt{T_{{\mathrm{11}}}} \mathord{ \rightarrow } \ottnt{T_{{\mathrm{12}}}}   \rangle^{ \ottnt{l'} } ~   \lambda \mathit{x} \mathord{:} \ottnt{T_{{\mathrm{31}}}} .~  \ottnt{e_{{\mathrm{1}}}}  $)] We
        step by \E{CastMerge} to:
        \[  \langle   \ottnt{T_{{\mathrm{31}}}} \mathord{ \rightarrow } \ottnt{T_{{\mathrm{32}}}}   \mathord{ \overset{   \mathcal{S}'  \cup  \mathcal{S}   \cup   \set{   \ottnt{T_{{\mathrm{11}}}} \mathord{ \rightarrow } \ottnt{T_{{\mathrm{12}}}}   }   }{\Rightarrow} }   \ottnt{T_{{\mathrm{21}}}} \mathord{ \rightarrow } \ottnt{T_{{\mathrm{22}}}}   \rangle^{ \ottnt{l} } ~   \lambda \mathit{x} \mathord{:} \ottnt{T_{{\mathrm{31}}}} .~  \ottnt{e_{{\mathrm{1}}}}   \]
        Let $\ottnt{e_{{\mathrm{2}}}} \, \in \,  \denot{ \ottnt{T_{{\mathrm{21}}}} } $. We step by \E{Unwrap}, observing that
        we can use cast congruence
        (Lemma~\ref{lem:heedfulcastcongruence}) to factor the domain
        and codomain casts, using the IH to handle $\mathcal{S}$ and
        $ \ottnt{T_{{\mathrm{21}}}} \mathord{ \rightarrow } \ottnt{T_{{\mathrm{22}}}} $ and the assumptions about $\ottnt{e'}$ to handle
        the rest.
      \end{itemize}      
    \end{itemize}
    \fi}
  \end{proof}
\end{lemma}

To be able to use our \textit{semantic} cast congruence lemma, we must
show that all well typed heedful \lambdah terms are in the relation we
define; this proof is standard.

\begin{lemma}[Strong normalization of heedful terms]
  \label{lem:heedfulnormalizes}
  ~

  \noindent
  \begin{itemize}
  \item\label{hn:tm} $ \Gamma   \vdash _{  \mathsf{H}  }  \ottnt{e}  :  \ottnt{T} $ implies $\Gamma  \models  \ottnt{e}  \ottsym{:}  \ottnt{T}$,
  \item\label{hn:ty} $ \mathord{  \vdash _{  \mathsf{H}  } }~ \ottnt{T} $ implies $\models  \ottnt{T}$, and
  \item\label{hn:ts} $ \mathord{  \vdash _{  \mathsf{H}  } }~ \mathcal{S}   \mathrel{\parallel}   \ottnt{T_{{\mathrm{1}}}}  \Rightarrow  \ottnt{T_{{\mathrm{2}}}} $ implies $\models  \mathcal{S}  \mathrel{\parallel}  \ottnt{T_{{\mathrm{1}}}}  \Rightarrow  \ottnt{T_{{\mathrm{2}}}}$.
  \end{itemize}
  \begin{proof}
    By mutual induction on the typing derivations.
    {\iffull
    \paragraph{Term typing \fbox{$ \Gamma   \vdash _{  \mathsf{H}  }  \ottnt{e}  :  \ottnt{T} $}}
    \begin{itemize}
    \item[\T{Var}] We know by assumption that $\sigma  \ottsym{(}  \mathit{x}  \ottsym{)} \, \in \,  \denot{ \ottnt{T} } $.
    \item[\T{Const}] Evaluation is by reflexivity; we find $ \mathsf{ty} ( \ottnt{k} )   \ottsym{=}  \ottnt{B}$ by assumption.
    \item[\T{Abs}] Let $\Gamma  \models  \sigma$. We must show that
      $ \lambda \mathit{x} \mathord{:} \ottnt{T_{{\mathrm{1}}}} .~  \sigma  \ottsym{(}  \ottnt{e_{{\mathrm{1}}}}  \ottsym{)}  \, \in \,  \denot{  \ottnt{T_{{\mathrm{1}}}} \mathord{ \rightarrow } \ottnt{T_{{\mathrm{2}}}}  } $. Let $\ottnt{e_{{\mathrm{2}}}} \, \in \,  \denot{ \ottnt{T_{{\mathrm{1}}}} } $. We
      must show that applying the abstraction to the result yields
      related values. If $\ottnt{e_{{\mathrm{2}}}}$ is blame we are done; if not, we
      step by \E{Beta}, to $ \sigma  \ottsym{(}  \ottnt{e_{{\mathrm{1}}}}  \ottsym{)}  [  \ottnt{e_{{\mathrm{2}}}} / \mathit{x}  ] $. But $ \Gamma , \mathit{x} \mathord{:} \ottnt{T_{{\mathrm{1}}}}   \models   \sigma  [  \ottnt{e_{{\mathrm{2}}}} / \mathit{x}  ] $, so we can apply IH (\ref{hn:tm}) on $\ottnt{e_{{\mathrm{1}}}}$.
    \item[\T{Op}] By IH (\ref{hn:tm}) on each argument, either one
      of the arguments goes to blame, we are done by \E{OpRaise}, or,
      all of the arguments normalize. We then reduce by \E{Op} on to
      have $\denot{ op } \, \ottsym{(}  \ottnt{k_{{\mathrm{1}}}}  \ottsym{,}  \dots  \ottsym{,}  \ottnt{k_{\ottmv{n}}}  \ottsym{)}$. We have assumed that the
      denotations of operations agree with their typings in
      \textit{all} modes, so then $\denot{ op } \, \ottsym{(}  \ottnt{k_{{\mathrm{1}}}}  \ottsym{,}  \dots  \ottsym{,}  \ottnt{k_{\ottmv{n}}}  \ottsym{)}$ produces a
      constant of appropriate base type (and, in fact, refinement) for
      $ \longrightarrow _{  \mathsf{H}  } $ in particular, and we are done.
    \item[\T{App}] Let $\Gamma  \models  \sigma$. We must show that
      $ \sigma  \ottsym{(}  \ottnt{e_{{\mathrm{1}}}}  \ottsym{)} ~ \sigma  \ottsym{(}  \ottnt{e_{{\mathrm{2}}}}  \ottsym{)}  \, \in \,  \denot{ \ottnt{T_{{\mathrm{2}}}} } $. But by IH (\ref{hn:tm}) on
      $\ottnt{e_{{\mathrm{1}}}}$ and $\ottnt{e_{{\mathrm{2}}}}$, we are done directly.
    \item[\T{Cast}] By Lemma~\ref{lem:heedfulsncast}, using IH
      (\ref{hn:ts}) on the type set and IH (\ref{hn:tm}) on the term.
    \item[\T{Blame}] By Lemma~\ref{lem:heedfulsnblame}.
    \item[\T{Check}] By IH (\ref{hn:tm}), we know that the active
      check reduces to a boolean or blame, which then reduces to blame
      or the appropriate constant $\ottnt{k}$.
    \end{itemize}
    
    \paragraph{Type well formedness \fbox{$ \mathord{  \vdash _{  \mathsf{H}  } }~ \ottnt{T} $}}
    \begin{itemize}
    \item[\WF{Base}] We can immediately see $  \mathsf{true}   [  \ottnt{k} / \mathit{x}  ]  \, \in \,  \denot{  \{ \mathit{x} \mathord{:}  \mathsf{Bool}  \mathrel{\mid}  \mathsf{true}  \}  } $ by reflexivity and definition of constants.
    \item[\WF{Refine}] By inversion, we know that $  \mathit{x} \mathord{:}  \{ \mathit{x} \mathord{:} \ottnt{B} \mathrel{\mid}  \mathsf{true}  \}     \vdash _{  \mathsf{H}  }  \ottnt{e}  :   \{ \mathit{x} \mathord{:}  \mathsf{Bool}  \mathrel{\mid}  \mathsf{true}  \}  $; by IH (\ref{hn:tm}), we find that
      $\sigma  \ottsym{(}  \ottnt{e}  \ottsym{)} \, \in \,  \denot{  \{ \mathit{x} \mathord{:}  \mathsf{Bool}  \mathrel{\mid}  \mathsf{true}  \}  } $, i.e., that $ \ottnt{e}  [  \ottnt{e_{{\mathrm{2}}}} / \mathit{x}  ]  \, \in \,  \denot{  \{ \mathit{x} \mathord{:}  \mathsf{Bool}  \mathrel{\mid}  \mathsf{true}  \}  } $ for all $\ottnt{e_{{\mathrm{2}}}} \, \in \,  \denot{  \{ \mathit{x} \mathord{:} \ottnt{B} \mathrel{\mid}  \mathsf{true}  \}  } $---which is
      what we needed to know.
    \item[\WF{Fun}] By IH (\ref{hn:ty}) on each of the types.
    \end{itemize}

    \paragraph{Type set well formedness \fbox{$ \mathord{  \vdash _{  \mathsf{H}  } }~ \mathcal{S}   \mathrel{\parallel}   \ottnt{T_{{\mathrm{1}}}}  \Rightarrow  \ottnt{T_{{\mathrm{2}}}} $}}
    \begin{itemize}
    \item[\WF{TypeSet}] By IH (\ref{hn:ty}) on each of the types.
    \end{itemize}
    \fi}
  \end{proof}
\end{lemma}

We define the logical relation in Figure~\reflr.
The main difference is that this relation is \textit{symmetric}:
classic and heedful \lambdah yield blame or values iff the other one
does, thought the blame labels may be different. The formulations are
otherwise the same, and the proof proceeds similarly---though heedful
\lambdah's more complicated cast merging leads to some more intricate
stepping in the cast lemma.

\begin{lemma}[Value relation relates only values]
  \label{lem:heedfullrvalue}
  If $ \ottnt{e_{{\mathrm{1}}}}   \sim _{  \mathsf{H}  }  \ottnt{e_{{\mathrm{2}}}}  :  \ottnt{T} $ then $ \mathsf{val} _{  \mathsf{C}  }~ \ottnt{e_{{\mathrm{1}}}} $ and $ \mathsf{val} _{  \mathsf{H}  }~ \ottnt{e_{{\mathrm{2}}}} $.
  \begin{proof}
    By induction on $\ottnt{T}$. We have $\ottnt{e_{{\mathrm{1}}}}  \ottsym{=}  \ottnt{e_{{\mathrm{2}}}} = \ottnt{k}$ when $\ottnt{T}  \ottsym{=}   \{ \mathit{x} \mathord{:} \ottnt{B} \mathrel{\mid} \ottnt{e} \} $ (and so we are done by \V{Const}). When $\ottnt{T}  \ottsym{=}   \ottnt{T_{{\mathrm{1}}}} \mathord{ \rightarrow } \ottnt{T_{{\mathrm{2}}}} $, we have the value derivations as assumptions.
  \end{proof}  
\end{lemma}

\begin{lemma}[Relation implies similarity]
  \label{lem:heedfullrsimilar}
  If $ \ottnt{T_{{\mathrm{1}}}}   \sim _{  \mathsf{H}  }  \ottnt{T_{{\mathrm{2}}}} $ then $\vdash  \ottnt{T_{{\mathrm{1}}}}  \mathrel{\parallel}  \ottnt{T_{{\mathrm{2}}}}$.
  \begin{proof}
    By induction on $\ottnt{T_{{\mathrm{1}}}}$, using \S{Refine} and \S{Fun}.
  \end{proof}
\end{lemma}

\begin{lemma}[Relating classic and heedful casts]
  \label{lem:heedfullrcast}
  If $ \ottnt{T_{{\mathrm{11}}}}   \sim _{  \mathsf{H}  }  \ottnt{T_{{\mathrm{21}}}} $ and $ \ottnt{T_{{\mathrm{12}}}}   \sim _{  \mathsf{H}  }  \ottnt{T_{{\mathrm{22}}}} $ and $\vdash  \ottnt{T_{{\mathrm{11}}}}  \mathrel{\parallel}  \ottnt{T_{{\mathrm{12}}}}$,
  then forall $ \ottnt{e_{{\mathrm{1}}}}   \sim _{  \mathsf{H}  }  \ottnt{e_{{\mathrm{2}}}}  :  \ottnt{T_{{\mathrm{21}}}} $, we have $  \langle  \ottnt{T_{{\mathrm{11}}}}  \mathord{ \overset{\bullet}{\Rightarrow} }  \ottnt{T_{{\mathrm{12}}}}  \rangle^{ \ottnt{l} } ~  \ottnt{e_{{\mathrm{1}}}}    \simeq _{  \mathsf{H}  }   \langle  \ottnt{T_{{\mathrm{21}}}}  \mathord{ \overset{\bullet}{\Rightarrow} }  \ottnt{T_{{\mathrm{22}}}}  \rangle^{ \ottnt{l'} } ~  \ottnt{e_{{\mathrm{2}}}}   :  \ottnt{T_{{\mathrm{22}}}} $.
  \begin{proof}
    By induction on the sum of the heights of $\ottnt{T_{{\mathrm{21}}}}$ and
    $\ottnt{T_{{\mathrm{22}}}}$.
    {\iffull
    By Lemma~\ref{lem:heedfullrsimilar}, we know that
    $\vdash  \ottnt{T_{{\mathrm{11}}}}  \mathrel{\parallel}  \ottnt{T_{{\mathrm{21}}}}$ and $\vdash  \ottnt{T_{{\mathrm{12}}}}  \mathrel{\parallel}  \ottnt{T_{{\mathrm{22}}}}$; by
    Lemma~\ref{lem:similaritytransitive}, we know that $\vdash  \ottnt{T_{{\mathrm{21}}}}  \mathrel{\parallel}  \ottnt{T_{{\mathrm{22}}}}$. We go by cases on $\ottnt{T_{{\mathrm{22}}}}$.
    The heedful term first steps by \E{TypeSet}, replacing its
    $ \bullet $ annotation with an empty set.
    \begin{itemize}
    \item[($\ottnt{T_{{\mathrm{22}}}}  \ottsym{=}   \{ \mathit{x} \mathord{:} \ottnt{B} \mathrel{\mid} \ottnt{e_{{\mathrm{22}}}} \} $)] It must be the case (by similarity)
      that all of the other types are also refinements. Moreover, it
      must be that case that $\ottnt{e_{{\mathrm{1}}}}  \ottsym{=}  \ottnt{e_{{\mathrm{2}}}} = \ottnt{k}$.

      Classic steps by \E{CheckNone}, while heedful steps by
      \E{CheckEmpty}. Since $ \ottnt{e_{{\mathrm{1}}}}   \sim _{  \mathsf{H}  }  \ottnt{e_{{\mathrm{2}}}}  :  \ottnt{T_{{\mathrm{21}}}}  =  \{ \mathit{x} \mathord{:} \ottnt{B} \mathrel{\mid} \ottnt{e_{{\mathrm{21}}}} \} $, we
      can find that $ \ottnt{e_{{\mathrm{1}}}}   \sim _{  \mathsf{H}  }  \ottnt{e_{{\mathrm{2}}}}  :   \{ \mathit{x} \mathord{:} \ottnt{B} \mathrel{\mid}  \mathsf{true}  \}  $ trivially. Then, since
      $  \{ \mathit{x} \mathord{:} \ottnt{B} \mathrel{\mid} \ottnt{e_{{\mathrm{12}}}} \}    \sim _{  \mathsf{H}  }   \{ \mathit{x} \mathord{:} \ottnt{B} \mathrel{\mid} \ottnt{e_{{\mathrm{22}}}} \}  $, we know that $  \ottnt{e_{{\mathrm{12}}}}  [  \ottnt{k} / \mathit{x}  ]    \simeq _{  \mathsf{H}  }   \ottnt{e_{{\mathrm{22}}}}  [  \ottnt{k} / \mathit{x}  ]   :   \{ \mathit{x} \mathord{:}  \mathsf{Bool}  \mathrel{\mid}  \mathsf{true}  \}  $.

      If $ \ottnt{e_{{\mathrm{12}}}}  [  \ottnt{k} / \mathit{x}  ]  \,  \longrightarrow ^{*}_{  \mathsf{C}  }  \,  \mathord{\Uparrow}  \ottnt{l'} $, then $ \ottnt{e_{{\mathrm{22}}}}  [  \ottnt{k} / \mathit{x}  ]  \,  \longrightarrow ^{*}_{  \mathsf{H}  }  \,  \mathord{\Uparrow}  \ottnt{l''} $, and both terms reduce to blame by \E{CheckInner}
      and \E{CheckRaise}---this completes the proof.
      If not, then both predicates reduce to a boolean together. If
      they reduce to $ \mathsf{false} $, then both terms eventually
      reduces to $ \mathord{\Uparrow}  \ottnt{l} $ via \E{CheckInner} and \E{CheckFail},
      and we are done.
      If they both go to $ \mathsf{true} $, then both sides step by
      \E{CheckInner} and \E{CheckOK} to yield $\ottnt{k}$, and we can find
      $ \ottnt{k}   \sim _{  \mathsf{H}  }  \ottnt{k}  :   \{ \mathit{x} \mathord{:} \ottnt{B} \mathrel{\mid} \ottnt{e_{{\mathrm{22}}}} \}  $ easily---we have a derivation for
      $ \ottnt{e_{{\mathrm{22}}}}  [  \ottnt{k} / \mathit{x}  ]  \,  \longrightarrow ^{*}_{  \mathsf{H}  }  \,  \mathsf{true} $ handy.

    \item[($\ottnt{T_{{\mathrm{22}}}}  \ottsym{=}   \ottnt{T_{{\mathrm{221}}}} \mathord{ \rightarrow } \ottnt{T_{{\mathrm{222}}}} $)] By Lemma~\ref{lem:heedfullrvalue},
      we know that $ \mathsf{val} _{  \mathsf{C}  }~ \ottnt{e_{{\mathrm{1}}}} $ and $ \mathsf{val} _{  \mathsf{H}  }~ \ottnt{e_{{\mathrm{2}}}} $. So the classic
      side is a value $\ottnt{e_{{\mathrm{11}}}}$ (by \V{ProxyC}), while the heedful
      side either steps by \E{CastMerge} to produce a function proxy
      $\ottnt{e_{{\mathrm{21}}}}$, or immediately has one, depending on the shape of
      $\ottnt{e_{{\mathrm{2}}}}$: an abstraction immediately yields a value by
      \V{ProxyH}, or \E{CastMerge} for a function proxy (again
      yielding a value by \V{ProxyF}).

      We must now show that $ \ottnt{e_{{\mathrm{11}}}}   \sim _{  \mathsf{H}  }  \ottnt{e_{{\mathrm{21}}}}  :   \ottnt{T_{{\mathrm{221}}}} \mathord{ \rightarrow } \ottnt{T_{{\mathrm{222}}}}  $, knowing
      that $ \ottnt{e_{{\mathrm{1}}}}   \sim _{  \mathsf{H}  }  \ottnt{e_{{\mathrm{2}}}}  :   \ottnt{T_{{\mathrm{211}}}} \mathord{ \rightarrow } \ottnt{T_{{\mathrm{212}}}}  $. Let $ \ottnt{e_{{\mathrm{12}}}}   \sim _{  \mathsf{H}  }  \ottnt{e_{{\mathrm{22}}}}  :  \ottnt{T_{{\mathrm{221}}}} $ be
      given. On the classic side, we step by \E{Unwrap} to find
      $ \langle  \ottnt{T_{{\mathrm{112}}}}  \mathord{ \overset{\bullet}{\Rightarrow} }  \ottnt{T_{{\mathrm{122}}}}  \rangle^{ \ottnt{l} } ~   (  \ottnt{e_{{\mathrm{1}}}} ~  (  \langle  \ottnt{T_{{\mathrm{121}}}}  \mathord{ \overset{\bullet}{\Rightarrow} }  \ottnt{T_{{\mathrm{111}}}}  \rangle^{ \ottnt{l} } ~  \ottnt{e_{{\mathrm{12}}}}  )   )  $. (Recall that the
      annotations are all empty.)

      We now go by cases on whether or not $\ottnt{e_{{\mathrm{2}}}}$ had to take a step
      to become a value:
      \begin{itemize}
      \item[(\V{ProxyH})] We have $\ottnt{e_{{\mathrm{21}}}}  \ottsym{=}   \langle   \ottnt{T_{{\mathrm{211}}}} \mathord{ \rightarrow } \ottnt{T_{{\mathrm{212}}}}   \mathord{ \overset{ \emptyset }{\Rightarrow} }   \ottnt{T_{{\mathrm{221}}}} \mathord{ \rightarrow } \ottnt{T_{{\mathrm{222}}}}   \rangle^{ \ottnt{l} } ~   \lambda \mathit{x} \mathord{:} \ottnt{T_{{\mathrm{211}}}} .~  \ottnt{e'_{{\mathrm{2}}}}  $ since $\ottnt{e_{{\mathrm{2}}}}  \ottsym{=}   \lambda \mathit{x} \mathord{:} \ottnt{T_{{\mathrm{211}}}} .~  \ottnt{e'_{{\mathrm{2}}}} $. We must show that:
        \[   \langle  \ottnt{T_{{\mathrm{112}}}}  \mathord{ \overset{\bullet}{\Rightarrow} }  \ottnt{T_{{\mathrm{122}}}}  \rangle^{ \ottnt{l} } ~   (  \ottnt{e_{{\mathrm{1}}}} ~  (  \langle  \ottnt{T_{{\mathrm{121}}}}  \mathord{ \overset{\bullet}{\Rightarrow} }  \ottnt{T_{{\mathrm{111}}}}  \rangle^{ \ottnt{l} } ~  \ottnt{e_{{\mathrm{12}}}}  )   )     \simeq _{  \mathsf{H}  }    (  \langle   \ottnt{T_{{\mathrm{211}}}} \mathord{ \rightarrow } \ottnt{T_{{\mathrm{212}}}}   \mathord{ \overset{ \emptyset }{\Rightarrow} }   \ottnt{T_{{\mathrm{221}}}} \mathord{ \rightarrow } \ottnt{T_{{\mathrm{222}}}}   \rangle^{ \ottnt{l} } ~   \lambda \mathit{x} \mathord{:} \ottnt{T_{{\mathrm{211}}}} .~  \ottnt{e'_{{\mathrm{2}}}}   )  ~ \ottnt{e_{{\mathrm{22}}}}   :  \ottnt{T_{{\mathrm{222}}}}  \]
        The heedful side steps by \E{Unwrap} with an empty type set,
        yielding $ \langle  \ottnt{T_{{\mathrm{212}}}}  \mathord{ \overset{\bullet}{\Rightarrow} }  \ottnt{T_{{\mathrm{222}}}}  \rangle^{ \ottnt{l} } ~   (   (  \lambda \mathit{x} \mathord{:} \ottnt{T_{{\mathrm{211}}}} .~  \ottnt{e'_{{\mathrm{2}}}}  )  ~  (  \langle  \ottnt{T_{{\mathrm{221}}}}  \mathord{ \overset{\bullet}{\Rightarrow} }  \ottnt{T_{{\mathrm{211}}}}  \rangle^{ \ottnt{l} } ~  \ottnt{e_{{\mathrm{22}}}}  )   )  $.
        By the IH, we know that $  \langle  \ottnt{T_{{\mathrm{121}}}}  \mathord{ \overset{\bullet}{\Rightarrow} }  \ottnt{T_{{\mathrm{111}}}}  \rangle^{ \ottnt{l} } ~  \ottnt{e_{{\mathrm{12}}}}    \simeq _{  \mathsf{H}  }   \langle  \ottnt{T_{{\mathrm{221}}}}  \mathord{ \overset{\bullet}{\Rightarrow} }  \ottnt{T_{{\mathrm{211}}}}  \rangle^{ \ottnt{l} } ~  \ottnt{e_{{\mathrm{22}}}}   :  \ottnt{T_{{\mathrm{211}}}} $. 
        If we get blame on both sides, we are done immediately by the
        appropriate \E{...Raise} rules.
        If not, we get values $ \ottnt{e'_{{\mathrm{12}}}}   \sim _{  \mathsf{H}  }  \ottnt{e'_{{\mathrm{22}}}}  :  \ottnt{T_{{\mathrm{211}}}} $. We know by
        assumption that $  \ottnt{e_{{\mathrm{1}}}} ~ \ottnt{e'_{{\mathrm{12}}}}    \simeq _{  \mathsf{H}  }   \ottnt{e_{{\mathrm{2}}}} ~ \ottnt{e'_{{\mathrm{22}}}}   :  \ottnt{T_{{\mathrm{212}}}} $; again, blame
        on finishes this case. So suppose both sides go to values
        $ \ottnt{e''_{{\mathrm{12}}}}   \sim _{  \mathsf{H}  }  \ottnt{e''_{{\mathrm{22}}}}  :  \ottnt{T_{{\mathrm{212}}}} $. By the IH, we know that
        $  \langle  \ottnt{T_{{\mathrm{112}}}}  \mathord{ \overset{\bullet}{\Rightarrow} }  \ottnt{T_{{\mathrm{122}}}}  \rangle^{ \ottnt{l} } ~  \ottnt{e''_{{\mathrm{12}}}}    \simeq _{  \mathsf{H}  }   \langle  \ottnt{T_{{\mathrm{212}}}}  \mathord{ \overset{\bullet}{\Rightarrow} }  \ottnt{T_{{\mathrm{222}}}}  \rangle^{ \ottnt{l} } ~  \ottnt{e''_{{\mathrm{22}}}}   :  \ottnt{T_{{\mathrm{222}}}} $, and
        we are done.

      \item[(\E{CastMerge})] We have $\ottnt{e_{{\mathrm{21}}}}  \ottsym{=}   \langle   \ottnt{T_{{\mathrm{31}}}} \mathord{ \rightarrow } \ottnt{T_{{\mathrm{32}}}}   \mathord{ \overset{  \mathcal{S}  \cup   \set{   \ottnt{T_{{\mathrm{211}}}} \mathord{ \rightarrow } \ottnt{T_{{\mathrm{212}}}}   }   }{\Rightarrow} }   \ottnt{T_{{\mathrm{221}}}} \mathord{ \rightarrow } \ottnt{T_{{\mathrm{222}}}}   \rangle^{ \ottnt{l} } ~   \lambda \mathit{x} \mathord{:} \ottnt{T_{{\mathrm{31}}}} .~  \ottnt{e'_{{\mathrm{2}}}}  $ since $\ottnt{e_{{\mathrm{2}}}}  \ottsym{=}   \langle   \ottnt{T_{{\mathrm{31}}}} \mathord{ \rightarrow } \ottnt{T_{{\mathrm{32}}}}   \mathord{ \overset{ \mathcal{S} }{\Rightarrow} }   \ottnt{T_{{\mathrm{211}}}} \mathord{ \rightarrow } \ottnt{T_{{\mathrm{212}}}}   \rangle^{ \ottnt{l'} } ~   \lambda \mathit{x} \mathord{:} \ottnt{T_{{\mathrm{31}}}} .~  \ottnt{e'_{{\mathrm{2}}}}  $.  
        We must show that:
        \[   \langle  \ottnt{T_{{\mathrm{112}}}}  \mathord{ \overset{\bullet}{\Rightarrow} }  \ottnt{T_{{\mathrm{122}}}}  \rangle^{ \ottnt{l} } ~   (  \ottnt{e_{{\mathrm{1}}}} ~  (  \langle  \ottnt{T_{{\mathrm{121}}}}  \mathord{ \overset{\bullet}{\Rightarrow} }  \ottnt{T_{{\mathrm{111}}}}  \rangle^{ \ottnt{l} } ~  \ottnt{e_{{\mathrm{12}}}}  )   )     \simeq _{  \mathsf{F}  }    (  \langle   \ottnt{T_{{\mathrm{31}}}} \mathord{ \rightarrow } \ottnt{T_{{\mathrm{32}}}}   \mathord{ \overset{  \mathcal{S}  \cup   \set{   \ottnt{T_{{\mathrm{211}}}} \mathord{ \rightarrow } \ottnt{T_{{\mathrm{212}}}}   }   }{\Rightarrow} }   \ottnt{T_{{\mathrm{221}}}} \mathord{ \rightarrow } \ottnt{T_{{\mathrm{222}}}}   \rangle^{ \ottnt{l} } ~   \lambda \mathit{x} \mathord{:} \ottnt{T_{{\mathrm{31}}}} .~  \ottnt{e'_{{\mathrm{2}}}}   )  ~ \ottnt{e_{{\mathrm{22}}}}   :  \ottnt{T_{{\mathrm{222}}}}  \]
        The right hand steps by \E{Unwrap},
        yielding $ \langle  \ottnt{T_{{\mathrm{32}}}}  \mathord{ \overset{   \mathsf{cod} ( \mathcal{S} )   \cup   \set{  \ottnt{T_{{\mathrm{212}}}}  }   }{\Rightarrow} }  \ottnt{T_{{\mathrm{222}}}}  \rangle^{ \ottnt{l} } ~   (   (  \lambda \mathit{x} \mathord{:} \ottnt{T_{{\mathrm{31}}}} .~  \ottnt{e'_{{\mathrm{2}}}}  )  ~  (  \langle  \ottnt{T_{{\mathrm{221}}}}  \mathord{ \overset{   \mathsf{dom} ( \mathcal{S} )   \cup   \set{  \ottnt{T_{{\mathrm{211}}}}  }   }{\Rightarrow} }  \ottnt{T_{{\mathrm{31}}}}  \rangle^{ \ottnt{l} } ~  \ottnt{e_{{\mathrm{22}}}}  )   )  $. We must show that this heedful term is related to
        the classic term $ \langle  \ottnt{T_{{\mathrm{112}}}}  \mathord{ \overset{\bullet}{\Rightarrow} }  \ottnt{T_{{\mathrm{122}}}}  \rangle^{ \ottnt{l} } ~   (  \ottnt{e_{{\mathrm{1}}}} ~  (  \langle  \ottnt{T_{{\mathrm{121}}}}  \mathord{ \overset{\bullet}{\Rightarrow} }  \ottnt{T_{{\mathrm{111}}}}  \rangle^{ \ottnt{l} } ~  \ottnt{e_{{\mathrm{12}}}}  )   )  $.

        We must now make a brief digression to examine the behavior of
        the cast that was eliminated by \E{CastMerge}.
        We know by the IH that $  \langle  \ottnt{T_{{\mathrm{121}}}}  \mathord{ \overset{\bullet}{\Rightarrow} }  \ottnt{T_{{\mathrm{111}}}}  \rangle^{ \ottnt{l} } ~  \ottnt{e_{{\mathrm{12}}}}    \simeq _{  \mathsf{H}  }   \langle  \ottnt{T_{{\mathrm{221}}}}  \mathord{ \overset{\bullet}{\Rightarrow} }  \ottnt{T_{{\mathrm{211}}}}  \rangle^{ \ottnt{l} } ~  \ottnt{e_{{\mathrm{22}}}}   :  \ottnt{T_{{\mathrm{211}}}} $, so both sides go to blame or to values $ \ottnt{e'_{{\mathrm{12}}}}   \sim _{  \mathsf{H}  }  \ottnt{e'_{{\mathrm{22}}}}  :  \ottnt{T_{{\mathrm{211}}}} $.  By Lemma~\ref{lem:heedfulcastcongruence}
        with $ \mathsf{dom} ( \mathcal{S} ) $ as the type set, we can find that
        $ \langle  \ottnt{T_{{\mathrm{211}}}}  \mathord{ \overset{\bullet}{\Rightarrow} }  \ottnt{T_{{\mathrm{31}}}}  \rangle^{ \ottnt{l} } ~  \ottnt{e'_{{\mathrm{22}}}}  \,  \longrightarrow ^{*}_{  \mathsf{H}  }  \, \ottnt{e''_{{\mathrm{22}}}}$ implies $ \langle  \ottnt{T_{{\mathrm{211}}}}  \mathord{ \overset{  \mathsf{dom} ( \mathcal{S} )  }{\Rightarrow} }  \ottnt{T_{{\mathrm{31}}}}  \rangle^{ \ottnt{l} } ~   (  \langle  \ottnt{T_{{\mathrm{221}}}}  \mathord{ \overset{\bullet}{\Rightarrow} }  \ottnt{T_{{\mathrm{211}}}}  \rangle^{ \ottnt{l} } ~  \ottnt{e_{{\mathrm{22}}}}  )   \,  \longrightarrow ^{*}_{  \mathsf{H}  }  \, \ottnt{e''_{{\mathrm{22}}}}$.
        But then we have that $ \langle  \ottnt{T_{{\mathrm{211}}}}  \mathord{ \overset{ \mathcal{S} }{\Rightarrow} }  \ottnt{T_{{\mathrm{31}}}}  \rangle^{ \ottnt{l} } ~   (  \langle  \ottnt{T_{{\mathrm{221}}}}  \mathord{ \overset{\bullet}{\Rightarrow} }  \ottnt{T_{{\mathrm{211}}}}  \rangle^{ \ottnt{l} } ~  \ottnt{e_{{\mathrm{22}}}}  )   \,  \longrightarrow _{  \mathsf{H}  }  \,  \langle  \ottnt{T_{{\mathrm{211}}}}  \mathord{ \overset{   \mathsf{dom} ( \mathcal{S} )   \cup   \set{  \ottnt{T_{{\mathrm{221}}}}  }   }{\Rightarrow} }  \ottnt{T_{{\mathrm{31}}}}  \rangle^{ \ottnt{l} } ~  \ottnt{e_{{\mathrm{22}}}} $, so we then know
        that $ \langle  \ottnt{T_{{\mathrm{221}}}}  \mathord{ \overset{   \mathsf{dom} ( \mathcal{S} )   \cup   \set{  \ottnt{T_{{\mathrm{221}}}}  }   }{\Rightarrow} }  \ottnt{T_{{\mathrm{31}}}}  \rangle^{ \ottnt{l} } ~  \ottnt{e_{{\mathrm{22}}}}  \,  \longrightarrow ^{*}_{  \mathsf{H}  }  \, \ottnt{e''_{{\mathrm{22}}}}$, just
        as if it were applied to $\ottnt{e'_{{\mathrm{22}}}}$.

        Now we can return to the meat of our proof. If
        $ \langle  \ottnt{T_{{\mathrm{121}}}}  \mathord{ \overset{\bullet}{\Rightarrow} }  \ottnt{T_{{\mathrm{111}}}}  \rangle^{ \ottnt{l} } ~  \ottnt{e_{{\mathrm{12}}}}  \,  \longrightarrow ^{*}_{  \mathsf{H}  }  \,  \mathord{\Uparrow}  \ottnt{l'} $, we are done---so must
        the heedful side (albeit possibly at a different blame
        label). If it reduces to a value $\ottnt{e'_{{\mathrm{12}}}}$, then we are left
        considering the term $ \langle  \ottnt{T_{{\mathrm{112}}}}  \mathord{ \overset{\bullet}{\Rightarrow} }  \ottnt{T_{{\mathrm{122}}}}  \rangle^{ \ottnt{l} } ~   (  \ottnt{e_{{\mathrm{1}}}} ~ \ottnt{e'_{{\mathrm{12}}}}  )  $ on the
        classic side. We know that $ \ottnt{e_{{\mathrm{1}}}}   \sim _{  \mathsf{H}  }  \ottnt{e_{{\mathrm{2}}}}  :  \ottnt{T_{{\mathrm{21}}}} $. Unfolding the
        definition of $\ottnt{e_{{\mathrm{2}}}}$, this means that $  \ottnt{e_{{\mathrm{1}}}} ~ \ottnt{e'_{{\mathrm{12}}}}    \simeq _{  \mathsf{F}  }   \langle  \ottnt{T_{{\mathrm{32}}}}  \mathord{ \overset{\bullet}{\Rightarrow} }  \ottnt{T_{{\mathrm{212}}}}  \rangle^{ \ottnt{l} } ~   (   (  \lambda \mathit{x} \mathord{:} \ottnt{T_{{\mathrm{31}}}} .~  \ottnt{e'_{{\mathrm{2}}}}  )  ~  (  \langle  \ottnt{T_{{\mathrm{211}}}}  \mathord{ \overset{\bullet}{\Rightarrow} }  \ottnt{T_{{\mathrm{31}}}}  \rangle^{ \ottnt{l} } ~  \ottnt{e'_{{\mathrm{22}}}}  )   )    :  \ottnt{T_{{\mathrm{212}}}} $.
        If the classic side produces blame, so must the heedful side
        and we are done, as indicated in the digression above. If not,
        then both sides produce values. For these terms to produce
        values, it must be the case that (a) the domain cast on the
        heedful side produces a value, (b) the heedful function
        produces a value given that input, and (c) the heedful
        codomain cast produces a value. Now, we know from our
        digression above that $ \langle  \ottnt{T_{{\mathrm{211}}}}  \mathord{ \overset{  \mathsf{dom} ( \mathcal{S} )  }{\Rightarrow} }  \ottnt{T_{{\mathrm{31}}}}  \rangle^{ \ottnt{l} } ~  \ottnt{e'_{{\mathrm{22}}}} $ and
        $ \langle  \ottnt{T_{{\mathrm{211}}}}  \mathord{ \overset{  \mathsf{dom} ( \mathcal{S} )  }{\Rightarrow} }  \ottnt{T_{{\mathrm{31}}}}  \rangle^{ \ottnt{l} } ~  \ottnt{e_{{\mathrm{22}}}} $ reduce to the exact same value,
        $\ottnt{e''_{{\mathrm{22}}}}$. So if $ \langle  \ottnt{T_{{\mathrm{32}}}}  \mathord{ \overset{  \mathsf{cod} ( \mathcal{S} )  }{\Rightarrow} }  \ottnt{T_{{\mathrm{212}}}}  \rangle^{ \ottnt{l} } ~   (   (  \lambda \mathit{x} \mathord{:} \ottnt{T_{{\mathrm{31}}}} .~  \ottnt{e'_{{\mathrm{2}}}}  )  ~  (  \langle  \ottnt{T_{{\mathrm{211}}}}  \mathord{ \overset{  \mathsf{dom} ( \mathcal{S} )  }{\Rightarrow} }  \ottnt{T_{{\mathrm{31}}}}  \rangle^{ \ottnt{l} } ~  \ottnt{e'_{{\mathrm{22}}}}  )   )   \,  \longrightarrow ^{*}_{  \mathsf{H}  }  \, \ottnt{e''_{{\mathrm{22}}}}$ then we can also
        see \[  \langle  \ottnt{T_{{\mathrm{32}}}}  \mathord{ \overset{  \mathsf{cod} ( \mathcal{S} )  }{\Rightarrow} }  \ottnt{T_{{\mathrm{212}}}}  \rangle^{ \ottnt{l} } ~   (   (  \lambda \mathit{x} \mathord{:} \ottnt{T_{{\mathrm{31}}}} .~  \ottnt{e'_{{\mathrm{2}}}}  )  ~  (  \langle  \ottnt{T_{{\mathrm{211}}}}  \mathord{ \overset{  \mathsf{dom} ( \mathcal{S} )  }{\Rightarrow} }  \ottnt{T_{{\mathrm{31}}}}  \rangle^{ \ottnt{l} } ~  \ottnt{e_{{\mathrm{22}}}}  )   )   \,  \longrightarrow ^{*}_{  \mathsf{H}  }  \,  \langle  \ottnt{T_{{\mathrm{32}}}}  \mathord{ \overset{\bullet}{\Rightarrow} }  \ottnt{T_{{\mathrm{212}}}}  \rangle^{ \ottnt{l} } ~  \ottnt{e_{{\mathrm{32}}}}   \longrightarrow ^{*}_{  \mathsf{H}  }  \ottnt{e''_{{\mathrm{22}}}}. \]

        We have shown that the domains and then the applied inner
        functions are equivalent.  It now remains to show that
        \[   \langle  \ottnt{T_{{\mathrm{112}}}}  \mathord{ \overset{\bullet}{\Rightarrow} }  \ottnt{T_{{\mathrm{122}}}}  \rangle^{ \ottnt{l} } ~  \ottnt{e''_{{\mathrm{11}}}}    \simeq _{  \mathsf{H}  }   \langle  \ottnt{T_{{\mathrm{32}}}}  \mathord{ \overset{   \mathsf{cod} ( \mathcal{S} )   \cup   \set{  \ottnt{T_{{\mathrm{212}}}}  }   }{\Rightarrow} }  \ottnt{T_{{\mathrm{222}}}}  \rangle^{ \ottnt{l} } ~   (   (  \lambda \mathit{x} \mathord{:} \ottnt{T_{{\mathrm{31}}}} .~  \ottnt{e'_{{\mathrm{2}}}}  )  ~  (  \langle  \ottnt{T_{{\mathrm{221}}}}  \mathord{ \overset{\bullet}{\Rightarrow} }  \ottnt{T_{{\mathrm{31}}}}  \rangle^{ \ottnt{l} } ~  \ottnt{e_{{\mathrm{22}}}}  )   )    :  \ottnt{T_{{\mathrm{222}}}}  \]
        We write the \textit{entire} heedful term to highlight the
        fact that we cannot freely apply congruence, but must instead
        carefully apply cast congruence
        (Lemma~\ref{lem:heedfulcastcongruence}) as we go.
        
        By the IH, we know that either $ \langle  \ottnt{T_{{\mathrm{112}}}}  \mathord{ \overset{\bullet}{\Rightarrow} }  \ottnt{T_{{\mathrm{122}}}}  \rangle^{ \ottnt{l} } ~  \ottnt{e''_{{\mathrm{11}}}} $ and
        $ \langle  \ottnt{T_{{\mathrm{212}}}}  \mathord{ \overset{\bullet}{\Rightarrow} }  \ottnt{T_{{\mathrm{222}}}}  \rangle^{ \ottnt{l} } ~  \ottnt{e''_{{\mathrm{22}}}} $ go to blame (perhaps with different
        labels) or to values. In the former case we are done; in the
        latter case, we already know that $ \langle  \ottnt{T_{{\mathrm{32}}}}  \mathord{ \overset{  \mathsf{cod} ( \mathcal{S} )  }{\Rightarrow} }  \ottnt{T_{{\mathrm{212}}}}  \rangle^{ \ottnt{l} } ~  \ottnt{e_{{\mathrm{32}}}}  \,  \longrightarrow ^{*}_{  \mathsf{H}  }  \, \ottnt{e''_{{\mathrm{22}}}}$, so we can apply cast congruence
        (Lemma~\ref{lem:heedfulcastcongruence}) with the empty type
        set to see that if $ \langle  \ottnt{T_{{\mathrm{212}}}}  \mathord{ \overset{\bullet}{\Rightarrow} }  \ottnt{T_{{\mathrm{222}}}}  \rangle^{ \ottnt{l} } ~  \ottnt{e''_{{\mathrm{22}}}}  \,  \longrightarrow ^{*}_{  \mathsf{H}  }  \, \ottnt{e'''_{{\mathrm{22}}}}$ then
        $ \langle  \ottnt{T_{{\mathrm{212}}}}  \mathord{ \overset{\bullet}{\Rightarrow} }  \ottnt{T_{{\mathrm{222}}}}  \rangle^{ \ottnt{l} } ~   (  \langle  \ottnt{T_{{\mathrm{32}}}}  \mathord{ \overset{  \mathsf{cod} ( \mathcal{S} )  }{\Rightarrow} }  \ottnt{T_{{\mathrm{212}}}}  \rangle^{ \ottnt{l} } ~  \ottnt{e_{{\mathrm{32}}}}  )   \,  \longrightarrow ^{*}_{  \mathsf{H}  }  \, \ottnt{e'''_{{\mathrm{22}}}}$.
        But we know that $ \langle  \ottnt{T_{{\mathrm{212}}}}  \mathord{ \overset{\bullet}{\Rightarrow} }  \ottnt{T_{{\mathrm{222}}}}  \rangle^{ \ottnt{l} } ~   (  \langle  \ottnt{T_{{\mathrm{32}}}}  \mathord{ \overset{  \mathsf{cod} ( \mathcal{S} )  }{\Rightarrow} }  \ottnt{T_{{\mathrm{212}}}}  \rangle^{ \ottnt{l} } ~  \ottnt{e_{{\mathrm{32}}}}  )   \,  \longrightarrow _{  \mathsf{H}  }  \,  \langle  \ottnt{T_{{\mathrm{32}}}}  \mathord{ \overset{   \mathsf{cod} ( \mathcal{S} )   \cup   \set{  \ottnt{T_{{\mathrm{212}}}}  }   }{\Rightarrow} }  \ottnt{T_{{\mathrm{222}}}}  \rangle^{ \ottnt{l} } ~  \ottnt{e_{{\mathrm{32}}}} $ deterministically,
        so we then we know that $ \langle  \ottnt{T_{{\mathrm{32}}}}  \mathord{ \overset{   \mathsf{cod} ( \mathcal{S} )   \cup   \set{  \ottnt{T_{{\mathrm{212}}}}  }   }{\Rightarrow} }  \ottnt{T_{{\mathrm{222}}}}  \rangle^{ \ottnt{l} } ~  \ottnt{e_{{\mathrm{32}}}}  \,  \longrightarrow ^{*}_{  \mathsf{H}  }  \, \ottnt{e'''_{{\mathrm{22}}}}$. Since $ \langle  \ottnt{T_{{\mathrm{32}}}}  \mathord{ \overset{   \mathsf{cod} ( \mathcal{S} )   \cup   \set{  \ottnt{T_{{\mathrm{212}}}}  }   }{\Rightarrow} }  \ottnt{T_{{\mathrm{222}}}}  \rangle^{ \ottnt{l} } ~   (   (  \lambda \mathit{x} \mathord{:} \ottnt{T_{{\mathrm{31}}}} .~  \ottnt{e'_{{\mathrm{2}}}}  )  ~  (  \langle  \ottnt{T_{{\mathrm{221}}}}  \mathord{ \overset{\bullet}{\Rightarrow} }  \ottnt{T_{{\mathrm{31}}}}  \rangle^{ \ottnt{l} } ~  \ottnt{e_{{\mathrm{22}}}}  )   )   \,  \longrightarrow ^{*}_{  \mathsf{H}  }  \,  \langle  \ottnt{T_{{\mathrm{32}}}}  \mathord{ \overset{\bullet}{\Rightarrow} }  \ottnt{T_{{\mathrm{222}}}}  \rangle^{ \ottnt{l} } ~  \ottnt{e_{{\mathrm{32}}}} $,
        we have shown that the classic term and heedful term reduce to
        values $ \ottnt{e'''_{{\mathrm{12}}}}   \sim _{  \mathsf{H}  }  \ottnt{e'''_{{\mathrm{22}}}}  :  \ottnt{T_{{\mathrm{222}}}} $, and we are done.
      \end{itemize}
    \end{itemize}
    \fi}
  \end{proof}
\end{lemma}

\begin{lemma}[Relating classic and heedful source programs]
  \label{lem:heedfullr}
  ~

  \noindent
  \begin{enumerate}
  \item \label{hlr:term} If $ \Gamma   \vdash _{  \mathsf{C}  }  \ottnt{e}  :  \ottnt{T} $ as a source program then
    $ \Gamma   \vdash   \ottnt{e}   \simeq _{  \mathsf{H}  }  \ottnt{e}  :  \ottnt{T} $.
  \item \label{hlr:type} If $ \mathord{  \vdash _{  \mathsf{C}  } }~ \ottnt{T} $ as a source program then $ \ottnt{T}   \sim _{  \mathsf{H}  }  \ottnt{T} $.
  \end{enumerate}
  \begin{proof}
    By mutual induction on the typing derivations.
    
    {\iffull
    \paragraph{Term typing \fbox{$ \Gamma   \vdash _{  \mathsf{C}  }  \ottnt{e}  :  \ottnt{T} $}}
    \begin{itemize}
    \item[\T{Var}] We know by assumption that $ \delta_{{\mathrm{1}}}  \ottsym{(}  \mathit{x}  \ottsym{)}   \sim _{  \mathsf{H}  }  \delta_{{\mathrm{2}}}  \ottsym{(}  \mathit{x}  \ottsym{)}  :  \ottnt{T} $.
    \item[\T{Const}] Since we are dealing with a source program, $\ottnt{T}  \ottsym{=}   \{ \mathit{x} \mathord{:} \ottnt{B} \mathrel{\mid}  \mathsf{true}  \} $. We have immediately that $ \mathsf{ty} ( \ottnt{k} )   \ottsym{=}  \ottnt{B}$ and
      $   \mathsf{true}   [  \ottnt{k} / \mathit{x}  ]    \simeq _{  \mathsf{H}  }    \mathsf{true}   [  \ottnt{k} / \mathit{x}  ]   :   \{ \mathit{x} \mathord{:}  \mathsf{Bool}  \mathrel{\mid}  \mathsf{true}  \}  $, so $ \ottnt{k}   \simeq _{  \mathsf{H}  }  \ottnt{k}  :   \{ \mathit{x} \mathord{:} \ottnt{B} \mathrel{\mid}  \mathsf{true}  \}  $.
    \item[\T{Abs}] Let $ \Gamma   \models _{  \mathsf{H}  }  \delta $. We must show that
      $  \lambda \mathit{x} \mathord{:} \ottnt{T_{{\mathrm{1}}}} .~  \delta_{{\mathrm{1}}}  \ottsym{(}  \ottnt{e_{{\mathrm{1}}}}  \ottsym{)}    \sim _{  \mathsf{H}  }   \lambda \mathit{x} \mathord{:} \ottnt{T_{{\mathrm{1}}}} .~  \delta_{{\mathrm{2}}}  \ottsym{(}  \ottnt{e_{{\mathrm{1}}}}  \ottsym{)}   :   \ottnt{T_{{\mathrm{1}}}} \mathord{ \rightarrow } \ottnt{T_{{\mathrm{2}}}}  $. Let $ \ottnt{e_{{\mathrm{2}}}}   \sim _{  \mathsf{H}  }  \ottnt{e'_{{\mathrm{2}}}}  :  \ottnt{T_{{\mathrm{1}}}} $. We must show that applying the abstractions to
      these values yields related values. Both sides step by \E{Beta},
      to $ \delta_{{\mathrm{1}}}  \ottsym{(}  \ottnt{e_{{\mathrm{1}}}}  \ottsym{)}  [  \ottnt{e_{{\mathrm{2}}}} / \mathit{x}  ] $ and $ \delta_{{\mathrm{2}}}  \ottsym{(}  \ottnt{e_{{\mathrm{1}}}}  \ottsym{)}  [  \ottnt{e'_{{\mathrm{2}}}} / \mathit{x}  ] $,
      respectively. But $  \Gamma , \mathit{x} \mathord{:} \ottnt{T_{{\mathrm{1}}}}    \models _{  \mathsf{H}  }   \delta  [  \ottnt{e_{{\mathrm{2}}}} , \ottnt{e'_{{\mathrm{2}}}} / \mathit{x}  ]  $, so we can
      apply IH (\ref{hlr:term}) on $\ottnt{e_{{\mathrm{1}}}}$ showing the two sides reduce to
      related results.
    \item[\T{Op}] By IH (\ref{hlr:term}) on each argument, either one
      of the arguments goes to blame (in both calculi), and we are done by
      \E{OpRaise}, or all of the arguments reduce to related
      values. Since $ \mathsf{ty} (\mathord{ \ottnt{op} }) $ is first order, these values must be
      related at refined base types, which means that they are in fact
      all equal constants. We then reduce by \E{Op} on both sides to
      have $\denot{ op } \, \ottsym{(}  \ottnt{k_{{\mathrm{1}}}}  \ottsym{,}  \dots  \ottsym{,}  \ottnt{k_{\ottmv{n}}}  \ottsym{)}$. We have assumed that the
      denotations of operations agree with their typings in
      \textit{all} modes, so then $\denot{ op } \, \ottsym{(}  \ottnt{k_{{\mathrm{1}}}}  \ottsym{,}  \dots  \ottsym{,}  \ottnt{k_{\ottmv{n}}}  \ottsym{)}$ satisfies
      the refinement for $ \longrightarrow _{  \mathsf{H}  } $ in particular, and we are done.
    \item[\T{App}]  Let $ \Gamma   \models _{  \mathsf{H}  }  \delta $. We must show that
      $  \delta_{{\mathrm{1}}}  \ottsym{(}  \ottnt{e_{{\mathrm{1}}}}  \ottsym{)} ~ \delta_{{\mathrm{1}}}  \ottsym{(}  \ottnt{e_{{\mathrm{2}}}}  \ottsym{)}    \simeq _{  \mathsf{H}  }   \delta_{{\mathrm{2}}}  \ottsym{(}  \ottnt{e_{{\mathrm{1}}}}  \ottsym{)} ~ \delta_{{\mathrm{2}}}  \ottsym{(}  \ottnt{e_{{\mathrm{2}}}}  \ottsym{)}   :  \ottnt{T_{{\mathrm{2}}}} $. But
      by IH (\ref{hlr:term}) on $\ottnt{e_{{\mathrm{1}}}}$ and $\ottnt{e_{{\mathrm{2}}}}$, we are done
      directly.
    \item[\T{Cast}] We know that the annotation is $ \bullet $, since
      we are dealing with a source term. Let $ \Gamma   \models _{  \mathsf{H}  }  \delta $. By IH
      (\ref{hlr:term}) on $\ottnt{e'}$, we know that $ \delta_{{\mathrm{1}}}  \ottsym{(}  \ottnt{e'}  \ottsym{)}   \simeq _{  \mathsf{H}  }  \delta_{{\mathrm{2}}}  \ottsym{(}  \ottnt{e'}  \ottsym{)}  :  \ottnt{T_{{\mathrm{1}}}} $, either $\delta_{{\mathrm{1}}}  \ottsym{(}  \ottnt{e'}  \ottsym{)} \,  \longrightarrow ^{*}_{  \mathsf{C}  }  \,  \mathord{\Uparrow}  \ottnt{l'} $ and
      $\delta_{{\mathrm{2}}}  \ottsym{(}  \ottnt{e'}  \ottsym{)} \,  \longrightarrow ^{*}_{  \mathsf{H}  }  \,  \mathord{\Uparrow}  \ottnt{l''} $ (and we are done) or
      $\delta_{{\mathrm{1}}}  \ottsym{(}  \ottnt{e'}  \ottsym{)}$ and $\delta_{{\mathrm{2}}}  \ottsym{(}  \ottnt{e'}  \ottsym{)}$ reduce to values $ \ottnt{e_{{\mathrm{1}}}}   \sim _{  \mathsf{H}  }  \ottnt{e_{{\mathrm{2}}}}  :  \ottnt{T_{{\mathrm{1}}}} $. By Lemma~\ref{lem:heedfullrcast} (using IH
      (\ref{hlr:type}) on the types), we know that $  \langle  \ottnt{T_{{\mathrm{1}}}}  \mathord{ \overset{\bullet}{\Rightarrow} }  \ottnt{T_{{\mathrm{2}}}}  \rangle^{ \ottnt{l} } ~  \ottnt{e_{{\mathrm{1}}}}    \sim _{  \mathsf{H}  }   \langle  \ottnt{T_{{\mathrm{1}}}}  \mathord{ \overset{\bullet}{\Rightarrow} }  \ottnt{T_{{\mathrm{2}}}}  \rangle^{ \ottnt{l} } ~  \ottnt{e_{{\mathrm{2}}}}   :  \ottnt{T_{{\mathrm{2}}}} $, so each side must reduce to a result
      $ \ottnt{e'_{{\mathrm{1}}}}   \sim _{  \mathsf{H}  }  \ottnt{e'_{{\mathrm{2}}}}  :  \ottnt{T_{{\mathrm{2}}}} $.
      We have cast congruence on the classic side straightforwardly,
      finding:
      \[  \langle  \ottnt{T_{{\mathrm{1}}}}  \mathord{ \overset{\bullet}{\Rightarrow} }  \ottnt{T_{{\mathrm{2}}}}  \rangle^{ \ottnt{l} } ~  \delta_{{\mathrm{1}}}  \ottsym{(}  \ottnt{e'}  \ottsym{)}  \,  \longrightarrow ^{*}_{  \mathsf{C}  }  \,  \langle  \ottnt{T_{{\mathrm{1}}}}  \mathord{ \overset{\bullet}{\Rightarrow} }  \ottnt{T_{{\mathrm{2}}}}  \rangle^{ \ottnt{l} } ~  \ottnt{e_{{\mathrm{1}}}}   \longrightarrow ^{*}_{  \mathsf{C}  }  \ottnt{e'_{{\mathrm{1}}}} \]
      On the heedful side, we can apply our derived cast congruence
      (Lemma~\ref{lem:heedfulcastcongruence}) to find that
      $ \langle  \ottnt{T_{{\mathrm{1}}}}  \mathord{ \overset{\bullet}{\Rightarrow} }  \ottnt{T_{{\mathrm{2}}}}  \rangle^{ \ottnt{l} } ~  \ottnt{e_{{\mathrm{2}}}}  \,  \longrightarrow ^{*}_{  \mathsf{H}  }  \, \ottnt{e'_{{\mathrm{2}}}}$ and $\delta_{{\mathrm{2}}}  \ottsym{(}  \ottnt{e'}  \ottsym{)} \,  \longrightarrow ^{*}_{  \mathsf{H}  }  \, \ottnt{e_{{\mathrm{2}}}}$ imply
      that $ \langle  \ottnt{T_{{\mathrm{1}}}}  \mathord{ \overset{\bullet}{\Rightarrow} }  \ottnt{T_{{\mathrm{2}}}}  \rangle^{ \ottnt{l} } ~  \delta_{{\mathrm{2}}}  \ottsym{(}  \ottnt{e'}  \ottsym{)}  \,  \longrightarrow ^{*}_{  \mathsf{H}  }  \, \ottnt{e'_{{\mathrm{2}}}}$.
    \item[\T{Blame}] Contradiction---doesn't appear in source
      programs. Though in fact it is in the relation, since
      $  \mathord{\Uparrow}  \ottnt{l}    \simeq _{  \mathsf{H}  }   \mathord{\Uparrow}  \ottnt{l'}   :  \ottnt{T} $ for any $\ottnt{l}$, $\ottnt{l'}$, and
      $\ottnt{T}$.
    \item[\T{Check}] Contradiction---doesn't appear in source
      programs.
    \end{itemize}
    
    \paragraph{Type well formedness \fbox{$ \mathord{  \vdash _{  \mathsf{C}  } }~ \ottnt{T} $}}
    \begin{itemize}
    \item[\WF{Base}] We can immediately see $   \mathsf{true}   [  \ottnt{k} / \mathit{x}  ]    \simeq _{  \mathsf{H}  }    \mathsf{true}   [  \ottnt{k} / \mathit{x}  ]   :   \{ \mathit{x} \mathord{:}  \mathsf{Bool}  \mathrel{\mid}  \mathsf{true}  \}  $ for any $ \ottnt{k}   \sim _{  \mathsf{H}  }  \ottnt{k}  :   \{ \mathit{x} \mathord{:} \ottnt{B} \mathrel{\mid}  \mathsf{true}  \}  $, i.e., any
      $\ottnt{k}$ such that $ \mathsf{ty} ( \ottnt{k} )   \ottsym{=}  \ottnt{B}$.
    \item[\WF{Refine}] By inversion, we know that $  \mathit{x} \mathord{:}  \{ \mathit{x} \mathord{:} \ottnt{B} \mathrel{\mid}  \mathsf{true}  \}     \vdash _{  \mathsf{C}  }  \ottnt{e}  :   \{ \mathit{x} \mathord{:}  \mathsf{Bool}  \mathrel{\mid}  \mathsf{true}  \}  $; by IH (\ref{hlr:term}), we find that
      $ \delta_{{\mathrm{1}}}  \ottsym{(}  \ottnt{e}  \ottsym{)}   \simeq _{  \mathsf{H}  }  \delta_{{\mathrm{2}}}  \ottsym{(}  \ottnt{e}  \ottsym{)}  :   \{ \mathit{x} \mathord{:}  \mathsf{Bool}  \mathrel{\mid}  \mathsf{true}  \}  $, i.e., that
      $  \ottnt{e}  [  \ottnt{e_{{\mathrm{1}}}} / \mathit{x}  ]    \simeq _{  \mathsf{H}  }   \ottnt{e}  [  \ottnt{e_{{\mathrm{2}}}} / \mathit{x}  ]   :   \{ \mathit{x} \mathord{:}  \mathsf{Bool}  \mathrel{\mid}  \mathsf{true}  \}  $ for all $ \ottnt{e_{{\mathrm{1}}}}   \sim _{  \mathsf{H}  }  \ottnt{e_{{\mathrm{2}}}}  :   \{ \mathit{x} \mathord{:} \ottnt{B} \mathrel{\mid}  \mathsf{true}  \}  $---which is what we needed to know.
    \item[\WF{Fun}] By IH (\ref{hlr:type}) on each of the types.
    \end{itemize}
    \fi}
  \end{proof}
\end{lemma}

We have investigated two alternatives to the formulation here: type
set optimization and invariants that clarify the role of type sets.

First, we can imagine a system that optimizes the type set of $ \langle  \ottnt{T_{{\mathrm{1}}}}  \mathord{ \overset{ \mathcal{S} }{\Rightarrow} }  \ottnt{T_{{\mathrm{2}}}}  \rangle^{ \ottnt{l} } ~     $ such that $\ottnt{T_{{\mathrm{1}}}}$ and $\ottnt{T_{{\mathrm{2}}}}$ don't appear in
$\mathcal{S}$---taking advantage of idempotence not only for the source type
(Lemma~\ref{lem:heedfulcastidempotence}) but also for the target
type. This change complicates the theory but doesn't give any stronger
theorems. Nevertheless, such an optimization would be a sensible
addition to an implementation.

Second, our proof relates source programs, which start with empty
annotations. In fact, all of the reasoning about type sets is encapsulated
in our proof cast congruence\iffull
(Lemma~\ref{lem:heedfulcastcongruence})\fi. 
We could define a function from heedful \lambdah to classic \lambdah
that unrolls type sets according to the $ \mathsf{choose} $ function. While
this proof would offer a direct understanding of heedful \lambdah
type sets in terms of the classic semantics, it wouldn't give us a
strong property---it degenerates to our proof in the empty type set
case.

\subsection{Relating  classic and eidetic manifest contracts}
\label{sec:eideticlr}

\fi}

\begin{lemma}[Idempotence of coercions]
  \label{lem:eideticcoercionidempotent}
  If $ \emptyset   \vdash _{  \mathsf{E}  }  \ottnt{k}  :   \{ \mathit{x} \mathord{:} \ottnt{B} \mathrel{\mid} \ottnt{e_{{\mathrm{1}}}} \}  $ and $ \mathord{  \vdash _{  \mathsf{E}  } }~  \mathsf{join} ( \ottnt{r_{{\mathrm{1}}}} , \ottnt{r_{{\mathrm{2}}}} )    \mathrel{\parallel}    \{ \mathit{x} \mathord{:} \ottnt{B} \mathrel{\mid} \ottnt{e_{{\mathrm{1}}}} \}   \Rightarrow   \{ \mathit{x} \mathord{:} \ottnt{B} \mathrel{\mid} \ottnt{e_{{\mathrm{2}}}} \}  $, then for all $ \mathsf{result} _{  \mathsf{E}  }~ \ottnt{e} $, we have
  $ \langle   \{ \mathit{x} \mathord{:} \ottnt{B} \mathrel{\mid} \ottnt{e_{{\mathrm{1}}}} \}   \mathord{ \overset{  \mathsf{join} ( \ottnt{r_{{\mathrm{1}}}} , \mathsf{drop} \, \ottsym{(}  \ottnt{r_{{\mathrm{2}}}}  \ottsym{,}   \{ \mathit{x} \mathord{:} \ottnt{B} \mathrel{\mid} \ottnt{e_{{\mathrm{1}}}} \}   \ottsym{)} )  }{\Rightarrow} }   \{ \mathit{x} \mathord{:} \ottnt{B} \mathrel{\mid} \ottnt{e_{{\mathrm{2}}}} \}   \rangle^{\bullet} ~  \ottnt{k}  \,  \longrightarrow ^{*}_{  \mathsf{E}  }  \, \ottnt{e}$ iff
  $ \langle   \{ \mathit{x} \mathord{:} \ottnt{B} \mathrel{\mid} \ottnt{e_{{\mathrm{1}}}} \}   \mathord{ \overset{  \mathsf{join} ( \ottnt{r_{{\mathrm{1}}}} , \ottnt{r_{{\mathrm{2}}}} )  }{\Rightarrow} }   \{ \mathit{x} \mathord{:} \ottnt{B} \mathrel{\mid} \ottnt{e_{{\mathrm{2}}}} \}   \rangle^{\bullet} ~  \ottnt{k}  \,  \longrightarrow ^{*}_{  \mathsf{E}  }  \, \ottnt{e}$.
  \begin{proof}
    By induction on their evaluation derivations: the only difference
    is that the latter derivation performs some extra checks that are
    implied by $ \ottnt{e_{{\mathrm{1}}}}  [  \ottnt{k} / \mathit{x}  ]  \,  \longrightarrow ^{*}_{  \mathsf{E}  }  \,  \mathsf{true} $---which we already know to
    hold.
  \end{proof}
\end{lemma}

As before, cast congruence is the key lemma in our proof---in this
case, the strongest property we have: reduction to identical results.

\begin{lemma}[Cast congruence (single step)]
  \label{lem:eideticcastcongruencesinglestep}
  If
  \begin{itemize}
  \item $ \emptyset   \vdash _{  \mathsf{E}  }  \ottnt{e_{{\mathrm{1}}}}  :  \ottnt{T_{{\mathrm{1}}}} $ and $ \mathord{  \vdash _{  \mathsf{E}  } }~ \ottnt{c}   \mathrel{\parallel}   \ottnt{T_{{\mathrm{1}}}}  \Rightarrow  \ottnt{T_{{\mathrm{2}}}} $ (and so
    $ \emptyset   \vdash _{  \mathsf{E}  }   \langle  \ottnt{T_{{\mathrm{1}}}}  \mathord{ \overset{ \ottnt{c} }{\Rightarrow} }  \ottnt{T_{{\mathrm{2}}}}  \rangle^{\bullet} ~  \ottnt{e_{{\mathrm{1}}}}   :  \ottnt{T_{{\mathrm{2}}}} $),
  \item $\ottnt{e_{{\mathrm{1}}}} \,  \longrightarrow _{  \mathsf{E}  }  \, \ottnt{e_{{\mathrm{2}}}}$ (and so $ \emptyset   \vdash _{  \mathsf{E}  }  \ottnt{e_{{\mathrm{2}}}}  :  \ottnt{T_{{\mathrm{1}}}} $),
  \end{itemize}
  then for all $ \mathsf{result} _{  \mathsf{E}  }~ \ottnt{e} $, we have $ \langle  \ottnt{T_{{\mathrm{1}}}}  \mathord{ \overset{ \ottnt{c} }{\Rightarrow} }  \ottnt{T_{{\mathrm{2}}}}  \rangle^{\bullet} ~  \ottnt{e_{{\mathrm{1}}}}  \,  \longrightarrow ^{*}_{  \mathsf{E}  }  \, \ottnt{e}$
  iff $ \langle  \ottnt{T_{{\mathrm{1}}}}  \mathord{ \overset{ \ottnt{c} }{\Rightarrow} }  \ottnt{T_{{\mathrm{2}}}}  \rangle^{\bullet} ~  \ottnt{e_{{\mathrm{2}}}}  \,  \longrightarrow ^{*}_{  \mathsf{E}  }  \, \ottnt{e}$.
  \begin{proof}
    By cases on the step taken to find $\ottnt{e_{{\mathrm{1}}}} \,  \longrightarrow _{  \mathsf{E}  }  \, \ottnt{e_{{\mathrm{2}}}}$.
    {\iffull
      
      There are two groups of reductions: straightforward merge-free
      reductions and merging reductions.
    In many cases, we simply show confluence, which implies the
    cotermination at identical values in our deterministic semantics
    (Lemma~\ref{lem:eideticdeterminism}).

    \paragraph{Merge-free reductions}
    In these cases, we apply \E{CoerceInner} and whatever rule derived
    $\ottnt{e_{{\mathrm{1}}}} \,  \longrightarrow _{  \mathsf{E}  }  \, \ottnt{e_{{\mathrm{2}}}}$ to find that $ \langle  \ottnt{T_{{\mathrm{1}}}}  \mathord{ \overset{ \ottnt{c} }{\Rightarrow} }  \ottnt{T_{{\mathrm{2}}}}  \rangle^{\bullet} ~  \ottnt{e_{{\mathrm{1}}}}  \,  \longrightarrow _{  \mathsf{E}  }  \,  \langle  \ottnt{T_{{\mathrm{1}}}}  \mathord{ \overset{ \ottnt{c} }{\Rightarrow} }  \ottnt{T_{{\mathrm{2}}}}  \rangle^{\bullet} ~  \ottnt{e_{{\mathrm{2}}}} $, i.e., $\ottnt{e}  \ottsym{=}   \langle  \ottnt{T_{{\mathrm{1}}}}  \mathord{ \overset{ \ottnt{c} }{\Rightarrow} }  \ottnt{T_{{\mathrm{2}}}}  \rangle^{\bullet} ~  \ottnt{e_{{\mathrm{2}}}} $.
    \begin{itemize}
    \item[(\E{Beta})] By \E{CoerceInner} and \E{Beta}.
    \item[(\E{Op})] By \E{CoerceInner} and \E{Op}.
    \item[(\E{Unwrap})] By \E{CoerceInner} and \E{Unwrap}.
    \item[(\E{AppL})] By \E{CoerceInner} with \E{AppL}.
    \item[(\E{AppR})] By \E{CoerceInner} with \E{AppR}.
    \item[(\E{AppRaiseL})] By \E{CoerceInner} with \E{AppRaiseL}; then
      by \E{CastRaise} on both sides.
    \item[(\E{AppRaiseR})] By \E{CoerceInner} with \E{AppRaiseR}; then
      by \E{CastRaise} on both sides.
    \item[(\E{Coerce})] By \E{CoerceInner} and \E{Coerce}.
    \item[(\E{StackDone})] By \E{CoerceInner} and \E{StackDone}.
    \item[(\E{StackPop})] By \E{CoerceInner} and \E{StackRaise}.
    \item[(\E{StackInner})] By \E{CoerceInner} and \E{StackRaise}.
    \item[(\E{StackRaise})] By \E{CoerceInner} and \E{StackRaise}.
    \item[(\E{CheckOK})] By \E{CoerceInner} and \E{CheckOK}.
    \item[(\E{CheckFail})] By \E{CoerceInner} and \E{CheckRaise}.
    \item[(\E{CheckFail})] By \E{CoerceInner} and \E{CheckRaise}.
    \item[(\E{OpInner})] By \E{CoerceInner} and \E{OpInner}.
    \item[(\E{OpRaise})] By \E{CoerceInner} and \E{OpRaise}.

    \end{itemize}

    \paragraph{Merging reductions}
    In these cases, some coercion in $\ottnt{e_{{\mathrm{1}}}}$ reduces when we step
    $\ottnt{e_{{\mathrm{1}}}} \,  \longrightarrow _{  \mathsf{E}  }  \, \ottnt{e_{{\mathrm{2}}}}$, but merges when we consider $ \langle  \ottnt{T_{{\mathrm{1}}}}  \mathord{ \overset{ \ottnt{c} }{\Rightarrow} }  \ottnt{T_{{\mathrm{2}}}}  \rangle^{\bullet} ~  \ottnt{e_{{\mathrm{1}}}} $. We must show that the merged term and $ \langle  \ottnt{T_{{\mathrm{1}}}}  \mathord{ \overset{ \ottnt{c} }{\Rightarrow} }  \ottnt{T_{{\mathrm{2}}}}  \rangle^{\bullet} ~  \ottnt{e_{{\mathrm{2}}}} $
    eventually meet at some common term $\ottnt{e}$.
    It's convenient to renumber the types and coercions, so we
    consider $ \langle  \ottnt{T_{{\mathrm{2}}}}  \mathord{ \overset{ \ottnt{c_{{\mathrm{2}}}} }{\Rightarrow} }  \ottnt{T_{{\mathrm{3}}}}  \rangle^{\bullet} ~  \ottnt{e_{{\mathrm{1}}}} $ where $\ottnt{e_{{\mathrm{1}}}} =  \langle  \ottnt{T_{{\mathrm{1}}}}  \mathord{ \overset{ \ottnt{c_{{\mathrm{1}}}} }{\Rightarrow} }  \ottnt{T_{{\mathrm{2}}}}  \rangle^{\bullet} ~  \ottnt{e}  \,  \longrightarrow _{  \mathsf{E}  }  \, \ottnt{e_{{\mathrm{2}}}}$ for some $\ottnt{e}$.
    \begin{itemize}
    \item[(\E{CoerceStack})] We have $\ottnt{e_{{\mathrm{1}}}}  \ottsym{=}   \langle  \ottnt{T_{{\mathrm{1}}}}  \mathord{ \overset{ \ottnt{r_{{\mathrm{1}}}} }{\Rightarrow} }  \ottnt{T_{{\mathrm{2}}}}  \rangle^{\bullet} ~  \ottnt{k} $ and
      $\ottnt{e_{{\mathrm{2}}}}  \ottsym{=}   \langle  \ottnt{T_{{\mathrm{2}}}} ,   \mathord{?}  ,  \ottnt{r_{{\mathrm{1}}}} ,  \ottnt{k} ,  \ottnt{k}  \rangle^{\bullet} $.
      Since our term is well typed, we know that $\ottnt{c_{{\mathrm{2}}}}$ is some
      $\ottnt{r_{{\mathrm{2}}}}$.
      In the original, unreduced term with $\ottnt{e_{{\mathrm{1}}}}$, we step to $ \langle  \ottnt{T_{{\mathrm{1}}}}  \mathord{ \overset{  \mathsf{join} ( \ottnt{r_{{\mathrm{1}}}} , \ottnt{r_{{\mathrm{2}}}} )  }{\Rightarrow} }  \ottnt{T_{{\mathrm{3}}}}  \rangle^{\bullet} ~  \ottnt{k} $ by \ECastMerge; we then step by
      \E{CoerceStack} to $ \langle  \ottnt{T_{{\mathrm{3}}}} ,   \mathord{?}  ,   \mathsf{join} ( \ottnt{r_{{\mathrm{1}}}} , \ottnt{r_{{\mathrm{2}}}} )  ,  \ottnt{k} ,  \ottnt{k}  \rangle^{\bullet} $. Call this term
      $\ottnt{e'_{{\mathrm{1}}}}$.

      In the reduced term with $\ottnt{e_{{\mathrm{2}}}}$, we have $ \langle  \ottnt{T_{{\mathrm{2}}}}  \mathord{ \overset{ \ottnt{r_{{\mathrm{2}}}} }{\Rightarrow} }  \ottnt{T_{{\mathrm{3}}}}  \rangle^{\bullet} ~   \langle  \ottnt{T_{{\mathrm{2}}}} ,   \mathord{?}  ,  \ottnt{r_{{\mathrm{1}}}} ,  \ottnt{k} ,  \ottnt{k}  \rangle^{\bullet}  $. Call this term $\ottnt{e'_{{\mathrm{2}}}}$.

      We must show that $\ottnt{e'_{{\mathrm{1}}}}$ reduces to a given result iff
      $\ottnt{e'_{{\mathrm{2}}}}$ does.

      The term $\ottnt{e'_{{\mathrm{1}}}}$ evaluates by running through the checks
      in $ \mathsf{join} ( \ottnt{r_{{\mathrm{1}}}} , \ottnt{r_{{\mathrm{2}}}} ) $, raising blame if a check fails or returning
      $\ottnt{k}$ if they all succeed.

      The term $\ottnt{e'_{{\mathrm{2}}}}$ evaluates by running through the checks
      in $\ottnt{r_{{\mathrm{1}}}}$, raising blame if a check fails or returning $\ottnt{k}$
      if they all succeed, eventually reducing to $ \langle  \ottnt{T_{{\mathrm{3}}}} ,   \mathord{?}  ,  \ottnt{r_{{\mathrm{2}}}} ,  \ottnt{k} ,  \ottnt{k}  \rangle^{\bullet} $
      in that case. This term, similarly, reduces to $\ottnt{k}$ or
      $ \mathord{\Uparrow}  \ottnt{l} $ if a given check fails.
      Note that types are preserved by $ \mathsf{join} $ left-to-right, so if
      there is a type that fails in $\ottnt{r_{{\mathrm{1}}}}$, it fails in $ \mathsf{join} ( \ottnt{r_{{\mathrm{1}}}} , \ottnt{r_{{\mathrm{2}}}} ) $, as well. So $\ottnt{e'_{{\mathrm{2}}}}$ fails in the first set of checks
      iff $\ottnt{e'_{{\mathrm{1}}}}$ fails in the first half of $ \mathsf{join} ( \ottnt{r_{{\mathrm{1}}}} , \ottnt{r_{{\mathrm{2}}}} ) $.
      
      Now we must consider those checks in $\ottnt{r_{{\mathrm{2}}}}$. It may be that
      there are types in $\ottnt{r_{{\mathrm{2}}}}$ that were subsumed by a check in $\ottnt{r_{{\mathrm{1}}}}$:
      these types are re-checked in $\ottnt{e'_{{\mathrm{2}}}}$ but not in
      $\ottnt{e'_{{\mathrm{1}}}}$, since the latter merges coercions while the former
      doesn't. We can show that these second checks are redundant by
      idempotence (Lemma~\ref{lem:eideticcoercionidempotent}),
      allowing us to conclude that $\ottnt{e'_{{\mathrm{1}}}}$ and $\ottnt{e'_{{\mathrm{2}}}}$ also behave
      the same on the second half of the checks, and therefore the two
      terms both go to the same blame label or to the same constant.

    \item[(\E{CoerceInner})] $\ottnt{e_{{\mathrm{1}}}}  \ottsym{=}   \langle  \ottnt{T_{{\mathrm{1}}}}  \mathord{ \overset{ \ottnt{c_{{\mathrm{1}}}} }{\Rightarrow} }  \ottnt{T_{{\mathrm{2}}}}  \rangle^{\bullet} ~  \ottnt{e'_{{\mathrm{1}}}} $ and $\ottnt{e_{{\mathrm{2}}}}  \ottsym{=}   \langle  \ottnt{T_{{\mathrm{1}}}}  \mathord{ \overset{ \ottnt{c_{{\mathrm{1}}}} }{\Rightarrow} }  \ottnt{T_{{\mathrm{2}}}}  \rangle^{\bullet} ~  \ottnt{e''_{{\mathrm{1}}}} $ such that $\ottnt{e_{{\mathrm{1}}}} \,  \longrightarrow _{  \mathsf{E}  }  \, \ottnt{e''_{{\mathrm{1}}}}$ and $\ottnt{e_{{\mathrm{1}}}}$ isn't a
      coercion term.
      Both sides reduce to the common term $ \langle  \ottnt{T_{{\mathrm{1}}}}  \mathord{ \overset{  \mathsf{join} ( \ottnt{c_{{\mathrm{1}}}} , \ottnt{c_{{\mathrm{2}}}} )  }{\Rightarrow} }  \ottnt{T_{{\mathrm{3}}}}  \rangle^{\bullet} ~  \ottnt{e''_{{\mathrm{1}}}} $.
      We step $ \langle  \ottnt{T_{{\mathrm{2}}}}  \mathord{ \overset{ \ottnt{c_{{\mathrm{2}}}} }{\Rightarrow} }  \ottnt{T_{{\mathrm{3}}}}  \rangle^{\bullet} ~   (  \langle  \ottnt{T_{{\mathrm{1}}}}  \mathord{ \overset{ \ottnt{c_{{\mathrm{1}}}} }{\Rightarrow} }  \ottnt{T_{{\mathrm{2}}}}  \rangle^{\bullet} ~  \ottnt{e'_{{\mathrm{1}}}}  )  $ first by
      \ECastMerge\ to $ \langle  \ottnt{T_{{\mathrm{1}}}}  \mathord{ \overset{  \mathsf{join} ( \ottnt{c_{{\mathrm{1}}}} , \ottnt{c_{{\mathrm{2}}}} )  }{\Rightarrow} }  \ottnt{T_{{\mathrm{3}}}}  \rangle^{\bullet} ~  \ottnt{e'} $, after which we can
      apply \E{CoerceInner}. We step $ \langle  \ottnt{T_{{\mathrm{2}}}}  \mathord{ \overset{ \ottnt{c_{{\mathrm{2}}}} }{\Rightarrow} }  \ottnt{T_{{\mathrm{3}}}}  \rangle^{\bullet} ~   (  \langle  \ottnt{T_{{\mathrm{1}}}}  \mathord{ \overset{ \ottnt{c_{{\mathrm{1}}}} }{\Rightarrow} }  \ottnt{T_{{\mathrm{2}}}}  \rangle^{\bullet} ~  \ottnt{e''_{{\mathrm{1}}}}  )  $ directly to $\ottnt{e}$ by \ECastMerge.

    \item[(\ECastMerge)] $\ottnt{e_{{\mathrm{1}}}}  \ottsym{=}   \langle  \ottnt{T_{{\mathrm{1}}}}  \mathord{ \overset{ \ottnt{c_{{\mathrm{1}}}} }{\Rightarrow} }  \ottnt{T_{{\mathrm{2}}}}  \rangle^{\bullet} ~   (  \langle  \ottnt{T_{{\mathrm{0}}}}  \mathord{ \overset{ \ottnt{c_{{\mathrm{0}}}} }{\Rightarrow} }  \ottnt{T_{{\mathrm{1}}}}  \rangle^{\bullet} ~  \ottnt{e'}  )  $
      and $\ottnt{e_{{\mathrm{2}}}}  \ottsym{=}   \langle  \ottnt{T_{{\mathrm{0}}}}  \mathord{ \overset{  \mathsf{join} ( \ottnt{c_{{\mathrm{0}}}} , \ottnt{c_{{\mathrm{1}}}} )  }{\Rightarrow} }  \ottnt{T_{{\mathrm{2}}}}  \rangle^{\bullet} ~  \ottnt{e'} $. The common term here is
      $ \langle  \ottnt{T_{{\mathrm{0}}}}  \mathord{ \overset{  \mathsf{join} (  \mathsf{join} ( \ottnt{c_{{\mathrm{0}}}} , \ottnt{c_{{\mathrm{1}}}} )  , \ottnt{c_{{\mathrm{2}}}} )  }{\Rightarrow} }  \ottnt{T_{{\mathrm{3}}}}  \rangle^{\bullet} ~  \ottnt{e'} $: we step the left-hand side
      by \ECastMerge\ twice, and $\ottnt{e_{{\mathrm{2}}}}$ only once.

    \item[(\E{CoerceRaise})] We have $\ottnt{e_{{\mathrm{1}}}}  \ottsym{=}   \langle  \ottnt{T_{{\mathrm{1}}}}  \mathord{ \overset{ \ottnt{c_{{\mathrm{1}}}} }{\Rightarrow} }  \ottnt{T_{{\mathrm{2}}}}  \rangle^{\bullet} ~   \mathord{\Uparrow}  \ottnt{l}  $
      and $\ottnt{e_{{\mathrm{2}}}}  \ottsym{=}   \mathord{\Uparrow}  \ottnt{l} $. Both reduce to the common term
      $ \mathord{\Uparrow}  \ottnt{l} $. The former first steps by \ECastMerge, and
      then both step by \E{CoerceRaise}.
    \end{itemize}
    \fi}
  \end{proof}
\end{lemma}

{\iffull
\begin{lemma}[Cast congruence]
  \label{lem:eideticcastcongruence}
  If
  \begin{itemize}
  \item $ \emptyset   \vdash _{  \mathsf{E}  }  \ottnt{e_{{\mathrm{1}}}}  :  \ottnt{T_{{\mathrm{1}}}} $ and $ \mathord{  \vdash _{  \mathsf{E}  } }~ \ottnt{c}   \mathrel{\parallel}   \ottnt{T_{{\mathrm{1}}}}  \Rightarrow  \ottnt{T_{{\mathrm{2}}}} $ (and so
    $ \emptyset   \vdash _{  \mathsf{E}  }   \langle  \ottnt{T_{{\mathrm{1}}}}  \mathord{ \overset{ \ottnt{c} }{\Rightarrow} }  \ottnt{T_{{\mathrm{2}}}}  \rangle^{\bullet} ~  \ottnt{e_{{\mathrm{1}}}}   :  \ottnt{T_{{\mathrm{2}}}} $),
  \item $\ottnt{e_{{\mathrm{1}}}} \,  \longrightarrow ^{*}_{  \mathsf{E}  }  \, \ottnt{e_{{\mathrm{2}}}}$ (and so $ \emptyset   \vdash _{  \mathsf{E}  }  \ottnt{e_{{\mathrm{2}}}}  :  \ottnt{T_{{\mathrm{1}}}} $),
  \end{itemize}
  then there exists an $\ottnt{e}$ such that $ \langle  \ottnt{T_{{\mathrm{1}}}}  \mathord{ \overset{ \ottnt{c} }{\Rightarrow} }  \ottnt{T_{{\mathrm{2}}}}  \rangle^{\bullet} ~  \ottnt{e_{{\mathrm{1}}}}  \,  \longrightarrow ^{*}_{  \mathsf{E}  }  \, \ottnt{e}$
  and $ \langle  \ottnt{T_{{\mathrm{1}}}}  \mathord{ \overset{ \ottnt{c} }{\Rightarrow} }  \ottnt{T_{{\mathrm{2}}}}  \rangle^{\bullet} ~  \ottnt{e_{{\mathrm{2}}}}  \,  \longrightarrow ^{*}_{  \mathsf{E}  }  \, \ottnt{e}$. Diagrammatically:
  %
  % \eideticcongruence
  \begin{center}
  \begin{tikzpicture}[description/.style={fill=white,inner sep=2pt},align at top]
    \matrix (m) [matrix of math nodes, row sep=4pt, nodes in empty cells,
                 text height=1.5ex, text depth=0.25ex]
    { 
      \ottnt{e_{{\mathrm{1}}}}           & & \ottnt{e_{{\mathrm{2}}}} \\
                       & \Downarrow & \\
       \langle  \ottnt{T_{{\mathrm{1}}}}  \mathord{ \overset{\bullet}{\Rightarrow} }  \ottnt{T_{{\mathrm{2}}}}  \rangle^{ \ottnt{l} } ~  \ottnt{e_{{\mathrm{1}}}}  & &  \langle  \ottnt{T_{{\mathrm{1}}}}  \mathord{ \overset{\bullet}{\Rightarrow} }  \ottnt{T_{{\mathrm{2}}}}  \rangle^{ \ottnt{l} } ~  \ottnt{e_{{\mathrm{2}}}}  \\[20pt]
      &  \mathsf{result} _{  \mathsf{E}  }~ \ottnt{e}  & \\
    };

    \path[->] (m-1-1) edge[E] (m-1-3);
    \path[->] (m-3-1) edge[E*] (m-4-2.north west);
    \path[->] (m-3-3) edge[E*] (m-4-2.north east);
  \end{tikzpicture}
  \end{center}
  \begin{proof}
    By induction on the derivation $\ottnt{e} \,  \longrightarrow ^{*}_{  \mathsf{E}  }  \, \ottnt{e_{{\mathrm{1}}}}$, using the
    single-step cast congruence
    (Lemma~\ref{lem:eideticcastcongruencesinglestep}).
  \end{proof}
\end{lemma}
\fi}

Our proof strategy is as follows: we show that the casts between
related types are applicative, and then we show that well typed source
programs in classic \lambdah are logically related to their
translation.
Our definitions are in Figure~\reflr. Our logical
relation is \textit{blame-exact}. Like our proofs relating forgetful and
heedful \lambdah to classic \lambdah, we use the space-efficient semantics
in the refinement case and use space-efficient type indices.

\begin{lemma}[Similar casts are logically related]
  \label{lem:eideticlrcast}
  If $ \ottnt{T_{{\mathrm{1}}}}   \sim _{  \mathsf{E}  }  \ottnt{T'_{{\mathrm{1}}}} $ and $ \ottnt{T_{{\mathrm{2}}}}   \sim _{  \mathsf{E}  }  \ottnt{T'_{{\mathrm{2}}}} $ and $ \ottnt{e_{{\mathrm{1}}}}   \sim _{  \mathsf{E}  }  \ottnt{e_{{\mathrm{2}}}}  :  \ottnt{T_{{\mathrm{1}}}} $, then
  $  \langle  \ottnt{T_{{\mathrm{1}}}}  \mathord{ \overset{\bullet}{\Rightarrow} }  \ottnt{T_{{\mathrm{2}}}}  \rangle^{ \ottnt{l} } ~  \ottnt{e_{{\mathrm{1}}}}    \simeq _{  \mathsf{E}  }   \langle  \ottnt{T'_{{\mathrm{1}}}}  \mathord{ \overset{\bullet}{\Rightarrow} }  \ottnt{T'_{{\mathrm{2}}}}  \rangle^{ \ottnt{l} } ~  \ottnt{e_{{\mathrm{2}}}}   :  \ottnt{T_{{\mathrm{2}}}} $.
  \begin{proof}
    By induction on the invariant relation, using coercion congruence
    in the function case when $\ottnt{e_{{\mathrm{2}}}}$ is a function proxy.
    {\iffull We always step first by \E{Coerce} on the right to
      $ \langle  \ottnt{T'_{{\mathrm{1}}}}  \mathord{ \overset{  \mathsf{coerce} ( \ottnt{T'_{{\mathrm{1}}}} , \ottnt{T'_{{\mathrm{2}}}} , \ottnt{l} )  }{\Rightarrow} }  \ottnt{T'_{{\mathrm{2}}}}  \rangle^{\bullet} ~  \ottnt{e_{{\mathrm{2}}}} $.
    \begin{itemize}
    \item[(\A{Refine})] Let $ \ottnt{e'_{{\mathrm{1}}}}   \sim _{  \mathsf{E}  }  \ottnt{e'_{{\mathrm{2}}}}  :   \{ \mathit{x} \mathord{:} \ottnt{B} \mathrel{\mid} \ottnt{e'_{{\mathrm{1}}}} \}  $, we
      know that $\ottnt{e'_{{\mathrm{1}}}}  \ottsym{=}  \ottnt{e'_{{\mathrm{2}}}} = \ottnt{k}$ such that $ \ottnt{e'_{{\mathrm{1}}}}  [  \ottnt{k} / \mathit{x}  ]  \,  \longrightarrow ^{*}_{  \mathsf{E}  }  \,  \mathsf{true} $.
      In classic \lambdah, we step by \E{CheckNone} to
      $ \langle   \{ \mathit{x} \mathord{:} \ottnt{B} \mathrel{\mid} \ottnt{e_{{\mathrm{2}}}} \}  ,   \ottnt{e_{{\mathrm{2}}}}  [  \ottnt{k} / \mathit{x}  ]  ,  \ottnt{k}  \rangle^{ \ottnt{l} } $; in eidetic \lambdah, we step by
      \E{Coerce} and then \E{CoerceStack} to
      $ \langle   \{ \mathit{x} \mathord{:} \ottnt{B} \mathrel{\mid} \ottnt{e'_{{\mathrm{2}}}} \}  ,   \mathord{?}  ,   \{ \mathit{x} \mathord{:} \ottnt{B} \mathrel{\mid} \ottnt{e'_{{\mathrm{2}}}} \}^{ \ottnt{l} }  ,  \ottnt{k} ,  \ottnt{k}  \rangle^{\bullet} $, and then by \E{StackPop} to 
      \[  \langle   \{ \mathit{x} \mathord{:} \ottnt{B} \mathrel{\mid} \ottnt{e'_{{\mathrm{2}}}} \}  ,   \mathord{\checkmark}  ,  \mathsf{nil} ,  \ottnt{k} ,   \langle   \{ \mathit{x} \mathord{:} \ottnt{B} \mathrel{\mid} \ottnt{e'_{{\mathrm{2}}}} \}  ,   \ottnt{e'_{{\mathrm{2}}}}  [  \ottnt{k} / \mathit{x}  ]  ,  \ottnt{k}  \rangle^{ \ottnt{l} }   \rangle^{\bullet}  \]
      Since $ \ottnt{k}   \sim _{  \mathsf{E}  }  \ottnt{k}  :   \{ \mathit{x} \mathord{:} \ottnt{B} \mathrel{\mid}  \mathsf{true}  \}  $ by definition and reflexivity of
      $ \longrightarrow ^{*}_{  \mathsf{E}  } $, we know that $  \ottnt{e_{{\mathrm{2}}}}  [  \ottnt{k} / \mathit{x}  ]    \simeq _{  \mathsf{E}  }   \ottnt{e'_{{\mathrm{2}}}}  [  \ottnt{k} / \mathit{x}  ]   :   \{ \mathit{x} \mathord{:}  \mathsf{Bool}  \mathrel{\mid}  \mathsf{true}  \}  $.
      If the predicates step to a blame label (the same one!), then
      both terms raise that label (by \E{CheckRaise}, with an added
      \E{StackRaise} on the right). Similarly, if the predicates go
      to $ \mathsf{false} $, then both sides raise $ \mathord{\Uparrow}  \ottnt{l} $ by
      \E{CheckFail} (followed by the same steps as for inner blame).
      Finally, if the predicates both go to $ \mathsf{true} $, then both
      checks return $\ottnt{k}$. After stepping by \E{StackDone} on the
      right, we find that both terms reduce to $\ottnt{k}$ and that
      $ \ottnt{e'_{{\mathrm{2}}}}  [  \ottnt{k} / \mathit{x}  ]  \,  \longrightarrow ^{*}_{  \mathsf{E}  }  \,  \mathsf{true} $.

    \item[(\A{Fun})] We have $  \ottnt{T_{{\mathrm{11}}}} \mathord{ \rightarrow } \ottnt{T_{{\mathrm{12}}}}    \sim _{  \mathsf{E}  }   \ottnt{T'_{{\mathrm{11}}}} \mathord{ \rightarrow } \ottnt{T'_{{\mathrm{12}}}}  $
      and $  \ottnt{T_{{\mathrm{21}}}} \mathord{ \rightarrow } \ottnt{T_{{\mathrm{22}}}}    \sim _{  \mathsf{E}  }   \ottnt{T'_{{\mathrm{21}}}} \mathord{ \rightarrow } \ottnt{T'_{{\mathrm{22}}}}  $. Let $ \ottnt{e_{{\mathrm{1}}}}   \sim _{  \mathsf{E}  }  \ottnt{e_{{\mathrm{2}}}}  :   \ottnt{T'_{{\mathrm{11}}}} \mathord{ \rightarrow } \ottnt{T'_{{\mathrm{12}}}}  $. The classic side is a value, by \V{ProxyC}. The
      eidetic \lambdah term is: \[  \langle   \ottnt{T'_{{\mathrm{11}}}} \mathord{ \rightarrow } \ottnt{T'_{{\mathrm{12}}}}   \mathord{ \overset{ \ottnt{c_{{\mathrm{1}}}}  \mapsto  \ottnt{c_{{\mathrm{2}}}} }{\Rightarrow} }   \ottnt{T'_{{\mathrm{21}}}} \mathord{ \rightarrow } \ottnt{T'_{{\mathrm{22}}}}   \rangle^{\bullet} ~  \ottnt{e_{{\mathrm{2}}}}  \] How this term steps depends on the shape of the value
      $\ottnt{e_{{\mathrm{2}}}}$: either $\ottnt{e_{{\mathrm{2}}}}$ is an abstraction $ \lambda \mathit{x} \mathord{:} \ottnt{T} .~  \ottnt{e} $ and we
      have a value by \V{ProxyE}, or it is a function proxy
      $ \langle   \ottnt{T_{{\mathrm{01}}}} \mathord{ \rightarrow } \ottnt{T_{{\mathrm{02}}}}   \mathord{ \overset{ \ottnt{c'_{{\mathrm{1}}}}  \mapsto  \ottnt{c'_{{\mathrm{2}}}} }{\Rightarrow} }   \ottnt{T'_{{\mathrm{11}}}} \mathord{ \rightarrow } \ottnt{T'_{{\mathrm{12}}}}   \rangle^{\bullet} ~   \lambda \mathit{x} \mathord{:} \ottnt{T_{{\mathrm{01}}}} .~  \ottnt{e}  $ and we step
      by \ECastMerge.
      \begin{itemize}
      \item[(\V{ProxyE})] We step to
        $ \langle   \ottnt{T'_{{\mathrm{11}}}} \mathord{ \rightarrow } \ottnt{T'_{{\mathrm{12}}}}   \mathord{ \overset{ \ottnt{c_{{\mathrm{1}}}}  \mapsto  \ottnt{c_{{\mathrm{2}}}} }{\Rightarrow} }   \ottnt{T'_{{\mathrm{21}}}} \mathord{ \rightarrow } \ottnt{T'_{{\mathrm{22}}}}   \rangle^{\bullet} ~  \ottnt{e_{{\mathrm{2}}}} $.  Let $ \ottnt{e'_{{\mathrm{1}}}}   \sim _{  \mathsf{E}  }  \ottnt{e'_{{\mathrm{2}}}}  :  \ottnt{T'_{{\mathrm{21}}}} $ be given. Both sides unwrap, giving us $ \langle  \ottnt{T_{{\mathrm{12}}}}  \mathord{ \overset{\bullet}{\Rightarrow} }  \ottnt{T_{{\mathrm{22}}}}  \rangle^{ \ottnt{l} } ~   (  \ottnt{e_{{\mathrm{1}}}} ~  (  \langle  \ottnt{T_{{\mathrm{21}}}}  \mathord{ \overset{\bullet}{\Rightarrow} }  \ottnt{T_{{\mathrm{11}}}}  \rangle^{ \ottnt{l} } ~  \ottnt{e'_{{\mathrm{1}}}}  )   )  $ on the classic side and $ \langle  \ottnt{T'_{{\mathrm{12}}}}  \mathord{ \overset{ \ottnt{c_{{\mathrm{2}}}} }{\Rightarrow} }  \ottnt{T'_{{\mathrm{22}}}}  \rangle^{\bullet} ~   (  \ottnt{e_{{\mathrm{2}}}} ~  (  \langle  \ottnt{T'_{{\mathrm{21}}}}  \mathord{ \overset{ \ottnt{c_{{\mathrm{1}}}} }{\Rightarrow} }  \ottnt{T'_{{\mathrm{12}}}}  \rangle^{\bullet} ~  \ottnt{e'_{{\mathrm{2}}}}  )   )  $. By the IH, the arguments
        are related and reduce to related results (by expansion via
        \E{Coerce} and the observation that $ \mathsf{coerce} ( \ottnt{T'_{{\mathrm{21}}}} , \ottnt{T'_{{\mathrm{11}}}} , \ottnt{l} )   \ottsym{=}  \ottnt{c_{{\mathrm{1}}}}$ ). Blame (at the same label!) aborts the computation. If
        the arguments produce values, then we apply our assumption
        that $ \ottnt{e_{{\mathrm{1}}}}   \sim _{  \mathsf{E}  }  \ottnt{e_{{\mathrm{2}}}}  :   \ottnt{T'_{{\mathrm{11}}}} \mathord{ \rightarrow } \ottnt{T'_{{\mathrm{12}}}}  $, so $  \ottnt{e_{{\mathrm{1}}}} ~ \ottnt{e'_{{\mathrm{1}}}}    \simeq _{  \mathsf{E}  }   \ottnt{e_{{\mathrm{2}}}} ~ \ottnt{e'_{{\mathrm{2}}}}   :  \ottnt{T'_{{\mathrm{12}}}} $. Again, blame (at the same label!)  aborts early. A
        value flows to the related codomain casts, and we are done
        by the IH and \E{Coerce}-expansion (observing
        $ \mathsf{coerce} ( \ottnt{T'_{{\mathrm{12}}}} , \ottnt{T'_{{\mathrm{22}}}} , \ottnt{l} )   \ottsym{=}  \ottnt{c_{{\mathrm{2}}}}$).
      \item[(\ECastMerge)] We step to $ \langle   \ottnt{T'_{{\mathrm{01}}}} \mathord{ \rightarrow } \ottnt{T'_{{\mathrm{02}}}}   \mathord{ \overset{  \mathsf{join} ( \ottnt{c_{{\mathrm{1}}}} , \ottnt{c'_{{\mathrm{1}}}} )   \mapsto   \mathsf{join} ( \ottnt{c'_{{\mathrm{2}}}} , \ottnt{c_{{\mathrm{2}}}} )  }{\Rightarrow} }   \ottnt{T'_{{\mathrm{21}}}} \mathord{ \rightarrow } \ottnt{T'_{{\mathrm{22}}}}   \rangle^{\bullet} ~   \lambda \mathit{x} \mathord{:} \ottnt{T'_{{\mathrm{01}}}} .~  \ottnt{e}  $. Let $ \ottnt{e'_{{\mathrm{1}}}}   \sim _{  \mathsf{E}  }  \ottnt{e'_{{\mathrm{2}}}}  :  \ottnt{T'_{{\mathrm{21}}}} $ be given. Both sides unwrap as above. On the classic
        side, we have the same term as before: $ \langle  \ottnt{T_{{\mathrm{12}}}}  \mathord{ \overset{\bullet}{\Rightarrow} }  \ottnt{T_{{\mathrm{22}}}}  \rangle^{ \ottnt{l} } ~   (  \ottnt{e_{{\mathrm{1}}}} ~  (  \langle  \ottnt{T_{{\mathrm{21}}}}  \mathord{ \overset{\bullet}{\Rightarrow} }  \ottnt{T_{{\mathrm{11}}}}  \rangle^{ \ottnt{l} } ~  \ottnt{e'_{{\mathrm{1}}}}  )   )  $. On the eidetic side, we have some extra
        coercions: $ \langle  \ottnt{T'_{{\mathrm{02}}}}  \mathord{ \overset{  \mathsf{join} ( \ottnt{c'_{{\mathrm{2}}}} , \ottnt{c_{{\mathrm{2}}}} )  }{\Rightarrow} }  \ottnt{T'_{{\mathrm{22}}}}  \rangle^{\bullet} ~   (   (  \lambda \mathit{x} \mathord{:} \ottnt{T'_{{\mathrm{01}}}} .~  \ottnt{e}  )  ~  (  \langle  \ottnt{T'_{{\mathrm{21}}}}  \mathord{ \overset{  \mathsf{join} ( \ottnt{c_{{\mathrm{1}}}} , \ottnt{c'_{{\mathrm{1}}}} )  }{\Rightarrow} }  \ottnt{T'_{{\mathrm{01}}}}  \rangle^{\bullet} ~  \ottnt{e'_{{\mathrm{1}}}}  )   )  $.
        We use coercion congruence to resolve these, and see that
        both terms behave the same.

        Considering the argument, we can factor it out to the term
        $ \langle  \ottnt{T'_{{\mathrm{11}}}}  \mathord{ \overset{ \ottnt{c'_{{\mathrm{1}}}} }{\Rightarrow} }  \ottnt{T'_{{\mathrm{01}}}}  \rangle^{\bullet} ~   (  \langle  \ottnt{T'_{{\mathrm{21}}}}  \mathord{ \overset{ \ottnt{c_{{\mathrm{1}}}} }{\Rightarrow} }  \ottnt{T'_{{\mathrm{11}}}}  \rangle^{\bullet} ~  \ottnt{e'_{{\mathrm{2}}}}  )  $. We know that
        $  \langle  \ottnt{T_{{\mathrm{21}}}}  \mathord{ \overset{\bullet}{\Rightarrow} }  \ottnt{T_{{\mathrm{11}}}}  \rangle^{ \ottnt{l} } ~  \ottnt{e'_{{\mathrm{1}}}}    \simeq _{  \mathsf{E}  }   \langle  \ottnt{T'_{{\mathrm{21}}}}  \mathord{ \overset{ \ottnt{c_{{\mathrm{1}}}} }{\Rightarrow} }  \ottnt{T'_{{\mathrm{11}}}}  \rangle^{\bullet} ~  \ottnt{e'_{{\mathrm{2}}}}   :  \ottnt{T'_{{\mathrm{11}}}} $ by the IH
        (with \E{Coerce}-expansion, observing that
        $ \mathsf{coerce} ( \ottnt{T'_{{\mathrm{21}}}} , \ottnt{T'_{{\mathrm{11}}}} , \ottnt{l} )   \ottsym{=}  \ottnt{c_{{\mathrm{1}}}}$), so they reduce to related
        results $ \ottnt{e''_{{\mathrm{1}}}}   \simeq _{  \mathsf{E}  }  \ottnt{e''_{{\mathrm{2}}}}  :  \ottnt{T'_{{\mathrm{11}}}} $. By coercion congruence
        (Lemma~\ref{lem:eideticcastcongruence}), we know that
        $ \langle  \ottnt{T'_{{\mathrm{11}}}}  \mathord{ \overset{ \ottnt{c'_{{\mathrm{1}}}} }{\Rightarrow} }  \ottnt{T'_{{\mathrm{01}}}}  \rangle^{\bullet} ~  \ottnt{e''_{{\mathrm{2}}}} $ and $ \langle  \ottnt{T'_{{\mathrm{11}}}}  \mathord{ \overset{ \ottnt{c'_{{\mathrm{1}}}} }{\Rightarrow} }  \ottnt{T'_{{\mathrm{01}}}}  \rangle^{\bullet} ~   (  \langle  \ottnt{T'_{{\mathrm{21}}}}  \mathord{ \overset{ \ottnt{c_{{\mathrm{1}}}} }{\Rightarrow} }  \ottnt{T'_{{\mathrm{11}}}}  \rangle^{\bullet} ~  \ottnt{e'_{{\mathrm{2}}}}  )  $ behave identically.
        We can make a similar observation in the codomain: factoring
        out to $ \langle  \ottnt{T'_{{\mathrm{12}}}}  \mathord{ \overset{ \ottnt{c_{{\mathrm{2}}}} }{\Rightarrow} }  \ottnt{T'_{{\mathrm{22}}}}  \rangle^{\bullet} ~   (  \langle  \ottnt{T'_{{\mathrm{02}}}}  \mathord{ \overset{ \ottnt{c'_{{\mathrm{2}}}} }{\Rightarrow} }  \ottnt{T'_{{\mathrm{12}}}}  \rangle^{\bullet} ~   (   (  \lambda \mathit{x} \mathord{:} \ottnt{T'_{{\mathrm{01}}}} .~  \ottnt{e}  )  ~  (  \langle  \ottnt{T'_{{\mathrm{11}}}}  \mathord{ \overset{ \ottnt{c'_{{\mathrm{1}}}} }{\Rightarrow} }  \ottnt{T'_{{\mathrm{01}}}}  \rangle^{\bullet} ~   (  \langle  \ottnt{T'_{{\mathrm{21}}}}  \mathord{ \overset{ \ottnt{c_{{\mathrm{1}}}} }{\Rightarrow} }  \ottnt{T'_{{\mathrm{11}}}}  \rangle^{\bullet} ~  \ottnt{e'_{{\mathrm{2}}}}  )   )   )   )  $, we know that
        this term is equivalent to $ \langle  \ottnt{T'_{{\mathrm{12}}}}  \mathord{ \overset{ \ottnt{c_{{\mathrm{2}}}} }{\Rightarrow} }  \ottnt{T'_{{\mathrm{22}}}}  \rangle^{\bullet} ~   (  \langle  \ottnt{T'_{{\mathrm{02}}}}  \mathord{ \overset{ \ottnt{c'_{{\mathrm{2}}}} }{\Rightarrow} }  \ottnt{T'_{{\mathrm{12}}}}  \rangle^{\bullet} ~   (   (  \lambda \mathit{x} \mathord{:} \ottnt{T'_{{\mathrm{01}}}} .~  \ottnt{e}  )  ~  (  \langle  \ottnt{T'_{{\mathrm{11}}}}  \mathord{ \overset{ \ottnt{c'_{{\mathrm{1}}}} }{\Rightarrow} }  \ottnt{T'_{{\mathrm{01}}}}  \rangle^{\bullet} ~  \ottnt{e''_{{\mathrm{2}}}}  )   )   )  $; by
        assumption, we know that $ \langle  \ottnt{T'_{{\mathrm{12}}}}  \mathord{ \overset{ \ottnt{c'_{{\mathrm{2}}}} }{\Rightarrow} }  \ottnt{T'_{{\mathrm{22}}}}  \rangle^{\bullet} ~   (   (  \lambda \mathit{x} \mathord{:} \ottnt{T'_{{\mathrm{01}}}} .~  \ottnt{e}  )  ~  (  \langle  \ottnt{T'_{{\mathrm{11}}}}  \mathord{ \overset{ \ottnt{c'_{{\mathrm{1}}}} }{\Rightarrow} }  \ottnt{T'_{{\mathrm{01}}}}  \rangle^{\bullet} ~  \ottnt{e''_{{\mathrm{2}}}}  )   )  $ is equivalent to $ \ottnt{e_{{\mathrm{1}}}} ~ \ottnt{e''_{{\mathrm{1}}}} $,
        so they both reduce to related results $ \ottnt{e'''_{{\mathrm{1}}}}   \simeq _{  \mathsf{E}  }  \ottnt{e'''_{{\mathrm{2}}}}  :  \ottnt{T'_{{\mathrm{12}}}} $. Now, by the IH on the codomain (along with \E{Coerce}
        expansion and the observation that $ \mathsf{coerce} ( \ottnt{T'_{{\mathrm{12}}}} , \ottnt{T'_{{\mathrm{22}}}} , \ottnt{l} )   \ottsym{=}  \ottnt{c_{{\mathrm{2}}}}$), we know that $  \langle  \ottnt{T_{{\mathrm{12}}}}  \mathord{ \overset{\bullet}{\Rightarrow} }  \ottnt{T_{{\mathrm{22}}}}  \rangle^{ \ottnt{l} } ~  \ottnt{e'''_{{\mathrm{1}}}}    \simeq _{  \mathsf{E}  }   \langle  \ottnt{T'_{{\mathrm{12}}}}  \mathord{ \overset{ \ottnt{c_{{\mathrm{2}}}} }{\Rightarrow} }  \ottnt{T'_{{\mathrm{22}}}}  \rangle^{\bullet} ~  \ottnt{e'''_{{\mathrm{2}}}}   :  \ottnt{T'_{{\mathrm{22}}}} $. We can apply coercion congruence again
        (Lemma~\ref{lem:eideticcastcongruence}) to see that the
        behavior on $\ottnt{e'''_{{\mathrm{2}}}}$ is the same as the behavior on the
        unreduced term.
      \end{itemize}
    \end{itemize}
    \fi}
  \end{proof}
\end{lemma}

\begin{lemma}[Relating classic and eidetic source programs]
  \label{lem:eideticlr}
  ~

  \noindent
  \begin{enumerate}
  \item \label{elr:term} If $ \Gamma   \vdash _{  \mathsf{C}  }  \ottnt{e}  :  \ottnt{T} $ as a source program then
    $ \Gamma   \vdash   \ottnt{e}   \simeq _{  \mathsf{E}  }  \ottnt{e}  :  \ottnt{T} $.
  \item \label{elr:type} If $ \mathord{  \vdash _{  \mathsf{C}  } }~ \ottnt{T} $ as a source program then $ \ottnt{T}   \sim _{  \mathsf{E}  }  \ottnt{T} $.
  \end{enumerate}
  \begin{proof}
    By mutual induction on the typing derivations.
    {\iffull
    \paragraph{Term typing \fbox{$ \Gamma   \vdash _{  \mathsf{C}  }  \ottnt{e}  :  \ottnt{T} $}}
    \begin{itemize}
    \item[\T{Var}] We have $\mathit{x}  \ottsym{=}  \mathit{x}$. We know by assumption
      that $ \delta_{{\mathrm{1}}}  \ottsym{(}  \mathit{x}  \ottsym{)}   \sim _{  \mathsf{E}  }  \delta_{{\mathrm{2}}}  \ottsym{(}  \mathit{x}  \ottsym{)}  :  \ottnt{T} $.
    \item[\T{Const}] We have $\ottnt{k}  \ottsym{=}  \ottnt{k}$. Since we are dealing
      with a source program, $\ottnt{T}  \ottsym{=}   \{ \mathit{x} \mathord{:} \ottnt{B} \mathrel{\mid}  \mathsf{true}  \} $. We have immediately
      that $ \mathsf{ty} ( \ottnt{k} )   \ottsym{=}  \ottnt{B}$ and $   \mathsf{true}   [  \ottnt{k} / \mathit{x}  ]    \simeq _{  \mathsf{E}  }    \mathsf{true}   [  \ottnt{k} / \mathit{x}  ]   :   \{ \mathit{x} \mathord{:}  \mathsf{Bool}  \mathrel{\mid}  \mathsf{true}  \}  $ by reflexivity of $ \longrightarrow ^{*}_{  \mathsf{C}  } $, so $ \ottnt{k}   \simeq _{  \mathsf{E}  }  \ottnt{k}  :   \{ \mathit{x} \mathord{:} \ottnt{B} \mathrel{\mid}  \mathsf{true}  \}  $.
    \item[\T{Abs}] Let $ \Gamma   \models _{  \mathsf{E}  }  \delta $. We must show that
      $  \lambda \mathit{x} \mathord{:} \ottnt{T_{{\mathrm{1}}}} .~  \delta_{{\mathrm{1}}}  \ottsym{(}  \ottnt{e_{{\mathrm{1}}}}  \ottsym{)}    \sim _{  \mathsf{E}  }   \lambda \mathit{x} \mathord{:} \ottnt{T_{{\mathrm{1}}}} .~  \delta_{{\mathrm{2}}}  \ottsym{(}  \ottnt{e_{{\mathrm{1}}}}  \ottsym{)}   :   \ottnt{T_{{\mathrm{1}}}} \mathord{ \rightarrow } \ottnt{T_{{\mathrm{2}}}}  $. Let $ \ottnt{e_{{\mathrm{2}}}}   \sim _{  \mathsf{E}  }  \ottnt{e'_{{\mathrm{2}}}}  :  \ottnt{T_{{\mathrm{1}}}} $. We must show that applying
      the abstractions to these values yields related values. Both
      sides step by \E{Beta}, to $ \delta_{{\mathrm{1}}}  \ottsym{(}  \ottnt{e_{{\mathrm{1}}}}  \ottsym{)}  [  \ottnt{e_{{\mathrm{2}}}} / \mathit{x}  ] $ and
      $ \delta_{{\mathrm{2}}}  \ottsym{(}  \ottnt{e_{{\mathrm{1}}}}  \ottsym{)}  [  \ottnt{e'_{{\mathrm{2}}}} / \mathit{x}  ] $, respectively. But $  \Gamma , \mathit{x} \mathord{:} \ottnt{T_{{\mathrm{1}}}}    \models _{  \mathsf{E}  }   \delta  [  \ottnt{e_{{\mathrm{2}}}} , \ottnt{e'_{{\mathrm{2}}}} / \mathit{x}  ]  $, so we can apply IH (\ref{elr:term}) on
      $\ottnt{e_{{\mathrm{1}}}}$ and $\ottnt{e_{{\mathrm{1}}}}$ to show the two sides reduce to
      related results.
    \item[\T{Op}] By IH (\ref{elr:term}) on each argument, either one
      of the arguments goes to blame (in both calculi), and we are done by
      \E{OpRaise}, or all of the arguments reduce to related
      values. Since $ \mathsf{ty} (\mathord{ \ottnt{op} }) $ is first order, these values must be
      related at refined base types, which means that they are in fact
      all equal constants. We then reduce by \E{Op} on both sides to
      have $\denot{ op } \, \ottsym{(}  \ottnt{k_{{\mathrm{1}}}}  \ottsym{,}  \dots  \ottsym{,}  \ottnt{k_{\ottmv{n}}}  \ottsym{)}$. We have assumed that the
      denotations of operations agree with their typings in
      \textit{all} modes, so then $\denot{ op } \, \ottsym{(}  \ottnt{k_{{\mathrm{1}}}}  \ottsym{,}  \dots  \ottsym{,}  \ottnt{k_{\ottmv{n}}}  \ottsym{)}$ satisfies
      the refinement for $ \longrightarrow _{  \mathsf{C}  } $ in particular, and we are done.
    \item[\T{App}] Let $ \Gamma   \models _{  \mathsf{E}  }  \delta $. We must show that
      $  \delta_{{\mathrm{1}}}  \ottsym{(}  \ottnt{e_{{\mathrm{1}}}}  \ottsym{)} ~ \delta_{{\mathrm{1}}}  \ottsym{(}  \ottnt{e_{{\mathrm{2}}}}  \ottsym{)}    \simeq _{  \mathsf{E}  }   \delta_{{\mathrm{2}}}  \ottsym{(}  \ottnt{e_{{\mathrm{1}}}}  \ottsym{)} ~ \delta_{{\mathrm{2}}}  \ottsym{(}  \ottnt{e_{{\mathrm{2}}}}  \ottsym{)}   :  \ottnt{T_{{\mathrm{2}}}} $. But by IH (\ref{elr:term}) on
      $\ottnt{e_{{\mathrm{1}}}}$ and $\ottnt{e_{{\mathrm{2}}}}$, we are done directly.
    \item[\T{Cast}] Let $ \Gamma   \models _{  \mathsf{E}  }  \delta $. By IH (\ref{elr:term}) on
      $\ottnt{e}$, we know that $ \delta_{{\mathrm{1}}}  \ottsym{(}  \ottnt{e}  \ottsym{)}   \simeq _{  \mathsf{E}  }  \delta_{{\mathrm{2}}}  \ottsym{(}  \ottnt{e}  \ottsym{)}  :  \ottnt{T_{{\mathrm{1}}}} $, either
      $\delta_{{\mathrm{1}}}  \ottsym{(}  \ottnt{e}  \ottsym{)} \,  \longrightarrow ^{*}_{  \mathsf{C}  }  \,  \mathord{\Uparrow}  \ottnt{l'} $ and $\delta_{{\mathrm{2}}}  \ottsym{(}  \ottnt{e}  \ottsym{)} \,  \longrightarrow ^{*}_{  \mathsf{E}  }  \,  \mathord{\Uparrow}  \ottnt{l'} $ (and we are done) or $\delta_{{\mathrm{1}}}  \ottsym{(}  \ottnt{e'}  \ottsym{)}$ and
      $\delta_{{\mathrm{2}}}  \ottsym{(}  \ottnt{e'}  \ottsym{)}$ reduce to values $ \ottnt{e_{{\mathrm{1}}}}   \sim _{  \mathsf{E}  }  \ottnt{e_{{\mathrm{2}}}}  :  \ottnt{T_{{\mathrm{1}}}} $. By
      Lemma~\ref{lem:eideticlrcast} (using IH (\ref{elr:type}) on the
      types), we know that $  \langle  \ottnt{T_{{\mathrm{1}}}}  \mathord{ \overset{\bullet}{\Rightarrow} }  \ottnt{T_{{\mathrm{2}}}}  \rangle^{ \ottnt{l} } ~  \ottnt{e_{{\mathrm{1}}}}    \sim _{  \mathsf{E}  }   \langle  \ottnt{T_{{\mathrm{1}}}}  \mathord{ \overset{\bullet}{\Rightarrow} }  \ottnt{T_{{\mathrm{2}}}}  \rangle^{ \ottnt{l} } ~  \ottnt{e_{{\mathrm{2}}}}   :  \ottnt{T_{{\mathrm{2}}}} $, so
      each side must reduce to a result $ \ottnt{e'_{{\mathrm{1}}}}   \sim _{  \mathsf{E}  }  \ottnt{e'_{{\mathrm{2}}}}  :  \ottnt{T_{{\mathrm{2}}}} $.
      We have cast congruence on the classic side straightforwardly,
      finding:
      \[  \langle  \ottnt{T_{{\mathrm{1}}}}  \mathord{ \overset{\bullet}{\Rightarrow} }  \ottnt{T_{{\mathrm{2}}}}  \rangle^{ \ottnt{l} } ~  \delta_{{\mathrm{1}}}  \ottsym{(}  \ottnt{e'}  \ottsym{)}  \,  \longrightarrow ^{*}_{  \mathsf{C}  }  \,  \langle  \ottnt{T_{{\mathrm{1}}}}  \mathord{ \overset{\bullet}{\Rightarrow} }  \ottnt{T_{{\mathrm{2}}}}  \rangle^{ \ottnt{l} } ~  \ottnt{e_{{\mathrm{1}}}}   \longrightarrow ^{*}_{  \mathsf{C}  }  \ottnt{e'_{{\mathrm{1}}}} \]
      On the heedful side, we can apply our derived cast congruence
      (Lemma~\ref{lem:eideticcastcongruence}) to find that
      $ \langle  \ottnt{T_{{\mathrm{1}}}}  \mathord{ \overset{\bullet}{\Rightarrow} }  \ottnt{T_{{\mathrm{2}}}}  \rangle^{ \ottnt{l} } ~  \ottnt{e_{{\mathrm{2}}}}  \,  \longrightarrow ^{*}_{  \mathsf{E}  }  \, \ottnt{e'_{{\mathrm{2}}}}$ and $\delta_{{\mathrm{2}}}  \ottsym{(}  \ottnt{e}  \ottsym{)} \,  \longrightarrow ^{*}_{  \mathsf{E}  }  \, \ottnt{e_{{\mathrm{2}}}}$ imply that $ \langle  \ottnt{T_{{\mathrm{1}}}}  \mathord{ \overset{\bullet}{\Rightarrow} }  \ottnt{T_{{\mathrm{2}}}}  \rangle^{ \ottnt{l} } ~  \delta_{{\mathrm{2}}}  \ottsym{(}  \ottnt{e}  \ottsym{)}  \,  \longrightarrow ^{*}_{  \mathsf{E}  }  \, \ottnt{e'_{{\mathrm{2}}}}$.
    \item[\T{Blame}] Contradiction---doesn't appear in source
      programs. Though in fact it is in the relation, since
      $  \mathord{\Uparrow}  \ottnt{l}    \simeq _{  \mathsf{E}  }   \mathord{\Uparrow}  \ottnt{l}   :  \ottnt{T} $ for any $\ottnt{l}$ and $\ottnt{T}$.
    \item[\T{Check}] Contradiction---doesn't appear in source
      programs.
    \end{itemize}
    
    \paragraph{Type well formedness \fbox{$ \mathord{  \vdash _{  \mathsf{C}  } }~ \ottnt{T} $}}
    \begin{itemize}
    \item[\WF{Base}] We can immediately see $   \mathsf{true}   [  \ottnt{k} / \mathit{x}  ]    \simeq _{  \mathsf{E}  }    \mathsf{true}   [  \ottnt{k} / \mathit{x}  ]   :   \{ \mathit{x} \mathord{:}  \mathsf{Bool}  \mathrel{\mid}  \mathsf{true}  \}  $ for any $ \ottnt{k}   \sim _{  \mathsf{E}  }  \ottnt{k}  :   \{ \mathit{x} \mathord{:} \ottnt{B} \mathrel{\mid}  \mathsf{true}  \}  $, i.e., any
      $\ottnt{k}$ such that $ \mathsf{ty} ( \ottnt{k} )   \ottsym{=}  \ottnt{B}$, since $ \longrightarrow ^{*}_{  \mathsf{C}  } $ is
      reflexive.
    \item[\WF{Refine}] By inversion, we know that $  \mathit{x} \mathord{:}  \{ \mathit{x} \mathord{:} \ottnt{B} \mathrel{\mid}  \mathsf{true}  \}     \vdash _{  \mathsf{C}  }  \ottnt{e}  :   \{ \mathit{x} \mathord{:}  \mathsf{Bool}  \mathrel{\mid}  \mathsf{true}  \}  $; by IH (\ref{elr:term}), we find that
      $ \delta_{{\mathrm{1}}}  \ottsym{(}  \ottnt{e}  \ottsym{)}   \simeq _{  \mathsf{E}  }  \delta_{{\mathrm{2}}}  \ottsym{(}  \ottnt{e}  \ottsym{)}  :   \{ \mathit{x} \mathord{:}  \mathsf{Bool}  \mathrel{\mid}  \mathsf{true}  \}  $, i.e., that
      $  \ottnt{e}  [  \ottnt{e_{{\mathrm{1}}}} / \mathit{x}  ]    \simeq _{  \mathsf{E}  }   \ottnt{e}  [  \ottnt{e_{{\mathrm{2}}}} / \mathit{x}  ]   :   \{ \mathit{x} \mathord{:}  \mathsf{Bool}  \mathrel{\mid}  \mathsf{true}  \}  $ for all $ \ottnt{e_{{\mathrm{1}}}}   \sim _{  \mathsf{E}  }  \ottnt{e_{{\mathrm{2}}}}  :   \{ \mathit{x} \mathord{:} \ottnt{B} \mathrel{\mid}  \mathsf{true}  \}  $---which is what we needed to know.
    \item[\WF{Fun}] By IH (\ref{elr:type}) on each of the types.
    \end{itemize}
    \fi}
   \end{proof}
\end{lemma}

\section{Proofs of bounds for space-efficiency}
\label{app:bounds}

This section contains our definitions for collecting types in a
program and the corresponding proof of bounded space consumption (for
all modes at once).

\begin{figure}
%  \sidebyside[.6][.36][t]
  {\hdr{Term type extraction}{\qquad \fbox{$ \mathsf{types} ( \ottnt{e} )  : \mathcal{P}(\ottnt{T})$}}
    \[ \begin{array}{r@{~}c@{~}l}
       \mathsf{types} ( \mathit{x} )  &=&  \emptyset  \\
       \mathsf{types} ( \ottnt{k} )  &=&  \emptyset  \\
       \mathsf{types} (  \lambda \mathit{x} \mathord{:} \ottnt{T} .~  \ottnt{e}  )  &=&   \mathsf{types} ( \ottnt{T} )   \cup   \mathsf{types} ( \ottnt{e} )   \\
       \mathsf{types} (  \langle  \ottnt{T_{{\mathrm{1}}}}  \mathord{ \overset{ \ottnt{a} }{\Rightarrow} }  \ottnt{T_{{\mathrm{2}}}}  \rangle^{ \ottnt{l} } ~  \ottnt{e}  )  &=&     \mathsf{types} ( \ottnt{T_{{\mathrm{1}}}} )   \cup   \mathsf{types} ( \ottnt{T_{{\mathrm{2}}}} )    \cup  {} \\  &  &   \mathsf{types} ( \ottnt{a} )    \cup   \mathsf{types} ( \ottnt{e} )   \\
       \mathsf{types} (  \ottnt{e_{{\mathrm{1}}}} ~ \ottnt{e_{{\mathrm{2}}}}  )  &=&   \mathsf{types} ( \ottnt{e_{{\mathrm{1}}}} )   \cup   \mathsf{types} ( \ottnt{e_{{\mathrm{2}}}} )   \\
       \mathsf{types} ( \ottnt{op}  \ottsym{(}  \ottnt{e_{{\mathrm{1}}}}  \ottsym{,}  \dots  \ottsym{,}  \ottnt{e_{\ottmv{n}}}  \ottsym{)} )  &=& \bigcup_{1 \le i \le n}  \mathsf{types} ( \ottnt{e_{\ottmv{i}}} )  \\
       \mathsf{types} (  \langle   \{ \mathit{x} \mathord{:} \ottnt{B} \mathrel{\mid} \ottnt{e_{{\mathrm{1}}}} \}  ,  \ottnt{e_{{\mathrm{2}}}} ,  \ottnt{k}  \rangle^{ \ottnt{l} }  )  &=&   \mathsf{types} (  \{ \mathit{x} \mathord{:} \ottnt{B} \mathrel{\mid} \ottnt{e_{{\mathrm{1}}}} \}  )   \cup   \mathsf{types} ( \ottnt{e_{{\mathrm{2}}}} )   \\
       \mathsf{types} (  \langle   \{ \mathit{x} \mathord{:} \ottnt{B} \mathrel{\mid} \ottnt{e_{{\mathrm{1}}}} \}  ,  \ottnt{s} ,  \ottnt{r} ,  \ottnt{k} ,  \ottnt{e}  \rangle^{\bullet}  )  &=& \\ 
      \multicolumn{3}{r}{   \mathsf{types} (  \{ \mathit{x} \mathord{:} \ottnt{B} \mathrel{\mid} \ottnt{e_{{\mathrm{1}}}} \}  )   \cup   \mathsf{types} ( \ottnt{r} )    \cup   \mathsf{types} ( \ottnt{e} )  } \\
       \mathsf{types} (  \mathord{\Uparrow}  \ottnt{l}  )  &=&  \emptyset 
    \end{array} \]}
  {\hdr{Type, type set, and coercion type extraction}{}
    \[ \begin{array}{rcl}
      \multicolumn{3}{l}{\fbox{$ \mathsf{types} ( \ottnt{T} )  : \mathcal{P}(\ottnt{T})$}} \\[1em]
       \mathsf{types} (  \{ \mathit{x} \mathord{:} \ottnt{B} \mathrel{\mid} \ottnt{e} \}  )  &=&   \set{   \{ \mathit{x} \mathord{:} \ottnt{B} \mathrel{\mid} \ottnt{e} \}   }   \cup   \mathsf{types} ( \ottnt{e} )   \\
       \mathsf{types} (  \ottnt{T_{{\mathrm{1}}}} \mathord{ \rightarrow } \ottnt{T_{{\mathrm{2}}}}  )  &=&    \set{   \ottnt{T_{{\mathrm{1}}}} \mathord{ \rightarrow } \ottnt{T_{{\mathrm{2}}}}   }   \cup  {} \\  &  &   \mathsf{types} ( \ottnt{T_{{\mathrm{1}}}} )    \cup   \mathsf{types} ( \ottnt{T_{{\mathrm{2}}}} )   \\[1em]
      \multicolumn{3}{l}{\fbox{$ \mathsf{types} ( \ottnt{a} )  : \mathcal{P}(\ottnt{T})$}} \\[1em]
       \mathsf{types} ( \bullet )  &=&  \emptyset  \\
\iffull       \mathsf{types} ( \mathcal{S} )  &=& \bigcup_{ \ottnt{T}  \in  \mathcal{S} }  \mathsf{types} ( \ottnt{T} )  \\[1em] \fi
       \mathsf{types} ( \mathsf{nil} )  &=&  \emptyset  \\
       \mathsf{types} (   \{ \mathit{x} \mathord{:} \ottnt{B} \mathrel{\mid} \ottnt{e} \}^{ \ottnt{l} }  , \ottnt{r}  )  &=&   \set{   \{ \mathit{x} \mathord{:} \ottnt{B} \mathrel{\mid} \ottnt{e} \}   }   \cup   \mathsf{types} ( \ottnt{r} )   \\
       \mathsf{types} ( \ottnt{c_{{\mathrm{1}}}}  \mapsto  \ottnt{c_{{\mathrm{2}}}} )  &=&   \mathsf{types} ( \ottnt{c_{{\mathrm{1}}}} )   \cup   \mathsf{types} ( \ottnt{c_{{\mathrm{2}}}} )  
    \end{array} \]}

  \hdr{Type height}{\qquad \fbox{$ \mathsf{height} ( \ottnt{T} ) $}}
  \[ \begin{array}{rcl}
     \mathsf{height} (  \{ \mathit{x} \mathord{:} \ottnt{B} \mathrel{\mid} \ottnt{e} \}  )  &=& 1 \\
     \mathsf{height} (  \ottnt{T_{{\mathrm{1}}}} \mathord{ \rightarrow } \ottnt{T_{{\mathrm{2}}}}  )  &=& 1 + \max_{i \in \set{1,2}}  \mathsf{height} ( \ottnt{T_{\ottmv{i}}} ) 
  \end{array} \]

  \caption{Type extraction and type height}
  \label{fig:types}
\end{figure}

We define a function collecting all of the distinct types that appear
in a program in Figure~\ref{fig:types}. If the type $\ottnt{T}  \ottsym{=}    \{ \mathit{x} \mathord{:}  \mathsf{Int}  \mathrel{\mid}  \mathit{x}  \mathrel{\ge}  \ottsym{0}  \}  \mathord{ \rightarrow }  \{ \mathit{y} \mathord{:}  \mathsf{Int}  \mathrel{\mid}  \mathit{y}  \mathrel{\ne}  \ottsym{0}  \}  $ appears in the program $\ottnt{e}$, then
$ \mathsf{types} ( \ottnt{e} ) $ includes the type $\ottnt{T}$ itself along with its
subparts $ \{ \mathit{x} \mathord{:}  \mathsf{Int}  \mathrel{\mid}  \mathit{x}  \mathrel{\ge}  \ottsym{0}  \} $ and $ \{ \mathit{y} \mathord{:}  \mathsf{Int}  \mathrel{\mid}  \mathit{y}  \mathrel{\ne}  \ottsym{0}  \} $.
\begin{lemma}
  \label{lem:typessubstitution}
  $ \mathsf{types} (  \ottnt{e}  [  \ottnt{e'} / \mathit{x}  ]  )  \subseteq   \mathsf{types} ( \ottnt{e} )   \cup   \mathsf{types} ( \ottnt{e'} )  $
  \begin{proof}
    By induction on $\ottnt{e}$.
    {\iffull
    \begin{itemize}
    \item[($\ottnt{e}  \ottsym{=}  \mathit{y}$)] $ \mathsf{types} ( \mathit{y} )  =  \emptyset $, so we must show
      that $  \mathsf{types} (  \mathit{y}  [  \ottnt{e'} / \mathit{x}  ]  )   \subseteq   \mathsf{types} ( \ottnt{e'} )  $. If $\mathit{x}  \neq  \mathit{y}$, then
      $ \mathsf{types} (  \mathit{y}  [  \ottnt{e'} / \mathit{x}  ]  )   \ottsym{=}   \mathsf{types} ( \mathit{y} )  =  \emptyset $, which is a subset
      of everything. If $\mathit{x}  \ottsym{=}  \mathit{y}$, then $ \mathsf{types} (  \mathit{x}  [  \ottnt{e'} / \mathit{x}  ]  )   \ottsym{=}   \mathsf{types} ( \ottnt{e'} ) $.
    \item[($\ottnt{e}  \ottsym{=}  \ottnt{k}$)] Immediate, since $\ottnt{k}$ is closed and
      $ \mathsf{types} ( \ottnt{k} )  =  \emptyset $.
    \item[($\ottnt{e}  \ottsym{=}   \lambda \mathit{y} \mathord{:} \ottnt{T} .~  \ottnt{e} $)] By the IH on $\ottnt{e}$ and the closure of
      $\ottnt{T}$.
    \item[($\ottnt{e}  \ottsym{=}   \langle  \ottnt{T_{{\mathrm{1}}}}  \mathord{ \overset{ \ottnt{a} }{\Rightarrow} }  \ottnt{T_{{\mathrm{2}}}}  \rangle^{ \ottnt{l} } ~  \ottnt{e} $)] By the IH on $\ottnt{e}$ and the
      closure of the types and annotations.
    \item[($\ottnt{e}  \ottsym{=}   \ottnt{e_{{\mathrm{1}}}} ~ \ottnt{e_{{\mathrm{2}}}} $)] By the IHs on $\ottnt{e_{{\mathrm{1}}}}$ and $\ottnt{e_{{\mathrm{2}}}}$.
    \item[($\ottnt{e}  \ottsym{=}  \ottnt{op}  \ottsym{(}  \ottnt{e_{{\mathrm{1}}}}  \ottsym{,}  \dots  \ottsym{,}  \ottnt{e_{\ottmv{n}}}  \ottsym{)}$)] By the IHs on each $\ottnt{e_{\ottmv{i}}}$.
    \item[($\ottnt{e}  \ottsym{=}   \langle   \{ \mathit{x} \mathord{:} \ottnt{B} \mathrel{\mid} \ottnt{e_{{\mathrm{1}}}} \}  ,  \ottnt{e_{{\mathrm{2}}}} ,  \ottnt{k}  \rangle^{ \ottnt{l} } $)] By the IH on $\ottnt{e_{{\mathrm{2}}}}$ and the
      closure of $ \{ \mathit{x} \mathord{:} \ottnt{B} \mathrel{\mid} \ottnt{e_{{\mathrm{1}}}} \} $---noting that $\ottnt{e_{{\mathrm{2}}}}$ is in fact
      closed in well typed terms.
    \item[($\ottnt{e}  \ottsym{=}   \langle   \{ \mathit{x} \mathord{:} \ottnt{B} \mathrel{\mid} \ottnt{e_{{\mathrm{1}}}} \}  ,  \ottnt{s} ,  \ottnt{r} ,  \ottnt{k} ,  \ottnt{e}  \rangle^{\bullet} $)] By the IH on
      $\ottnt{e}$---though all of the terms are closed.
    \item[($\ottnt{e}  \ottsym{=}   \mathord{\Uparrow}  \ottnt{l} $)] Immediate since $ \mathord{\Uparrow}  \ottnt{l} $ is
      closed and $ \mathsf{types} (  \mathord{\Uparrow}  \ottnt{l}  )  =  \emptyset $.
    \end{itemize}
    \fi}
  \end{proof}
\end{lemma}

{\iffull
\begin{lemma}
  \label{lem:typesmerge}
  $    \mathsf{types} ( \ottnt{T_{{\mathrm{1}}}} )   \cup   \mathsf{types} (  \mathsf{merge} _{ \ottnt{m} }( \ottnt{T_{{\mathrm{1}}}} , \ottnt{a_{{\mathrm{1}}}} , \ottnt{T_{{\mathrm{2}}}} , \ottnt{a_{{\mathrm{2}}}} , \ottnt{T_{{\mathrm{3}}}} )  )    \cup   \mathsf{types} ( \ottnt{T_{{\mathrm{3}}}} )    \subseteq       \mathsf{types} ( \ottnt{T_{{\mathrm{1}}}} )   \cup   \mathsf{types} ( \ottnt{a_{{\mathrm{1}}}} )    \cup   \mathsf{types} ( \ottnt{T_{{\mathrm{2}}}} )    \cup   \mathsf{types} ( \ottnt{a_{{\mathrm{2}}}} )    \cup   \mathsf{types} ( \ottnt{T_{{\mathrm{3}}}} )   $
  \begin{proof}
    By cases on each, but observing that the merged annotation is
    always no bigger than the original, and that the type $\ottnt{T_{{\mathrm{2}}}}$
    may or may not vanish.
  \end{proof}
\end{lemma}
\fi}

\begin{lemma}
  \label{lem:typesdom}
  $  \mathsf{types} (  \mathsf{dom} ( \ottnt{a} )  )   \subseteq   \mathsf{types} ( \ottnt{a} )  $
  \begin{proof}
    This property is trivial when $\ottnt{a}  \ottsym{=}  \bullet$.

    {\iffull
    When the annotation is a type set, for $ \mathsf{dom} ( \mathcal{S} ) $ to be
    defined, every type in $\mathcal{S}$ must be a function type. So:
    \[ \begin{array}{rcl}
       \mathsf{types} ( \mathcal{S} )  &=& \bigcup_{ \ottnt{T}  \in  \mathcal{S} }  \mathsf{types} ( \ottnt{T} )  \\
      &=& \bigcup_{  \ottnt{T_{{\mathrm{1}}}} \mathord{ \rightarrow } \ottnt{T_{{\mathrm{2}}}}   \in  \mathcal{S} }  \mathsf{types} (  \ottnt{T_{{\mathrm{1}}}} \mathord{ \rightarrow } \ottnt{T_{{\mathrm{2}}}}  )  \\
      &=& \bigcup_{  \ottnt{T_{{\mathrm{1}}}} \mathord{ \rightarrow } \ottnt{T_{{\mathrm{2}}}}   \in  \mathcal{S} }    \set{   \ottnt{T_{{\mathrm{1}}}} \mathord{ \rightarrow } \ottnt{T_{{\mathrm{2}}}}   }   \cup  {} \\  &  &   \mathsf{types} ( \ottnt{T_{{\mathrm{1}}}} )    \cup   \mathsf{types} ( \ottnt{T_{{\mathrm{2}}}} )   \\
      &\supseteq& \bigcup_{  \ottnt{T_{{\mathrm{1}}}} \mathord{ \rightarrow } \ottnt{T_{{\mathrm{2}}}}   \in  \mathcal{S} }  \mathsf{types} ( \ottnt{T_{{\mathrm{1}}}} )  \\
      &=& \bigcup_{ \ottnt{T}  \in   \mathsf{types} (  \mathsf{dom} ( \mathcal{S} )  )  }  \mathsf{types} ( \ottnt{T} )  \\
      &=&  \mathsf{types} (  \mathsf{dom} ( \mathcal{S} )  ) 
    \end{array} \]
    \fi}

    Immediate when $\ottnt{a}  \ottsym{=}  \ottnt{c_{{\mathrm{1}}}}  \mapsto  \ottnt{c_{{\mathrm{2}}}}$.
  \end{proof}
\end{lemma}

\begin{lemma}
  \label{lem:typescod}
  $  \mathsf{types} (  \mathsf{cod} ( \ottnt{a} )  )   \subseteq   \mathsf{types} ( \ottnt{a} )  $
  \begin{proof}
    {\iffull
    This property is trivial when $\ottnt{a}  \ottsym{=}  \bullet$.

    When the annotation is a type set, for $ \mathsf{cod} ( \ottnt{a} ) $ to be
    defined, every type in $\ottnt{a}$ must be a function type. So:
    \[ \begin{array}{rcl}
       \mathsf{types} ( \mathcal{S} )  &=& \bigcup_{ \ottnt{T}  \in  \mathcal{S} }  \mathsf{types} ( \ottnt{T} )  \\
      &=& \bigcup_{  \ottnt{T_{{\mathrm{1}}}} \mathord{ \rightarrow } \ottnt{T_{{\mathrm{2}}}}   \in  \mathcal{S} }  \mathsf{types} (  \ottnt{T_{{\mathrm{1}}}} \mathord{ \rightarrow } \ottnt{T_{{\mathrm{2}}}}  )  \\
      &=& \bigcup_{  \ottnt{T_{{\mathrm{1}}}} \mathord{ \rightarrow } \ottnt{T_{{\mathrm{2}}}}   \in  \mathcal{S} }    \set{   \ottnt{T_{{\mathrm{1}}}} \mathord{ \rightarrow } \ottnt{T_{{\mathrm{2}}}}   }   \cup   \mathsf{types} ( \ottnt{T_{{\mathrm{1}}}} )    \cup   \mathsf{types} ( \ottnt{T_{{\mathrm{2}}}} )   \\
      &\supseteq& \bigcup_{  \ottnt{T_{{\mathrm{1}}}} \mathord{ \rightarrow } \ottnt{T_{{\mathrm{2}}}}   \in  \mathcal{S} }  \mathsf{types} ( \ottnt{T_{{\mathrm{2}}}} )  \\
      &=& \bigcup_{ \ottnt{T}  \in   \mathsf{types} (  \mathsf{cod} ( \mathcal{S} )  )  }  \mathsf{types} ( \ottnt{T} )  \\
      &=&  \mathsf{types} (  \mathsf{cod} ( \mathcal{S} )  ) 
    \end{array} \]

    Immediate when $\ottnt{a}  \ottsym{=}  \ottnt{c_{{\mathrm{1}}}}  \mapsto  \ottnt{c_{{\mathrm{2}}}}$.
    \else
    Similar to Lemma~\ref{lem:typesdom}.
    \fi}
  \end{proof}
\end{lemma}

\begin{lemma}[Coercing types doesn't introduce types]
  \label{lem:typescoerce}
  ~
  
  \noindent
  $  \mathsf{types} (  \mathsf{coerce} ( \ottnt{T_{{\mathrm{1}}}} , \ottnt{T_{{\mathrm{2}}}} , \ottnt{l} )  )   \subseteq    \mathsf{types} ( \ottnt{T_{{\mathrm{1}}}} )   \cup   \mathsf{types} ( \ottnt{T_{{\mathrm{2}}}} )   $
  \begin{proof}
    By induction on $\ottnt{T_{{\mathrm{1}}}}$ and $\ottnt{T_{{\mathrm{2}}}}$.
    When they are refinements, we have the coercion just being
    $ \{ \mathit{x} \mathord{:} \ottnt{B} \mathrel{\mid} \ottnt{e_{{\mathrm{2}}}} \}^{ \ottnt{l} } $. When they are functions, by the IH.
  \end{proof}
\end{lemma}

\begin{lemma}[Dropping types doesn't introduce types]
  \label{lem:typesdrop}
  ~

  \noindent
  $  \mathsf{types} ( \mathsf{drop} \, \ottsym{(}  \ottnt{r}  \ottsym{,}   \{ \mathit{x} \mathord{:} \ottnt{B} \mathrel{\mid} \ottnt{e} \}   \ottsym{)} )   \subseteq   \mathsf{types} ( \ottnt{r} )  $
  \begin{proof}
    By induction on $\ottnt{r}$.
    \begin{itemize}
      \item[($\ottnt{r}  \ottsym{=}  \mathsf{nil}$)] The two sides are immediately equal.
      \item[($\ottnt{r}  \ottsym{=}    \{ \mathit{x} \mathord{:} \ottnt{B} \mathrel{\mid} \ottnt{e'} \}^{ \ottnt{l} }  , \ottnt{r'} $)] If $ \{ \mathit{x} \mathord{:} \ottnt{B} \mathrel{\mid} \ottnt{e'} \}  \, \not \supset \,  \{ \mathit{x} \mathord{:} \ottnt{B} \mathrel{\mid} \ottnt{e} \} $, then the
        two are identical. If not, then we have $  \mathsf{types} ( \ottnt{r'} )   \subseteq   \mathsf{types} ( \ottnt{r} )  $ by the IH.
    \end{itemize}
  \end{proof}
\end{lemma}

\begin{lemma}[Coercion merges don't introduce types]
  \label{lem:typescoercion}
  ~
  
  \noindent
  $  \mathsf{types} (  \mathsf{join} ( \ottnt{r_{{\mathrm{1}}}} , \ottnt{r_{{\mathrm{2}}}} )  )   \subseteq    \mathsf{types} ( \ottnt{r_{{\mathrm{1}}}} )   \cup   \mathsf{types} ( \ottnt{r_{{\mathrm{2}}}} )   $
  \begin{proof}
    By induction on $\ottnt{r_{{\mathrm{1}}}}$.
    \begin{itemize}
      \item[($\ottnt{r_{{\mathrm{1}}}}  \ottsym{=}  \mathsf{nil}$)] The two sides are immediately equal.
      \item[($\ottnt{r_{{\mathrm{1}}}}  \ottsym{=}    \{ \mathit{x} \mathord{:} \ottnt{B} \mathrel{\mid} \ottnt{e} \}^{ \ottnt{l} }  , \ottnt{r'_{{\mathrm{1}}}} $)] Using
        Lemma~\ref{lem:typesdrop}, we find:
        \[ \begin{array}{r@{~}c@{~}l}
           \mathsf{types} (  \mathsf{join} ( \ottnt{r_{{\mathrm{1}}}} , \ottnt{r_{{\mathrm{2}}}} )  )  &=&   \set{   \{ \mathit{x} \mathord{:} \ottnt{B} \mathrel{\mid} \ottnt{e} \}   }   \cup  {} \\  &  &   \mathsf{types} (  \mathsf{join} ( \ottnt{r'_{{\mathrm{1}}}} , \mathsf{drop} \, \ottsym{(}  \ottnt{r_{{\mathrm{2}}}}  \ottsym{,}   \{ \mathit{x} \mathord{:} \ottnt{B} \mathrel{\mid} \ottnt{e} \}   \ottsym{)} )  )   \\
          &\subseteq&    \set{   \{ \mathit{x} \mathord{:} \ottnt{B} \mathrel{\mid} \ottnt{e} \}   }   \cup   \mathsf{types} ( \ottnt{r'_{{\mathrm{1}}}} )    \cup  {} \\  &  &   \mathsf{types} ( \mathsf{drop} \, \ottsym{(}  \ottnt{r_{{\mathrm{2}}}}  \ottsym{,}   \{ \mathit{x} \mathord{:} \ottnt{B} \mathrel{\mid} \ottnt{e} \}   \ottsym{)} )   \\
          &\subseteq&    \set{   \{ \mathit{x} \mathord{:} \ottnt{B} \mathrel{\mid} \ottnt{e} \}   }   \cup   \mathsf{types} ( \ottnt{r'_{{\mathrm{1}}}} )    \cup   \mathsf{types} ( \ottnt{r_{{\mathrm{2}}}} )   \\
          &=&    \mathsf{types} ( \ottnt{r_{{\mathrm{1}}}} )   \cup   \mathsf{types} ( \ottnt{r_{{\mathrm{2}}}} )  
        \end{array} \]
    \end{itemize}
  \end{proof}
\end{lemma}

\begin{lemma}[Reduction doesn't introduce types]
  \label{lem:typesreduction}
  If $\ottnt{e} \,  \longrightarrow _{ \ottnt{m} }  \, \ottnt{e'}$ then $  \mathsf{types} ( \ottnt{e'} )   \subseteq   \mathsf{types} ( \ottnt{e} )  $.
  \begin{proof}
    By induction on the step taken.
    {\iffull
    \paragraph{Shared rules}
    \begin{itemize}
    \item[(\E{Beta})] \[ \begin{array}{rclr}
         \mathsf{types} (   (  \lambda \mathit{x} \mathord{:} \ottnt{T} .~  \ottnt{e_{{\mathrm{12}}}}  )  ~ \ottnt{e_{{\mathrm{2}}}}  )  &=&    \mathsf{types} ( \ottnt{T} )   \cup   \mathsf{types} ( \ottnt{e_{{\mathrm{12}}}} )    \cup   \mathsf{types} ( \ottnt{e_{{\mathrm{2}}}} )   & \\
        &\supseteq&   \mathsf{types} ( \ottnt{e_{{\mathrm{12}}}} )   \cup   \mathsf{types} ( \ottnt{e_{{\mathrm{2}}}} )   & \text{(Lemma~\ref{lem:typessubstitution})} \\
        &=&  \mathsf{types} (  \ottnt{e_{{\mathrm{12}}}}  [  \ottnt{e_{{\mathrm{2}}}} / \mathit{x}  ]  ) 
      \end{array} \]
    \item[(\E{Op})] \[ \begin{array}{rclr}
         \mathsf{types} ( \ottnt{op}  \ottsym{(}  \ottnt{e_{{\mathrm{1}}}}  \ottsym{,} \, ... \, \ottsym{,}  \ottnt{e_{\ottmv{n}}}  \ottsym{)} )  &=& \bigcup_{1 \le i \le n}  \mathsf{types} ( \ottnt{e_{\ottmv{i}}} )  & \\
        &\supseteq&  \emptyset  & \\
        &=&  \mathsf{types} ( \ottnt{k} )  & \text{(operations are first order)} \\
        &=&  \mathsf{types} ( \denot{ op } \, \ottsym{(}  \ottnt{e_{{\mathrm{1}}}}  \ottsym{,} \, ... \, \ottsym{,}  \ottnt{e_{\ottmv{n}}}  \ottsym{)} ) 
      \end{array} \]
    \item[(\E{Unwrap})] \[ \begin{array}{@{~~}cl}
        \multicolumn{2}{l}{ \mathsf{types} (  \langle   \ottnt{T_{{\mathrm{11}}}} \mathord{ \rightarrow } \ottnt{T_{{\mathrm{12}}}}   \mathord{ \overset{ \ottnt{a} }{\Rightarrow} }   \ottnt{T_{{\mathrm{21}}}} \mathord{ \rightarrow } \ottnt{T_{{\mathrm{22}}}}   \rangle^{ \ottnt{l} } ~   \ottnt{e_{{\mathrm{1}}}} ~ \ottnt{e_{{\mathrm{2}}}}   ) } \\
        =&      \mathsf{types} (  \ottnt{T_{{\mathrm{11}}}} \mathord{ \rightarrow } \ottnt{T_{{\mathrm{12}}}}  )   \cup   \mathsf{types} (  \ottnt{T_{{\mathrm{21}}}} \mathord{ \rightarrow } \ottnt{T_{{\mathrm{22}}}}  )    \cup   \mathsf{types} ( \ottnt{a} )    \cup   \mathsf{types} ( \ottnt{e_{{\mathrm{1}}}} )    \cup   \mathsf{types} ( \ottnt{e_{{\mathrm{2}}}} )    \\
        =&          \set{   \ottnt{T_{{\mathrm{11}}}} \mathord{ \rightarrow } \ottnt{T_{{\mathrm{12}}}}   }   \cup   \mathsf{types} ( \ottnt{T_{{\mathrm{11}}}} )    \cup   \mathsf{types} ( \ottnt{T_{{\mathrm{12}}}} )    \cup   \set{   \ottnt{T_{{\mathrm{21}}}} \mathord{ \rightarrow } \ottnt{T_{{\mathrm{22}}}}   }    \cup   \mathsf{types} ( \ottnt{T_{{\mathrm{21}}}} )    \cup   \mathsf{types} ( \ottnt{T_{{\mathrm{22}}}} )    \cup   \mathsf{types} ( \ottnt{a} )    \cup   \mathsf{types} ( \ottnt{e_{{\mathrm{1}}}} )    \cup   \mathsf{types} ( \ottnt{e_{{\mathrm{2}}}} )    \\
        \supseteq&        \mathsf{types} ( \ottnt{T_{{\mathrm{11}}}} )   \cup   \mathsf{types} ( \ottnt{T_{{\mathrm{12}}}} )    \cup   \mathsf{types} ( \ottnt{T_{{\mathrm{21}}}} )    \cup   \mathsf{types} ( \ottnt{T_{{\mathrm{22}}}} )    \cup   \mathsf{types} ( \ottnt{a} )    \cup   \mathsf{types} ( \ottnt{e_{{\mathrm{1}}}} )    \cup   \mathsf{types} ( \ottnt{e_{{\mathrm{2}}}} )    \\
        \supseteq&         \mathsf{types} ( \ottnt{T_{{\mathrm{11}}}} )   \cup   \mathsf{types} ( \ottnt{T_{{\mathrm{12}}}} )    \cup   \mathsf{types} ( \ottnt{T_{{\mathrm{21}}}} )    \cup   \mathsf{types} ( \ottnt{T_{{\mathrm{22}}}} )    \cup   \mathsf{types} (  \mathsf{dom} ( \ottnt{a} )  )    \cup   \mathsf{types} (  \mathsf{cod} ( \ottnt{a} )  )    \cup   \mathsf{types} ( \ottnt{e_{{\mathrm{1}}}} )    \cup   \mathsf{types} ( \ottnt{e_{{\mathrm{2}}}} )   \\
        \multicolumn{2}{r}{\text{(Lemmas~\ref{lem:typesdom} and~\ref{lem:typescod})} } \\
        =&  \mathsf{types} (  \langle  \ottnt{T_{{\mathrm{12}}}}  \mathord{ \overset{  \mathsf{cod} ( \ottnt{a} )  }{\Rightarrow} }  \ottnt{T_{{\mathrm{22}}}}  \rangle^{ \ottnt{l} } ~   (  \ottnt{e_{{\mathrm{1}}}} ~  (  \langle  \ottnt{T_{{\mathrm{21}}}}  \mathord{ \overset{  \mathsf{dom} ( \ottnt{a} )  }{\Rightarrow} }  \ottnt{T_{{\mathrm{11}}}}  \rangle^{ \ottnt{l} } ~  \ottnt{e_{{\mathrm{2}}}}  )   )   ) 
      \end{array} \]
    \item[(\E{CheckNone})]  \[ \begin{array}{rclr}
         \mathsf{types} (  \langle   \{ \mathit{x} \mathord{:} \ottnt{B} \mathrel{\mid} \ottnt{e_{{\mathrm{1}}}} \}   \mathord{ \overset{ \bullet }{\Rightarrow} }   \{ \mathit{x} \mathord{:} \ottnt{B} \mathrel{\mid} \ottnt{e_{{\mathrm{2}}}} \}   \rangle^{ \ottnt{l} } ~  \ottnt{k}  )  &=&     \mathsf{types} (  \{ \mathit{x} \mathord{:} \ottnt{B} \mathrel{\mid} \ottnt{e_{{\mathrm{1}}}} \}  )   \cup   \mathsf{types} ( \bullet )    \cup   \mathsf{types} (  \{ \mathit{x} \mathord{:} \ottnt{B} \mathrel{\mid} \ottnt{e_{{\mathrm{2}}}} \}  )    \cup   \mathsf{types} ( \ottnt{k} )   & \\
        &\supseteq&   \mathsf{types} (  \{ \mathit{x} \mathord{:} \ottnt{B} \mathrel{\mid} \ottnt{e_{{\mathrm{2}}}} \}  )   \cup   \mathsf{types} ( \ottnt{k} )   & \\
        &=&    \set{   \{ \mathit{x} \mathord{:} \ottnt{B} \mathrel{\mid} \ottnt{e_{{\mathrm{2}}}} \}   }   \cup   \mathsf{types} ( \ottnt{e_{{\mathrm{2}}}} )    \cup   \mathsf{types} ( \ottnt{k} )   & \\
        &=&    \mathsf{types} (  \{ \mathit{x} \mathord{:} \ottnt{B} \mathrel{\mid} \ottnt{e_{{\mathrm{2}}}} \}  )   \cup   \mathsf{types} ( \ottnt{e_{{\mathrm{2}}}} )    \cup   \mathsf{types} ( \ottnt{k} )   & \\
        &\supseteq&   \mathsf{types} (  \{ \mathit{x} \mathord{:} \ottnt{B} \mathrel{\mid} \ottnt{e_{{\mathrm{2}}}} \}  )   \cup   \mathsf{types} (  \ottnt{e_{{\mathrm{2}}}}  [  \ottnt{k} / \mathit{x}  ]  )   & \text{(Lemma~\ref{lem:typessubstitution})} \\
        &=&  \mathsf{types} (  \{ \mathit{x} \mathord{:} \ottnt{B} \mathrel{\mid} \ottnt{e_{{\mathrm{2}}}} \}  )  & \text{since $ \mathsf{types} ( \ottnt{k} )  =  \bullet $}\\
        &=&  \mathsf{types} (  \langle   \{ \mathit{x} \mathord{:} \ottnt{B} \mathrel{\mid} \ottnt{e_{{\mathrm{2}}}} \}  ,   \ottnt{e_{{\mathrm{2}}}}  [  \ottnt{k} / \mathit{x}  ]  ,  \ottnt{k}  \rangle^{ \ottnt{l} }  ) 
      \end{array} \]
    \item[(\E{CheckOK})]  \[ \begin{array}{rclr}
         \mathsf{types} (  \langle   \{ \mathit{x} \mathord{:} \ottnt{B} \mathrel{\mid} \ottnt{e} \}  ,   \mathsf{true}  ,  \ottnt{k}  \rangle^{ \ottnt{l} }  )  &=&   \mathsf{types} (  \{ \mathit{x} \mathord{:} \ottnt{B} \mathrel{\mid} \ottnt{e} \}  )   \cup   \mathsf{types} (  \mathsf{true}  )   & \\
        &\supseteq&  \emptyset  & \\
        &=&  \mathsf{types} ( \ottnt{k} ) 
      \end{array} \]
    \item[(\E{CheckFail})]  \[ \begin{array}{rclr}
         \mathsf{types} (  \langle   \{ \mathit{x} \mathord{:} \ottnt{B} \mathrel{\mid} \ottnt{e} \}  ,   \mathsf{false}  ,  \ottnt{k}  \rangle^{ \ottnt{l} }  )  &=&   \mathsf{types} (  \{ \mathit{x} \mathord{:} \ottnt{B} \mathrel{\mid} \ottnt{e} \}  )   \cup   \mathsf{types} (  \mathsf{false}  )   & \\
        &\supseteq&  \emptyset  & \\
        &=&  \mathsf{types} (  \mathord{\Uparrow}  \ottnt{l}  ) 
      \end{array} \]
    \item[(\E{AppL})]  \[ \begin{array}{rclr}
         \mathsf{types} (  \ottnt{e_{{\mathrm{1}}}} ~ \ottnt{e_{{\mathrm{2}}}}  )  &=&   \mathsf{types} ( \ottnt{e_{{\mathrm{1}}}} )   \cup   \mathsf{types} ( \ottnt{e_{{\mathrm{2}}}} )   & \\
        &\supseteq&   \mathsf{types} ( \ottnt{e'_{{\mathrm{1}}}} )   \cup   \mathsf{types} ( \ottnt{e_{{\mathrm{2}}}} )   & \text{(IH)} \\
        &=&  \mathsf{types} (  \ottnt{e'_{{\mathrm{1}}}} ~ \ottnt{e_{{\mathrm{2}}}}  ) 
      \end{array} \]
    \item[(\E{AppR})]  \[ \begin{array}{rclr}
         \mathsf{types} (  \ottnt{e_{{\mathrm{1}}}} ~ \ottnt{e_{{\mathrm{2}}}}  )  &=&   \mathsf{types} ( \ottnt{e_{{\mathrm{1}}}} )   \cup   \mathsf{types} ( \ottnt{e_{{\mathrm{2}}}} )   & \\
        &\supseteq&   \mathsf{types} ( \ottnt{e_{{\mathrm{1}}}} )   \cup   \mathsf{types} ( \ottnt{e'_{{\mathrm{2}}}} )   & \text{(IH)} \\
        &=&  \mathsf{types} (  \ottnt{e_{{\mathrm{1}}}} ~ \ottnt{e'_{{\mathrm{2}}}}  ) 
      \end{array} \]
    \item[(\E{OpInner})]  \[ \begin{array}{rclr}
         \mathsf{types} ( \ottnt{op}  \ottsym{(}  \ottnt{e_{{\mathrm{1}}}}  \ottsym{,}  \dots  \ottsym{,}  \ottnt{e_{{\ottmv{i}-1}}}  \ottsym{,}  \ottnt{e_{\ottmv{i}}}  \ottsym{,}  \dots  \ottsym{,}  \ottnt{e_{\ottmv{n}}}  \ottsym{)} )  &=& \bigcup_{1 \le i \le n}  \mathsf{types} ( \ottnt{e_{\ottmv{i}}} )  & \\
        &\supseteq& \bigcup_{1 \le j \le i}  \mathsf{types} ( \ottnt{e_{\ottmv{j}}} )  \cup  \mathsf{types} ( \ottnt{e'_{\ottmv{i}}} )  \cup \bigcup_{i+1 \le j \le n}  \mathsf{types} ( \ottnt{e_{\ottmv{j}}} )  & \text{(IH)} \\
        &=&  \mathsf{types} ( \ottnt{op}  \ottsym{(}  \ottnt{e_{{\mathrm{1}}}}  \ottsym{,}  \dots  \ottsym{,}  \ottnt{e_{{\ottmv{i}-1}}}  \ottsym{,}  \ottnt{e'_{\ottmv{i}}}  \ottsym{,}  \dots  \ottsym{,}  \ottnt{e_{\ottmv{n}}}  \ottsym{)} ) 
      \end{array} \]
    \item[(\E{CheckInner})]  \[ \begin{array}{rclr}
         \mathsf{types} (  \langle   \{ \mathit{x} \mathord{:} \ottnt{B} \mathrel{\mid} \ottnt{e_{{\mathrm{1}}}} \}  ,  \ottnt{e_{{\mathrm{2}}}} ,  \ottnt{k}  \rangle^{ \ottnt{l} }  )  &=&   \mathsf{types} (  \{ \mathit{x} \mathord{:} \ottnt{B} \mathrel{\mid} \ottnt{e_{{\mathrm{1}}}} \}  )   \cup   \mathsf{types} ( \ottnt{e_{{\mathrm{2}}}} )    & \\
        &\supseteq&   \mathsf{types} (  \{ \mathit{x} \mathord{:} \ottnt{B} \mathrel{\mid} \ottnt{e_{{\mathrm{1}}}} \}  )   \cup   \mathsf{types} ( \ottnt{e'_{{\mathrm{2}}}} )   & \text{(IH)} \\
        &=&  \mathsf{types} (  \langle   \{ \mathit{x} \mathord{:} \ottnt{B} \mathrel{\mid} \ottnt{e_{{\mathrm{1}}}} \}  ,  \ottnt{e'_{{\mathrm{2}}}} ,  \ottnt{k}  \rangle^{ \ottnt{l} }  ) 
      \end{array} \]
    \item[(\E{AppRaiseL})]  \[ \begin{array}{rclr}
         \mathsf{types} (   \mathord{\Uparrow}  \ottnt{l}  ~ \ottnt{e_{{\mathrm{2}}}}  )  &=&   \mathsf{types} (  \mathord{\Uparrow}  \ottnt{l}  )   \cup   \mathsf{types} ( \ottnt{e_{{\mathrm{2}}}} )  & \\
        &\supseteq&  \emptyset  & \\
        &=&  \mathsf{types} (  \mathord{\Uparrow}  \ottnt{l}  ) 
      \end{array} \]
    \item[(\E{AppRaiseR})]  \[ \begin{array}{rclr}
         \mathsf{types} (  \ottnt{e_{{\mathrm{1}}}} ~  \mathord{\Uparrow}  \ottnt{l}   )  &=&   \mathsf{types} ( \ottnt{e_{{\mathrm{1}}}} )   \cup   \mathsf{types} (  \mathord{\Uparrow}  \ottnt{l}  )   & \\
        &\supseteq&  \emptyset  & \\
        &=&  \mathsf{types} (  \mathord{\Uparrow}  \ottnt{l}  ) 
      \end{array} \]
    \item[(\E{CastRaise})]  \[ \begin{array}{rclr}
         \mathsf{types} (  \langle  \ottnt{T_{{\mathrm{1}}}}  \mathord{ \overset{ \mathcal{S} }{\Rightarrow} }  \ottnt{T_{{\mathrm{2}}}}  \rangle^{ \ottnt{l} } ~   \mathord{\Uparrow}  \ottnt{l'}   )  &=&     \mathsf{types} ( \ottnt{T_{{\mathrm{1}}}} )   \cup   \mathsf{types} ( \ottnt{T_{{\mathrm{2}}}} )    \cup   \mathsf{types} ( \mathcal{S} )    \cup   \mathsf{types} (  \mathord{\Uparrow}  \ottnt{l'}  )   & \\
        &\supseteq&  \emptyset  & \\
        &=&  \mathsf{types} (  \mathord{\Uparrow}  \ottnt{l'}  ) 
      \end{array} \]
    \item[(\E{OpRaise})]  \[ \begin{array}{rclr}
         \mathsf{types} ( \ottnt{op}  \ottsym{(}  \ottnt{e_{{\mathrm{1}}}}  \ottsym{,}  \dots  \ottsym{,}  \ottnt{e_{{\ottmv{i}-1}}}  \ottsym{,}   \mathord{\Uparrow}  \ottnt{l}   \ottsym{,}  \dots  \ottsym{,}  \ottnt{e_{\ottmv{n}}}  \ottsym{)} )  &=& 
        \bigcup_{1 \le j \le i}  \mathsf{types} ( \ottnt{e_{\ottmv{j}}} )  \cup  \mathsf{types} (  \mathord{\Uparrow}  \ottnt{l}  )  \cup 
        \bigcup_{i+1 \le j \le n}  \mathsf{types} ( \ottnt{e_{\ottmv{j}}} ) & \\
        &\supseteq&  \emptyset  & \\
        &=&  \mathsf{types} (  \mathord{\Uparrow}  \ottnt{l}  ) 
      \end{array} \]
    \item[(\E{CheckRaise})]  \[ \begin{array}{rclr}
         \mathsf{types} (  \langle   \{ \mathit{x} \mathord{:} \ottnt{B} \mathrel{\mid} \ottnt{e} \}  ,   \mathord{\Uparrow}  \ottnt{l}  ,  \ottnt{k}  \rangle^{ \ottnt{l'} }  )  &=&   \mathsf{types} (  \{ \mathit{x} \mathord{:} \ottnt{B} \mathrel{\mid} \ottnt{e} \}  )   \cup   \mathsf{types} (  \mathord{\Uparrow}  \ottnt{l}  )   & \\
        &\supseteq&  \emptyset  & \\
        &=&  \mathsf{types} (  \mathord{\Uparrow}  \ottnt{l}  ) 
      \end{array} \]
    \end{itemize}
      
    \paragraph{Classic \lambdah rules}
    \begin{itemize}
    \item[(\E{CastInnerC})]  \[ \begin{array}{rclr}
         \mathsf{types} (  \langle  \ottnt{T_{{\mathrm{1}}}}  \mathord{ \overset{ \bullet }{\Rightarrow} }  \ottnt{T_{{\mathrm{2}}}}  \rangle^{ \ottnt{l} } ~  \ottnt{e}  )  &=& 
            \mathsf{types} ( \ottnt{T_{{\mathrm{1}}}} )   \cup   \mathsf{types} ( \ottnt{T_{{\mathrm{2}}}} )    \cup   \mathsf{types} ( \bullet )    \cup   \mathsf{types} ( \ottnt{e} )   & \\
        &\supseteq&     \mathsf{types} ( \ottnt{T_{{\mathrm{1}}}} )   \cup   \mathsf{types} ( \ottnt{T_{{\mathrm{2}}}} )    \cup   \mathsf{types} ( \bullet )    \cup   \mathsf{types} ( \ottnt{e'} )   & \text{(IH)} \\
        &=&  \mathsf{types} (  \langle  \ottnt{T_{{\mathrm{1}}}}  \mathord{ \overset{ \bullet }{\Rightarrow} }  \ottnt{T_{{\mathrm{2}}}}  \rangle^{ \ottnt{l} } ~  \ottnt{e'}  ) 
      \end{array} \]
    \end{itemize} 

    \paragraph{Efficient \lambdah rules (\ifpopl$\ottnt{m}  \ottsym{=}   \mathsf{E} $\else$\ottnt{m}  \neq   \mathsf{C} $\fi)}
    \begin{itemize}
    \item[(\ECastInner)]  \[ \begin{array}{rclr}
         \mathsf{types} (  \langle  \ottnt{T_{{\mathrm{2}}}}  \mathord{ \overset{ \ottnt{a} }{\Rightarrow} }  \ottnt{T_{{\mathrm{3}}}}  \rangle^{ \ottnt{l} } ~  \ottnt{e_{{\mathrm{2}}}}  )  &=& 
            \mathsf{types} ( \ottnt{T_{{\mathrm{2}}}} )   \cup   \mathsf{types} ( \ottnt{T_{{\mathrm{3}}}} )    \cup   \mathsf{types} ( \ottnt{a} )    \cup   \mathsf{types} ( \ottnt{e_{{\mathrm{2}}}} )   & \\
        &\supseteq&     \mathsf{types} ( \ottnt{T_{{\mathrm{2}}}} )   \cup   \mathsf{types} ( \ottnt{T_{{\mathrm{3}}}} )    \cup   \mathsf{types} ( \ottnt{a} )    \cup   \mathsf{types} ( \ottnt{e'_{{\mathrm{2}}}} )   & \text{(IH)} \\
        &=&  \mathsf{types} (  \langle  \ottnt{T_{{\mathrm{2}}}}  \mathord{ \overset{ \ottnt{a} }{\Rightarrow} }  \ottnt{T_{{\mathrm{3}}}}  \rangle^{ \ottnt{l} } ~  \ottnt{e'_{{\mathrm{2}}}}  ) 
      \end{array} \]
    \item[(\ECastMerge)]  \[ \begin{array}{rclr}
         \mathsf{types} (  \langle  \ottnt{T_{{\mathrm{2}}}}  \mathord{ \overset{ \ottnt{a_{{\mathrm{2}}}} }{\Rightarrow} }  \ottnt{T_{{\mathrm{3}}}}  \rangle^{ \ottnt{l} } ~   (  \langle  \ottnt{T_{{\mathrm{1}}}}  \mathord{ \overset{ \ottnt{a_{{\mathrm{1}}}} }{\Rightarrow} }  \ottnt{T_{{\mathrm{2}}}}  \rangle^{ \ottnt{l'} } ~  \ottnt{e_{{\mathrm{2}}}}  )   )  &=& 
               \mathsf{types} ( \ottnt{T_{{\mathrm{2}}}} )   \cup   \mathsf{types} ( \ottnt{T_{{\mathrm{3}}}} )    \cup   \mathsf{types} ( \ottnt{a_{{\mathrm{2}}}} )    \cup  {} \\  &  &   \mathsf{types} ( \ottnt{T_{{\mathrm{1}}}} )    \cup   \mathsf{types} ( \ottnt{T_{{\mathrm{2}}}} )    \cup   \mathsf{types} ( \ottnt{a_{{\mathrm{1}}}} )    \cup   \mathsf{types} ( \ottnt{e_{{\mathrm{2}}}} )   & \\
        &\supseteq&     \mathsf{types} ( \ottnt{T_{{\mathrm{1}}}} )   \cup   \mathsf{types} ( \ottnt{T_{{\mathrm{3}}}} )    \cup  {} \\  &  &   \mathsf{types} (  \mathsf{merge} _{ \ottnt{m} }( \ottnt{T_{{\mathrm{1}}}} , \ottnt{a_{{\mathrm{1}}}} , \ottnt{T_{{\mathrm{2}}}} , \ottnt{a_{{\mathrm{2}}}} , \ottnt{T_{{\mathrm{3}}}} )  )    \cup   \mathsf{types} ( \ottnt{e_{{\mathrm{2}}}} )   & \text{(Lemma~\ref{lem:typesmerge}))}\\
        &=&  \mathsf{types} (  \langle  \ottnt{T_{{\mathrm{1}}}}  \mathord{ \overset{  \mathsf{merge} _{ \ottnt{m} }( \ottnt{T_{{\mathrm{1}}}} , \ottnt{a_{{\mathrm{1}}}} , \ottnt{T_{{\mathrm{2}}}} , \ottnt{a_{{\mathrm{2}}}} , \ottnt{T_{{\mathrm{3}}}} )  }{\Rightarrow} }  \ottnt{T_{{\mathrm{3}}}}  \rangle^{ \ottnt{l} } ~  \ottnt{e_{{\mathrm{2}}}}  ) 
      \end{array} \]
    \end{itemize} 

    \paragraph{Heedful \lambdah rules}
    \begin{itemize}
    \item[(\E{TypeSet})] \[ \begin{array}{rclr}
         \mathsf{types} (  \langle  \ottnt{T_{{\mathrm{1}}}}  \mathord{ \overset{ \bullet }{\Rightarrow} }  \ottnt{T_{{\mathrm{2}}}}  \rangle^{ \ottnt{l} } ~  \ottnt{e}  )  &=&    \mathsf{types} ( \ottnt{T_{{\mathrm{1}}}} )   \cup   \mathsf{types} ( \ottnt{T_{{\mathrm{2}}}} )    \cup   \mathsf{types} ( \ottnt{e} )   \\
        &&=  \mathsf{types} (  \langle  \ottnt{T_{{\mathrm{1}}}}  \mathord{ \overset{ \emptyset }{\Rightarrow} }  \ottnt{T_{{\mathrm{2}}}}  \rangle^{ \ottnt{l} } ~  \ottnt{e}  ) 
        \end{array} \]
    \item[(\E{CheckEmpty})]  \[ \begin{array}{rclr}
         \mathsf{types} (  \langle   \{ \mathit{x} \mathord{:} \ottnt{B} \mathrel{\mid} \ottnt{e_{{\mathrm{1}}}} \}   \mathord{ \overset{ \emptyset }{\Rightarrow} }   \{ \mathit{x} \mathord{:} \ottnt{B} \mathrel{\mid} \ottnt{e_{{\mathrm{2}}}} \}   \rangle^{ \ottnt{l} } ~  \ottnt{k}  )  &=&     \mathsf{types} (  \{ \mathit{x} \mathord{:} \ottnt{B} \mathrel{\mid} \ottnt{e_{{\mathrm{1}}}} \}  )   \cup   \mathsf{types} ( \emptyset )    \cup   \mathsf{types} (  \{ \mathit{x} \mathord{:} \ottnt{B} \mathrel{\mid} \ottnt{e_{{\mathrm{2}}}} \}  )    \cup   \mathsf{types} ( \ottnt{k} )   & \\
        &\supseteq&   \mathsf{types} (  \{ \mathit{x} \mathord{:} \ottnt{B} \mathrel{\mid} \ottnt{e_{{\mathrm{2}}}} \}  )   \cup   \mathsf{types} ( \ottnt{k} )   & \\
        &=&    \set{   \{ \mathit{x} \mathord{:} \ottnt{B} \mathrel{\mid} \ottnt{e_{{\mathrm{2}}}} \}   }   \cup   \mathsf{types} ( \ottnt{e_{{\mathrm{2}}}} )    \cup   \mathsf{types} ( \ottnt{k} )   & \\
        &=&    \mathsf{types} (  \{ \mathit{x} \mathord{:} \ottnt{B} \mathrel{\mid} \ottnt{e_{{\mathrm{2}}}} \}  )   \cup   \mathsf{types} ( \ottnt{e_{{\mathrm{2}}}} )    \cup   \mathsf{types} ( \ottnt{k} )   & \\
        &\supseteq&   \mathsf{types} (  \{ \mathit{x} \mathord{:} \ottnt{B} \mathrel{\mid} \ottnt{e_{{\mathrm{2}}}} \}  )   \cup   \mathsf{types} (  \ottnt{e_{{\mathrm{2}}}}  [  \ottnt{k} / \mathit{x}  ]  )   & \text{(Lemma~\ref{lem:typessubstitution})} \\
        &=&  \mathsf{types} (  \{ \mathit{x} \mathord{:} \ottnt{B} \mathrel{\mid} \ottnt{e_{{\mathrm{2}}}} \}  )  & \text{since $ \mathsf{types} ( \ottnt{k} )  =  \emptyset $}\\
        &=&  \mathsf{types} (  \langle   \{ \mathit{x} \mathord{:} \ottnt{B} \mathrel{\mid} \ottnt{e_{{\mathrm{2}}}} \}  ,   \ottnt{e_{{\mathrm{2}}}}  [  \ottnt{k} / \mathit{x}  ]  ,  \ottnt{k}  \rangle^{ \ottnt{l} }  ) 
      \end{array} \]
    \item[(\E{CheckSet})]  \[ \begin{array}{rcl}
         \mathsf{types} (  \langle   \{ \mathit{x} \mathord{:} \ottnt{B} \mathrel{\mid} \ottnt{e_{{\mathrm{1}}}} \}   \mathord{ \overset{ \mathcal{S} }{\Rightarrow} }   \{ \mathit{x} \mathord{:} \ottnt{B} \mathrel{\mid} \ottnt{e_{{\mathrm{3}}}} \}   \rangle^{ \ottnt{l} } ~  \ottnt{k}  )  &=& 
            \mathsf{types} (  \{ \mathit{x} \mathord{:} \ottnt{B} \mathrel{\mid} \ottnt{e_{{\mathrm{1}}}} \}  )   \cup   \mathsf{types} (  \{ \mathit{x} \mathord{:} \ottnt{B} \mathrel{\mid} \ottnt{e_{{\mathrm{3}}}} \}  )    \cup   \mathsf{types} ( \mathcal{S} )    \cup   \mathsf{types} ( \ottnt{k} )    \\
        &=&    \mathsf{types} (  \{ \mathit{x} \mathord{:} \ottnt{B} \mathrel{\mid} \ottnt{e_{{\mathrm{1}}}} \}  )   \cup   \mathsf{types} (  \{ \mathit{x} \mathord{:} \ottnt{B} \mathrel{\mid} \ottnt{e_{{\mathrm{3}}}} \}  )    \cup   \mathsf{types} ( \mathcal{S} )    \\
        &=&     \mathsf{types} (  \{ \mathit{x} \mathord{:} \ottnt{B} \mathrel{\mid} \ottnt{e_{{\mathrm{1}}}} \}  )   \cup   \mathsf{types} (  \{ \mathit{x} \mathord{:} \ottnt{B} \mathrel{\mid} \ottnt{e_{{\mathrm{3}}}} \}  )    \cup  {} \\  &  &   \mathsf{types} (  \mathcal{S}  \setminus   \{ \mathit{x} \mathord{:} \ottnt{B} \mathrel{\mid} \ottnt{e_{{\mathrm{2}}}} \}   )    \cup   \mathsf{types} (  \{ \mathit{x} \mathord{:} \ottnt{B} \mathrel{\mid} \ottnt{e_{{\mathrm{2}}}} \}  )   \\
        \multicolumn{3}{r}{\text{($  \{ \mathit{x} \mathord{:} \ottnt{B} \mathrel{\mid} \ottnt{e_{{\mathrm{2}}}} \}   \in  \mathcal{S} $)}} \\
        &\supseteq&      \mathsf{types} (  \{ \mathit{x} \mathord{:} \ottnt{B} \mathrel{\mid} \ottnt{e_{{\mathrm{1}}}} \}  )   \cup   \mathsf{types} (  \{ \mathit{x} \mathord{:} \ottnt{B} \mathrel{\mid} \ottnt{e_{{\mathrm{3}}}} \}  )    \cup  {} \\  &  &   \mathsf{types} (  \mathcal{S}  \setminus   \{ \mathit{x} \mathord{:} \ottnt{B} \mathrel{\mid} \ottnt{e_{{\mathrm{2}}}} \}   )    \cup   \mathsf{types} (  \{ \mathit{x} \mathord{:} \ottnt{B} \mathrel{\mid} \ottnt{e_{{\mathrm{2}}}} \}  )    \cup   \mathsf{types} (  \ottnt{e_{{\mathrm{2}}}}  [  \ottnt{k} / \mathit{x}  ]  )   \\
        \multicolumn{3}{r}{\text{(Lemma~\ref{lem:typessubstitution} and $ \mathsf{types} ( \ottnt{k} )  =  \emptyset $)}}\\
        &=&  \mathsf{types} (  \langle   \{ \mathit{x} \mathord{:} \ottnt{B} \mathrel{\mid} \ottnt{e_{{\mathrm{2}}}} \}   \mathord{ \overset{  \mathcal{S}  \setminus   \{ \mathit{x} \mathord{:} \ottnt{B} \mathrel{\mid} \ottnt{e_{{\mathrm{2}}}} \}   }{\Rightarrow} }   \{ \mathit{x} \mathord{:} \ottnt{B} \mathrel{\mid} \ottnt{e_{{\mathrm{3}}}} \}   \rangle^{ \ottnt{l} } ~   \langle   \{ \mathit{x} \mathord{:} \ottnt{B} \mathrel{\mid} \ottnt{e_{{\mathrm{2}}}} \}  ,   \ottnt{e_{{\mathrm{2}}}}  [  \ottnt{k} / \mathit{x}  ]  ,  \ottnt{k}  \rangle^{ \ottnt{l} }   ) 
      \end{array} \]
    \end{itemize}

    \paragraph{Eidetic \lambdah rules}
    \begin{itemize}
    \item[(\E{Coerce})] \[ \begin{array}{rclr}
         \mathsf{types} (  \langle  \ottnt{T_{{\mathrm{1}}}}  \mathord{ \overset{\bullet}{\Rightarrow} }  \ottnt{T_{{\mathrm{2}}}}  \rangle^{ \ottnt{l} } ~  \ottnt{e}  )  &=&    \mathsf{types} ( \ottnt{T_{{\mathrm{1}}}} )   \cup   \mathsf{types} ( \ottnt{T_{{\mathrm{2}}}} )    \cup   \mathsf{types} ( \ottnt{e} )   & \\
        &=&     \mathsf{types} (  \mathsf{coerce} ( \ottnt{T_{{\mathrm{1}}}} , \ottnt{T_{{\mathrm{2}}}} , \ottnt{l} )  )   \cup   \mathsf{types} ( \ottnt{T_{{\mathrm{1}}}} )    \cup   \mathsf{types} ( \ottnt{T_{{\mathrm{2}}}} )    \cup   \mathsf{types} ( \ottnt{e} )   & \text{(Lemma~\ref{lem:typescoerce})} \\
        &=&  \mathsf{types} (  \langle  \ottnt{T_{{\mathrm{1}}}}  \mathord{ \overset{  \mathsf{coerce} ( \ottnt{T_{{\mathrm{1}}}} , \ottnt{T_{{\mathrm{2}}}} , \ottnt{l} )  }{\Rightarrow} }  \ottnt{T_{{\mathrm{2}}}}  \rangle^{\bullet} ~  \ottnt{e}  )  &
      \end{array} \]
    \item[(\E{CoerceStack})] \[ \begin{array}{rclr}
         \mathsf{types} (  \langle   \{ \mathit{x} \mathord{:} \ottnt{B} \mathrel{\mid} \ottnt{e_{{\mathrm{1}}}} \}   \mathord{ \overset{ \ottnt{r} }{\Rightarrow} }   \{ \mathit{x} \mathord{:} \ottnt{B} \mathrel{\mid} \ottnt{e_{{\mathrm{2}}}} \}   \rangle^{\bullet} ~  \ottnt{k}  )  &=&    \mathsf{types} (  \{ \mathit{x} \mathord{:} \ottnt{B} \mathrel{\mid} \ottnt{e_{{\mathrm{1}}}} \}  )   \cup   \mathsf{types} (  \{ \mathit{x} \mathord{:} \ottnt{B} \mathrel{\mid} \ottnt{e_{{\mathrm{2}}}} \}  )    \cup   \mathsf{types} ( \ottnt{r} )   & \\
        &\supseteq&   \mathsf{types} (  \{ \mathit{x} \mathord{:} \ottnt{B} \mathrel{\mid} \ottnt{e_{{\mathrm{2}}}} \}  )   \cup   \mathsf{types} ( \ottnt{r} )   & \\
        &=&  \langle   \{ \mathit{x} \mathord{:} \ottnt{B} \mathrel{\mid} \ottnt{e_{{\mathrm{2}}}} \}  ,   \mathord{?}  ,  \ottnt{r} ,  \ottnt{k} ,  \ottnt{k}  \rangle^{\bullet}  &
      \end{array} \]
    \item[(\E{StackDone})] \[ \begin{array}{rclr}
         \mathsf{types} (  \langle   \{ \mathit{x} \mathord{:} \ottnt{B} \mathrel{\mid} \ottnt{e} \}  ,   \mathord{\checkmark}  ,  \mathsf{nil} ,  \ottnt{k} ,  \ottnt{k}  \rangle^{\bullet}  )  &=&  \mathsf{types} (  \{ \mathit{x} \mathord{:} \ottnt{B} \mathrel{\mid} \ottnt{e} \}  )  & \\
        &\supseteq&  \emptyset  & \\
        &=&  \mathsf{types} ( \ottnt{k} )  &
      \end{array} \]
    \item[(\E{StackPop})] \[ \begin{array}{rclr}
         \mathsf{types} (  \langle   \{ \mathit{x} \mathord{:} \ottnt{B} \mathrel{\mid} \ottnt{e} \}  ,  \ottnt{s} ,  \ottsym{(}    \{ \mathit{x} \mathord{:} \ottnt{B} \mathrel{\mid} \ottnt{e'} \}^{ \ottnt{l} }  , \ottnt{r}   \ottsym{)} ,  \ottnt{k} ,  \ottnt{k}  \rangle^{\bullet}  )  &=&   \mathsf{types} (  \{ \mathit{x} \mathord{:} \ottnt{B} \mathrel{\mid} \ottnt{e} \}  )   \cup   \mathsf{types} (   \{ \mathit{x} \mathord{:} \ottnt{B} \mathrel{\mid} \ottnt{e'} \}^{ \ottnt{l} }  , \ottnt{r}  )   & \\
        &=&     \mathsf{types} (  \{ \mathit{x} \mathord{:} \ottnt{B} \mathrel{\mid} \ottnt{e} \}  )   \cup   \mathsf{types} (  \{ \mathit{x} \mathord{:} \ottnt{B} \mathrel{\mid} \ottnt{e'} \}  )    \cup   \mathsf{types} ( \ottnt{r} )    \cup   \mathsf{types} ( \ottnt{e'} )   & \\
        &=&     \mathsf{types} (  \{ \mathit{x} \mathord{:} \ottnt{B} \mathrel{\mid} \ottnt{e} \}  )   \cup   \mathsf{types} (  \{ \mathit{x} \mathord{:} \ottnt{B} \mathrel{\mid} \ottnt{e'} \}  )    \cup   \mathsf{types} ( \ottnt{r} )    \cup   \mathsf{types} (  \ottnt{e'}  [  \ottnt{k} / \mathit{x}  ]  )   & \text{(Lemma~\ref{lem:typessubstitution})} \\
        &=&  \mathsf{types} (  \langle   \{ \mathit{x} \mathord{:} \ottnt{B} \mathrel{\mid} \ottnt{e} \}  ,   \ottnt{s}  \vee ( \ottnt{e}  =  \ottnt{e'} )  ,  \ottnt{r} ,  \ottnt{k} ,   \langle   \{ \mathit{x} \mathord{:} \ottnt{B} \mathrel{\mid} \ottnt{e'} \}  ,   \ottnt{e'}  [  \ottnt{k} / \mathit{x}  ]  ,  \ottnt{k}  \rangle^{ \ottnt{l} }   \rangle^{\bullet}  )  &
      \end{array} \]
    \item[(\E{StackInner})] \[ \begin{array}{rclr}
         \mathsf{types} (  \langle   \{ \mathit{x} \mathord{:} \ottnt{B} \mathrel{\mid} \ottnt{e} \}  ,  \ottnt{s} ,  \ottnt{r} ,  \ottnt{k} ,  \ottnt{e'}  \rangle^{\bullet}  )  &=&    \mathsf{types} (  \{ \mathit{x} \mathord{:} \ottnt{B} \mathrel{\mid} \ottnt{e} \}  )   \cup   \mathsf{types} ( \ottnt{r} )    \cup   \mathsf{types} ( \ottnt{e'} )   & \\
        &\supseteq&     \mathsf{types} (  \{ \mathit{x} \mathord{:} \ottnt{B} \mathrel{\mid} \ottnt{e} \}  )   \cup   \mathsf{types} ( \ottnt{r} )    \cup   \mathsf{types} ( \ottnt{e''} )   & \text{(IH)} \\
        &=&  \mathsf{types} (  \langle   \{ \mathit{x} \mathord{:} \ottnt{B} \mathrel{\mid} \ottnt{e} \}  ,  \ottnt{s} ,  \ottnt{r} ,  \ottnt{k} ,  \ottnt{e''}  \rangle^{\bullet}  )  &
      \end{array} \]
    \item[(\E{StackRaise})] \[ \begin{array}{rclr}
         \mathsf{types} (  \langle   \{ \mathit{x} \mathord{:} \ottnt{B} \mathrel{\mid} \ottnt{e} \}  ,   \mathord{\checkmark}  ,  \mathsf{nil} ,  \ottnt{k} ,  \ottnt{k}  \rangle^{\bullet}  )  &=&  \mathsf{types} (  \{ \mathit{x} \mathord{:} \ottnt{B} \mathrel{\mid} \ottnt{e} \}  )  & \\
        &\supseteq&  \emptyset  & \\
        &=&  \mathsf{types} ( \ottnt{k} )  &
      \end{array} \]
    \end{itemize} 
    \fi}
  \end{proof}
\end{lemma}

\end{document}